\documentclass[12pt]{article}
\pdfoutput=1
\usepackage{amsmath,amssymb,amsthm,mathtools,bm,mathdots,mathrsfs,xfrac}
\usepackage{graphicx,float,array,multirow,multicol,rotfloat,caption,subcaption,booktabs}
\usepackage{tabularx}
\usepackage{enumerate,parskip,adjustbox}
\usepackage[shortlabels]{enumitem}
\usepackage[normalem]{ulem}
\usepackage{xcolor}
\usepackage{pdflscape}
\usepackage{refcount}
\usepackage{rotating}
\usepackage{empheq}

\usepackage{geometry}
\newif\ifsaSerifFont
\saSerifFonttrue 

\ifsaSerifFont
  \newcommand{\safont}{\rmfamily}
\else
  \newcommand{\safont}{\sffamily}
\fi

\usepackage{dsfont}
\usepackage{pifont}
\DeclareSymbolFont{largesymbolsA}{U}{jkpexa}{m}{n}
\SetSymbolFont{largesymbolsA}{bold}{U}{jkpexa}{bx}{n}
\DeclareMathSymbol{\varprod}{\mathop}{largesymbolsA}{16}

\usepackage[vcentermath]{youngtab}
\usepackage[boxsize=.8em]{ytableau}
\usepackage[all]{xy}
\usepackage{tikz,tikz-cd}
\usetikzlibrary{positioning,arrows.meta,calc,fit,backgrounds,arrows,shapes.misc,decorations.pathmorphing,decorations.markings,decorations.pathreplacing,matrix,patterns}
\usepackage{pgfplots}
\pgfplotsset{compat=1.18}
\usepgfplotslibrary{fillbetween}

\usepackage{blkarray}

\tikzset{
    scale cd/.style={every label/.append style={scale=#1}, cells={nodes={scale=#1}}},
    gauge/.style={rounded rectangle, draw=black!100, thick, minimum size=2mm}, 
    gaugeD/.style={rounded rectangle, draw=black!100,double,thick,minimum size=2mm},  
    empty/.style={rounded rectangle, draw=white!100, thick, minimum size=2mm}, 
    flavor/.style={rectangle, draw=black!100, thick, minimum size=2mm},
    flavorD/.style={rectangle, draw=black!100, double,thick, minimum size=2mm},
    node/.style={circle, thick, draw=black!100,fill=white!100,  minimum size=2mm, inner sep=0pt},
    sqnode/.style={rectangle, thick, draw=black!100,fill=white!100,  minimum size=2mm, inner sep=0pt},
    sonode/.style={circle, thick, draw=black!100,fill=red!100,  minimum size=2mm, inner sep=0pt},
    spnode/.style={circle, thick, draw=black!100,fill=blue!100,  minimum size=2mm, inner sep=0pt},
    fnode/.style={rectangle, thick, draw=black!100,fill=white!100,  minimum size=2mm, inner sep=0pt},
    tnode/.style={rounded rectangle, outer sep=0pt, thick, minimum size=2mm},
    brace/.style={decoration={brace, mirror},decorate},
    ncbar angle/.initial=90,
    ncbar/.style={
        to path=(\tikztostart)
        -- ($(\tikztostart)!#1!\pgfkeysvalueof{/tikz/ncbar angle}:(\tikztotarget)$)
        -- ($(\tikztotarget)!($(\tikztostart)!#1!\pgfkeysvalueof{/tikz/ncbar angle}:(\tikztotarget)$)!\pgfkeysvalueof{/tikz/ncbar angle}:(\tikztostart)$)
        -- (\tikztotarget)
    },
    ncbar/.default=0.5cm,
    square left brace/.style={ncbar=0.2cm},
    square right brace/.style={ncbar=-0.2cm},
    snake it/.style={decorate, decoration={snake, amplitude=.4mm, segment length=2mm, post length=0mm,pre length=0mm}}
}

\usepackage[numbers,sort&compress]{natbib}
\usepackage{csquotes}
\usepackage{hhline}
\usepackage{longtable}
\usepackage{todonotes}
\graphicspath{{figs/}}

\usepackage[hidelinks,linktocpage]{hyperref}
\usepackage[capitalize,sort]{cleveref}
\usepackage{orcidlink}

\hypersetup{
	pdftitle={Solver Agent: An Agentic Framework for Mathematics and Theoretical Physics, Applied to F-theory Uplifts of O3-planes and S-folds},
	pdfauthor={\textcopyright\ Eliott Morgensztern, Cesar Fierro Cota, Alessandro Mininno},
	pdfsubject={hep-th, cs.AI, math.AG},
	pdfcreator={pdfLaTeX},
	pdfproducer={LaTeX},
	pdfkeywords={},
	colorlinks=true
  ,urlcolor=blue
  ,anchorcolor=blue
  ,citecolor=blue
  ,filecolor=blue
  ,linkcolor=blue
  ,menucolor=blue
  ,linktocpage=true
}

\crefname{figure}{Figure}{Figures}
\crefname{table}{Table}{Tables}
\crefname{definition}{Definition}{Definitions}
\crefname{proposition}{Proposition}{Propositions}
\crefname{claim}{Claim}{Claims}
\crefname{conjecture}{Conjecture}{Conjectures}

\numberwithin{equation}{section}
\allowdisplaybreaks[1]
\makeatletter
\def\myrotate{\ifodd\c@page\else+\fi 90}
\g@addto@macro{\landscape}{\PLS@Rotate{\myrotate}}
\makeatother

\newcommand{\Lpagenumber}{\ifdim\textwidth=\linewidth\else\bgroup
	\dimendef\margin=0 
	\ifodd\value{page}\margin=\oddsidemargin
	\else\margin=\evensidemargin
	\fi
	\raisebox{\dimexpr -3\topmargin-\headheight-\headsep-0.5\linewidth}[0pt][0pt]{%
		\rlap{\hspace{\dimexpr \margin+\textheight+3\footskip}%
			\llap{\rotatebox{90}{\hspace{-4.5cm}\thepage\hfill}}}}%
	\egroup\fi}
\AddToHook{shipout/background}{\Lpagenumber}%

\theoremstyle{plain}
\newtheorem*{thm*}{Theorem}

\DeclareMathOperator{\SO}{SO}

\newcommand{\CC}{\mathbb{C}}
\newcommand{\PP}{\mathbb{P}}
\newcommand{\RR}{\mathbb{R}}
\newcommand{\ZZ}{\mathbb{Z}}

\newcommand{\ID}{\mathds{1}}

\newcommand{\coma}{\ensuremath{\,,\quad}}
\newcommand{\fstop}{\ensuremath{\,.}}

\newcommand{\IX}{{X}}
\newcommand{\IY}{{Y}}

\newcommand{\mailref}[1]{\href{mailto:#1}{\color{black}\nolinkurl{#1}}}

\renewcommand{\bar}[1]{\overline{#1}}

\renewcommand{\tilde}[1]{\widetilde{#1}}

\definecolor{saAccent}{HTML}{2B53E0}      
\definecolor{saAssumption}{HTML}{1F7FC4}  
\definecolor{saDerivation}{HTML}{7C4ED0}  
\definecolor{saResult}{HTML}{C451A6}      
\definecolor{saSymbolic}{HTML}{2B53E0}    
\definecolor{saNumerical}{HTML}{167D4F}   
\definecolor{saCalabi--Yau}{HTML}{0891B2}          
\definecolor{saReference}{HTML}{5B6275}   
\definecolor{saVerification}{HTML}{B3811A}
\definecolor{saCorrection}{HTML}{C52941}  
\definecolor{saFinal}{HTML}{C46F1A}       
\definecolor{saText}{HTML}{14171F}
\definecolor{saMuted}{HTML}{5B6275}
\definecolor{saPrompt}{HTML}{5B6275}    
\definecolor{saSurface}{HTML}{F8F9FB}

\newcommand{\saTag}[1]{\small\bfseries\color{#1!80!black}}

\tikzset{
  saBox/.style={
    draw=#1!55, fill=#1!6, rounded corners=2.2pt, thick,
    text=saText, font=\small, align=center,
    minimum height=8.5mm, inner xsep=2.6mm, inner ysep=1.8mm},
  saBoxWide/.style={saBox=#1, minimum width=34mm},
  saLabel/.style={font=\footnotesize, text=saMuted},
  saEdge/.style={-{Stealth[length=2.2mm]}, thick, draw=saMuted!80},
  saEdgeAccent/.style={-{Stealth[length=2.2mm]}, thick, draw=#1!70},
  saGroup/.style={draw=saMuted!35, dashed, rounded corners=3pt, inner sep=3mm},
}

\newcommand{\saHead}[2]{{\bfseries\color{#1!72!black}#2}}
\newcommand{\saSub}[1]{{\footnotesize\color{saMuted}#1}}
\tikzset{
  saCard/.style 2 args={
    fill=white, draw=#1!45, line width=.9pt, rounded corners=5.5pt,
    text=saText, font=\small\safont, align=left,
    minimum height=11mm, inner xsep=4.2mm, inner ysep=2.1mm,
    minimum width=#2,
    text width=\dimexpr#2-8.4mm\relax,
    path picture={\fill[#1!85!black, rounded corners=.55mm]
      ([xshift=1.2mm, yshift=1.6mm]path picture bounding box.south west)
      rectangle
      ([xshift=2.3mm, yshift=-1.6mm]path picture bounding box.north west);}},
  saCardEm/.style 2 args={saCard={#1}{#2}, fill=#1!5, draw=#1!60, line width=.9pt},
  saPanel/.style={fill=#1!6, draw=#1!35, line width=.6pt,
    rounded corners=7pt, inner xsep=3.2mm, inner ysep=3.2mm},
  saChip/.style={fill=#1!10, draw=#1!45, line width=.5pt, rounded corners=3.5pt,
    inner xsep=2.2mm, inner ysep=1.1mm, font=\fontsize{7}{8}\selectfont\safont,
    text=#1!55!black},
  saFlow/.style={{Circle[length=1.1mm]}-{Stealth[length=2.6mm, width=2.2mm, round]},
    line width=.9pt, draw=#1!70, rounded corners=5pt,
    shorten >=1.3mm, shorten <=1.3mm},
}

\usepackage{footnotehyper}
\usepackage[most]{tcolorbox}
\newcommand{\saCalloutBar}[3]{%
  \path[fill=#1!85!black, rounded corners=.55mm]
    ([xshift=1.7mm, yshift=#2]interior.south west) rectangle
    ([xshift=2.8mm, yshift=#3]interior.north west);

  \ifdim#2=0pt
    \path[fill=#1!85!black]
      ([xshift=1.7mm]interior.south west) rectangle
      ([xshift=2.8mm,yshift=.6mm]interior.south west);
  \fi

  \ifdim#3=0pt
    \path[fill=#1!85!black]
      ([xshift=1.7mm,yshift=-.6mm]interior.north west) rectangle
      ([xshift=2.8mm]interior.north west);
  \fi
}
\tcbset{saCallout/.style={
    enhanced, breakable, lines before break=3,
    colback=#1!4, colframe=#1!45, boxrule=.9pt,
    arc=8pt, outer arc=8pt,
    left=7mm, right=4mm, top=3.2mm, bottom=3.2mm,
    underlay unbroken={\saCalloutBar{#1}{2.6mm}{-2.6mm}},
    underlay first={\saCalloutBar{#1}{0mm}{-2.6mm}},
    underlay middle={\saCalloutBar{#1}{0mm}{0mm}},
    underlay last={\saCalloutBar{#1}{2.6mm}{0mm}},
    before skip=3.5mm, after skip=3.5mm}}
\newtcolorbox{saBox}[2]{saCallout=#1, fontupper=\small,
  before upper={\setlength{\parskip}{2mm}\setlength{\parindent}{0pt}%
    {\safont\bfseries\normalsize\color{#1!72!black}#2}%
    \par\vspace{1.4mm}}}

\newcounter{safncount}
\newtoks\safntoks
\makeatletter
\newcommand{\safnadd}[2]{%
  \global\safntoks=\expandafter{\the\safntoks \footnotetext[#1]{#2}}%
}
\newcommand{\safnfootnote}[1]{%
  \stepcounter{footnote}%
  \setcounter{safncount}{\value{footnote}}%
  \footnotemark[\value{safncount}]%
  \expandafter\safnadd\expandafter{\the\value{safncount}}{#1}%
}
\makeatother

\NewTColorBox[
  auto counter,
  number within=section,
  crefname={Prompt}{Prompts},
  Crefname={Prompt}{Prompts}
]{saPromptBox}{ O{} }{%
  saCallout=saPrompt,
  fontupper=\small,
  title/.store in=\satitle,
  title={},
  #1,
  before={\par\addvspace{0.5\baselineskip}\noindent},
  after={\the\safntoks\global\safntoks={}},
  before upper={%
    \setlength{\parskip}{2mm}%
    \setlength{\parindent}{0pt}%
    \let\footnote\safnfootnote
    \def\empty{}%
    {\safont\bfseries\normalsize\color{saPrompt!72!black}%
      Prompt \thetcbcounter
      \ifx\satitle\empty\else\space(\satitle)\fi
    }%
    \par\vspace{1.4mm}%
  }
}

\newcounter{saresult}

\crefname{saresult}{Result}{Results}
\Crefname{saresult}{Result}{Results}
\crefname{saResultBox}{Result}{Results}
\Crefname{saResultBox}{Result}{Results}

\NewTColorBox{saResultBox}{ O{} }{%
  saCallout=saFinal,
  fontupper=\small,
  prompt/.store in=\saprompt,
  prompt={},
  title/.store in=\satitle,
  title={},
  label/.store in=\salabel,
  label={},
  #1,
  before={\noindent},
  after={\the\safntoks\global\safntoks={}},
  phantom={%
    \xdef\thesaresult{\getrefnumber{\saprompt}}%
    \refstepcounter{saresult}%
    \def\empty{}%
    \ifx\salabel\empty
    \else
      \label{\salabel}%
    \fi
  },
  before upper={%
    \setlength{\parskip}{2mm}%
    \setlength{\parindent}{0pt}%
    \let\footnote\safnfootnote
    \def\empty{}%
    {\safont\bfseries\normalsize\color{saFinal!72!black}%
      Result \thesaresult
      \ifx\satitle\empty\else\space(\satitle)\fi
    }%
    \par\vspace{1.4mm}%
  }
}

\NewTColorBox{saResultBoxManipulated}{ O{} }{%
  saCallout=saFinal,
  fontupper=\small,
  prompt/.store in=\saprompt,
  prompt={},
  title/.store in=\satitle,
  title={},
  label/.store in=\salabel,
  label={},
  #1,
  before={\noindent},
  after={\the\safntoks\global\safntoks={}},
  phantom={%
    \xdef\thesaresult{\getrefnumber{\saprompt}}%
    \refstepcounter{saresult}%
    \def\empty{}%
    \ifx\salabel\empty
    \else
      \label{\salabel}%
    \fi
  },
  before upper={%
    \setlength{\parskip}{2mm}%
    \setlength{\parindent}{0pt}%
    \let\footnote\safnfootnote
    \def\empty{}%
    {\safont\bfseries\normalsize\color{saFinal!72!black}%
      Result \thesaresult
      \ifx\satitle\empty\else\space(\satitle)\fi
      {} (Manipulated)%
    }%
    \par\vspace{1.4mm}%
  }
}

\NewTColorBox{saResultBoxCondensed}{ O{} }{%
  saCallout=saFinal,
  fontupper=\small,
  prompt/.store in=\saprompt,
  prompt={},
  title/.store in=\satitle,
  title={},
  label/.store in=\salabel,
  label={},
  #1,
  before={\noindent},
  after={\the\safntoks\global\safntoks={}},
  phantom={%
    \xdef\thesaresult{\getrefnumber{\saprompt}}%
    \refstepcounter{saresult}%
    \def\empty{}%
    \ifx\salabel\empty
    \else
      \label{\salabel}%
    \fi
  },
  before upper={%
    \setlength{\parskip}{2mm}%
    \setlength{\parindent}{0pt}%
    \let\footnote\safnfootnote
    \def\empty{}%
    {\safont\bfseries\normalsize\color{saFinal!72!black}%
      Result \thesaresult
      \ifx\satitle\empty\else\space(\satitle)\fi
      \space(Condensed)%
    }%
    \par\vspace{1.4mm}%
  }
}

\definecolor{saTheorem}{HTML}{168A8A}

\NewTColorBox[
  auto counter,
  crefname={Theorem}{Theorems},
  Crefname={Theorem}{Theorems}
]{saTheoremBox}{ O{} }{%
  before={\addvspace{\baselineskip}\noindent},
  after={\addvspace{\baselineskip}\the\safntoks\global\safntoks={}},
  saCallout=saTheorem, fontupper=\small,
  #1,
  before upper={%
    \setlength{\parskip}{2mm}%
    \setlength{\parindent}{0pt}%
    \let\footnote\safnfootnote
    {\safont\bfseries\normalsize\color{saTheorem!72!black}%
      Theorem \thetcbcounter}%
    \par\vspace{1.4mm}%
  }
}

\begin{document}

\begin{titlepage}
 \begin{center}
{\LARGE\bfseries \texttt{Solver Agent}: an Agentic AI Framework for
Theoretical Physics Computations\\[10pt]
{\Large\bfseries Applied to
F-theory Uplifts of O3-planes and S-folds}}
\vspace{0.4cm}

{\large Eliott Morgensztern\,\orcidlink{0009-0002-6333-4537},$^1$ Cesar Fierro Cota\,\orcidlink{0000-0003-2788-5921},$^1$ and Alessandro Mininno\,\orcidlink{0000-0002-9593-0440}$^2$}\\
\vspace{.6cm}

{$^1$ Sorbonne Universit\'e, CNRS, Laboratoire de Physique Th\'eorique et Hautes Energies,\\ Campus Pierre et Marie Curie, 4 place Jussieu, F-75005, Paris, France}\\
\vspace{.1cm}
{$^2$ Department of Physics, University of Wisconsin--Madison,\\1150 University Avenue, Madison, WI 53706, USA}\par
\vspace{.2cm}

\scalebox{0.8}{\tt \mailref{eliott@morgensztern.com}, \mailref{fierrocota@lpthe.jussieu.fr}, \mailref{mininno@physics.wisc.edu}}
\vspace{0.3cm}

\end{center}

\begin{abstract}
\noindent We introduce \texttt{Solver Agent}, an AI framework based on large language models for calculations and proofs in mathematics and theoretical physics. The solution process is tracked through a persistent ledger that records assumptions, derivations, and computations. A central agent delegates tasks to specialized sub-agents, while independent agents verify both intermediate steps and the final result. This setup improves the traceability, reproducibility, and verification of computer-assisted calculations.

\noindent Applying \texttt{Solver Agent}, 
we study global F-theory uplifts of Type IIB orientifolds and their S-fold generalizations. We establish sufficient conditions for Weierstrass models over projective threefolds with terminal $\mathbb{Z}_k$ quotient singularities ($k\in\{2,3,4,6\}$) to give $\mathbb{Q}$-factorial projective elliptically fibered Calabi--Yau fourfolds with isolated Gorenstein terminal quotient singularities. These geometries realize O3-planes and S-folds, where local D3-brane probes of the latter yield four-dimensional $\mathcal{N}=3$ superconformal field theories. Using stringy invariants, we derive fixed-point contributions to Hodge data and Euler characteristics, and show that these Euler corrections determine the localized 
D3-brane charges required for tadpole cancellation. We illustrate these results using toric hypersurface constructions, where a single three-dimensional polytope determines both the Type IIB Calabi--Yau threefold and the F-theory base; here, the orientifold double cover naturally forms a bisection of an alternative genus-one-fibered uplift with discrete $\mathbb{Z}_2$ gauge symmetry. Finally, we provide methods for toric computations and four-form flux analysis in four-dimensional $\mathcal{N}=1$ compactifications with non-abelian gauge sectors.
\end{abstract}

\vspace{1cm}
\vfill 
\end{titlepage}

\tableofcontents
\bigskip\medskip
\hrule
\bigskip\bigskip

\section{Introduction}
\label{sec:intro}

\subsubsection*{Summary of the Results}

\begin{figure}[!htp]
  \centering
  \renewcommand{\safont}{\sffamily}
  \begin{tikzpicture}[saLabel/.style={font=\footnotesize\sffamily, text=saMuted},scale=0.99]
    \node[saCardEm={saMuted}{36mm}] (user) at (0,0)
      {\saHead{saMuted}{Researcher}\\[-1pt]\saSub{human in the loop}};
    \node[saCardEm={saAccent}{36mm}] (solver) at (5.4,0)
      {\saHead{saAccent}{Main solver}\\[-1pt]\saSub{coordinating agent}};

    \node[saCard={saSymbolic}{43mm}] (sym) at (12.4,2.2)
      {\saHead{saSymbolic}{Symbolic}\\[-1pt]\saSub{computer algebra}};
    \node[saCard={saNumerical}{43mm}] (num) at (12.4,0.75)
      {\saHead{saNumerical}{Numerical}\\[-1pt]\saSub{evaluation, plots}};
    \node[saCard={saCalabi--Yau}{43mm}] (Calabi--Yau) at (12.4,-0.75)
      {\saHead{saCalabi--Yau}{Calabi--Yau analysis}\\[-1pt]\saSub{geometric data}};
    \node[saCard={saReference}{43mm}] (ref) at (12.4,-2.2)
      {\saHead{saReference}{Reference lookup}\\[-1pt]\saSub{literature search}};

    \node[saCard={saMuted}{47mm}, densely dashed] (sandbox) at (12.4,-4.5)
      {\saHead{saMuted}{Sandbox}\\[-1pt]\saSub{isolated code environment}};

    \node[saCardEm={saDerivation}{68mm}] (ledger) at (5.4,-4.05)
      {\saHead{saDerivation}{Ledger}\\[-1pt]
       \saSub{persistent record of assumptions, steps,}\\[-2pt]
       \saSub{computations, verdicts, conclusions}};

    \node[saCard={saVerification}{57mm}] (stepver) at (2.2,-7.1)
      {\saHead{saVerification}{Step verification}\\[-1pt]
       \saSub{subset of entries re-examined}};
    \node[saCard={saVerification}{57mm}] (fullver) at (8.6,-7.1)
      {\saHead{saVerification}{Full-solution verification}\\[-1pt]
       \saSub{complete record re-examined}};

    \begin{scope}[on background layer]
      \node[saPanel=saAccent, fit=(sym)(num)(Calabi--Yau)(ref)] (compgroup) {};
      \node[saPanel=saVerification, fit=(stepver)(fullver)] (vergroup) {};
    \end{scope}
    \node[saChip=saAccent] at (compgroup.north) {SPECIALIZED SUB-AGENTS};
    \node[saChip=saVerification] at (vergroup.south) {VERIFICATION AGENTS};

    \draw[saFlow=saMuted] ([yshift=2.2mm]user.east) --
      node[saLabel, above=.4mm, align=center] {problem and\\[-3pt]follow-ups}
      ([yshift=2.2mm]solver.west);
    \draw[saFlow=saFinal] ([yshift=-2.2mm]solver.west) --
      node[saLabel, below=.4mm] {answer} ([yshift=-2.2mm]user.east);

    \draw[saFlow=saAccent] (solver.east) --
      node[saLabel, below=.4mm, align=center] {delegated\\[-3pt]tasks}
      (compgroup.west |- solver.east);

    \draw[saFlow=saDerivation] ([xshift=-2.6mm]solver.south) --
      node[saLabel, left=.4mm, pos=0.795] {append entries}
      ([xshift=-2.6mm]ledger.north);
    \draw[saFlow=saDerivation] ([xshift=2.6mm]ledger.north) --
      node[saLabel, right=.4mm, pos=0.22] {read record}
      ([xshift=2.6mm]solver.south);

    \draw[saFlow=saSymbolic] ([yshift=-12mm]compgroup.west) -|
      (9.4,-4.05) -- (ledger.east);
    \node[saLabel, anchor=east, align=right] at (9.4,-1.8)
      {recorded results\\[-3pt]code and output};

    \draw[saFlow=saMuted] ([xshift=-2.6mm]compgroup.south) --
      node[saLabel, left=.4mm] {run code} ([xshift=-2.6mm]sandbox.north);
    \draw[saFlow=saMuted] ([xshift=2.6mm]sandbox.north) --
      node[saLabel, right=.4mm, pos=0.43] {output} ([xshift=2.6mm]compgroup.south);

    \draw[saFlow=saVerification] (10.74,0 |- vergroup.north) --
      node[saLabel, left=.6mm, align=right]
      {independent\\[-3pt]re-computations} (10.74,0 |- sandbox.south);
    \draw[saFlow=saVerification] (11.26,0 |- sandbox.south) --
      node[saLabel, right=.4mm] {output} (11.26,0 |- vergroup.north);

    \draw[saFlow=saVerification] ([xshift=-2.6mm]ledger.south) --
      node[saLabel, left=.4mm] {review} ([xshift=-2.6mm]vergroup.north);
    \draw[saFlow=saVerification] ([xshift=2.6mm]vergroup.north) --
      node[saLabel, right=.4mm] {verdicts} ([xshift=2.6mm]ledger.south);

    \draw[saFlow=saVerification]
      ([xshift=-14mm]solver.south) |-
      node[saLabel, above=.4mm, pos=0.75] {request verification}
      (0.9,-1.8) -- (0.9,0 |- vergroup.north);
  \end{tikzpicture}
  \caption{Schematic overview of \texttt{Solver Agent} architecture. A coordinating
  agent --- main solver --- decomposes the problem and records its work in a persistent ledger;
  specialized sub-agents run their code in an isolated sandbox and record the
  code and output in the ledger; independent verification agents read the
  ledger, re-run computations in their own sandboxed sessions, and append
  their assessments. The final answer is released to the researcher only
  after the recorded solution has been reviewed as a whole.}
  \label{fig:architecture}
\end{figure}
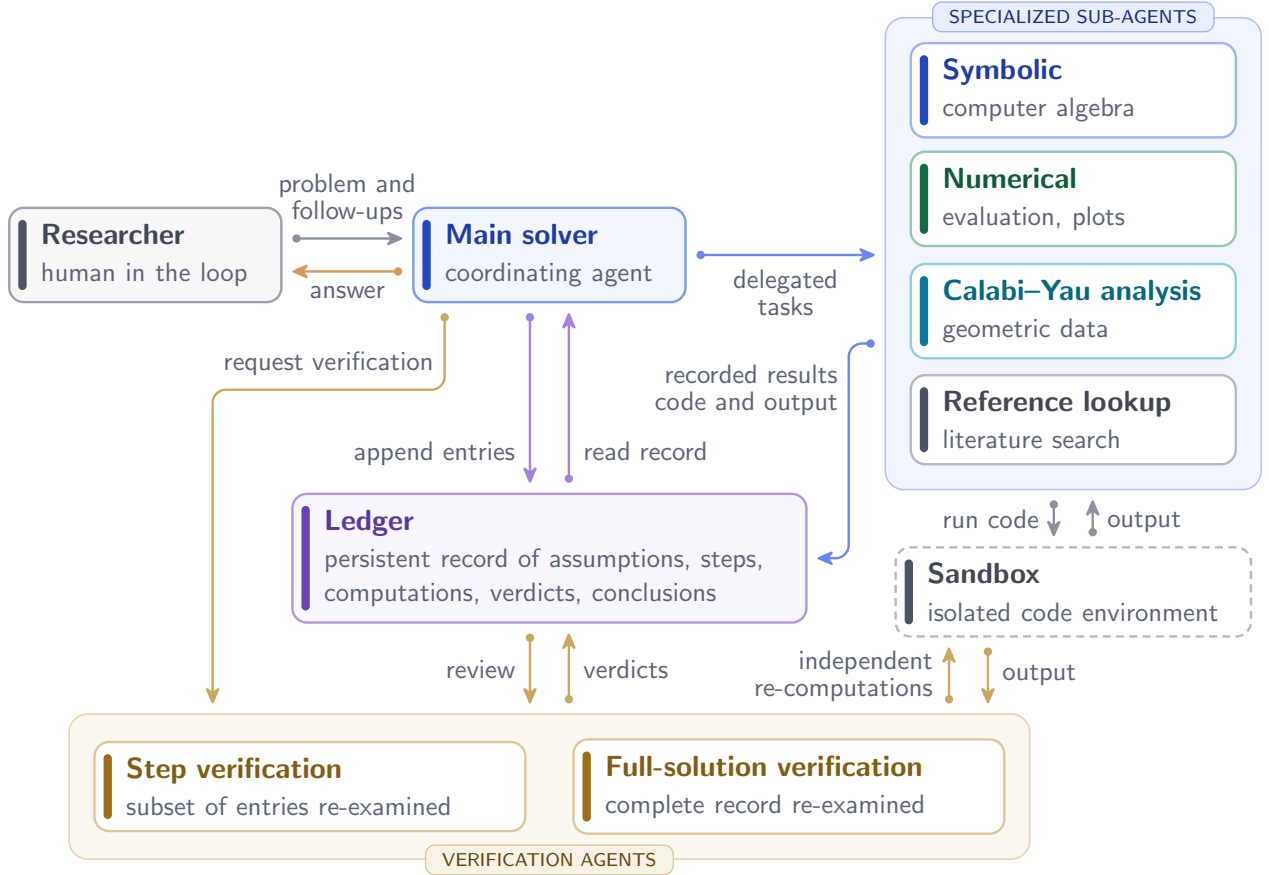

In this work, we introduce \texttt{Solver Agent}, a large language model-based framework for proofs and computations in mathematics and theoretical physics. The system automatically splits the solution process, delegating tasks to \emph{agents} --- specialized language model instances with access to external programs such as symbolic, numerical, and other domain-specific algorithms or literature search engines  --- and allows for both automatic and human verification of each individual step.
Figure \ref{fig:architecture} presents the architecture of the framework.

The \href{https://github.com/starrfree/solver-agent}{full source code},\footnote{Public Git repository under \href{https://github.com/starrfree/solver-agent/blob/main/LICENSE}{AGPL-3.0} license: \href{https://github.com/starrfree/solver-agent}{github.com/starrfree/solver-agent}} including the core engine that orchestrates the agents --- the backend --- an interface for using the system, visualizing and verifying the proof process, as well as exporting sessions for repeatability --- the frontend --- is open-source,\footnote{\texttt{Solver Agent} requires an external API subject to usage fees.} along with a complete \href{https://github.com/starrfree/solver-agent\#installation}{installation guide} and a \href{https://github.com/starrfree/solver-agent\#using-solver-agent}{user guide}. For a detailed presentation of \texttt{Solver Agent}, we refer the reader to Section \ref{sec:solver-agent}.

\texttt{Solver Agent} has been used in this work to derive new results in string theory, and more specifically in F-theory. It has proven to be a powerful tool to significantly accelerate calculations, but more importantly, it was fundamental for the proof of Theorem \ref{thm:weierstrass_proof}, and for the explicit constructions of the elliptically fibered Calabi--Yau fourfolds with terminal singularities. 

In fact, from the physical point of view, the main result of this paper is a controlled global embedding of O3-planes and non-perturbative S-fold sectors in compact F-theory. Noting that isolated fixed points of O3 orientifold involutions descend to terminal $\mathbb{Z}_2$ quotient singularities of the threefold base, we generalize this structure to terminal $\mathbb{Z}_k$ quotient singularities to construct S-folds. Together with this geometric embedding, we provide a prescription for computing the induced D3-brane charges from the stringy invariants of the singular Calabi--Yau fourfold; in particular, the resulting correction to the Euler characteristic of the fourfold is the one necessary to match the prediction from the tadpole, as we show in explicit examples in \cref{sec:ExampleO3,sec:Ftheory-SfoldUplifts}. We illustrate these properties through toric trilayer constructions, wherein a three-dimensional reflexive polytope determines both the Type IIB Calabi--Yau threefold and the F-theory base. In this framework, the orientifold double cover is realized as a bisection of an auxiliary genus-one-fibered Calabi--Yau fourfold, yielding an alternative F-theory uplift with a discrete $\mathbb{Z}_2$ gauge symmetry that holds independently of a Sen limit. Finally, we develop accessible methods for detailed toric computations (including the extraction of intersection data) and for the systematic analysis of four-form fluxes in four-dimensional $\mathcal{N}=1$ compactifications with non-abelian gauge sectors.

Throughout the paper, the reader will encounter {\color{saPrompt}\textbf{Prompt}} and {\color{saFinal}\textbf{Result}} boxes, as we summarize in Table \ref{tab:prompts-results}. These contain, respectively, inputs to and potentially lightly manipulated or condensed outputs of \texttt{Solver Agent}. The exact sessions are available for reproducibility \href{https://storage.googleapis.com/solver-agent-sessions/solver-agent-f-theory-uplifts/index.html}{here}. Each is in a separate folder with the full ledger, generated files, system version, and model used in different formats. The most human-readable format is \texttt{report.html}, the raw data is in \texttt{ledger.json}, and for conversion to a printable document, \texttt{report.tex}. When we specify that the {\color{saFinal}\textbf{Result}} boxes are ``condensed,'' we mean that we have trimmed the otherwise verbose output given by \texttt{Solver Agent}. Analogously, a {\color{saFinal}\textbf{Result}} box is ``manipulated'' if we have rephrased some of the content of \texttt{Solver Agent} to improve readability without changing its meaning. This was necessary because some prompts were divided into sub-prompts while being reported all together in a single box.

\begin{table}[!htp]
    \centering
    \begin{tabular}{c|l|c|c}
        \textbf{Section} & \multicolumn{1}{c|}{\textbf{Purpose}} & \textbf{Prompt} & \textbf{Result} \\
        \hhline{=|=|=|=}
        \multirow{3}{*}{Section \ref{ssec:o3_terminal}} 
        & Check terminal singularities & \ref{prmpt:terminal_p31112} & \ref{res:terminal_p31112} \\
        & Count terminal singularities & \ref{prmpt:terminal_weighted} & \ref{res:terminal_weighted} \\
        & Identify local groups & \ref{prmpt:local_groups_weighted} & \ref{res:local_groups_weighted} \\
        \hline
        Section \ref{ssec:weierstrass_terminal}
        & Prove Theorem \ref{thm:weierstrass_proof} & \ref{prmpt:weierstrass_proof} & \ref{res:weierstrass_proof} \\
        \hline
        \multirow{3}{*}{Section \ref{sec:ftheory_corrections_2}} 
        & Stringy Euler definition & \ref{prmpt:StringyEuler1} & \ref{res:StringyEuler1} \\
        & Compute stringy Euler & \ref{prmpt:StringyEuler2} & \ref{res:StringyEuler2} \\
        & Evaluate stringy correction & \ref{prmpt:StringyEuler3} & \ref{res:StringyEuler3} \\
        \hline
        Section \ref{sec:orientifold-in-genus1} 
        & Analyze reflexive polytope & \ref{prmpt:ks_data} & \ref{res:ks_data} \\
        \hline
        \multirow{7}{*}{Section \ref{sec:X18Uplift}} 
        & Check non-Fano threefold & \ref{prmpt:NHC1} & \ref{res:NHC1} \\
        & Construct toric fan & \ref{prmpt:NHC2} & \ref{res:NHC2} \\
        & Find interior points & \ref{prmpt:NHC3} & \ref{res:NHC3} \\
        & Find compatible FRST & \ref{prmpt:NHC4} & \ref{res:NHC4} \\
        & Identify divisor basis & \ref{prmpt:NHC5} & \ref{res:NHC5} \\
        & Compute Chern classes & \ref{prmpt:computechiofY4} & \ref{res:computechiofY4} \\
        & Vertical flux constraints & \ref{prmpt:FluxConstraints} & \ref{res:FluxConstraints} \\
        \hline
        \multirow{6}{*}{Section \ref{sec:CYExamples-SFoldsGauge}} 
        & Project lattice points & \ref{prmpt:Boss1} & \ref{res:Boss1} \\
        & Find boundary point & \ref{prmpt:Boss2} & \ref{res:Boss2} \\
        & Compute singular locus & \ref{prmpt:Boss3} & \ref{res:Boss3} \\
        & Identify interior points & \ref{prmpt:Boss4} & \ref{res:Boss4} \\
        & Check compatible FRST & \ref{prmpt:Boss5} & \ref{res:Boss5} \\
        & Determine singular locus & \ref{prmpt:Boss6} & \ref{res:Boss6} \\
    \end{tabular}
    \caption{Overview of prompts provided to \texttt{Solver Agent} and their corresponding results, categorized by section and purpose.}
    \label{tab:prompts-results}
\end{table}

\subsubsection*{Overview}
Artificial intelligence has progressively evolved from data analysis to formal mathematical problem-solving. Scaled reasoning architectures now directly tackle intractable problems in mathematical physics. For example, large language models have been integrated with the Lean proof assistant to formally verify precise finite-time singularity bounds for the generalized Korteweg--de Vries equation~\cite{Lefkowitz_2026, anthropic2024claude, de2015lean, Benander_2025}, to conjecture closed-form expressions for single-minus gluon tree amplitudes~\cite{guevara2026singleminus}, to formally construct the free bosonic quantum field theory in four-dimensional Euclidean spacetime~\cite{douglas2026formalizationqft}, and to tackle Millennium problems (see, e.g., the recent solution to the Navier--Stokes existence and smoothness problem \cite{OpenAI_Navier_2026}).

These results demonstrate that machine learning can be applied well beyond the big data analysis and pattern recognition paradigms that have already proven useful in theoretical physics. Early works successfully utilized neural networks and reinforcement learning to scan the string landscape, from exploring intersecting D-brane configurations~\cite{He:2017aed, Halverson:2019tkf, Ruehle:2020jrk, Loges:2021hvn, Loges:2022mao} to sampling string vacua using generative models and genetic algorithms~\cite{Cole:2021nnt, Yip:2025hon, Walden:2025cpf}. More recently, machine learning algorithms have been utilized to compute numerical Calabi--Yau metrics and advance string model building~\cite{Brodie:2019dfx, He:2020lbz, Anderson:2020hux, Douglas:2021ces, Larfors:2021pbb, Constantin:2021for, Larfors:2022nep, Halverson:2023ndu, Anderson:2023viv, Ebelt:2023clh, Lust:2026mys}. Parallel efforts have integrated interactive theorem provers, such as Lean, with language agents to formalize high-energy physics workflows and automate theorem proving~\cite{Krippendorf:2025mhp, Krippendorf:2026pou, Gu:2026zbf}. Recently, researchers have also successfully trained transformers to representatively generate fine, regular, and star triangulations\footnote{The FRST conditions restrict the triangulation at the polytope level: ``fine'' demands that every integral point in the dual polytope is a vertex of a simplex; ``regular'' requires the triangulation to follow from the projection of a higher-dimensional convex hull; and ``star'' ensures that the origin is a vertex of all simplices.} (FRSTs) of 4D reflexive polytopes that give rise to smooth Calabi--Yau (CY) threefolds in toric ambient space \cite{Yip:2025hon, Arnal:2026zyo} or to explore heterotic line bundle standard models~\cite{Yip:2026jhw}, confirming that sequence modeling architectures can efficiently parse massive datasets within string phenomenology.

The introduction of large language models~\cite{OpenAI:2023ktj} has motivated the automation of symbolic and code-assisted proofs. However, the application of standalone language models in theoretical physics remains restricted, as open-ended physical problems require long-horizon, multi-step derivations. Although current models are capable of executing formal tasks, they typically fail on open-ended derivations by generating incorrect intermediate steps or reaching correct conclusions via flawed logic. To address these limitations, recent methodologies have incorporated reasoning distillation and process reward models to evaluate intermediate physical steps~\cite{Pan_2025,Cai__2025}, and specialized fine-tuning has been applied to reasoning models in quantum field theory~\cite{Woodward:2026abc}. Furthermore, language agents have been utilized to evaluate tree amplitudes, calculate loop integrals, and automate high-energy workflows~\cite{lu2025languageagentsphysics,guevara2026singleminus,Schwartz:2026ekw,Shih:2026lmy,Shih:2026jfe,Plehn:2026gxv,Agrawal:2026lvg}. To mitigate logical degradation, multi-agent architectures, verifier-gated language models, and structured evaluation frameworks have been proposed to isolate interaction variables and refine theoretical reasoning~\cite{jaiswal2024mora,xu2025multiagentphysicist,Yu:2026ogu,Niarchos:2026abc}. Nevertheless, these models currently lack the intrinsic capacity for rigorous inspection. Applying neural architectures to complex, multi-stage physical theories necessitates a structured environment that enforces reproducibility and maintains a persistent ledger of assumptions.

To overcome the reliability limitations of standalone language models in these tasks, it is necessary to deploy an architecture that decomposes long derivations, formally verifies intermediate algebraic steps, maintains a record of all underlying assumptions, and allows inspection of all individual steps of the derivation: we introduce \texttt{Solver Agent}, an agentic framework~\cite{Yao:2022react,Wu:2023autogen} for automating and organizing computer-assisted calculations in mathematics and theoretical physics. A coordinating language-model agent decomposes a problem and delegates computational tasks to specialized sub-agents, while a persistent ledger records everything happening during the calculation or proof: the assumptions, derivations, code, corrections, verification verdicts, and conclusions.

\texttt{Solver Agent} is not intended to provide a formal certificate of correctness. Rather, it supplies an automated, inspectable, and reproducible workflow in which errors can be localized, computations can be rerun, assumptions can be revised explicitly, and the evidential basis of each conclusion can be examined by the researcher. In this work, we present the framework and use it as a computational tool underlying our results.  

In particular, we use it to derive new results in F-theory.
In the context of string phenomenology, Type IIB string theory compactifications on Calabi--Yau orientifolds with D-branes and fluxes yield four-dimensional effective field theories (EFTs) with minimal supersymmetry \cite{Dasgupta:1999ss,Gukov:1999ya,Giddings:2001yu,Grana:2005jc,Douglas:2006es,Blumenhagen:2006ci,Ibanez:2012zz}. In these setups, classical superpotentials fix complex structure moduli via Calabi--Yau periods \cite{Hosono:1993qy,Morrison:1991cd,Giddings:2001yu,Demirtas:2019sip,Demirtas:2023als}, while K\"ahler moduli are stabilized via perturbative and non-perturbative corrections \cite{Kachru:2003aw,Balasubramanian:2005zx,Cicoli:2021dhg,McAllister:2023vgy}. However, constructing consistent vacua in this perturbative Type IIB framework requires introducing specific configurations of D7-branes and worldvolume fluxes by hand to cancel orientifold tadpoles \cite{Blumenhagen:2006ci,Denef:2008wq,Braun:2008ua,Collinucci:2008pf,Collinucci:2008sq,Grimm:2011dj,Crino:2022zjk}. Furthermore, the resulting geometry typically possesses quotient singularities at the orientifold fixed loci that remain unresolved perturbatively \cite{Garcia-Etxebarria:2015wns,Weigand:2018rez,Apruzzi:2020pmv,Heckman:2020svr}. We therefore turn to F-theory \cite{Vafa:1996xn}, which provides the non-perturbative completion of this framework by compactifying on an elliptically fibered Calabi--Yau fourfold~\cite{Sen:1996vd,Sen:1997gv,Aluffi:2009tm,Esole:2012tf}. By geometrizing the varying Type IIB axio-dilaton and the monodromies of $[p,q]$-7-branes, F-theory inherently absorbs these physical consistency conditions into the topology of the fourfold. In this way, the deformation moduli of D7-branes become complex structure moduli of the fourfold, meaning the entire classical superpotential is captured by the fourfold periods and quantized 4-form fluxes \cite{Vafa:1996xn,Morrison:1996na,Morrison:1996pp,Braun:2011zm,Weigand:2018rez}. This geometric translation provides a natural foundation for constructing Standard Model-like phenomenological models \cite{Cvetic:2001nr,Cvetic:2001tj,Ibanez:2001nd,Marchesano:2004xz,Blumenhagen:2005mu,Ibanez:2006da,Blumenhagen:2009up,Beasley:2008dc,Beasley:2008kw,Marsano:2009gv,Grimm:2009yu,Donagi:2008ca,Mayrhofer:2014laa,Cvetic:2019gnh,Cvetic:2022fnv,Marchesano:2022qbx,Marchesano:2024gul}. More recently, F-theory uplifts of Type IIB orientifolds have been used
to investigate how quantum corrections affect classical infinite-distance
limits~\cite{Kaufmann:2026fli, Kaufmann:2026mha}, with a complementary Type IIA/M-theory
analysis providing further insight into these obstructions~\cite{Kaufmann:2026tsy}.

Although this non-perturbative map is conceptually well-understood, explicitly constructing the elliptically fibered Calabi--Yau fourfolds from general Type IIB orientifolds remains difficult \cite{Collinucci:2008zs,Blumenhagen:2009up,Collinucci:2009uh,Aluffi:2009tm,Blumenhagen:2009yv,Esole:2012tf}. However, a systematic algorithm to compute these fourfold uplifts for Calabi--Yau threefold hypersurfaces in toric varieties was proposed in \cite{Hassfeld:2026rzd}. In this work, we use \texttt{Solver Agent} to both prove mathematical statements and perform explicit computations.

\subsubsection*{Structure of the Paper}

The paper is structured as follows. In Section \ref{sec:solver-agent}, we introduce \texttt{Solver Agent}, schematically described in Figure \ref{fig:architecture}. There, we explain what an agentic framework is, how the \texttt{Solver Agent} software system is designed, and its scope. Since our application of \texttt{Solver Agent} is in the context of F-theory uplifts of Type IIB compactifications on CY threefolds with orientifolds and fluxes, in Section \ref{sec:typeIIBCalabi--Yau} we set our conventions, reviewing how to construct orientifolds and compute the D7- and D3-brane tadpoles that determine the topology of the elliptically fibered Calabi--Yau fourfold. From Section \ref{sec:Codim3Sing}, we start using \texttt{Solver Agent} to prove geometrical statements that help us construct the uplift to F-theory of a given compactification. There, we also discuss terminal quotient singularities and how they can be described on the CY fourfold. Section \ref{ssec:weierstrass_terminal} in particular contains the explicit construction of such uplifts in the corresponding global Weierstrass model of the CY fourfold, and the first theorem (\cref{thm:weierstrass_proof}) proven entirely by \texttt{Solver Agent} in \cref{res:weierstrass_proof}. In Section \ref{sec:Ftheoryuplift}, we focus on F-theory uplifts for CY threefolds as toric hypersurfaces. We first review the general procedure and then, in Section \ref{sec:orientifold-in-genus1}, we rephrase it as uplifts to genus-one-fibered CY fourfolds. Many examples are provided, and intermediate results are solved by \texttt{Solver Agent}, as summarized in Table \ref{tab:prompts-results}. In Section \ref{sec:conclusions}, we conclude and discuss possible outlooks. Finally, in Appendix \ref{sec:KS}, we briefly review the construction of CY threefold orientifolds as hypersurfaces in toric ambient space.

\section{\texorpdfstring{\texttt{Solver Agent}}{Solver Agent} System}
\label{sec:solver-agent}

\texttt{Solver Agent} is a software system designed to support technically demanding
calculations in mathematical and theoretical physics. It organizes the work of
a large language model around an explicit persistent record of the solution
process, and it delegates substantive computation to external programs
whose inputs and outputs are stored alongside the reasoning that motivated
them. The system has been used, in the work reported here, to carry out and
document computations arising in F-theory and type~IIB compactifications,
including the evaluation of Chern classes, topological invariants, and
intersection data on Calabi--Yau geometries.

The purpose of this section is to describe the system in detail: how it operates, what kinds of problems it is intended to address, and how its outputs are meant to be read and checked. We first recall
what distinguishes an agent-based workflow from the ordinary conversational use of
a language model in Section~\ref{sec:agent-workflows}, and then describe the
architecture of the system in Section~\ref{sec:architecture}. The three
central components of this system are: a
structured ledger that records everything happening during the solving process, including problem statements, assumptions, computations, verifications, and
conclusions (Section~\ref{sec:ledger}); an explicit mechanism for stating and
revising assumptions to eliminate any ambiguity in the statements (Section~\ref{sec:assumptions}); and two complementary
verification procedures that review individual steps and the complete candidate final solution (Section~\ref{sec:verification}). We then describe the integration of both general-purpose and domain-specific computational tools in Section~\ref{sec:tools}, the way a researcher interacts with the system and
the artifacts it produces in Section~\ref{sec:interaction}, and discuss the intended scope of the system and the design choices regarding traceability and reproducibility in Section~\ref{sec:discussion}.

\texttt{Solver Agent} should be understood as a methodology for organizing and automating computer-assisted calculations.
The system is intended to make the individual steps of a calculation
explicit, inspectable, and repeatable; it does not establish mathematical
correctness, and its outputs are subject to the same scrutiny as any other computer-assisted result.

\subsection{Language Models and Agent-based Workflows}
\label{sec:agent-workflows}

The most familiar mode of use of large language models (LLMs)~\cite{OpenAI:2023ktj} is
through a conversational interface: the user poses a question, the model
answers in free-form prose, and the exchange continues turn by turn. Modern chat interfaces sometimes allow the model to access a limited selection of external tools, such as web search and code execution environments. This mode
is designed to be useful for answering any question across a broad range of fields.
For mathematics or theoretical physics domains, specific technical calculations may be imposed by the conversational format, which enforces a
particular structure on the work. In chat interfaces, the state of the calculation resides in the
conversation transcript, which mixes digressions, reasoning, and corrections. Intermediate derivations happen within the model's own generated text or usually inaccessible internal reasoning, rather than by dedicated software; and there is no
built-in notion of a step that can be individually referenced, revised, or
re-examined at a later time. One usually has to redo most of the steps in order to trust the generated answer.

An \emph{agent-based} workflow~\cite{Yao:2022react} arranges the same underlying capability
differently. The language model runs 
inside a loop in which it may repeatedly invoke external programs, commonly called \emph{tools}, and observe their results before deciding how to
proceed. A tool may be a deterministic program or a specialized LLM. These may include computer-algebra routines, numerical solvers, literature-search interfaces, and any other program exposed to the model through
a declared calling convention. The state of the work, moreover, need not
reside in the conversation itself: it can be held in an external data
structure that the model reads from and writes to through the same tool
mechanism, and that persists and can be checked at any point.

The underlying model, and thus intrinsic knowledge, is the same in the two modes, as they both rely on the same core LLM; the difference is in the workflow and structure. In an agent-based
system, each specific task is rooted in a different model instance, the \emph{agent} --- specialized by its prompt, context, and available tools. One agent may be formulating tasks, interpreting outputs, and organizing the argument while delegating code writing and verification to other agents. Because the record is explicit, separate review procedures can be applied
to it after the fact. None of this makes the outputs of the model reliable by
themselves; a tool can be invoked with the wrong input, and a well-organized
record can document a flawed argument, but it creates points of attachment
for the bookkeeping and review practices that long calculations ordinarily
require.

\texttt{Solver Agent} is an implementation of such an agentic system specialized in
mathematical and theoretical physics. The remainder of this section describes
its components in detail.

\subsection{Overall Architecture and Workflow}
\label{sec:architecture}

The system consists of a small number of cooperating components, shown in
Figure~\ref{fig:architecture}. The workflow is as follows: a researcher submits a problem statement in
ordinary mathematical prose, together with optional configuration choices
discussed below. The problem is handed to a coordinating LLM instance, the \emph{main solver}, which is responsible for organizing the
solution: it decomposes the problem, records its interpretation and plan,
requests computations, and eventually proposes an answer. The main instance delegates substantive computations. Algebraic manipulations,
numerical evaluations, geometric analysis, and literature lookups are
 handled by specialized agents~\cite{Wu:2023autogen}, each of which executes code in
an isolated environment and returns its results along with the code that
produced them.

All of this activity is recorded in a single persistent data structure, the
\emph{ledger}, described in Section~\ref{sec:ledger}. The ledger is stored in
a database, independently of the conversation, and is the medium through
which the components communicate: the main solver appends its assumptions,
derivation steps, and conclusions; the computational sub-agents append the
tasks they were given and the outputs they produced; and two separate
verification processes (see Section~\ref{sec:verification}) read the ledger and
append their assessments to it. The answer that the main solver proposes is
released to the researcher only after the second of these verification
processes has reviewed the recorded solution as a whole.

\paragraph{The architecture is LLM-agnostic.} Each of the roles just described --- the
main solver, the computational sub-agents, and the verifiers --- is a language
model invoked through a uniform interface, and any model capable of
structured tool calling can be used. Different roles may use different models. However, the quality of the resulting
work naturally depends on the choice of models.

\subsection{The Solution Ledger}
\label{sec:ledger}

The central data structure of the system is the ledger: an append-only,
typed record of the solution process. The ledger is a sequence of discrete \emph{entries}, each with a
declared type, a status, a short summary, a detailed body, and a list of
references to the earlier entries on which it depends. The ledger is intended
to play the role of a research notebook: it states what is being assumed, what has
been computed, on what grounds each conclusion rests, and which earlier material has been revised.
\begin{table}[!htp]
    \centering
    \renewcommand{\tabularxcolumn}[1]{m{#1}}
    \renewcommand{\arraystretch}{1.35}
    \begin{tabularx}{\textwidth}{l|c|X}
        \textbf{Entry type} & \textbf{Written by} & \multicolumn{1}{c}{\textbf{Content}} \\
        \hhline{=|=|=}
        Assumption & main solver & interpretive choices, definitions, notation, conventions, domains of validity \\
        Derivation & main solver & reasoning steps and plans, in mathematical prose \\
        Result & main solver & interpretation of a computation, with an explicit dependence on it \\
        \hline
        Symbolic computation & sub-agent & computer-algebra task, code, and output \\
        Numerical computation & sub-agent & numerical task, code, output, and generated files \\
        Calabi--Yau analysis & sub-agent & geometric quantities from domain-specific software \\
        Reference lookup & sub-agent & sourced statements retrieved from the literature \\
        \hline
        Verification & verifier & verdict and justification for a step or for the whole solution \\
        Correction & main solver & replacement for a step that was rejected or superseded \\
        Final answer & main solver & the proposed solution, pending full verification \\
        Follow-up & researcher & clarifications, amendments to the problem statement, or subsequent problem \\
    \end{tabularx}
    \caption{Entry types of the solution ledger. Each entry carries a status
(accepted, rejected, or superseded), a summary, a detailed body, and
references to the entries on which it depends.}
    \label{tab:entry-types}
\end{table}

Table~\ref{tab:entry-types} lists the entry types. Three of them carry the
mathematical argument: \emph{assumption} entries record interpretive and
technical hypotheses (see Section~\ref{sec:assumptions}); \emph{derivation}
entries record reasoning and plans in mathematical prose; and \emph{result}
entries record the interpretation of a computation. Four types are written by
the computational sub-agents rather than by the main solver, and they preserve the
exact task submitted, the code executed, and the output produced. The
remaining types support review and revision: \emph{verification} entries
record the verdicts of the two review processes, \emph{correction} entries
record replacements for steps found to be defective, \emph{final answer}
entries record the proposed solution, and \emph{follow-up} entries record
additional input supplied by the researcher after the initial problem
statement.

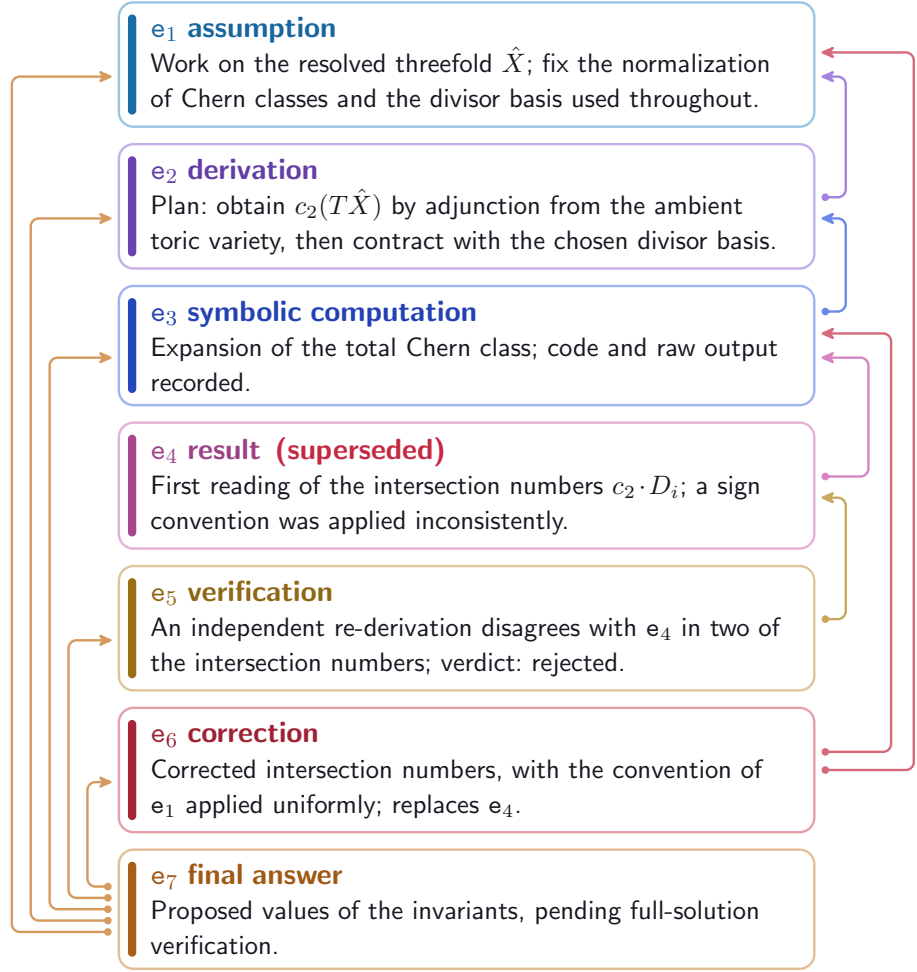
\begin{figure}[!htp]
\centering
\renewcommand{\safont}{\sffamily}
\begin{tikzpicture}[node distance=2.mm,
    saLabel/.style={font=\footnotesize\safont, text=saMuted}]
  \tikzset{
    card/.style={saCard={#1}{92mm}, minimum height=0pt, inner ysep=1.9mm,
      font=\footnotesize\safont},
    dep/.style={{Circle[length=1mm]}-{Stealth[length=2.2mm, width=1.9mm, round]},
      line width=.9pt, draw=#1!70, rounded corners=3pt,
      shorten <=.8mm, shorten >=.8mm},
  }
  \node[card=saAssumption] (e1)
    {{\saTag{saAssumption}$\texttt{e}_1$ assumption}\\
     Work on the resolved threefold $\hat{X}$; fix the normalization of
     Chern classes and the divisor basis used throughout.};
  \node[card=saDerivation, below=of e1] (e2)
    {{\saTag{saDerivation}$\texttt{e}_2$ derivation}\\
     Plan: obtain $c_2(T\hat{X})$ by adjunction from the ambient toric
     variety, then contract with the chosen divisor basis.};
  \node[card=saSymbolic, below=of e2] (e3)
    {{\saTag{saSymbolic}$\texttt{e}_3$ symbolic computation}\\
     Expansion of the total Chern class; code and raw output recorded.};
  \node[card=saResult, below=of e3] (e4)
    {{\saTag{saResult}$\texttt{e}_4$ result \,{\color{saCorrection}(superseded)}}\\
     First reading of the intersection numbers $c_2\!\cdot\!D_i$; a sign
     convention was applied inconsistently.};
  \node[card=saVerification, below=of e4] (e5)
    {{\saTag{saVerification}$\texttt{e}_5$ verification}\\
     An independent re-derivation disagrees with $\texttt{e}_4$ in two of the
     intersection numbers; verdict: rejected.};
  \node[card=saCorrection, below=of e5] (e6)
    {{\saTag{saCorrection}$\texttt{e}_6$ correction}\\
     Corrected intersection numbers, with the convention of $\texttt{e}_1$ applied
     uniformly; replaces $\texttt{e}_4$.};
  \node[card=saFinal, below=of e6] (e7)
    {{\saTag{saFinal}$\texttt{e}_7$ final answer}\\
     Proposed values of the invariants, pending full-solution verification.};

  \draw[dep=saDerivation]   ([yshift=1.2mm]e2.east) -- ++(4mm,0)
    |- ([yshift=-1.6mm]e1.east);
  \draw[dep=saSymbolic]     ([yshift=4.5mm]e3.east) -- ++(4mm,0)
    |- ([yshift=-1.6mm]e2.east);
  \draw[dep=saResult]       ([yshift=1.2mm]e4.east) -- ++(7mm,0)
    |- ([yshift=-1.6mm]e3.east);
  \draw[dep=saVerification] ([yshift=1.2mm]e5.east) -- ++(4mm,0)
    |- ([yshift=-1.6mm]e4.east);
  \draw[dep=saCorrection]   ([yshift=2.4mm]e6.east) -- ++(10mm,0)
    |- ([yshift=1.6mm]e3.east);
  \draw[dep=saCorrection]   (e6.east)               -- ++(13mm,0)
    |- ([yshift=1.6mm]e1.east);

  \draw[dep=saFinal]        ([yshift=3mm]e7.west)   -- ++(-4mm,0)
    |- ([yshift=-1.6mm]e6.west);
  \draw[dep=saFinal]        ([yshift=1.5mm]e7.west) -- ++(-6.5mm,0)
    |- ([yshift=-1.6mm]e5.west);
  \draw[dep=saFinal]        (e7.west)               -- ++(-9mm,0)
    |- ([yshift=-1.6mm]e3.west);
  \draw[dep=saFinal]        ([yshift=-1.5mm]e7.west) -- ++(-11.5mm,0)
    |- ([yshift=-1.6mm]e2.west);
  \draw[dep=saFinal]        ([yshift=-3mm]e7.west)  -- ++(-14mm,0)
    |- ([yshift=-1.6mm]e1.west);

  \path ([xshift=-14mm]e1.west) ([xshift=14mm]e1.east);
\end{tikzpicture}
\caption{Schematic example of a ledger. A rejected result ($\texttt{e}_4$) remains in the record with its
verification verdict ($\texttt{e}_5$), and its replacement ($\texttt{e}_6$) declares its own
dependencies. The proposed answer ($\texttt{e}_7$) depends on every entry except the superseded one ($\texttt{e}_4$). Arrows indicate declared dependencies; later entries point to the earlier entries they rely on.}
\label{fig:ledger}
\end{figure}

When a step turns out to be wrong or is refined, its status
is changed to \emph{rejected} or \emph{superseded}, the replacement is
appended as a new entry, and both remain visible in the record. The history of
a calculation, including unsuccessful attempts, is therefore preserved. Furthermore, entries declare their dependencies explicitly. A result
entry must reference the computation it interprets; a derivation may reference
the assumptions and results it uses; the final answer references the entries
that support it. The ledger can thus be seen as a
directed graph, in which one can trace any conclusion backward through the
results, computations, and assumptions that led to it. Figure~\ref{fig:ledger}
shows a small example of a ledger in schematic form.

\subsection{Explicit Treatment of Assumptions and Ambiguities}
\label{sec:assumptions}

A problem submitted to \texttt{Solver Agent} may contain ambiguities in the sense that not everything is fully specified, convention choices may be made tacitly, and edge cases may silently be ignored.
The assumption-entry mechanism is intended to lift such ambiguities. Before deriving anything, the main solver is required to record its
reading of the problem as a set of assumption entries: the interpretation of
the statement, the definitions and notation to be used, the conventions
adopted, and the domain in which the results are claimed to hold. When the
problem statement admits more than one reasonable reading, the system is
instructed to explicitly write an assumption, visible in the ledger from that point
on. Later steps that rely
on an interpretive choice reference the corresponding assumption entry in
their dependency lists so that the consequences of each choice can be traced
through the record. This allows, in particular, the researcher to verify that the system solves the problem it is intended to solve. The verification system is also
instructed to check that no assumptions were overlooked or strengthened without an explicit ledger entry.

The mechanism does not, of course, guarantee that every tacit choice is
caught---an assumption can only be scrutinized once it has been written
down, and the system may still fail to recognize that a choice was made. But it reduces the risk of such a tacit choice.

\subsection{Verification Procedures}
\label{sec:verification}

Two review processes operate on the ledger. Both are carried out by language
models that are separate from the main solver: each runs with its own
instructions, sees only the material it is given to review, and has access to
its own computational tools so that it can re-derive quantities without
relying on the computations it is checking.

\paragraph{Step Verification.}
The first process reviews a single subset of entries. It is invoked selectively, at the
main solver initiative, for steps considered pivotal by the main solver. The verifier receives the entry under
review together with the full text of the entries it declares as
dependencies and nothing else. It is instructed to restate the claim in its
own terms, to examine whether the stated dependencies actually support it,
and to attempt an independent check: re-deriving the result by another route
where feasible or testing it symbolically or numerically on special cases
and boundary cases. The outcome is a verification entry containing a verdict,
accepted or rejected, together with a justification and, if one was found,
an explicit counterexample. A rejection changes
the reviewed entry status to rejected, and the main solver is
required to supersede it by appending a correction entry and redoing any work that
depended on it.

\paragraph{Full-solution Verification.}
The second process reviews the recorded solution as a whole. When the main solver considers
the problem solved, it records its proposed solution as a final-answer entry
and submits the entire ledger --- the problem statement, every entry in order,
and the candidate answer --- for review. The full-solution verifier works
through the dependency graph from the answer back toward the problem
statement. It checks that the assumptions are recorded, that each derivation and result is supported by the entries it cites,
that no step of the argument is missing from the record, that superseded
material is not silently relied upon, and it repeats independent spot checks
of key quantities with its own tools. Its verdict is again recorded in the
ledger: either the solution is verified, or it is rejected together with a list of issues naming the entry concerned,
and the correction required.

The independence of the verification systems from the main solver, favored by the separation of context, instructions, and tools, is intended to reduce the risk that the reviewer simply reproduces the
solver's reasoning. But, of course, it is not perfect; review can fail to notice a
defect just as the original derivation did.

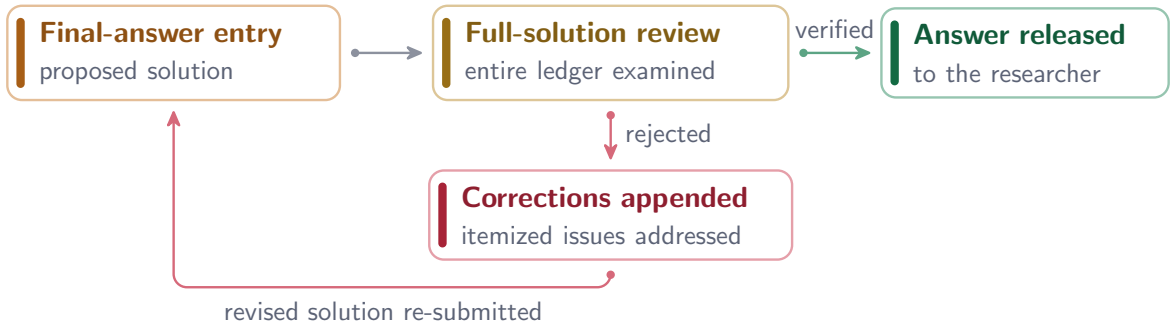
\begin{figure}[!htp]
  \centering
  \renewcommand{\safont}{\sffamily}
  \begin{tikzpicture}[node distance=7mm and 12mm,
      saLabel/.style={font=\footnotesize\sffamily, text=saMuted}]
    \node[saCard={saFinal}{44mm}] (cand)
      {\saHead{saFinal}{Final-answer entry}\\[-1pt]\saSub{proposed solution}};
    \node[saCard={saVerification}{47mm}, right=of cand] (review)
      {\saHead{saVerification}{Full-solution review}\\[-1pt]\saSub{entire ledger examined}};
    \node[saCard={saNumerical}{38mm}, right=of review] (release)
      {\saHead{saNumerical}{Answer released}\\[-1pt]\saSub{to the researcher}};
    \node[saCard={saCorrection}{48mm}, below=9mm of review] (fix)
      {\saHead{saCorrection}{Corrections appended}\\[-1pt]\saSub{itemized issues addressed}};

    \draw[saFlow=saMuted] (cand) -- (review);
    \draw[saFlow=saNumerical] (review) --
      node[saLabel, above=.4mm] {verified} (release);
    \draw[saFlow=saCorrection] (review) --
      node[saLabel, right=.4mm] {rejected} (fix);
    \draw[saFlow=saCorrection] (fix.south) -- ++(0,-3.5mm) -|
      node[saLabel, pos=0.26, below=.4mm] {revised solution re-submitted}
      (cand.south);
  \end{tikzpicture}
  \caption{Finalization of a solution. The proposed answer is withheld until
  the full-solution review returns a favorable verdict; a rejection blocks
  release and returns a list of required corrections. Any later
  change to the solution invalidates the verdict and requires a further
  review.}
  \label{fig:gate}
\end{figure}

\subsection{Computational Tools and Domain-specific Extensions}
\label{sec:tools}

Computation in the system is carried out through tools: programs exposed to
the language models through a declared calling convention. A tool is
specified by a typed interface describing the arguments it accepts, together
with a handler that performs the work; the model requests an invocation by
producing a structured call, and the handler's response is returned to it as
data. Every invocation is persisted as a
ledger entry of the corresponding type, recording the task as it was posed,
the code that was executed, the output that was produced, and any files
generated along the way.

Two general-purpose computational tools are always available. The symbolic tool carries out computer-algebra tasks --- series expansions, simplification,
polynomial and tensor manipulations --- in an isolated Python environment based
on \texttt{SymPy} \cite{Sympy}, with an optional bridge to the Wolfram engine. The numerical tool provides floating-point evaluation, linear
algebra, integration, and plotting through the standard scientific Python
libraries, with the option of compiling C++ programs when a computation is
too intensive for Python code. Plots and data files it produces are stored as artifacts attached to the corresponding ledger entry that the agents and the researcher can access.

Alongside these, the system allows for the addition of domain-specific tools, and the
applications reported in this paper rely on one: a Calabi--Yau
analysis tool built on the \textsc{Calabi--Yau Tools} package~\cite{Demirtas:2022hqf}, which gives the system
access to standard operations on toric Calabi--Yau geometries: handling
reflexive polytopes from the Kreuzer--Skarke list \cite{Kreuzer:2000xy}, constructing
triangulations, and computing Hodge numbers, intersection numbers, etc. These quantities often recur in F-theory and type~IIB compactification problems, and obtaining them from an established package is useful in this context. Another optional tool is a reference-retrieval tool that searches
the literature on the web for stated results and returns sourced excerpts.

Because the source code of \texttt{Solver Agent} is publicly available, one can add custom tools that use their own code or domain-specific packages to solve particular problems.

\subsection{Interaction with the System}
\label{sec:interaction}

A calculation begins with a problem statement, the \emph{prompt}, written in ordinary
mathematical prose, with formulas in \LaTeX{} where convenient. At this point, the
researcher also makes the tool choices mentioned earlier. While the system works, the ledger is displayed as a live timeline: entries
appear as they are appended. A complementary view renders the ledger as its dependency graph, which is a quick way to see how a conclusion is supported. The researcher may intervene at any point; a message sent mid-run is recorded as a follow-up entry and treated as an amendment to the problem.

A completed calculation yields several artifacts. First, the
ledger, which contains the assumptions, the argument, every
computation with its code and output, the verification verdicts, and the
answer; it can be read in the interface.
Files produced along the way (plots, tables, datasets) are stored with
their generating entries and can be accessed individually. Lastly,
the system can generate, on request, a self-contained
prose account of the solution, composed from the ledger, with the computations summarized and cited. It provides a convenient and readable summary of the ledger.

\subsection{Traceability, Reproducibility, and Scope}
\label{sec:discussion}

\texttt{Solver Agent} is built to leverage the ever increasing knowledge, coding, and computation capabilities of large language models for mathematics and theoretical physics while reducing mistakes and ensuring all derivations are fully verifiable and reproducible.

Traceability is provided by the structure of the ledger, which keeps track of every assumption, derivation, computation, verification, and conclusion.
One should emphasize that all reasoning
components of the system are language models, which can inherently make mistakes and hallucinate results. What the system offers is a framework for machine-assisted calculation and proof in which a researcher can identify which assumption or
reasoning step supports a conclusion and check the relevant code. This is a
stronger standard than a free-form generated answer, but it is not a formal
proof, the ledger does not assign formal logical semantics to its entries, and
an accepted verification verdict is not a machine-checked certificate of
truth. 
\texttt{Solver Agent} has nevertheless proven to be more reliable than usual LLM interfaces for both avoiding declaring a result it does not have and identifying errors or inconsistencies in the prompts. The latter has turned out to be useful for grasping subtleties in the results we obtained, especially when applying them to examples.

To ensure reproducibility, users of \texttt{Solver Agent} are advised to make all interactions with the system publicly available, including the full ledger and version of the system, in the format of their choice. The interface provides a way to easily export them in various formats. All sessions used in this work are available \href{https://storage.googleapis.com/solver-agent-sessions/solver-agent-f-theory-uplifts/index.html}{here}.

A recent complementary approach was presented in \cite{douglas2026formalizationqft}. The authors construct
the massive Gaussian free field in four-dimensional Euclidean spacetime in
Lean~\cite{deMoura:2021lean4} and prove that it satisfies the Glimm--Jaffe/Osterwalder--Schrader
axioms. In the current version, the proof contains no unproven axioms and is
checked using Lean and its mathematical library \textsc{Mathlib}~\cite{mathlib:2020}. The language
models used in that work assist in writing and organizing the formal
development.

This type of agentic use has a different goal from \texttt{Solver Agent}. A Lean
formalization translates definitions and proofs into a formal language with
precise axiomatic logical semantics and provides a very strong guarantee once the
formal statement itself has been accepted. \texttt{Solver Agent}
, instead, works at the level of mathematical and physical calculations, sometimes in an exploratory way, as they
are ordinarily carried out by researchers: in prose, with symbolic and
numerical computations, specialized software, literature results, convention choices, and assumptions that may not have a completely rigorous formulation.
It therefore applies to a different class of tasks, including those in
which some physical inputs are heuristic, effective, convention-dependent, or
supported by numerical evidence. Its verification agents can independently
rederive and cross-check such steps, but they cannot provide the logical
guarantee of a Lean proof. The two approaches are complementary.

We now turn to an application of \texttt{Solver Agent} to a set of string theory problems related to F-theory uplift constructions and S-folds.

\section{Review of Type IIB Compactifications on Calabi--Yau Orientifolds}
\label{sec:typeIIBCalabi--Yau}

This section provides a brief summary of the construction of Calabi--Yau orientifolds for Type IIB compactifications with fluxes. A comprehensive review of this topic is beyond the scope of this work; instead, the reader is referred to the standard textbooks and reviews \cite{Grana:2005jc, Blumenhagen:2006ci, Douglas:2006es, Denef:2008wq, Weigand:2010wm, Ibanez:2012zz, Maharana:2012tu, Blumenhagen:2013fgp, Weigand:2018rez, Marchesano:2022qbx,Marchesano:2024gul}.

\subsection{Orientifolding Calabi--Yau threefolds}
\label{sec:KSorientifolds}

The orientifold projection in Type II compactification is specified by an involution $\iota$ that acts on the K\"ahler form $J$ and the holomorphic top form $\Omega_3$ of the Calabi--Yau threefold~\cite{Sen:1997gv,Giddings:2001yu,Blumenhagen:2006ci,Ibanez:2012zz}.\footnote{The orientifold projection in Type IIB compactifications involves the world-sheet parity reversal $\Omega_p$ combined with a holomorphic involution $\iota$ of the Calabi--Yau threefold. Specifically, the orientifold generator is $\mathcal{O}_\text{or} = \Omega_p (-1)^{F_L} \iota$ for O3/O7-planes and $\mathcal{O}_\text{or} = \Omega_p \iota$ for O5/O9-planes, where $F_L$ is the left-moving spacetime fermion number.} In the case of Type IIB compactification, the involutions are
\begin{equation}
    \begin{cases}
        \iota^* J = J\coma \iota^* \Omega_3 = -\Omega_3 & \text{for O3/O7-planes}\,, \\
        \iota^* J = J\coma \iota^* \Omega_3 = \Omega_3 & \text{for O5/O9-planes}\,.
    \end{cases}
\end{equation}
This action splits the cohomology groups $H^{p,q}(\IX_3)$ into even and odd eigenspaces under the pull-back of the involution, decomposing the K\"ahler and complex structure moduli spaces into invariant and anti-invariant sectors (with dimensions $h^{1,1}_{\pm}$ and $h^{2,1}_{\pm}$). The dimensions of the eigenspaces are determined by~\cite{Camara:2003ku,Grana:2003ek,Grimm:2004uq,Crino:2022zjk}:
\begin{equation}
h^{2,1}_+ - h^{2,1}_- = 1 - S_O + (h^{1,1}_+ - h^{1,1}_-) - \frac{1}{2} \chi_{\text{fix}}\,,
\end{equation}
where $h^{1,1} = h^{1,1}_+ + h^{1,1}_-$ and $h^{2,1} = h^{2,1}_+ + h^{2,1}_-$ by construction, so that we obtain 
\begin{equation}
h^{2,1}_+ = \frac{1}{2} \left( h^{2,1} + 1 - S_O + h^{1,1}_+ - h^{1,1}_- - \frac{1}{2} \chi_{\text{fix}} \right)\coma h^{2,1}_- = h^{2,1} - h^{2,1}_+\,.
\end{equation}
The parameter $S_O$ encodes the choice of the orientifold projection; setting $S_O = +1$ corresponds to the O3/O7-plane configurations, while $S_O = -1$ defines O5/O9-planes. The total Euler characteristic of the fixed loci, $\chi_{\text{fix}}$, decomposes into the sum of the Euler characteristics $\chi(S)$ of the codimension-one fixed surfaces $S$ and the number of  isolated fixed points $N_{O3}$. These components are evaluated using the intersection data of the Calabi--Yau, i.e.~\cite{Morrison:1996na,Morrison:1996pp,Denef:2008wq,Weigand:2010wm}:
\begin{equation}
\chi_{\text{fix}} = \sum_{\text{O7}} \left( \int_{\IX_3} [S]^3 + \int_{\IX_3} c_2(T\IX_3) \wedge [S] \right) + \sum_{\text{O3}} \int_{\IX_3} [D_{i_1}] \wedge [D_{i_2}] \wedge [D_{i_3}]\,.
\label{eq:EulerFix}
\end{equation}
However, we must note that the above formulas assume that the Calabi--Yau $\IX_3$ is smooth everywhere. While $\IX_3$ is smooth generically, the orientifold base $B_3 = \IX_3/\iota$ possesses quotient singularities at the locations of the O3-planes. Furthermore, when the Calabi--Yau $\IX_3$ itself inherits singularities from the ambient toric space, the naive evaluation of $\chi_{\text{fix}}$ via the smooth intersection formulas in~\eqref{eq:EulerFix} fails to capture the topological data. Since we rely on this computation of $\chi_{\text{fix}}$ and the Lefschetz fixed point theorem to determine $h^{2,1}_+$ and $h^{2,1}_-$, an incorrect (and often fractional) $\chi_{\text{fix}}$ leads to fractional values for $h^{2,1}_+$ and $h^{2,1}_-$. This does not imply an inconsistent theory~\cite{Gendler:2022qof,Jefferson:2022ssj}; rather, it serves as a signal that the formulas must be corrected to properly account for the singularities \cite{Collinucci:2014taa}. Discovering the effects of quotient singularities on the Calabi--Yau orientifold bases (and their F-theory uplifts) is precisely one of the purposes of this work. 

\subsection{Comments on Divisor Cohomologies}
\label{sec:divisorcohomologies}

In the literature, the divisors are classified based on their Hodge numbers. A divisor $D$ is said to be \textit{rigid} if it cannot be continuously deformed within the Calabi--Yau $X_3$; therefore, all its normal bundle global sections vanish and $h^{2,0}(D) = 0$ \cite{Witten:1996bn,Blumenhagen:2009qh,Blumenhagen:2012kz}. In the context of string phenomenology, one also encounters a \textit{completely rigid} divisor that additionally satisfies $h^{1,0}(D) = 0$, meaning it is free of any moduli, including Wilson lines \cite{Gao:2013pra}. The Hodge numbers of a completely rigid divisor $D_{\text{rig}}\subset X_3$ are 
\begin{equation}
 h^{0,0}(D_{\text{rig}}) = 1, \quad h^{1,0}(D_{\text{rig}})= 0,\quad h^{2,0}(D_{\text{rig}})= 0, \quad h^{1,1}(D_{\text{rig}})>0 \fstop
\end{equation} 
For this reason, rigid divisors that are not completely rigid, i.e., $h^{1,0}(D)>0$, are called \textit{Wilson divisors} \cite{Gao:2013pra}. Finally, divisors with $h^{1,0}(D)=0$ but $h^{2,0}(D)>0$ are \textit{non-rigid}; hence, they are deformable, but they do not admit Wilson lines; such divisors are also classified as \textit{K3 divisors} \cite{Gao:2013pra}.

In moduli stabilization, generating a non-perturbative superpotential $W_\text{np} \supset A_\text{Pf} e^{-2\pi T_i}$ to stabilize K\"ahler moduli~\cite{Giddings:2001yu,Kachru:2003aw,Denef:2008wq,Cicoli:2023opf} requires the wrapped ED3-instanton to carry exactly two neutral fermionic zero modes~\cite{Witten:1996bn,Katz:1996fh,Blumenhagen:2006xt,Blumenhagen:2009qh} (see also \cite{Caraffi:2026wzk} for a recent discussion on zero-mode constraints in poly-instanton configurations). A completely rigid divisor automatically satisfies this requirement, as it lacks any moduli that would otherwise generate extra zero modes. While instantons wrapping rigid (Wilson) divisors can also contribute if these extra zero modes are lifted by background fluxes~\cite{Blumenhagen:2007bn,Bianchi:2011qh,Palti:2020qlc} or the orientifold projection~\cite{Polchinski:1995mt,Gimon:1996rq,Polchinski:1998rq,Polchinski:1998rr}, complete rigidity remains the simplest prerequisite for instanton contributions. An identical restriction applies when the non-perturbative superpotential is generated by gaugino condensation on a stack of D7-branes~\cite{Katz:1996fh,Katz:1996th,Diaconescu:1998ua}: any geometric deformations or Wilson lines manifest as massless adjoint matter in the four-dimensional EFT, preventing condensation unless the wrapped divisors are completely rigid geometrically~\cite{Morrison:2012np,Morrison:2014lca} or these modes are rendered massive by background fluxes~\cite{Camara:2004jj,Gomis:2005wc,Jockers:2005zy}.

\subsection{Tadpole Cancellation and Fluxes}
\label{sec:tadpole_fluxes}

In the following, we focus on O7/O3 orientifold configurations and review the mechanisms for canceling the tadpole contributions induced by these O-planes \cite{Sethi:1996es, Uranga:2000xp, Giddings:2001yu}.\footnote{For applications to global model building and comprehensive reviews of tadpole constraints in Type IIB and F-theory compactifications, see e.g., \cite{Marchesano:2004yq, Grana:2005jc, Blumenhagen:2006ci, Blumenhagen:2008zz, Denef:2008wq, Weigand:2010wm, Ibanez:2012zz}.} Specifically, when an O7-plane wraps a divisor $S$ (with local coordinate $x_S$) of a Calabi--Yau $\IX_3$, it carries a local D7-brane charge of $-8[S]$. This tadpole is canceled locally by introducing D7-branes on appropriate divisors. While various configurations can achieve this, we restrict our analysis to the two most common geometric resolutions \cite{Sen:1996vd, Sen:1997gv, Collinucci:2008pf, Braun:2008ua}.

If the divisor admits complex structure deformations (hence it is not rigid), one can recombine the branes into a Whitney umbrella wrapping the homology class $8[S]$ defined by the polynomial \cite{Aluffi:2007sx,Braun:2008ua,Collinucci:2008pf,Collinucci:2008sq,Collinucci:2008zs,Aluffi:2009tm} \begin{equation}
\label{eq:WU_def}
    \eta^2 - x_S^2 \psi = 0\coma
\end{equation}
where $x_S$ is the local coordinate of the O7-plane locus $S$, and $\eta$ and $\psi$ are sections of $\mathcal{O}(4S)$ and $\mathcal{O}(6S)$, respectively. This configuration does not support any non-abelian gauge group. Alternatively, one can place four D7-branes and their four orientifold images on top of the O7-plane locus $x_S^8 = 0$. This configuration generates an $\SO(8)$ gauge group \cite{Polchinski:1995mt,Gimon:1996rq,Sen:1997gv,Collinucci:2008pf}, and it is the typical configuration whenever the O7-plane wraps a rigid divisor.\footnote{More precisely, for a rigid divisor $S$ in a Calabi--Yau threefold, the normal bundle is typically negative, which implies that $h^0(\mathcal{O}(k S)) = 1$ for all $k \geq 1$. Consequently, there are no other effective representatives in the homology class $8[S]$, making it impossible to wrap the D7-branes on any surface other than $S$ itself to cancel the local tadpole.} In this case, the D7-branes cannot be recombined into a local Whitney umbrella, leaving them bound to the O7-plane locus \cite{Blumenhagen:2008zz,Cicoli:2011qg,Morrison:2012np}. However, if the homology class $8[S]$ can be decomposed into a linear combination of other effective divisors, one can, in principle, cancel the tadpole by wrapping the D7-branes on these other divisors instead. In such cases, the realized gauge groups depend on the specific action of the orientifold on these new divisors, allowing one to realize other gauge group configurations.

Once the D7-branes are placed, they can support worldvolume gauge fluxes $\mathcal{F}_2 = F_2 -B_2$, where $F_2$ is the field strength of the worldvolume gauge theory and $B_2$ is the Neveu--Schwarz-Neveu--Schwarz (NSNS)  2-form potential restricted to the D7-brane worldvolume. It is known that, by the Freed--Witten (FW) anomaly \cite{Minasian:1997mm,Freed:1999vc,Berasaluce-Gonzalez:2012awn}, the gauge flux requires $$F_2 + \frac{1}{2}c_1(TS_{D7})\in H^2(S_{D7},\ZZ)\coma$$ where $S_{D7}$ is the divisor wrapped by the D7-brane. If the divisor is Spin (i.e., $c_1(TS_{D7})$ is an even cohomology class), then $F_2$ can be chosen to be integral and trivial. In such a scenario, if $B_2$ is also trivial on the worldvolume of the brane, the total worldvolume gauge flux $\mathcal{F}_2$ vanishes. On the other hand, if $S_{D7}$ is non-Spin, the FW anomaly requires a half-integer flux for $F_2$, leading to a generally non-zero $\mathcal{F}_2$. For an $\SO(8)$ stack, while the D7-branes cannot be separated from the O7-plane, turning on a non-trivial gauge flux $\mathcal{F}_2$ breaks the $\mathrm{SO}(8)$ gauge group (e.g., to $\mathrm{U}(4)$). Thus, preserving the full $\mathrm{SO}(8)$ symmetry requires $\mathcal{F}_2 = 0$. Conversely, for a Whitney brane, the gauge flux $F_2$ is constrained by the line bundles $\mathcal{O}(4S)$ and $\mathcal{O}(6S)$ of the sections $\eta$ and $\psi$ defining the Whitney umbrella~\cite{Braun:2008ua,Collinucci:2008pf,Collinucci:2008sq}. While one can sometimes choose a $B_2$ field to cancel this flux, $B_2$ is a global closed string field whose pullback generally cannot cancel local worldvolume fluxes across multiple branes simultaneously. Thus, non-trivial fluxes $\mathcal{F}_2$ are generically necessary.

We can now consider the D3-tadpole condition, which receives contributions from the localized O3-planes, the induced D3-brane charges of the D7/O7 configuration, the $F_3$ and $H_3$ fluxes, and the D7-brane worldvolume fluxes. Because after the orientifold action, the resulting quotient base $B_3 = \IX_3/\iota$ is singular at the orientifold fixed loci, we evaluate the tadpole in terms of the intersection data of the original Calabi--Yau $\IX_3$.\footnote{The difference in convention is the reason why, in some string theory literature, the D3-tadpole is sometimes expressed in terms of $B_3$, leading to a factor of $2$ in \eqref{eq:tadpole_full}.} The total tadpole condition is\footnote{We assume a single O7-plane and its corresponding D7-brane configuration for simplicity. In the presence of multiple O7-planes and D7-brane stacks, their contributions to the D3-tadpole sum over each component.} 
\begin{equation}
\label{eq:tadpole_full}
2n_{D3} + \underbrace{\frac{1}{2} \int_{\IX_3} F_3 \wedge H_3 + \frac{1}{2} \int_{S_{D7}} \mathrm{tr}(\mathcal{F}_2 \wedge \mathcal{F}_2)}_{N_{\text{flux}}} = \underbrace{\frac{N_{O3}}{2} + \frac{\chi(S)}{6} + \frac{\chi(S_{D7})}{24}}_{N_{D3}}\,.
\end{equation}
Here, $n_{D3}$ is the number of D3-branes in $B_3$, so that $2n_{D3}$ is their contribution evaluated in the double cover $\IX_3$. Analogously to the D7-tadpole, the O3- and O7-planes carry negative D3-brane charges. While the D7-branes also carry induced D3-brane charges that partially compensate for the O7-planes, the net charge of the localized sources $N_{D3}$ remains negative. This net negative charge acts as a budget that must be canceled by introducing positive contributions: the mobile D3-branes ($2n_{D3}$) and the background fluxes ($N_{\text{flux}}$). Since both of these contributions are non-negative in a supersymmetric vacuum, the tadpole constraint effectively bounds the flux vacua via the inequality $N_{\text{flux}}\leq N_{D3}$ \cite{Bena:2020xrh,Braun:2020jrx,Plauschinn:2021hkp,Gao:2022fdi,Grana:2022dfw,Lust:2022mhk}.

Depending on the D7-brane configuration chosen to cancel the O7-tadpole, the contributions to $N_{D3}$ and to $N_\text{flux}$ coming from the D7-branes will vary. The geometric contribution from the D7-branes depends on their Euler characteristic, which, for a generic smooth divisor $S_{D7}$, can be computed by integrating its top Chern class using the adjunction formula:
\begin{equation}
\label{eq:adjunction_chi}
\chi(S_{D7}) = \int_{\IX_3} \left( [S_{D7}]^3 + c_2(T\IX_3) \wedge [S_{D7}] \right)\,.
\end{equation}
Since global D7-tadpole cancellation demands that the total D7-brane homology class satisfies $[S_{D7_{\text{tot}}}] = 8[S]$, the distinct D7-brane configurations chosen to cancel the O7-tadpole will substitute differently into \eqref{eq:adjunction_chi}. 

For instance, for the $\mathrm{SO}(8)$ stack, four D7-branes and their four orientifold images wrap the O7-plane locus $S$; thus, their contribution to the Euler characteristic is simply eight times that of $S$:
\begin{equation}
\label{eq:chiD7_SO8}
\chi(S_{D7_{\mathrm{SO}(8)}}) = 8 \chi(S) = \int_{\IX_3} \left( 8 [S]^3 + 8 c_2(T\IX_3) \wedge [S] \right)\,.
\end{equation}
Substituting this into \eqref{eq:tadpole_full}, we recover the standard result~\cite{Sethi:1996es,Sen:1996vd,Sen:1997gv}:
\begin{equation}
\label{eq:D3chSO8}
N_{D3}^{\text{SO}(8)} = \frac{N_{O3}}{2} + \frac{\chi(S)}{6} + \frac{\chi(S_{D7_{\mathrm{SO}(8)}})}{24} = \frac{N_{O3}}{2} + \frac{\chi(S)}{2}\,.
\end{equation}
Correspondingly, the worldvolume flux on this stack adds $\frac{1}{2}\int_{S}\mathrm{tr}(\mathcal{F}_2\wedge\mathcal{F}_2)$ to $N_{\text{flux}}$~\cite{Douglas:1995bn,Green:1996dd,Minasian:1997mm}.\footnote{Note that the trace over the fundamental representation of the gauge group accounts for the multiplicity of the branes in the stack.}

A more complicated situation is that of the Whitney umbrella, which is defined algebraically by equation \eqref{eq:WU_def}. The total homology class wrapped by the brane is thus $8[S]$. To evaluate the D3-brane tadpole \eqref{eq:tadpole_full}, we must first compute the geometric contribution $\chi(S_{D7_{\text{WU}}})$. The Euler characteristic of a generic smooth divisor in the class $8[S]$ is given by \eqref{eq:adjunction_chi}, i.e. 
\begin{equation}
\chi^{\text{sm}}(S_{D7_{\text{WU}}}) = \int_{\IX_3} \left( (8[S])^3 + c_2(T\IX_3) \wedge [8S] \right) = \int_{\IX_3} \left( 512 [S]^3 + 8 c_2(T\IX_3) \wedge [S] \right)\,.
\end{equation}
Unlike a generic divisor, the Whitney umbrella \eqref{eq:WU_def} contains a non-trivial singular locus: a double curve at $x_S=\eta=0$ (of homology class $4[S]^2$) and pinch points at $x_S=\eta=\psi=0$ (of homology class $24[S]^3$) \cite{Aluffi:2007sx,Collinucci:2008pf}. This alters the topology of the brane, and the actual Euler characteristic of $S_{D7_{\text{WU}}}$ is corrected by $\Delta \chi$ \cite{Aluffi:2007sx,Collinucci:2008pf}, i.e.
\begin{equation}
    \chi(S_{D7_{\text{WU}}}) = \chi^{\text{sm}}(S_{D7_{\text{WU}}}) +  \Delta \chi\fstop
\end{equation}

One way to compute $\chi(S_{D7_{\text{WU}}})$ is to consider the resolved surface $\Sigma_\text{WU}$ to be a smooth complete intersection within a $\mathbb{P}^1$-bundle $P = \mathbb{P}(\mathcal{O} \oplus \mathcal{O}(2S))$ over $\IX_3$~\cite{Aluffi:2007sx,Collinucci:2008pf}. The hyperplane class $H$ of the bundle satisfies $H^2 = 2H[S]$, and $\Sigma_\text{WU}$ is the intersection of two divisors $D_1 = 2H - 2[S]$ and $D_2 = 3H + [S]$, such that $D_1 \wedge D_2 = 6H^2 - 4H[S] - 2[S]^2$ simplifies to $8H[S] - 2[S]^2$. Pushing this down to the base yields $8[S]$.

In this configuration, $\chi(\Sigma_\text{WU}) \equiv \chi(S_{D7_{\text{WU}}})$ is given by the integral of the top Chern class of $\Sigma_\text{WU}$. The total Chern class is obtained via the adjunction formula $c(T\Sigma_\text{WU}) = c(TP)/c(N_{\Sigma_\text{WU}})$. Since $c_1(T\IX_3) = 0$, the tangent bundle of $P$ has total Chern class 
$$c(TP) = (1+c_2(T\IX_3))(1+2H-2[S]+H^2-2H[S])=(1+c_2(T\IX_3))(1+2H-2[S])\coma$$
while the normal bundle $N_{\Sigma_\text{WU}} = \mathcal{O}(D_1) \oplus \mathcal{O}(D_2)$ has total Chern class 
$$c(N_{\Sigma_\text{WU}}) = (1+2H-2[S])(1+3H+[S])\fstop$$ 
The ratio leaves:
\begin{equation}
    c(T\Sigma_\text{WU}) = \frac{1+c_2(T\IX_3)}{1+3H+[S]}\fstop
\end{equation}
Expanding and substituting $H^2 = 2H[S]$, the second Chern class evaluates to:
\begin{equation}
    [c(T\Sigma_\text{WU})]_2 = c_2(T\IX_3) + 24H[S] + [S]^2\fstop
\end{equation}
Integrating $[c(T\Sigma_\text{WU})]_2 \wedge D_1 \wedge D_2$ over the four-dimensional bundle $P$, we obtain the Euler characteristic:
\begin{equation}
\begin{split}
    \chi(\Sigma_\text{WU}) &= \int_P (8H[S] - 2[S]^2) \wedge (c_2(T\IX_3) + 24H[S] + [S]^2) \\
    &= \int_{\IX_3} \left( 344 [S]^3 + 8 c_2(T\IX_3) \wedge [S] \right)\,,
\end{split}
\end{equation}
The geometric Euler characteristic for the Whitney umbrella is thus~\cite{Collinucci:2008pf}:
\begin{equation}
\label{eq:chiD7_WU}
\chi(S_{D7_{\text{WU}}}) \equiv \chi(\Sigma_\text{WU}) = \int_{\IX_3} \left( 344 [S]^3 + 8 c_2(T\IX_3) \wedge [S] \right)\,.
\end{equation}
Substituting this contribution into \eqref{eq:tadpole_full}, the tadpole evaluated in the double cover becomes:
\begin{equation}
\label{eq:D3WD7geom_bare}
N_{D3}^{\text{WD7}} = \frac{N_{O3}}{2} + \frac{\chi(S)}{6} + \frac{\chi(S_{D7_{\text{WU}}})}{24} = \frac{N_{O3}}{2} + \frac{29}{2}\int_{\IX_3} [S]^3 + \frac{1}{2}\int_{\IX_3} [S]\wedge c_2(T\IX_3)\,.
\end{equation}
The full D3-brane tadpole \eqref{eq:tadpole_full} also requires evaluating the worldvolume flux contribution $N_{\text{flux}}$. For the Whitney umbrella, this contribution is necessarily non-trivial even when the worldvolume fluxes ($\mathcal{F}_2=0$) vanish. Because the divisor wrapped by the brane is singular at the pinch points (where the two local sheets of the D7-brane, defined by the branches of \eqref{eq:WU_def}, undergo a non-trivial monodromy), defining a consistent gauge bundle requires a half-integral worldvolume flux. This flux is determined by the line bundles $\mathcal{O}(4S)$ and $\mathcal{O}(6S)$ of the sections defining the Whitney umbrella~\cite{Collinucci:2008pf,Braun:2011zm}. The total flux contribution to the tadpole in \eqref{eq:tadpole_full} combines this mandatory flux with any continuous flux $F_2$:
\begin{equation}
\label{eq:D3WD7flux}
N_{\text{flux}}^{\text{WD7}} = \int_{\IX_3} [S] \wedge (3[S]-2F_2+2B_2)\wedge (3[S] + 2F_2 - 2B_2)\,.
\end{equation}
Expanding this expression yields $9 \int_{\IX_3} [S]^3 - 4 \int_{\IX_3} [S] \wedge (F_2-B_2)^2$. Both terms are non-negative and contribute positively to $N_{\text{flux}}$; consequently, they consume part of $N_{D3}^{\text{WD7}}$.

\subsection{F-theory Uplift}

It is possible to translate the perturbative type IIB compactification on a Calabi--Yau orientifold that we described in the previous sections into the F-theory  language~\cite{Vafa:1996xn,Morrison:1996na,Morrison:1996pp,Sen:1996vd,Sen:1997gv,Denef:2008wq,Collinucci:2008zs,Blumenhagen:2009up,Collinucci:2009uh,Aluffi:2009tm,Weigand:2018rez}. Such a configuration corresponds to F-theory compactified on a Calabi--Yau fourfold $\IY_4$ given by the elliptic fibration over a base $B_3$.

\paragraph{Sen Limit Motivation.}
\label{sec:sen_limit}

In order to construct the Calabi--Yau $\IY_4$, we recall that, in the F-theory framework~\cite{Vafa:1996xn}, the axio-dilaton $\tau = C_0 + i e^{-\phi}$, where $C_0$ is the Ramond--Ramond (RR) zero-form and $\phi$ is the string dilaton, becomes the complex structure modulus of an auxiliary elliptic curve fibered over the physical spacetime. The compactification space is an elliptically fibered Calabi--Yau fourfold $\pi: \IY_4 \to B_3$, where the physical compactification base $B_3 = \IX_3/\iota$ is the orientifold quotient of the Calabi--Yau threefold.

Sen's weak coupling limit \cite{Sen:1996vd,Sen:1997gv} establishes the connection between the F-theory fourfold $\IY_4$ and the Type IIB Calabi--Yau orientifold $\IX_3$. The elliptic fibration is described by a Weierstrass model
\begin{equation}
    P_W \equiv y^2 - x^3 - f x z^4 - g z^6 = 0\coma
\end{equation}
where $[x:y:z]$ are homogeneous coordinates of the ambient $\mathbb{P}_{[2,3,1]}$ fiber. To ensure $\IY_4$ is Calabi--Yau, the sections $f$ and $g$ must transform as sections of the line bundles $\mathcal{O}(-4K_{B_3})$ and $\mathcal{O}(-6K_{B_3})$, respectively~\cite{Morrison:1996na,Morrison:1996pp}. The locations of the Type IIB seven-branes correspond to the discriminant locus $\Delta_W = 4f^3 + 27g^2 = 0$, where the elliptic fiber degenerates \cite{Bershadsky:1996nh}.

The sections $f$ and $g$ are parametrized by $c$ as
\begin{equation}
    f = -3h^2 + c\eta\,, \qquad g = -2h^3 + c h \eta - \frac{c^2}{12}\psi\fstop
\end{equation}
Here, $h \in \Gamma(B_3, \mathcal{O}(-2K_{B_3}))$, while $\eta \in \Gamma(B_3, \mathcal{O}(-4K_{B_3}))$ and $\psi \in \Gamma(B_3, \mathcal{O}(-6K_{B_3}))$. Substituting this parametrization into the discriminant yields
\begin{equation}
    \Delta_W = -9 c^2 h^2 \left( \eta^2 - h \psi \right) + \mathcal{O}(c^3)\fstop
\end{equation}
In the weak coupling limit $c \to 0$, the discriminant locus factorizes into two components. The component $h = 0$ (appearing squared) corresponds to the location of the O7-plane in the base $B_3$. The canonical class of the base must satisfy $K_{B_3} = -\frac{1}{2}[h]$, ensuring that the double cover branched over the O7-plane locus remains a Calabi--Yau threefold.

The Type IIB Calabi--Yau threefold $\IX_3$ is defined algebraically over the base $B_3$ by introducing a local coordinate $\xi$ and considering the hypersurface~\cite{Sen:1997gv}
\begin{equation}
    \xi^2 = h\fstop
\end{equation}
This space admits the orientifold involution $\sigma: \xi \to -\xi$. The fixed-point locus of this involution is $\xi = 0$, which identifies $\xi$ as the local coordinate $x_S$ of the O7-plane divisor $S$ introduced in Section~\ref{sec:tadpole_fluxes}. This identification explicitly sets $h = x_S^2$. 
The second component of the discriminant locus, $\eta^2 - h \psi = 0$, corresponds to the D7-brane configuration in the base $B_3$. Uplifting this locus to the Calabi--Yau threefold $\IX_3$ by substituting $h = x_S^2$ yields the recombined Whitney umbrella equation
\begin{equation}
    \eta^2 - x_S^2 \psi = 0\coma
\end{equation}
where $\eta$ and $\psi$ are now understood as sections of $\mathcal{O}(4S)$ and $\mathcal{O}(6S)$ on $\IX_3$~\cite{Braun:2008ua,Collinucci:2008pf,Collinucci:2008sq}. This matches the configuration described in Section~\ref{sec:tadpole_fluxes} that globally cancels the O7-plane tadpole.

If the O7-plane wraps a rigid divisor, the sections $\eta$ and $\psi$ are forced to factorize as $\eta = \tilde{\eta} h^2$ and $\psi= \tilde{\psi} h^3$ and the Whitney umbrella polynomial becomes  $\eta^2 - h \psi = h^4(\tilde{\eta}^2 - \tilde{\psi})$. This $h^4$ component merges with the O7-plane locus $h^2$, forcing the total discriminant to vanish to degree six ($\Delta \propto h^6$). This yields the $I_0^*$ singularity characteristic of an $\mathrm{SO}(8)$ stack, where four D7-branes and their four orientifold images sit on top of the O7-plane.

\subsection{F-theory Tadpole Cancellation Conditions  and Corrections}
\label{sec:ftheory_corrections}

The D3-brane tadpole condition in F-theory takes the form~\cite{Becker:1996gj,Sethi:1996es}:
\begin{equation}
\label{eq:FtheoryTadpole}
n_{D3} + \frac{1}{2} \int_{\IY_4} G_4 \wedge G_4 = \frac{\chi_\text{st}(\IY_4)}{24} \equiv \frac{N_{D3}}{2}\,,
\end{equation}
where $G_4$ is the four-form flux on the Calabi--Yau fourfold $\IY_4$, corresponding to the $N_\text{flux}/2$ contribution in \eqref{eq:tadpole_full}, and $\chi_\text{st}(\IY_4)$ is the Euler characteristic of $Y_4$. However, we note that in defining $N_{D3}/2 = \chi_\text{st}(\IY_4)/24$, F-theory includes contributions coming from the presence of O3-planes in the Type IIB description.

As we will see in Section~\ref{sec:Codim3Sing}, the base $B_3$ contains terminal quotient singularities corresponding to the orientifold quotient of O3-plane fixed points~\cite{morrison1984terminal,anno2003four,Garcia-Etxebarria:2015wns}, which are inherited by the uplifted Calabi--Yau fourfold $\IY_4$. If one naively ignores them and computes the Euler characteristic from the top Chern class of the resulting Calabi--Yau fourfold, the result will fail to match the value expected by the tadpole condition~\cite{Sethi:1996es,Dasgupta:1996yh,Dasgupta:1999ss,Collinucci:2008zs,Collinucci:2008pf}. More precisely, the actual $\chi_\text{st}(\IY_4)$ that one should use to match the tadpole contribution should be what is known as the ``stringy'' Euler characteristic \cite{Batyrev:1997hj}, which, in the case of singularities arising only from O3-planes, is found to be
\begin{equation}
   \chi_\text{st}(Y_4) = \int_{Y_4} c_4(TY_4) + 6N_{O3} \fstop 
\label{eq:Euler4fold}
\end{equation}
In Section \ref{sec:Codim3Sing}, we will actually prove \eqref{eq:Euler4fold} in full generality, obtaining the expression for $\chi_\text{st}(Y_4)$ also when other types of terminal singularities arise.

In \cite{Jefferson:2022ssj}, an analogous correction was proposed for the vertical divisor $\overline{D}$ in $\IY_4$, which corresponds to the uplift of rigid toric divisors $\hat{D}$ hosting ED3-instantons \cite{Witten:1996bn}. In order to understand it, let us first compute the naive Euler characteristic of $\overline{D}$ by integrating its third Chern class over the ambient fourfold. For the generic smooth Weierstrass model, $\overline{D}$ is embedded in the projective bundle $\pi_P: P = \mathbb{P}(\mathcal{O} \oplus \mathcal{O}(2L) \oplus \mathcal{O}(3L)) \to B_3$ over the base $B_3$, where $L = c_1(TB_3)$. We denote by $\hat{D}_B$ the divisor in the base $B_3$ such that $\pi_\text{or}^* \hat{D}_B = \hat{D}$, where $\pi_\text{or}: \IX_3 \to B_3$ is the orientifold quotient map. Applying the adjunction formula yields the total Chern class
\begin{equation}
c(T\overline{D}) = \frac{c(TB_3)(1+H)(1+H+2L)(1+H+3L)}{(1+3H+6L)(1+\hat{D}_B)}\,,
\end{equation}
where $H$ is the hyperplane class of the $\mathbb{P}^2$ fiber, and $3H+6L$ is the class of the Weierstrass divisor. Expanding this expression and pushing forward to the base $B_3$ using the intersection relations of the projective bundle ($\pi_{P*} H^2 = 1$, $\pi_{P*} H^3 = -5L$), we obtain the integrated contribution evaluated on $B_3$:
\begin{equation}
\label{eq:chi_WU_base}
\chi_{\text{naive}}^{\text{WU}}(\overline{D}) = \int_{B_3} \left( -12 \hat{D}_B^2\cdot L - 60 \hat{D}_B \cdot L^2 \right) \,.
\end{equation}
To express this result in terms of the double cover Calabi--Yau threefold $\IX_3$, we utilize $\pi_\text{or}: \IX_3 \to B_3$, branched over the O7-plane locus $S_B$ (with class $[S_B] = 2L$ in $B_3$). The canonical bundle relation $K_{\IX_3} = \pi_\text{or}^* K_{B_3} + S$ \cite[Proposition 2.3]{hartshorne1977algebraic} implies $\pi_\text{or}^* L = S$. Accounting for the standard double cover pullback $\int_{B_3} \alpha = \frac{1}{2} \int_{\IX_3} \pi_\text{or}^* \alpha$, \eqref{eq:chi_WU_base} becomes
\begin{equation}
\label{eq:chi_WU_X3}
\chi_{\text{naive}}^{\text{WU}}(\overline{D}) = -6 \int_{\IX_3} \left( \hat{D}^2 \cdot  S + 5 \hat{D} \cdot  S^2 \right) \,.
\end{equation}

For the $\mathrm{SO}(8)$ configuration, the presence of the D7-branes on top of the O7-plane locus $S_B = 2L$ induces an $I_0^*$ singularity in the Weierstrass model. Resolving this singularity requires a sequence of four crepant blowups along $S_B$. To determine the Euler characteristic of the resolved geometry, one needs to evaluate the pushforwards of the total Chern class of the resolved threefold $\overline{D}$ down to the base surface $\hat{D}_B$, as studied in \cite{Esole:2017qeh,Esole:2018tuz,Esole:2018bmf,Esole:2017kyr,Esole:2019asj,Esole:2019hgr,Esole:2019ocl,Esole:2020alo}. When applied to the pushforward of Chern classes of blowups along complete intersections \cite{Aluffi:2010cbu}, one has
\begin{equation}\label{eq:cvarpioverlineD}
\tilde{\pi}_* (c(T\overline{D})) = 12 \frac{L + 3 S_B\cdot L - 2 S_B^2}{(1+S_B)(1+6L - 4S_B)} c(T\hat{D}_B) \,,
\end{equation}
where $\tilde{\pi}_*$ denotes the composition of the successive blowdowns and the projective bundle projection. Using \eqref{eq:cvarpioverlineD}, one can compute the Euler characteristic of $\overline{D}$ restricted to $\hat{D}_B$:
\begin{equation}
\chi_{\text{naive}}^{\text{SO(8)}}(\overline{D}) = 12 \int_{\hat{D}_B} \left( c_1(T\hat{D}_B) \cdot L - 6L^2 + 6 L\cdot S_B - 2S_B^2 \right) \,.
\end{equation}
By using that $c_1(T\hat{D}_B) = L - \hat{D}_B$, $S_B = 2L$, and by pulling back the integral to the double cover Calabi--Yau threefold $\IX_3$ via $\pi_\text{or}^*(\hat{D}_B) = \hat{D}$ and $\pi_\text{or}^*(L) = S$, we obtain
\begin{equation}\label{eq:chi_SO8_X3}
\begin{split}
    \chi_{\text{naive}}^{\text{SO(8)}}(\overline{D}) &= 12 \int_{\hat{D}_B} \left( -5L^2 - \hat{D}_B \cdot L + 6 L\cdot (2L) - 2(2L)^2 \right) \\
    & = -6 \int_{\IX_3} \left( \hat{D}^2 \cdot  S + \hat{D} \cdot  S^2 \right) \,.
\end{split}
\end{equation}
While the previous computations capture purely classical effects, these Euler characteristics receive corrections due to stringy orbifold effects from the presence of terminal singularities, in a manner similar to how they modify the Euler characteristic of the fourfold in \eqref{eq:Euler4fold}. In fact, the authors of \cite{Jefferson:2022ssj} proposed
\begin{equation}
\chi(\overline{D}) = \chi_{\text{naive}}(\overline{D}) + 2 N_{O3}(\hat{D})\,.
\label{eq:KimJefferson}
\end{equation}
This expression is valid for any type of terminal singularities, but the authors of \cite{Jefferson:2022ssj} were interested only in those arising from O3-planes. However, if we compute the ``stringy'' Euler characteristic for the vertical divisors in the presence of O3-planes, as we will show in Section \ref{sec:Codim3Sing}, the expression becomes
\begin{equation}
    \chi_{\text{st}}(\overline{D}) = \chi_{\text{naive}}(\overline{D})+ 6 N_{O3}(\hat{D})\fstop
\end{equation}

\section{F-theory Uplifts with  Codimension-3 Singularities}
\label{sec:Codim3Sing}

In F-theory on elliptically fibered Calabi--Yau fourfolds, degenerations of the fiber over codimension-one, -two, and -three loci encode gauge enhancements, charged matter, and Yukawa points, and are usually handled by crepant resolutions. $\mathbb{Q}$-factorial terminal singularities admit no such resolution, yet still carry physical information. On the Jacobian of a genus one fibration, they are the geometric manifestation of a discrete gauge symmetry~\cite{Arras:2016evy, Grassi:2018rva}, while isolated terminal cyclic quotient singularities in codimension-three have been proposed to mark the location of perturbative O3-planes in the Sen limit, and more generally of S-folds~\cite{Garcia-Etxebarria:2015wns}. The latter are non-perturbative generalizations of O3-planes, whose worldvolume theories realize four-dimensional $\mathcal{N}=3$ Superconformal Field Theories (SCFTs). Our aim here is to give a geometric characterization of these singularities in F-theory uplift constructions.

In this section, we rely on \texttt{Solver Agent} for mathematical computations and proofs. We begin in Section~\ref{ssec:o3_terminal} by introducing the Reid--Tai criterion for identifying terminal quotient singularities on a threefold base for F-theory. In Section~\ref{ssec:weierstrass_terminal}, we present the general construction of global Weierstrass models over bases with terminal quotient singularities. Theorem~\ref{thm:weierstrass_proof} describes how these singularities lift to realize O3-planes and S-folds. This theorem is proved by \texttt{Solver Agent} through Prompt~\ref{prmpt:weierstrass_proof}, whose output is reformulated and presented in Result~\ref{res:weierstrass_proof}. Finally, in Section~\ref{sec:ftheory_corrections_2}, we introduce stringy invariants, which incorporate corrections to topological quantities such as Hodge numbers and the Euler characteristic. In the setting of Theorem~\ref{thm:weierstrass_proof}, we derive through Prompts~\ref{prmpt:StringyEuler1},~\ref{prmpt:StringyEuler2}, and~\ref{prmpt:StringyEuler3} the correction formulas presented in Results~\ref{res:StringyEuler2} and~\ref{res:StringyEuler3}, which we interpret in terms of localized D3-brane charges associated with O3-planes and S-folds. Through Prompt~\ref{prmpt:StringyEuler4}, we also prove Jefferson--Kim's conjecture~\cite{Jefferson:2022ssj} in the general setting of Theorem~\ref{thm:weierstrass_proof} and present in Result~\ref{res:StringyEuler4} a refinement incorporating stringy corrections from isolated terminal quotient singularities.

\subsection{Terminal Cyclic  Quotient Singularities on the Base}
\label{ssec:o3_terminal}

 We now fix the geometric setting. Our starting point is a $\mathbb{Q}$-factorial threefold $B_3$ with terminal cyclic quotient singularities, which serves as the base of the F-theory uplift to an elliptically fibered Calabi--Yau fourfold~\cite{Grassi:2026ept,Taylor:2025gnp}. The Reid--Tai criterion provides the local test that singles out precisely these singularities, and hence the candidate O3-plane and S-fold loci~\cite{Garcia-Etxebarria:2015wns, Weigand:2018rez,  Apruzzi:2020pmv, Heckman:2020svr}.

To analyze these geometries, we apply the Reid--Tai criterion, which detects terminal 
cyclic quotient singularities. Any cyclic action on $\mathbb{C}^n$ can be diagonalized, 
so we may take $U=\mathbb{C}^n/G$ with $G=\langle g\rangle$ acting as
\begin{equation}
g:(x_1,\ldots,x_n)\longmapsto
\left(\zeta^{a_1}x_1,\ldots,\zeta^{a_n}x_n\right)\coma
\end{equation}
where $\zeta=e^{2\pi i/k}$, $k=|G|$ and $0\leq a_j<k$. Following the common shorthand, 
we denote the corresponding quotient singularity by $\tfrac{1}{k}(a_1,\ldots,a_n)$. 
The age of $g$ is defined as
\begin{equation}
\operatorname{age}(g):=\frac{1}{k}\left(a_1+\cdots+a_n\right)\coma
\label{eq:age}
\end{equation}
and likewise for any other element $h\in G$, with its exponents again taken in the same 
range. The Reid--Tai criterion then states that $\mathbb{C}^n/G$ is terminal if and only 
if $\operatorname{age}(h)>1$ for every non-identity element $h\in 
G$~\cite{Kollr2008ExercisesIT, Morrison2008TerminalQS}. Assuming in addition that $G$ 
contains no quasi-reflections, the singularity at the image of the origin is isolated 
precisely when all eigenvalues of $g$ are primitive $k$-th roots of 
unity~\cite{Morrison2008TerminalQS}. 

To illustrate how terminal cyclic quotient singularities arise in compact geometries, 
we consider the simple base example discussed in~\cite{Taylor:2025gnp}. The 
\texttt{Solver Agent} prompt below identifies the terminal singularities of a weighted 
projective space.

\begin{saPromptBox}[label={prmpt:terminal_p31112}]
Consider the weighted projective space $\mathbb{P}^3_{[1,1,1,2]}$. Compute the
number of $\mathbb{Q}$-terminal singularities this geometry possesses.
\end{saPromptBox}

\begin{saResultBoxCondensed}[label=res:terminal_p31112, prompt=prmpt:terminal_p31112]The unique singular point is $[0:0:0:1]$, locally of terminal type
$\tfrac12(1,1,1)$. Hence the number of $\mathbb{Q}$-terminal singularities
is $1$.
\end{saResultBoxCondensed}

Let us now discuss in detail the result obtained by \texttt{Solver Agent} in 
\cref{res:terminal_p31112}. The weighted projective space $\mathbb{P}^3_{[1,1,1,2]}$, 
which may be regarded as a $\mathbb{Z}_2$-quotient of $\mathbb{P}^3$, contains a single 
isolated orbifold singularity located at the point $[0:0:0:1]$ in homogeneous 
coordinates $[u_0:u_1:u_2:u_3]$. In the corresponding local chart $U_3$, setting 
$u_3=1$ leaves a residual $\mathbb{Z}_2\subset\mathbb{C}^*$ acting on the affine 
coordinates as
\begin{equation}
g:(u_0,u_1,u_2)\longmapsto(-u_0,-u_1,-u_2)\fstop
\label{eq:propmt1}
\end{equation}
Hence $U_3\cong\mathbb{C}^3/\mathbb{Z}_2$, of type $\tfrac{1}{2}(1,1,1)$, and since the 
origin is the unique fixed point of $g$, the singularity is isolated. As $g$ is the only 
non-identity element of the group and $\operatorname{age}(g)=\tfrac{3}{2}$, the 
Reid--Tai criterion is satisfied, and the origin is an isolated terminal quotient 
singularity. As we will see in Section~\ref{sec:ExampleO3}, this singular point corresponds to an O3-plane fixed point in the corresponding 
Type IIB orientifold picture.

Similarly, we use \texttt{Solver Agent} to search for terminal cyclic quotient singularities in the remaining simplest weighted projective threefolds arising as toric varieties from three-dimensional reflexive polytopes~\cite{Kreuzer:1998vb, Taylor:2025gnp}.

\begin{saPromptBox}[label={prmpt:terminal_weighted}]
Count the number of $\mathbb{Q}$-terminal singularities in the weighted
projective spaces $\mathbb{P}^3_{[w_1,w_2,w_3,w_4]}$, for
\begin{equation}
  \begin{split}
      (w_1,w_2,w_3,w_4)\in &\left\{  
(1,1,2,2),\ (1,1,1,3),\ (1,1,2,3),\ (1,1,2,4),\right.\\
&\,\, \left. (1,1,3,4),\ (1,1,3,5),\ (1,1,4,6)\right\}\fstop
  \end{split}
\end{equation}
\end{saPromptBox}

\begin{saResultBoxCondensed}[label=res:terminal_weighted, prompt=prmpt:terminal_weighted]We determine the number of isolated quotient singularities by examining the fixed loci of the weighted projective group actions. Applying the Reid--Tai criterion to the corresponding local coordinate charts evaluates the terminality of each isolated fixed point. For the given sequence of weight vectors, the total counts of $\mathbb{Q}$-factorial terminal singularities are exactly $(0,0,2,0,1,1,0)$.
\end{saResultBoxCondensed}

\begin{saPromptBox}[label={prmpt:local_groups_weighted}]
For each of the $\mathbb{Q}$-terminal geometries in the list, write the local group corresponding to the $\mathbb{Q}$-terminal singularities.
\end{saPromptBox}

\begin{saResultBoxCondensed}[label=res:local_groups_weighted, prompt=prmpt:local_groups_weighted]We denote a local cyclic quotient action by $\sigma_k:\frac1k(a_1, a_2, a_3) \equiv (k | a_1, a_2, a_3)$. 
Evaluating the residual group actions at the singular loci of the three geometries containing terminal points yields:
\begin{equation*}
\begin{array}{c|c}
  w_1,w_2,w_3,w_4 & \text{Local groups} \\
\hline
1,1,2,3 & (2| 1,1,1);\, (3| 1,1,2) \\
1,1,3,4 & (4| 1,1,3) \\
1,1,3,5 & (3| 1,1,2)
\end{array}
\end{equation*}
\end{saResultBoxCondensed}

We asked \texttt{Solver Agent} in \cref{prmpt:terminal_weighted} to count terminal singularities for a set of weighted projective spaces known to contain local quotient singularities of the form $\mathbb{C}^3 / \mathbb{Z}_k$. It determined the local group action and applied the Reid--Tai criterion to each isolated fixed point. It successfully proved that only $\mathbb{P}^3_{[1,1,2,3]}$, $\mathbb{P}^3_{[1,1,3,4]}$, and $\mathbb{P}^3_{[1,1,3,5]}$ contain terminal quotient singularities, identifying their precise local structure in \cref{prmpt:local_groups_weighted,res:local_groups_weighted,res:terminal_weighted}.

\subsection{Global Weierstrass Model with Gorenstein Singularities}
\label{ssec:weierstrass_terminal}

Having characterized the isolated terminal quotient singularities of the base geometry, 
we now describe how they uplift in the corresponding global Weierstrass model. The upshot 
is that we will be dealing with elliptically fibered Calabi--Yau fourfolds carrying 
$\mathbb{Q}$-factorial terminal quotient singularities over fixed points of cyclic 
groups $\mathbb{Z}_k$ with $k\in\{2,3,4,6\}$, the orders compatible with an automorphism 
of the elliptic fiber. We further show how the presence of these singularities corrects 
the standard formulae for certain topological invariants. 

In the general uplift construction that follows, we assume that the terminal quotient 
singularities do not lie on divisors supporting Kodaira degenerations at codimension 
one. This is motivated by the Type IIB orientifold picture, where O3-planes and O7-planes 
sit at separate loci. While no such perturbative description is available in the presence 
of S-folds, we consider its natural non-perturbative counterpart: locally 
on the base, the terminal quotient singularities are taken to lie away from the 
discriminant locus. 
This is realized concretely in the following Weierstrass model construction: 

\begin{saTheoremBox}[label={thm:weierstrass_proof}]
Let $B_3$ be a projective threefold with terminal quotient singularities that are  
analytically of the form $\mathbb{C}^3/\mathbb{Z}_k$ with $ k \in \{2,3,4,6\}$, and suppose that
\begin{equation}
h^i(B_3,\mathcal{O}_{B_3}) = 0 \coma 0<i \leq 3 \fstop 
\end{equation}
Let $\pi : Y_4 \to B_3$ be the Weierstrass model given by the hypersurface
\begin{equation}
\IY_4:\left\{y^2-\left(x^3+fxz^4+gz^6\right)=0\right\}
\subset
\mathbb{P}_{[2,3,1]}^2\left(\mathcal{E}\right)  \coma 
\end{equation}
where  $\mathcal{E} = \mathcal{O}_{B_3}(-2K_{B_3})\oplus \mathcal{O}_{B_3}(-3K_{B_3}) \oplus\mathcal{O}_{B_3}$,
\begin{equation}
f \in \Gamma (B , \mathcal{O}_{B_3}(-4K_{B_3})) \coma g \in \Gamma (B_3 , \mathcal{O}_{B_3}(-6K_{B_3})) \coma
\end{equation}
and $\Delta_W = 4f^3 + 27g^2$ denotes the discriminant. Here we assume that $-4K_{B_3}$ and $-6K_{B_3}$ are globally generated. 
Then, the following holds:
\begin{enumerate}
 \item $\IY_4$ is a 
 normal projective Calabi--Yau fourfold that is  elliptically fibered over $B_3$.  
 Here we consider the Calabi--Yau criterion $\omega_{\IY_4} \cong \mathcal{O}_{\IY_4}$ and $h^i(\IY_4,\mathcal{O}_{\IY_4}) = 0$ for $0<i<4$. 
\item For generic $(f,g)$ one has
\begin{equation}
  \{\Delta_W=0\} \cap \operatorname{Sing}(B_3)=\varnothing .
\end{equation}
Thus, for every $p\in\operatorname{Sing}(B_3)$ with local uniformizing cover $ q:\mathbb{C}^3\to U_p\cong\mathbb{C}^3/\mathbb{Z}_k,$ 
the fiber $ E_p:=\pi_0^{-1}(0)$ 
of the corresponding local Weierstrass model
$\pi_0:\IY_4^{\mathrm{loc}}\to\mathbb{C}^3$ is a smooth elliptic curve carrying the induced
order-$k$ automorphism. Moreover, $\IY_4$ is smooth over $B^{\mathrm{sm}}_3 := B_3 \backslash \operatorname{Sing}(B_3)$, and
$\operatorname{Sing}(\IY_4)$ is therefore the finite set of images of the points
of $E_p$ with non-trivial stabilizer
\begin{equation} \mathbb{Z}_{\ell}\subseteq\mathbb{Z}_k \,,  \qquad
  \ell\in\{2,3,4,6\}\,. 
\end{equation}
At such a point, $\IY_4$ is analytically of the form 
$  \mathbb{C}^4/\mathbb{Z}_{\ell}$, 
where the generator acts on local coordinates by
\begin{equation}
  z_a\mapsto e^{2\pi i v^a}z_a \coma
  \qquad
  v=\frac{1}{\ell}(1,1,-1,-1)\coma
\end{equation}
with $z_1,z_2,z_3$ coordinates along the base directions and $z_4$ along the
elliptic fiber. 
These singularities are isolated, Gorenstein and terminal. In particular, for
generic $(f,g)$ the fourfold $\IY_4$ is $\mathbb{Q}$-factorial. 
We call \emph{O3-plane points} the singularities analytically of the form
$\mathbb{C}^4/\mathbb{Z}_2$, and \emph{S-fold points} those analytically of the
form $\mathbb{C}^4/\mathbb{Z}_{\ell}$ with
$\ell\in\{3,4,6\}$.

\item Let $\Sigma\subset B^{\mathrm{sm}}_3$ be a smooth divisor and suppose that the
Weierstrass model has a fixed split Kodaira type $\mathsf T\notin\{I_0,I_1,II\}$
along $\Sigma$. Assume moreover that $ \Delta_W\cap\operatorname{Sing}(B_3)=\varnothing$.
Then the isolated terminal quotient singularities lying over $\operatorname{Sing}(B_3)$ are
precisely those described in~2. and are unaffected by the Kodaira degeneration
along $\Sigma$. Over $\Sigma$, the singular locus of $\IY_4$ contains a surface
$S_\Sigma$ lying over $\Sigma$ such that, at a general point of $S_\Sigma$,
the singularity transverse to $\Sigma$ is the ADE surface singularity
associated with $\mathsf T$. These singularities are canonical but not
terminal. Assuming that all higher-codimension enhancement loci remain minimal
and introduce no non-canonical singularities, $\IY_4$ has canonical singularities
and is not terminal.
 
\item Suppose that the singularities supported over $\Sigma$ admit a
projective crepant partial resolution
$ \rho:\hat \IY_4\rightarrow \IY_4 $
which extracts the Cartan divisors associated with $\mathsf T$ and resolves all
non-terminal singularities over $\Sigma$, while remaining an isomorphism near
$\pi^{-1}(\operatorname{Sing}(B_3))$. Then $\hat \IY_4$ is again a $\mathbb Q$-factorial
Calabi--Yau fourfold with terminal singularities, and the isolated terminal
quotient singularities described in~2. are left unchanged.

\end{enumerate}

\end{saTheoremBox}

Proving such a statement using \texttt{Solver Agent} requires converting it to a {\color{saPrompt}{\textbf{Prompt}}}. Asking \texttt{Solver Agent} directly to prove the theorem is generally not the best approach. Indeed, this introduces a significant bias toward the theorem being true: the system might provide a (flawed) proof even if the theorem is false. Instead, it is better to ask more neutrally about the conditions under which the theorem holds—without stating them oneself; this allows for a quick, retrospective check of whether these conditions match those identified by the researcher or provide a refinement thereof.
\begin{saPromptBox}[label={prmpt:weierstrass_proof}]
Let $B_3$ be a complex projective threefold with $\mathbb{Q}$-factorial terminal cyclic quotient singularities in codimension 3. Let
$\pi \colon Y_4 \to B_3$ be the global Weierstrass model over $B_3$. Assume $B_3$ contains a rigid divisor $\Sigma \subset B_3$ that is a component of the discriminant $\Delta(\pi)$, so that the fibration $\pi$ develops a Kodaira degeneration in codimension one over $\Sigma$. Let $\rho \colon \hat{Y}_4 \to Y_4$ be the partial resolution of $Y_4$ that resolves precisely this codimension-one Kodaira degeneration, and set $\hat{\pi} \coloneqq \pi \circ \rho \colon \hat{Y}_4 \to B_3$. Find the conditions on $B_3$ such that the following holds:
\begin{enumerate}
    \item $\pi \colon Y_4 \to B_3$ is an elliptically fibered Calabi--Yau fourfold whose singular locus consists of finitely many $\mathbb{Q}$-factorial terminal cyclic quotient singularities, related to a cyclic group $\mathbb{Z}_k$ with $k \in \{2,3,4,6\}$.
    \item The same holds for $\hat{\pi} \colon \hat{Y}_4 \to B_3$: it is again an elliptically fibered Calabi--Yau fourfold with singularities of exactly the type described in part 1. In particular, $\rho$ is crepant and neither introduces new quotient singularities nor removes the existing ones.
\end{enumerate}
\end{saPromptBox}

The proofs below follow one principle: away from the quotient points, we use the ordinary Weierstrass model, and near a quotient point, we compute on the smooth uniformizing cover, track the character of the finite group, and descend with the resulting invariant geometric data.

\paragraph{The Global Model and the Local Quotient Description.}

Notice that $\mathcal{L} = \mathcal{O}_{B_3}(-K_{B_3})$ is a rank-one reflexive sheaf
that is not invertible at the orbifold points, as $-K_{B_3}$ is
$\mathbb{Q}$-Cartier but not Cartier there. Given that
$B_3^{\mathrm{sm}}$ is smooth, the
restriction $\mathcal{L}|_{B_3^{\mathrm{sm}}}$ is a line bundle, and the weighted
projective bundle above is therefore defined a priori only over
$B_3^{\mathrm{sm}}$.

To extend this description over all of $B_3$, we write
\begin{equation}
\mathcal{L}^{[m]}:=\mathcal{O}_{B_3}(-mK_{B_3})
\end{equation}
for the reflexive powers of $\mathcal{L}$, following the generalized
Weierstrass construction of \cite{NAKAYAMA1988405, Nakayama2002Global}.
Locally, one passes to a finite cover on which $\mathcal{L}$ becomes
invertible, constructs the ordinary Weierstrass model there, and then takes
the corresponding finite quotient. These local quotient models patch together
to give a global Weierstrass model $\pi:\IY_4\to B_3 $.
Since the ordinary local Weierstrass models are projective and projectivity
descends under finite quotients, $\pi$ is projective.

Over $B_3^{\mathrm{sm}}$, this construction reduces to the ordinary Weierstrass
hypersurface in $$\mathbb P_{[2,3,1]}
(\mathcal{L}^2\oplus\mathcal{L}^3 \oplus \mathcal{O}_{B_3})\fstop$$
Near a point $p\in\operatorname{Sing}(B_3)$, we then consider an analytic uniformizing
quotient
\begin{equation}
  q:\mathbb C^3\to U_p\cong\mathbb C^3/\mathbb Z_k\fstop
\end{equation}
On the cover, the reflexive powers of $\mathcal{L}$ become equivariantly
trivial line bundles, and the corresponding local model is the ordinary
equivariant Weierstrass model
\begin{equation}
\pi_0:Y_4^{\mathrm{loc}}\to\mathbb C^3\coma
  Y|_{U_p}\cong Y_4^{\mathrm{loc}}/\mathbb Z_k\fstop
\end{equation}
We shall use this quotient description only to analyze the geometry at the
orbifold points.

Throughout, by an \emph{elliptic fibration} we mean a projective morphism
with a section whose generic fiber is a smooth genus one curve. The induced
quotient morphism need not be flat at an orbifold point, and flatness will not
be used in what follows.

\paragraph{Characters.}
Let $\sigma_k$ generate $\mathbb{Z}_k$, acting on the local smooth cover of the base by
\begin{equation}
  \sigma_k : (u_1,u_2,u_3) \longmapsto
  (\zeta^{a_1}u_1, \zeta^{a_2}u_2, \zeta^{a_3}u_3) \coma
  \zeta = e^{2\pi i/k} \coma
  s \equiv a_1 + a_2 + a_3 \pmod{k} \,.
\end{equation}
The holomorphic three-form $\Omega = \mathrm{d}u_1 \wedge \mathrm{d}u_2 \wedge
\mathrm{d}u_3$ satisfies $\sigma_k^{*}\Omega = \zeta^{s}\Omega$, so the local
anti-canonical generator $\Omega^{-1}$ has character $-s$. Fiber
coordinates carry the character opposite to that of the generator of the line
bundle on which they are coordinates~\cite{Schuett:2009pej}. Since $x$ and $y$ are the fiber coordinates of $\mathcal{L}^{2}$ and
$\mathcal{L}^{3}$,
\begin{equation}
  x \mapsto \zeta^{2s}x \coma
  y \mapsto \zeta^{3s}y \coma
  z \mapsto z \,.
\end{equation}
Writing the pulled-back Weierstrass coefficients as
$q^{*}f = \widetilde f \, \Omega^{-4}$ and $q^{*}g = \widetilde g \,
\Omega^{-6}$, with $\widetilde f, \widetilde g$ holomorphic on $\mathbb{C}^3$,
the pullback of a section downstairs is $\mathbb{Z}_k$-invariant, which in turn gives
\begin{equation}
  \widetilde f(\sigma_k u) = \zeta^{4s} \widetilde f(u) \coma
  \widetilde g(\sigma_k u) = \zeta^{6s} \widetilde g(u) \,.
\end{equation}
Every monomial of the Weierstrass polynomial therefore has character $6s$. Thus,  the hypersurface 
equation is semi-invariant and its zero locus is $\mathbb{Z}_k$-invariant.

One piece of classification input is used in the following. After choosing a generator and permuting coordinates, a
three-dimensional terminal cyclic quotient singularity may be written as
\begin{equation}
  \frac1k(1,a,k-a)\coma
   \gcd(a,k)=1\fstop
\end{equation}
For this choice,
\begin{equation}
  s\equiv1+a+(k-a)\equiv1\pmod{k}\fstop
  \label{eq:s-equiv-one}
\end{equation}
For $k\in\{2,3,4,6\}$, one has, moreover, $a\equiv\pm1\pmod{k}$; thus, the 
quotient is equivalent, after permuting coordinates, to
$\frac1k(1,1,-1)$.

\paragraph{The Quotient Points Lie off the Discriminant.}
For a given fixed $p \in \operatorname{Sing}(B_3)$, by \eqref{eq:s-equiv-one}, the local lifts
$\widetilde f, \widetilde g$ have characters $4$ and $6$ modulo $k$, so a
non-zero constant term is allowed precisely when the corresponding character is
trivial:
\begin{center}
\begin{tabular}{ccl}
\toprule
$k$ & trivial character & consequence \\
\midrule
$2$ & $\widetilde f$ and $\widetilde g$ & $f(p), g(p)$ may both be non-zero \\
$3$ & $\widetilde g$ & $g(p) \neq 0$ suffices for $\Delta_W(p) \neq 0$ \\
$4$ & $\widetilde f$ & $f(p) \neq 0$ suffices for $\Delta_W(p) \neq 0$ \\
$6$ & $\widetilde g$ & $g(p) \neq 0$ suffices for $\Delta_W(p) \neq 0$ \\
\bottomrule
\end{tabular}
\end{center}
When the character is trivial, the corresponding reflexive power is locally
free at $p$, and global generation allows one to choose a section whose local
lift has a non-zero constant term. Thus, for each
$p\in\operatorname{Sing}(B_3)$, the condition $\Delta_W(p)\neq0$ can be
satisfied. Since $\operatorname{Sing}(B_3)$ is finite, a general pair $(f,g)$
satisfies these non-vanishing conditions simultaneously, and hence
\begin{equation}
\{\Delta_W=0\}\cap\operatorname{Sing}(B_3)=\varnothing \fstop
\end{equation}
The list $\{2,3,4,6\}$ is not accidental. Since $\sigma_k$ fixes the zero
section, whenever the fiber over the local smooth cover is elliptic, it induces
an automorphism fixing its origin, whose possible orders are
$1,2,3,4$ and $6$. Equivalently, for any other $k$, both constant terms are
forbidden by equivariance, forcing $\Delta_W(p)=0$.

\begin{saResultBoxManipulated}[label=res:weierstrass_proof, prompt=prmpt:weierstrass_proof]

\begin{enumerate}[label=\textbf{1.\alph*.}, wide=0pt, topsep=1ex]

\item \textbf{Normality and the Elliptic Generic Fiber.} 
        
We first recall a standard property of ordinary Weierstrass models over a
smooth base. Let $W \to V$ be an ordinary Weierstrass model with $V$ a smooth
and irreducible variety, and suppose that its discriminant locus is a proper subset of
$V$. By~\cite[Lemma~5.1.1]{Nakayama2002Global}, $W \to V$ is a flat
Gorenstein morphism, is smooth away from the discriminant locus, and every
fiber is an irreducible plane cubic, either smooth, nodal, or cuspidal. Since
$V$ is smooth, the total space $W$ is Gorenstein and hence
Cohen--Macaulay.

If the discriminant locus is empty, then $W$ is smooth. Otherwise, it is a
divisor in $V$. Since $W \to V$ is flat, a point at which the fiber is smooth
is also a smooth point of the morphism, and hence a smooth point of $W$.
Therefore, $\operatorname{Sing}(W)$ is contained in the locus of singular
points of the fibers. As every singular Weierstrass cubic has a unique
singular point, this locus is finite over the discriminant locus, and hence
$ \dim \operatorname{Sing}(W)\leq \dim V - 1 = \dim W - 2 $. 
Thus, $W$ is regular in codimension one, while the Cohen--Macaulay property
implies Serre's condition $S_2$~\cite[Tag~0342]{StacksProject}. Serre's criterion~\cite[Tag~031S]{StacksProject} therefore gives that $W$
is normal.

Applying this with $V=B_3^{\mathrm{sm}}$ shows that $Y_4$ is normal away from the
orbifold points, while applying it with $V=\mathbb C^3$ shows that the local
model $Y_4^{\mathrm{loc}}$ is normal and Cohen--Macaulay. Since
\begin{equation}
  Y_4|_{U_p}\cong Y_4^{\mathrm{loc}}/\mathbb Z_k
\end{equation}
and finite quotients of normal varieties are normal, $Y_4$ is normal also at
the orbifold points. 

Setting $z=0$ in the Weierstrass equation gives $y^2=x^3$, which determines
the unique point $[1{:}1{:}0]\in\mathbb P^2_{[2,3,1]}$. This point is fixed by the
induced $\mathbb Z_k$-action. Indeed, using the weighted projective
identification
\begin{equation}
  [x{:}y{:}z]
  =
  [\lambda^2x{:}\lambda^3y{:}\lambda z] \coma
\end{equation}
and choosing $\lambda=\zeta^{-s}$, one obtains
\begin{equation}
  [\zeta^{2s}{:}\zeta^{3s}{:}0]
  =
  [1{:}1{:}0] \fstop
\end{equation}
Hence, the zero section of the local model is equivariant and descends to
\begin{equation}
  S_0=\{z=0\}\cong B_3 \fstop
\end{equation}

Finally, by assumption, the discriminant locus is a proper subset of $B_3$.
The generic point of $B_3$ therefore lies outside it, so the generic fiber is
a smooth genus one curve. Thus $\pi:Y_4\to B_3$ is an elliptic fibration in the
sense fixed above.

\item \textbf{$\omega_{Y_4}\cong\mathcal{O}_{Y_4}$.} 

Over $B_3^{\mathrm{sm}}$, ordinary adjunction for the Weierstrass model gives
\begin{equation}
  \omega_{Y_4}
  \cong
  \pi^*(\omega_B\otimes\mathcal{L})
  \cong
  \mathcal{O}_{Y_4}\coma
\end{equation}
since $\mathcal{L}=\mathcal{O}_{B_3}(-K_{B_3})$. 
It remains to check that this trivialization extends across the orbifold
points. Near a given $p\in\operatorname{Sing}(B_3)$, we work on the local smooth cover,
where the ordinary Weierstrass model
$\pi_0:Y_4^{\mathrm{loc}}\to\mathbb {C}^3
$
has a canonical generator
\begin{equation}
\Omega_{Y_4^{\mathrm{loc}}}
  =
\Omega_{}\wedge\frac{dx}{y} \fstop
\end{equation}
Under the generator $\sigma_k$ of $\mathbb Z_k$, we have 
\begin{equation}
\sigma_k^*\Omega=\zeta^s\Omega\coma 
\sigma_k^*\!\left(\frac{dx}{y}\right)
  =
  \zeta^{2s-3s}\frac{dx}{y}
  =
  \zeta^{-s}\frac{dx}{y}.
\end{equation}
Hence, the two characters cancel 
\begin{equation}
\sigma_k^*\Omega_{Y_4^{\mathrm{loc}}}
  =
\Omega_{Y_4^{\mathrm{loc}}} \fstop
\end{equation}
Thus $\Omega_{Y_4^{\mathrm{loc}}}$ is $\mathbb Z_k$-invariant and descends to a
trivialization of the canonical sheaf on the quotient away from the
higher-codimension fixed locus. Since $Y_4$ is normal and both $\omega_{Y_4}$ and
$\mathcal{O}_{Y_4}$ are rank-one reflexive sheaves, this isomorphism extends uniquely
across that~\cite[Tag~0EBJ]{StacksProject}. Therefore
\begin{equation}
\omega_{Y_4}\cong\mathcal{O}_{Y_4}\fstop
\end{equation}

\item \textbf{Vanishing of $h^i(Y_4,\mathcal{O}_{Y_4})$.} 

Let $\mathcal{L}_0$ denote the line bundle on the local smooth cover
$\mathbb{C}^3$ corresponding to $\mathcal{L}$. For the ordinary Weierstrass model
$\pi_0:Y_4^{\mathrm{loc}}\to\mathbb{C}^3$, the standard direct image calculation gives
\begin{equation}
\pi_{0*}\mathcal{O}_{Y_4^{\mathrm{loc}}}\cong\mathcal{O}_{\mathbb{C}^3} \coma 
R^1\pi_{0*}\mathcal{O}_{Y_4^{\mathrm{loc}}}
\cong\mathcal{L}_0^{-1}
\cong\omega_{\mathbb{C}^3}\fstop
\end{equation}
Here, the first two identifications follow from cohomology and base change, together with Grothendieck--Serre duality. See, for instance~\cite[Section 6.2]{Bartocci:2009tcf}. Since the fibers of $\pi_0$ are
one-dimensional~\cite[Chap.~III, Cor.~11.2]{hartshorne1977algebraic}
\begin{equation}
R^i\pi_{0*}\mathcal{O}_{Y_4^{\mathrm{loc}}}=0\coma
   i\geq2 \fstop
\end{equation}
These identifications are compatible with the $\mathbb{Z}_k$-action. In particular,
the generator $\Omega$ of $\omega_{\mathbb{C}^3}$ has character $+s$. Fiber-wise,
on a smooth elliptic fiber $E$, the differential $dx/y$ generates
$H^0(E,\omega_E)$ and has character $-s$; thus, its Serre dual
$H^1(E,\mathcal{O}_E)$ has character $+s$, in agreement with
$\omega_{\mathbb{C}^3}$.

Let   $Q:Y_4^{\mathrm{loc}}\to Y_4|_{U_p}$ 
be the finite quotient map near $p\in\operatorname{Sing}(B_3)$. Since
$\pi\circ Q=q\circ\pi_0$, the pushforwards along $Q$ and $q$ are exact,
and taking $\mathbb{Z}_k$-invariants is exact over $\mathbb{C}$. Therefore, 
\begin{equation}
R^i\pi_*\mathcal{O}_{Y_4}\big|_{U_p}
  \cong
\left(q_*R^i\pi_{0*}\mathcal{O}_{Y_4^{\mathrm{loc}}}\right)^{\mathbb{Z}_k}\fstop
\end{equation}
For $i=0$, this gives
\begin{equation}
\pi_*\mathcal{O}_{Y_4}\big|_{U_p}
  \cong
\left(q_*\mathcal{O}_{\mathbb{C}^3}\right)^{\mathbb{Z}_k}
  \cong
\mathcal{O}_{U_p}\coma
\end{equation}
while for $i=1$ one obtains
\begin{equation}
R^1\pi_*\mathcal{O}_{Y_4}\big|_{U_p}
  \cong
\left(q_*\omega_{\mathbb{C}^3}\right)^{\mathbb{Z}_k}
  \cong
  \omega_{U_p}\coma
\end{equation}
where the last isomorphism is the standard description of the dualizing
sheaf of a finite quotient. Together with the ordinary Weierstrass
calculation over $B_3^{\mathrm{sm}}$~\cite{Bartocci:2009tcf, hartshorne1977algebraic}, these local identifications give globally
\begin{equation}
\pi_*\mathcal{O}_{Y_4}\cong\mathcal{O}_{B_3}\coma
R^1\pi_*\mathcal{O}_{Y_4}\cong\omega_{B_3}\coma
R^i\pi_*\mathcal{O}_{Y_4}=0 \coma\text{where } i\geq2 \fstop
\label{eq:directIm}
\end{equation}

We now apply the Leray spectral sequence, following the standard calculation
for elliptically fibered Calabi--Yau fourfolds. See, for example~\cite[Appendix~A.1]{Anderson:2015yzz}. Quotient singularities are Cohen--Macaulay, so $\omega_{B_3}$ is a dualizing sheaf and Serre duality gives
\begin{equation}
H^p(B_3,\omega_{B_3})
  \cong
  H^{3-p}(B_3,\mathcal{O}_{B_3})^\vee\coma
\end{equation}
without requiring $\omega_{B_3}$ to be invertible. The hypothesis on the
cohomology of $B_3$ therefore yields
\begin{equation}
H^p(B_3,\omega_{B_3})=0\coma   0\leq p<3
  \coma
H^3(B_3,\omega_{B_3})\cong\mathbb{C}\fstop
\end{equation}
By~(\ref{eq:directIm}), the Leray spectral sequence has only the two rows
\begin{equation}
E_2^{p,0}=H^p(B_3,\mathcal{O}_{B_3})
  \coma
E_2^{p,1}=H^p(B_3,\omega_{B_3})\fstop
\end{equation}
Every term contributing to total degree $1$, $2$, or $3$ already vanishes on
the $E_2$-page. Hence
\begin{equation}
h^i(Y_4,\mathcal{O}_{Y_4})=0\coma 0<i<4\fstop
\end{equation}
Moreover $E_2^{0,0}\cong\mathbb{C}$ and
$E_2^{3,1}\cong\mathbb{C}$ survive for degree reasons, giving
$h^0(Y_4,\mathcal{O}_{Y_4})=h^4(Y_4,\mathcal{O}_{Y_4})=1$. This completes the proof of 1.
    \end{enumerate}

    \begin{enumerate}[label=\textbf{2.\alph*.}, ref={2.\alph*}, wide=0pt, topsep=1ex]
    
        \item\label{itm:th2a} \textbf{Smoothness over $B_3^{\mathrm{sm}}$ for General Coefficients.} 
        
        Over $B_3^{\mathrm{sm}}$, the Weierstrass model is an ordinary Weierstrass
fibration. Since $\mathcal{L}^4$ and $\mathcal{L}^6$ are globally generated,
a general pair $(f,g)$ defines a generic Weierstrass model. Such a generic
Weierstrass model over a smooth base has smooth total space 
\cite[Section 3.2]{Weigand:2018rez}. Hence $Y_4|_{B_3^{\mathrm{sm}}}$ is smooth
for a general pair $(f,g)$.

\item \textbf{The Terminal Quotient Singularities above $\operatorname{Sing}(B_3)$.} 

Let us fix $p\in\operatorname{Sing}(B_3)$ and use the local quotient model introduced
above. By~(\ref{eq:s-equiv-one}), we may choose $U_p$ sufficiently small such that
$U_p\cap\Delta_W=\varnothing$. Hence
$\pi_0:Y_4^{\mathrm{loc}}\to\mathbb{C}^3$ is smooth near the elliptic fiber $
  E_p:=\pi_0^{-1}(0)$. 
The three base directions have weights $(1,1,-1)$, while the character
calculation above gives
\begin{equation}
\sigma_k^\ast:\frac{dx}{y}\mapsto\zeta^{-1}\frac{dx}{y},
\end{equation}
so the tangent direction along $E_p$ has weight $-1$.

Let $w_0\in E_p$ have non-trivial stabilizer
\begin{equation}
G_{w_0}:=\operatorname{Stab}(w_0)\cong\mathbb{Z}_{\ell}
  \coma \ell\mid k\coma
\end{equation}
and set $w=[w_0]\in Y_4$. The local quotient germ is
\begin{equation}
(Y_4,w)\cong(Y_4^{\mathrm{loc}},w_0)/G_{w_0}\fstop
\end{equation}
Since $\pi_0$ is smooth at $w_0$, its differential gives a $G_{w_0}$-equivariant exact sequence
\begin{equation}
  0\longrightarrow T_{w_0}E_p
  \longrightarrow T_{w_0}Y_4^{\mathrm{loc}}
\xrightarrow{\,d\pi_0\,}
  T_0\mathbb{C}^3
  \longrightarrow0\fstop
\end{equation}
Because $G_{w_0}$ is finite, this sequence splits $G_{w_0}$-equivariantly, and hence
\begin{equation}
T_{w_0}Y_4^{\mathrm{loc}}
\cong T_0\mathbb{C}^3\oplus T_{w_0}E_p \fstop
\end{equation}
 If $\sigma_{\ell}$ denotes a generator of $G_{w_0}$,
its four tangent weights are therefore
$(1,1,-1,-1)$ modulo $\ell$.
By analytic linearization of the finite group action, the germ $(Y_4,w)$ is
thus analytically isomorphic to
$\mathbb{C}^4/\mathbb{Z}_{\ell}$, 
where the generator acts by
\begin{equation}
\sigma_\ell  : z_a\mapsto e^{2\pi i v^a}z_a\coma
  v=\frac{1}{\ell}(1,1,-1,-1)\fstop
  \label{eq:Spoint}
\end{equation}
After reordering the coordinates, this representation is the direct sum of
two faithful $\operatorname{SL}(2,\mathbb{C})$ representations. Hence
\cite[Theorem~2.4 (ii)]{morrison1984terminal} implies that the quotient singularity
is isolated, Gorenstein and terminal.

Finally, the fixed-point set of every non-trivial automorphism of $E_p$ is
finite. Since $\operatorname{Sing}(B_3)$ is finite and, by \ref{itm:th2a}, $Y_4$ is smooth
over $B_3^{\mathrm{sm}}$ for generic $(f,g)$, these are all the singular
points of $Y_4$.

\item \textbf{$\mathbb{Q}$-factoriality.} 

Let $w=[w_0]\in Y_4$ be one of the quotient singularities described above, with
stabilizer 
$G_{w_0}\cong\mathbb{Z}_{\ell}$. 
We choose a sufficiently small $G_{w_0}$-invariant neighborhood $V\subset Y_4^{\mathrm{loc}}$ of
$w_0$, so that $V/G_{w_0}$ is a neighborhood of $w$ in $Y_4$, and denote the
quotient map by
$Q_{G_{w_0}}:V\to V/G_{w_0}$. 
Let $D$ be a Weil divisor on $V/G_{w_0}$. Since $V$ is smooth, its pullback is
Cartier and hence locally principal, i.e.
\begin{equation}
Q_{G_{w_0}}^*D=\operatorname{div}_V(h)\coma  h\in K(V)^\times\fstop
\end{equation}
Let us then form the norm
\begin{equation}
N_{G_{w_0}}(h):=\prod_{g\in G_{w_0}}g^*h\coma
\end{equation}
which is $G_{w_0}$-invariant and therefore descends to a rational function
  $\bar{h}\in K(V/G_{w_0})^\times.$
Moreover,
\begin{equation}
\operatorname{div}_V(N_{G_{w_0}}(h))
  =
  \vert G_{w_0}\vert  Q_{G_{w_0}}^*D
  =
Q_{G_{w_0}}^*\operatorname{div}_{V/G_{w_0}}(\bar{h})\fstop
\end{equation}
Comparing coefficients of prime divisors under the finite quotient map,
the vanishing of the pullback implies
\begin{equation}
\operatorname{div}_{V/G_{w_0}}(\overline{h})
  =
  \vert G_{w_0}\vert D\fstop
\end{equation}
Since the left-hand side is a principal divisor, $|G_{w_0}|D$ is Cartier.
Thus every Weil divisor is $\mathbb{Q}$-Cartier, and hence $V/G_{w_0}$ is
$\mathbb{Q}$-factorial. Since the remaining points of $Y_4$ are smooth,
$Y_4$ is $\mathbb{Q}$-factorial.

    \end{enumerate}

\begin{enumerate}[label=\textbf{\arabic*.}, start=3, wide=0pt, topsep=1ex]

\item \textbf{Decoupled singularities.}

Since $\Sigma\subset B_3^{\mathrm{sm}}$ and
$  \{\Delta_W =0\}\cap\operatorname{Sing}(B_3)=\varnothing$,
the Kodaira degeneration along $\Sigma$ is disjoint from the orbifold points
of the base. Hence the local quotient models above
$\operatorname{Sing}(B_3)$ are exactly those analyzed in 2., and their isolated
terminal quotient singularities are unchanged.

Now let $b\in\Sigma$ be a general point and let 
$(t,u,v)$ be local coordinates on $B_3^{\mathrm{sm}}$ centered at $b$ such that
$ \Sigma=\{t=0\}$.
Restricting the Weierstrass model to a smooth curve transverse to $\Sigma$,
obtained by fixing $u$ and $v$, gives a Weierstrass surface over the
$t$-disc with Kodaira fiber of type $\mathsf T$. For
$\mathsf T\notin\{I_0,I_1,II\}$, the corresponding Weierstrass surface has a
rational double point of the ADE type associated with $\mathsf T$.
Thus, as $b$ varies in $\Sigma$, these singular points sweep out a surface
$ S_\Sigma\subset\operatorname{Sing}(Y_4)$
lying over $\Sigma$, and the singularity transverse to $S_\Sigma$ at a
general point is the corresponding ADE surface singularity.

Rational double points are canonical, and their minimal resolutions are
crepant. Consequently, resolving the transverse ADE singularity produces
exceptional divisors over $S_\Sigma$ with discrepancy zero. Hence the
singularities along $S_\Sigma$ are canonical but not terminal.

The singularities above $\operatorname{Sing}(B_3)$ are terminal by 2., and
therefore canonical. Assuming that all higher-codimension enhancement loci
remain minimal and introduce no non-canonical singularities, every
singularity of $Y_4$ is canonical. Since $Y_4$ is singular along $S_\Sigma$ with
zero discrepancy exceptional divisors, it is not terminal.

\item \textbf{Crepant partial resolution and the Calabi--Yau property.}

Since $\rho$ is projective and $Y_4$ is projective, $\hat{Y}_4$ is
projective. Moreover, $\rho$ is crepant, so
\begin{equation}
K_{\hat{Y}_4}=\rho^*K_{Y_4}\fstop
\end{equation}
Since $\omega_{Y_4}\cong\mathcal{O}_{Y_4}$ by 1., it follows that $\omega_{\hat{Y}_4}\cong\mathcal{O}_{\hat{Y}_4}$.

By assumption, $\rho$ resolves all non-terminal singularities supported over
$\Sigma$. Away from this locus it is an isomorphism, except possibly near the
isolated quotient singularities lying over $\operatorname{Sing}(B_3)$, where it
is an isomorphism by hypothesis. These quotient singularities are therefore
unchanged and remain terminal by 2. Thus $\hat{Y}_4$ has only terminal
singularities.

It remains to check the cohomology. By 3., $Y_4$ has canonical singularities,
while $\hat{Y}_4$ has terminal singularities; in particular, both have
rational singularities. For the proper birational morphism $\rho$ this gives
\begin{equation}
\rho_*\mathcal{O}_{\hat{Y}_4}\cong\mathcal{O}_{Y_4}\coma
R^i\rho_*\mathcal{O}_{\hat{Y}_4}=0\coma\text{where }
   i>0\fstop
\end{equation}
The Leray spectral sequence therefore yields
\begin{equation}
H^i(\hat{Y}_4,\mathcal{O}_{\hat{Y}_4})
  \cong
H^i(Y,\mathcal{O}_{Y_4})\coma
\end{equation}
and hence, by 1.,
\begin{equation}
h^i(\hat{Y}_4,\mathcal{O}_{\hat{Y}_4})=0\coma 0<i<4\fstop
\end{equation}
Thus $\hat{Y}_4$ is again a Calabi--Yau fourfold in the sense used above.

Finally, $\mathbb{Q}$-factoriality is local. Over the resolved locus above
$\Sigma$, $\hat{Y}_4$ is smooth and hence factorial. Near
$\pi^{-1}(\operatorname{Sing}(B_3))$, the morphism $\rho$ is an isomorphism,
so $\hat{Y}_4$ has the same $\mathbb{Q}$-factorial quotient singularities
as in 2. Away from these loci, $\hat{Y}_4$ is isomorphic to the smooth
locus of $Y_4$. Hence $\hat{Y}_4$ is $\mathbb{Q}$-factorial.

\end{enumerate}

\end{saResultBoxManipulated}

\subsection{Stringy Corrections from O3-planes and S-folds Singularities}
\label{sec:ftheory_corrections_2}

Having characterized the main properties of global F-theory uplift compactifications containing O3-planes or S-folds, we now turn to the study of topological invariants of physical interest for these geometries. We invoke \texttt{Solver Agent} to prove the Euler characteristic formulae~(\ref{eq:Euler4fold}) and~(\ref{eq:KimJefferson}) discussed in Section~\ref{sec:ftheory_corrections}, as well as their generalizations.

 To describe the Euler characteristic of geometries with local orbifold singularities, one must distinguish among several inequivalent notions appearing in the literature. For our purposes, the relevant invariant is the \emph{stringy Euler characteristic}, or more generally the stringy $E$-polynomial and its associated stringy Hodge numbers. These invariants incorporate the additional contributions localized at the quotient singularities and provide the appropriate framework for extending the Hodge-theoretic relations predicted by mirror symmetry to singular Calabi--Yau varieties
~\cite{batyrev1996strong,Batyrev:1997hj}. In the following, we use this framework to determine the stringy corrections associated with O3-plane and S-fold singularities in our setup, and thereby compare the relevant topological data in the Type IIB orientifold and F-theory descriptions.

We now recall the definition of the stringy $E$-function for a normal $\mathbb{Q}$-Gorenstein complex variety $X$ with at worst log-terminal singularities.  Let $\rho:\widetilde X\to X$ be a log resolution whose exceptional locus is a simple-normal-crossing divisor
$D=\cup_{i\in I}D_i$, where 
$I=\{1,\ldots,r\}$. 
Then, the canonical divisors are related by
\begin{equation}
K_{\widetilde X}
=
\rho^{*}K_X+\sum_{i\in I}a_iD_i\coma
\end{equation}
where the $a_i$ are the discrepancies. Since $X$ has at worst log-terminal singularities, they satisfy $a_i>-1$. 
For every subset $J\subseteq I$, define
\begin{equation}
D_J
:=
\begin{cases} \bigcap_{j\in J} D_j &\text{ if } J\neq \varnothing \\ \widetilde{X} &\text{ if }  J = \varnothing  \end{cases} \coma \text{and} 
\qquad 
D_J^{\circ}
:=
D_J\setminus\bigcup_{i\in I\setminus J}D_i\fstop
\end{equation}
The varieties $D_J^\circ$ are the locally closed strata determined by the exceptional divisor $D$. The stringy $E$-function is then defined by
\begin{equation}
E_{\mathrm{st}}(X;u,v)
:=
\sum_{J\subseteq I}
E(D_J^{\circ};u,v)
\prod_{j\in J}
\frac{uv-1}{(uv)^{a_j+1}-1}\coma
\end{equation}
where $E(D_J^\circ;u,v)$ denotes the Hodge--Deligne polynomial computed using compactly supported cohomology, and the empty product is understood to be equal to one. This expression is independent of the chosen log resolution~\cite{Batyrev:1997hj}.

In the setting considered here, where $X$ is projective and has only Gorenstein quotient singularities, the stringy $E$-function is a polynomial. Its coefficients define the stringy Hodge numbers through
\begin{equation}
E_{\mathrm{st}}(X;u,v)
=
\sum_{p,q}
(-1)^{p+q}h_{\mathrm{st}}^{p,q}(X)u^pv^q
\end{equation}
and the stringy Euler characteristic is given by
\begin{equation}
\chi_{\mathrm{st}}(X)
:=
\lim_{u,v\to 1}E_{\mathrm{st}}(X;u,v)
=
E_{\mathrm{st}}(X;1,1)\fstop
\end{equation}
The final equality holds because $E_{\mathrm{st}}(X;u,v)$ is a polynomial in the present setting~\cite{batyrev1996strong, Batyrev:1997hj}. Finally, if $X$ is smooth and projective, then
\begin{equation}
E_{\mathrm{st}}(X;u,v)=E(X;u,v)\coma 
h_{\mathrm{st}}^{p,q}(X)=h^{p,q}(X)\coma
\end{equation}
so the stringy invariants reduce to the ordinary Hodge-theoretic invariants.

We now specialize to our setting of a $\mathbb{Q}$-factorial elliptically fibered Calabi--Yau fourfold $\pi:\hat Y_4\to B_3$ with only isolated Gorenstein terminal singularities. Let $\hat Y_4^{\mathrm{sm}}:=\hat Y_4\setminus \operatorname{Sing}(\hat{Y}_4)$ and $\mu:\tilde Y_4\to\hat Y_4$ be a log resolution with exceptional divisors $D_i$ and discrepancies $a_i>0$, indexed by $i\in I$. Using the strata defined above, we obtain
\begin{equation}
E_{\mathrm{st}}(\hat Y_4;u,v)
=
E(\hat Y_4^{\mathrm{sm}};u,v)
+
\sum_{s\in \operatorname{Sing}(\hat{Y}_4)}
\sum_{\varnothing\neq J\subseteq I}
E(D_J^\circ\cap\mu^{-1}(s);u,v)
\prod_{j\in J}
\frac{uv-1}{(uv)^{a_j+1}-1}\fstop
\end{equation}
We are interested in the correction to the ordinary Hodge--Deligne polynomial contributed by the singular points. By additivity and the identity $E({s};u,v)=1$, this correction is
\begin{equation}
E_{\mathrm{st}}(\hat Y_4;u,v)-E(\hat Y_4;u,v)
=
\sum_{s}
\left(
\sum_{\varnothing\neq J\subseteq I}
E(D_J^\circ\cap\mu^{-1}(s);u,v)
\prod_{j\in J}
\frac{uv-1}{(uv)^{a_j+1}-1}
-1
\right)\fstop
\label{eq:DiffEuler}
\end{equation}

With this in mind, we now apply \texttt{Solver Agent} to derive the corrections to the tadpole cancellation condition discussed in Section~\ref{sec:ftheory_corrections} that arise from the stringy invariants of the singularities associated with O3-planes and S-fold points:

\begin{saPromptBox}[label={prmpt:StringyEuler1}]

Let $\hat{Y}_4$ be a projective $\mathbb Q$-factorial Calabi--Yau fourfold
equipped with an elliptic fibration
$\pi:\hat{Y}_4 \to B_3$. 
Assume that the only singularities of $\hat{Y}_4$ are isolated Gorenstein
terminal quotient singularities of the form
\begin{equation*}
s \in\operatorname{Sing}(\hat{Y}_4) \coma
(\hat{Y}_4,s)\cong \mathbb C^4/\mathbb Z_{\ell_s} \coma
\ell_s\in\{2,3,4,6\}\fstop
\end{equation*}
A generator of $\mathbb Z_{\ell_s}$ acts by
\begin{equation*}
\sigma_{\ell_s}:z_a\mapsto e^{2\pi i v^a}z_a\coma 
v=\frac{1}{\ell_s}(1,1,-1,-1)\coma 
a=1,\ldots,4\fstop
\end{equation*}

Suppose that $B_3$ is a projective threefold with isolated terminal
quotient singularities of order $k\in\{2,3,4,6\}$ and that every
singular point of $\hat{Y}_4$ lies over $\operatorname{Sing}(B_3)$. For a point
$b\in\operatorname{Sing}(B_3)$ of order $k$, assume that the multiset of
stabilizer orders of the singular points lying over $b$ is
\begin{equation}
\left\{
\ell_s
\;\middle|\;
s\in\operatorname{Sing}(\hat{Y}_4)\,, \pi(s)=b
\right\}_{\mathrm{mult}}
=
\begin{cases}
\{2,2,2,2\}\coma & k=2\coma\\
\{3,3,3\}\coma   & k=3\coma\\
\{2,4,4\}\coma   & k=4\coma\\
\{2,3,6\}\coma   & k=6\fstop
\end{cases}
\label{eq:stabilizer-profiles}
\end{equation}

 Compute the age of every non-identity element of
$\mathbb Z_{\ell_s}$ and hence determine the local contribution of
$s$ to
\begin{equation}
E_{\mathrm{st}}(\hat{Y}_4;u,v)-E(\hat{Y}_4;u,v) \fstop
\end{equation}
Here $E(\hat{Y}_4;u,v)$ is the ordinary Hodge--Deligne polynomial of the
coarse space $\hat{Y}_4$, while the stringy $E$-function is, in the present
setting, the polynomial
\begin{equation*}
E_{\mathrm{st}}(\hat{Y}_4;u,v)
=
\sum_{p,q}(-1)^{p+q}h_{\mathrm{st}}^{p,q}(\hat{Y}_4)u^pv^q\fstop
\end{equation*}
Identify the Hodge bidegree of the corresponding twisted-sector
contributions.

\end{saPromptBox}

\begin{saResultBoxManipulated}
[label=res:StringyEuler1, prompt=prmpt:StringyEuler1]

Although the individual exceptional strata depend on the resolution, their weighted sum (\ref{eq:DiffEuler}) does not. Thus (\ref{eq:DiffEuler})  depends only on each  singularity germ and can be evaluated using
$$
(\hat Y_4,s)\cong
\bigl(\mathbb C^4/G_s,0\bigr)\coma s\in \operatorname{Sing}(\hat{Y}_4)\coma G_s:=\mathbb Z_{\ell_s}\fstop
$$
The Hodge realization of the local McKay correspondence gives~\cite[Theorem 5.1]{DenefLoeserMcKay}
\begin{equation}
F_s(u,v)
:= \sum_{\varnothing\neq J\subseteq I}
E(D_J^\circ\cap\mu^{-1}(s);u,v)
\prod_{j\in J}
\frac{uv-1}{(uv)^{a_j+1}-1}=
\sum_{[g]\in\operatorname{Conj}(G_s)}
(uv)^{\operatorname{age}(g)},
\end{equation}
where the sum runs over conjugacy classes and $\operatorname{age}(g)$ is the sum of the fractional eigenvalue weights of $g$. 
Let $\sigma$ generate $G_s$, acting with weights
$\frac1{\ell_s}(1,1,-1,-1)$. For every nonidentity element
$\sigma^j$, $1\leq j\leq\ell_s-1$, the fractional weights are
\begin{equation}
\left(
\frac{j}{\ell_s},\frac{j}{\ell_s},
1-\frac{j}{\ell_s},1-\frac{j}{\ell_s}
\right),
\end{equation}
and hence $\operatorname{age}(\sigma^j)=2$. 
None of these weights vanishes, so the fixed locus of each nonidentity element is the origin. Since \(G_s\) is abelian, its conjugacy classes are its individual elements. The identity has age zero, giving
\begin{equation}
F_s(u,v)=1+(\ell_s-1)(uv)^2,
\qquad
\boxed{\Delta_s(u,v):=F_s(u,v)-1=(\ell_s-1)u^2v^2.}
\end{equation}

Thus singularities of orders $2,3,4,6$ contribute respectively $1,2,3,5$ copies of $u^2v^2$. Each corresponding point sector starts in Hodge bidegree $(0,0)$; its age-two shift places the contribution in bidegree $(2,2)$, with positive sign. 
Finally, using the assumed stabilizer profiles above a singular base point $b\in \operatorname{Sing}(B_3)$, the total correction is
\begin{equation}
\Delta_b(u,v)
:=
\sum_{\substack{s\in S\\\pi(s)=b}}\Delta_s(u,v)
=
\begin{cases}
4u^2v^2,&k=2,\quad \{\ell_s\}=\{2,2,2,2\},\\
6u^2v^2,&k=3,\quad \{\ell_s\}=\{3,3,3\},\\
7u^2v^2,&k=4,\quad \{\ell_s\}=\{2,4,4\},\\
8u^2v^2,&k=6,\quad \{\ell_s\}=\{2,3,6\}.
\end{cases}
\end{equation}
Therefore, if \(N_k\) denotes the number of order-\(k\) singular points of \(B_3\),
\begin{equation}
\boxed{
E_{\mathrm{st}}(\hat Y_4;u,v)-E(\hat Y_4;u,v)
=
(4N_2+6N_3+7N_4+8N_6)\,u^2v^2.
}
\end{equation}

\end{saResultBoxManipulated}

\begin{saPromptBox}[label={prmpt:StringyEuler2}]

Define the stringy Euler characteristic by
\begin{equation*}
\chi_{\mathrm{st}}(\hat{Y}_4)
:=
E_{\mathrm{st}}(\hat{Y}_4;1,1)
=
\sum_{p,q}(-1)^{p+q}h_{\mathrm{st}}^{p,q}(\hat{Y}_4)\fstop
\end{equation*}
Compare it with the Euler--Satake characteristic
\begin{equation}
\chi_{\mathrm{ES}}(\mathcal{Y}_4)
:=
\int_{\mathcal{Y}_4}c_4(T\mathcal{Y}_4)\coma
\end{equation}
where $\mathcal{Y}_4$ is the canonical smooth Deligne--Mumford stack
associated with $\hat{Y}_4$. Determine a function $\varepsilon(\ell)$ such
that
\begin{equation}
\chi_{\mathrm{st}}(\hat{Y}_4)
=
\int_{\mathcal{Y}_4}c_4(T\mathcal{Y}_4)
+
\sum_{s\in\operatorname{Sing}(\hat{Y}_4)}
\varepsilon(\ell_s)\fstop
\label{eq:correctionGoal}
\end{equation}
\end{saPromptBox}

\begin{saResultBoxManipulated}
[label=res:StringyEuler2, prompt=prmpt:StringyEuler2]

 Setting $u=v=1$ in the result of Prompt~\ref{res:StringyEuler1} gives
$\chi_{\mathrm{st}}(\hat Y_4)
=
\chi(\hat Y_4)
+
\sum_{s\in \operatorname{Sing}(\hat{Y}_4)}(\ell_s-1)$.
Since $\hat Y_4$ is projective, its ordinary and compactly supported Euler characteristics agree. Additivity of the compactly supported Euler characteristic therefore gives
\begin{equation}
\chi(\hat Y_4)
=
\chi_{\mathrm{c}}(\hat Y_4^{\mathrm{sm}})
+
\sum_{s\in \operatorname{Sing}(\hat{Y}_4)}1\coma
\end{equation}
where $\chi_c(\hat{Y}_4^{\mathrm{sm}})$ denotes the compactly supported Euler characteristic of the smooth locus.

We now compare this with the canonical smooth Deligne--Mumford stack
$p:\mathcal Y_4\to\hat Y_4$.
Over $\hat Y_4^{\mathrm{sm}}$, the map $p$ is an isomorphism.
Over each singular point $s\in \operatorname{Sing}(\hat{Y}_4)$, the reduced stacky stratum is the residual gerbe~\cite{Sharpe:2001bs}
\begin{equation}
BG_s=[\operatorname{pt}/G_s]\coma
G_s = \mathbb{Z}_{\ell_s}\fstop
\end{equation}
The Euler--Satake characteristic is additive with respect to this finite stratification and assigns to $BG_s$ the fractional weight
$\chi_{\mathrm{ES}}(BG_s)
=
1/\ell_s$. 
This follows from the standard normalization of proper pushforward in the rational Chow theory of Deligne--Mumford stacks, which divides the fundamental class by the order of the generic stabilizer~\cite[Section 2.1]{AbramovichGraberVistoli}. Hence
\begin{equation}
\chi_{\mathrm{ES}}(\mathcal Y_4)
=
\chi_c(\hat Y_4^{\mathrm{sm}})
+
\sum_{s\in S}\frac{1}{\ell_s}\fstop
\end{equation}
Because $\mathcal Y_4$ is a smooth proper complex Deligne--Mumford stack of complex dimension four, the orbifold Gauss--Bonnet theorem identifies the preceding weighted Euler characteristic with the Chern number~\cite{SatakeGaussBonnet, Behrend:2005wzg}:
\begin{equation}
\chi_{\mathrm{ES}}(\mathcal Y_4)
=
\int_{\mathcal Y_4}c_4(T\mathcal Y_4).
\end{equation}
Comparing the stringy and Euler--Satake expressions, we obtain
\begin{equation}
\boxed{
\chi_{\mathrm{st}}(\hat Y_4)
=
\int_{\mathcal Y_4}c_4(T\mathcal Y_4)
+
\sum_{s\in\operatorname{Sing}(\hat Y_4)}
\left(
\ell_s-\frac{1}{\ell_s}
\right)
}\fstop
\label{eq:stringyTerminal}
\end{equation}
Therefore,
\begin{equation}
\boxed{
\varepsilon(\ell)
=
\ell-\frac{1}{\ell}
}\fstop
\end{equation}

\end{saResultBoxManipulated}

\begin{saPromptBox}[label={prmpt:StringyEuler3}]
Let $b\in\operatorname{Sing}(B_3)$ be a point of order $k$.
Using the stabilizer profiles in~\eqref{eq:stabilizer-profiles},
determine the total correction
\begin{equation}
\delta(k)
:=
\sum_{\substack{s\in\operatorname{Sing}(\hat{Y}_4)\\ \pi(s)=b}}
\varepsilon(\ell_s)\fstop
\end{equation}
Evaluate $\delta(k)$ for $k\in\{2,3,4,6\}$ and express the result as
a uniform function of $k$.

\end{saPromptBox}

\begin{saResultBox}
[label=res:StringyEuler3, prompt=prmpt:StringyEuler3]

From the result in Prompt~\ref{res:StringyEuler2}, each isolated terminal quotient singularity $s$ of order
$\ell_s$ contributes
$\varepsilon(\ell_s)
=
\ell_s-\frac{1}{\ell_s}$.
The stabilizer profiles list, with multiplicity, the orders of the
singular points lying over a fixed base singularity $b$. Their
contributions therefore add as follows:
\begin{equation}
\begin{aligned}
\delta(2)
&=
4\varepsilon(2)
=
4\left(2-\frac{1}{2}\right)
=
6\coma
\\
\delta(3)
&=
3\varepsilon(3)
=
3\left(3-\frac{1}{3}\right)
=
8\coma
\\
\delta(4)
&=
\varepsilon(2)+2\varepsilon(4)
=
\frac{3}{2}
+
2\left(\frac{15}{4}\right)
=
9\coma
\\
\delta(6)
&=
\varepsilon(2)+\varepsilon(3)+\varepsilon(6)
=
\frac{3}{2}
+
\frac{8}{3}
+
\frac{35}{6}
=
10\fstop
\end{aligned}
\end{equation}
For the allowed orders $k\in\{2,3,4,6\}$, these values admit the
uniform expression
\begin{equation}
\boxed{
\delta(k)
=
12\left(1-\frac{1}{k}\right)
}\coma
k\in\{2,3,4,6\}\fstop
\label{eq:delta}
\end{equation}

Using the expression for $\delta(k)$, we may equivalently group
the corrections according to the singular points of the base:
\begin{equation}
\boxed{
\chi_{\mathrm{st}}(\hat Y_4)
=
\int_{\mathcal Y_4} c_4(T\mathcal Y_4)
+
12\sum_{b\in\operatorname{Sing}(B_3)}
\left(1-\frac{1}{k_b}\right)
}\fstop
\end{equation}
Here $k_b\in\{2,3,4,6\}$ denotes the order of the local cyclic group
defining the terminal quotient singularity at
$b\in\operatorname{Sing}(B_3)$.

\end{saResultBox}

The physical meaning of the stringy correction becomes particularly
transparent from the M2/D3-brane tadpole cancellation condition. For the quotient
singularities considered here, the fixed-point contribution to the
curvature induced tadpole is captured by the stringy Euler
characteristic. In the absence of additional discrete torsion, the
tadpole condition therefore takes the form
\begin{equation}
n_{\mathrm{D3}}
+
\frac{1}{2}\int_{\mathcal Y_4}G_4\wedge G_4
=
\frac{\chi_{\mathrm{st}}(\hat Y_4)}{24} = \frac{1}{24}
\int_{\mathcal Y_4}c_4(T\mathcal Y_4)
+
\frac{1}{24}
\sum_{s\in\operatorname{Sing}(\hat Y_4)}
\left(
\ell_s-\frac{1}{\ell_s}
\right),
\end{equation}
where the flux integral is understood in the orbifold or stack sense~\cite{Sharpe:2001bs}.
Consequently, a fourfold singularity located on
$\mathbb{C}^4/\mathbb{Z}_{\ell_s}$ carries the curvature induced
M2-brane charge
\begin{equation}
Q_{\mathrm{M2}}(s)
=
-\frac{1}{24}
\left(
\ell_s-\frac{1}{\ell_s}
\right),
\end{equation}
in agreement with~\cite[Eq.~(3.8)]{Bergman:2009zh}. Summing over the
singular points lying above
$b\in\operatorname{Sing}(B_3)$ gives the geometric D3-brane charge of
the corresponding O3-plane or S-fold point,
\begin{equation}
Q_{\mathrm{D3}}(b)
=
-\frac{\delta(k_b)}{24}
=
-\frac{1}{2}
\left(
1-\frac{1}{k_b}
\right),
\end{equation}
as in~\cite{Apruzzi:2020pmv}. For
$k_b=2$, this reproduces the charge $-1/4$ of an
$\mathrm{O3}^{-}$-plane~\cite{Hanany:2000fq}.

As an independent check for formula~(\ref{eq:delta}), we use the orbifold Euler
characteristic~\cite{Dixon:1985jw}. Let $b\in\operatorname{Sing}(B_3)$
have order $k$ and local group $\mathbb{Z}_k$, and take $b$
away from the discriminant locus, so that $E_b\cong T^2$ is smooth.
The local model 
$(\mathbb C^3\times E_b)/\mathbb{Z}_k$,
with $\mathbb{Z}_k$ acting diagonally, is a chart to which the orbifold formula
applies even though $\hat Y_4$ need not be a global quotient. Working
with compactly supported Euler characteristics, $\chi_c(\mathbb
C^3\times E_b)=0$ since $\chi(E_b)=0$, so the identity pair drops out
and the local gap between the stringy and Euler--Satake
characteristics comes entirely from twisted sectors
\begin{equation}
\delta(k)=\frac{1}{k}\sum_{\substack{g,h\in G_b\\(g,h)\neq(1,1)}}
\chi_c\!\left((\mathbb C^3\times E_b)^{g,h}\right).
\end{equation}
Any nontrivial $\langle g,h\rangle$ fixes only the origin of $\mathbb
C^3$, leaving $\{0\}\times E_b^{\langle g,h\rangle}$. Since elliptic
curve automorphisms of order $2,3,4,6$ have $4,3,2,1$ fixed points,
counting ordered pairs by the subgroup they generate gives
\begin{equation}
\begin{aligned}
\delta(2)&=\tfrac{1}{2}(3\cdot4)=6, &\qquad
\delta(3)&=\tfrac{1}{3}(8\cdot3)=8,\\
\delta(4)&=\tfrac{1}{4}(3\cdot4+12\cdot2)=9, &
\delta(6)&=\tfrac{1}{6}(3\cdot4+8\cdot3+24\cdot1)=10 .
\end{aligned}
\end{equation}
This reproduces the stabilizer profiles
in~\eqref{eq:stabilizer-profiles} and confirms
$\delta(k)=12(1-1/k)$ by fixed points. Additivity over
$b\in\operatorname{Sing}(B_3)$ then recovers the global expression.

Using \texttt{Solver Agent}, we now prove the Jefferson--Kim conjecture~\cite[Eq. (1.3)]{Jefferson:2022ssj} in a broader setting that includes S-folds and establish an analogous result incorporating corrections from stringy invariants:

\begin{saPromptBox}[label={prmpt:StringyEuler4}]
 Let
$p:\mathcal{Y}_4\to \hat{Y}_4$ be the canonical smooth Deligne--Mumford stack
associated with $\hat{Y}_4$. Let $D\subset B_3$ be a divisor such that the
vertical pullback 
$\bar D:=\pi^*D$ 
is a reduced normal divisor in $\hat{Y}_4$, and assume that the induced stack
divisor
$\mathcal{D}:=p^{-1}(\bar D)\subset\mathcal{Y}_4$ 
is smooth. Since $p$ is an isomorphism over $\hat{Y}_4^{\mathrm{sm}} = \hat{Y}_4 \setminus \operatorname{Sing}(\hat{Y}_4)$, the
divisor $\bar D$ is smooth away from
$\bar D\cap\operatorname{Sing}(\hat{Y}_4)$. In each local quotient chart, the
inverse image of $\bar D$ is a smooth invariant hypersurface.

Identify the induced three-dimensional quotient singularities of
$\bar D$ at the points of
$\bar D\cap\operatorname{Sing}(\hat{Y}_4)$ and verify that they are terminal,
and hence klt, so that the stringy Euler characteristic of
$\bar D$ is defined. Consider
\begin{equation}
\chi_{\mathrm{ES}}(\mathcal D)
:=
\int_{\mathcal D}c_3(T\mathcal D)\,,
\qquad
\chi(\bar D)
:=
E(\bar D;1,1)\,,
\qquad
\chi_{\mathrm{st}}(\bar D)
:=
\lim_{u,v\to1}E_{\mathrm{st}}(\bar D;u,v).
\end{equation}

Using the local quotient contributions and the stabilizer profiles
in~\eqref{eq:stabilizer-profiles}, determine the corrections
$\eta_b(D)$ and $\theta_b(D)$ such that
\begin{equation}
\chi_{\mathrm{ES}}(\mathcal D)
=
\chi(\bar D)
+
\sum_{b\in\operatorname{Sing}(B_3)}\eta_b(D),
\end{equation}
and
\begin{equation}
\chi_{\mathrm{ES}}(\mathcal D)
=
\chi_{\mathrm{st}}(\bar D)
+
\sum_{b\in\operatorname{Sing}(B_3)}\theta_b(D)\,.
\end{equation}
Set $\eta_b(D)=\theta_b(D)=0$ when $b\notin D$. For $b\in D$,
evaluate these corrections for
$k_b\in\{2,3,4,6\}$ and express the results uniformly in terms of
$k_b$.
\end{saPromptBox}

\begin{saResultBox}[label=res:StringyEuler4, prompt=prmpt:StringyEuler4]

Let $s\in\bar D\cap\operatorname{Sing}(\hat{Y}_4)$ be a singular point with
stabilizer group $G_s=\mathbb Z_{\ell_s}$. In the local smooth cover
$\mathbb C^4$, the generator acts with weights 
$\frac{1}{\ell_s}(1,1,-1,-1)$.
Since the lift of $\bar D$ is a smooth invariant hypersurface, its
tangent space at the origin is an invariant hyperplane. Restricting
the ambient representation to this hyperplane gives
\begin{equation}
\frac{1}{\ell_s}(1,1,-1)
\qquad\text{or}\qquad
\frac{1}{\ell_s}(1,-1,-1).
\end{equation}
These two representations are equivalent after replacing the
generator by its inverse and permuting the coordinates. We may
therefore write the induced quotient singularity as
\begin{equation}
(\bar D,s )
\cong
\mathbb C^3\big/G_s,
\qquad
\frac{1}{\ell_s}(1,1,\ell_s-1).
\end{equation}
For $\ell_s\in\{2,3,4,6\}$, the possible singularity types are thus
\begin{equation}
\frac{1}{2}(1,1,1),
\qquad
\frac{1}{3}(1,1,2),
\qquad
\frac{1}{4}(1,1,3),
\qquad
\frac{1}{6}(1,1,5).
\end{equation}
For $1\leq j\leq\ell_s-1$, the fractional weights of the element
$\sigma^j$ are 
$
\left(
\frac{j}{\ell_s},
\frac{j}{\ell_s},
1-\frac{j}{\ell_s}
\right)$,
and hence
\begin{equation}
\operatorname{age}(\sigma^j)
=
1+\frac{j}{\ell_s}
>
1.
\end{equation}
None of the fractional weights vanishes, so every nonidentity element
fixes only the origin. The induced quotient singularity is therefore
isolated and terminal by the age criterion. In particular, it is log
terminal, and the stringy Euler characteristic of $\bar D$ is
well-defined. Notice that these three-dimensional quotient singularities are
$\mathbb Q$-Gorenstein of index $\ell_s$, but non-Gorenstein, as
reflected by their fractional ages.

We now compare the three Euler characteristics locally. The coarse
singular point contributes $1$ to $\chi(\bar D)$, whereas the
corresponding residual gerbe contributes $1/\ell_s$ to
$\chi_{\mathrm{ES}}(\mathcal D)$. Therefore,
$\eta_{\mathrm{loc}}(\ell_s)
=
\frac{1}{\ell_s}-1$, where $\eta_{\mathrm{loc}}$ denotes the local contribution of a
quotient point of order $\ell_s$ to
$\chi_{\mathrm{ES}}-\chi$. 
On the other hand, the local McKay correspondence gives
\begin{equation}
\left(
E_{\mathrm{st}}(\bar D;u,v)-E(\bar D;u,v)
\right)_s
=
\sum_{j=1}^{\ell_s-1}
(uv)^{1+j/\ell_s}.
\end{equation}
Setting $u=v=1$ gives $\ell_s-1$. Thus the total local stringy Euler
weight of the point is
$1+(\ell_s-1)=\ell_s$, 
and consequently 
$\theta_{\mathrm{loc}}(\ell_s)
=
\frac{1}{\ell_s}-\ell_s$.

For $b\in D$, summing these local contributions over the stabilizer
profiles in~\eqref{eq:stabilizer-profiles} gives
\begin{equation}
\begin{array}{c|cccc}
k_b & 2 & 3 & 4 & 6 \\ \hline
\eta_b(D)   & -2 & -2 & -2 & -2 \\
\theta_b(D)& -6 & -8 & -9 & -10
\end{array}
\end{equation}
and therefore
\begin{equation}
\boxed{
\eta_b(D)=-2,
\qquad
\theta_b(D)
=
-12\left(1-\frac{1}{k_b}\right)
}
\qquad
\text{for } b\in D.
\end{equation}
For $b\notin D$, both corrections vanish. Equivalently,
\begin{equation}
\eta_b(D)
=
\begin{cases}
-2, & b\in D,\\
0,  & b\notin D,
\end{cases}
\qquad
\theta_b(D)
=
\begin{cases}
-12\left(1-\dfrac{1}{k_b}\right), & b\in D,\\
0, & b\notin D.
\end{cases}
\end{equation}

\end{saResultBox}

\section{F-theory Uplifts for Calabi--Yau Toric Hypersurfaces}
\label{sec:Ftheoryuplift}

In this section, we illustrate through explicit examples both how F-theory uplifts can be realized as toric hypersurfaces and how~\texttt{Solver Agent}  integrates \texttt{CYTools} to perform physically relevant computations in a user-friendly manner. We construct orientifold uplifts as well as toric Calabi--Yau fourfolds containing S-fold points with no perturbative global realization. Together, these two classes of examples provide explicit realizations of the settings described in Theorem~\ref{thm:weierstrass_proof}, and we show that their respective tadpole cancellation conditions receive corrections determined by stringy invariants described in Section~\ref{sec:ftheory_corrections_2}.

We begin in Section~\ref{sec:trilayer1} by introducing our main method for constructing Calabi--Yau threefolds admitting orientifold involutions, called trilayer polytope construction~\cite{Hassfeld:2026rzd},  which takes a three-dimensional polytope as input. In Section~\ref{sec:orientifold-in-genus1}, we show that such a polytope defines a three-dimensional toric variety whose double cover is a Calabi--Yau threefold admitting an orientifold involution; this double cover can also be identified with a bisection of an associated genus one fibered Calabi--Yau fourfold. The proof presented there is restricted to cases without Kodaira degenerations giving rise to gauge enhancement. In Section~\ref{sec:ExampleO3}, we present the simplest trilayer constructions of Type IIB orientifold uplifts. Section~\ref{sec:Ftheory-SfoldUplifts} provides an analogous compact Calabi--Yau fourfold containing both O3-planes and S-fold points, but admitting no global Type IIB orientifold description. In Section~\ref{sec:X18Uplift}, we study an orientifold uplift that gives rise to a non-Higgsable cluster with gauge algebra $\mathfrak{so}(8)$, for which our interpretation in terms of the trilayer polytope construction remains valid. We then provide a detailed implementation of our methods for computing the intersection data of a smooth resolution of this geometry and use this data to study non-abelian gauge fluxes in F-theory. Finally, in Section~\ref{sec:CYExamples-SFoldsGauge} we present a non-trivial example of Theorem~\ref{thm:weierstrass_proof} based on a threefold base whose only singularities are terminal $\mathbb{Z}_4$ quotient singularities. As an uplift of this base, we consider a tuned Weierstrass model with gauge algebra $\mathfrak{su}(6)$ containing both O3-planes and S-fold points.

\subsection{F-theory Uplifts of Orientifolds from Trilayer Polytopes}
\label{sec:trilayer1}

Our aim is to uplift a given Calabi--Yau threefold $\IX_3$ to an elliptically fibered Calabi--Yau fourfold $\pi:\IY_4\to B_3$ by constructing a Weierstrass model over the orientifold quotient base $B_3=\IX_3/\iota$. 

A simple class of Calabi--Yau orientifolds $\IX_3$ can be obtained
through the \textit{trilayer polytope} construction formalized
in~\cite{Hassfeld:2026rzd}. In this construction, $\IX_3$ is realized
as a Calabi--Yau  hypersurface in a four-dimensional ambient toric
variety $\mathcal{A}_{\mathrm{HM}}$, determined by a suitable
triangulation of the lattice polytope
\begin{equation}
  \Delta_{\mathrm{HM}}
=\operatorname{Conv}\bigl\{(v,1),(0,-1)
\mid v\in\operatorname{Vert}(\Delta_3)\bigr\}\coma
\end{equation}
where $\Delta_3$ is a three-dimensional lattice polytope and
$\Delta_{\mathrm{HM}}$ is required to be reflexive. 

A simpler subclass
is obtained by choosing $\Delta_3$ among the 4319 three-dimensional
reflexive polytopes classified by Kreuzer and Skarke~\cite{Kreuzer:1998vb}. For this subclass, the origin is the only lattice point of
$\Delta_{\mathrm{HM}}$ with coordinates of the form $(v_1, v_2, v_3, 0)$. Consequently,
the orientifold quotient $B_3=\IX_3/\iota$ is itself a toric
threefold, and the generic Weierstrass model over $B_3$ has no associated non-Higgsable cluster~\cite[Section~2.6]{Hassfeld:2026rzd}. 
In these cases, the associated F-theory uplift $\IY_4^s$ for the corresponding orientifold  is realized as a complete intersection in the ambient toric sixfold $V_6^s$, as defined in Section~\ref{sec:systematic_F_theory_uplifts}. For these specific models, $B_3$ has been shown to be isomorphic to a toric variety \cite{Hassfeld:2026rzd}, $\PP_{\Delta_3}$, related to the three-dimensional lattice polytope $\Delta_3$ defining the trilayer structure. Thus, the resulting F-theory uplift can be realized through a generic Weierstrass model $\pi : \IY_4 \to  B_3$ as a toric hypersurface, which constitutes a smooth Calabi--Yau fourfold. A large set of examples constructed in this manner can be found in~\cite{Klemm:1996ts}.

We now consider a more general class of trilayer polytopes for which the input three-dimensional lattice polytope $\Delta_3$ need not be reflexive. An example is the non-reflexive base polytope associated with $\mathbb{P}^3_{[1,1,1,2]}$, discussed in~\cite{Taylor:2025gnp}. The corresponding four-dimensional polytope $\Delta_{\mathrm{HM}}$ is reflexive precisely when $0\in\operatorname{int}(\Delta_3)$ and $2\Delta_3^*$ are lattice polytopes. When the origin is the only lattice point of $\Delta_{\mathrm{HM}}$ with vanishing fourth coordinate, the orientifold quotient $B_3=\IX_3/\iota$ is a toric threefold~\cite[Section~2.6]{Hassfeld:2026rzd}. Suitable non-reflexive choices of $\Delta_3$ can yield $\mathbb{Q}$-factorial bases with isolated terminal $\mathbb{Z}_2$ quotient singularities. These types of singularities lift to O3-planes fixed points on the corresponding Calabi--Yau double cover $X_3$.

Let $\Delta_{\mathrm{HM}}$ be a trilayer polytope determined by a three-dimensional polytope $\Delta_3$, and assume that its associated toric base $B_3$ has no non-Higgsable clusters. In the next section, we show that the Calabi--Yau double cover $X_3\to B_3$ is isomorphic to a bisection of an auxiliary genus one fibered Calabi--Yau fourfold over $B_3$. Using this identification and the intersection theory of $B_3$, we express the topological invariants of the trilayer orientifold $X_3$ in terms of data on $B_3$. With this in mind, we construct the F-theory uplift of $X_3$ directly as a toric hypersurface defined by a generic Weierstrass model $\pi:Y_4\to B_3$. We illustrate this construction with two examples in Section~\ref{sec:ExampleO3}.

In Section~\ref{sec:X18Uplift}, we treat an example with a non-Higgsable cluster. There, we consider a Calabi--Yau orientifold $X_3$ whose quotient $B_3=X_3/\iota$ is a non-Fano toric threefold. We also construct its corresponding Weierstrass model as a toric hypersurface.

\subsection{Orientifolds as Bisections of Genus One Fibrations}
\label{sec:orientifold-in-genus1}

We now express the intersection data of the Calabi--Yau orientifolds $\IX_3$, their orientifold quotients $B_3$, and their elliptically fibered Calabi--Yau fourfold uplifts $\IY_4$ in terms of a single input. In our setting, each orientifold trilayer construction is specified by the three-dimensional reflexive polytope $\Delta_3$, or equivalently, by the associated toric variety $\mathbb{P}_{\Delta_3}$. We therefore seek to determine the intersection data of both $\IX_3$ and $\IY_4$ directly from that of $\mathbb{P}_{\Delta_3}$. For $\IY_4$, this is achieved through adjunction formula computations applied to its Weierstrass model realization~\cite{Klemm:1996ts}. An analogous computation for $\IX_3$ can be performed by introducing an auxiliary genus one fibration.

The key step is to engineer an auxiliary genus one fibered Calabi--Yau fourfold $\varpi:\widetilde{\IY}_4\to B_3$ that admits no global section but possesses a bisection $S_0$. This bisection defines a double cover of the base and is identified with the Calabi--Yau orientifold $S_0 \cong \IX_3$. We construct this geometry as a Calabi--Yau toric hypersurface in a toric projective bundle $\mathbb{P}^2_{[1,1,2]}(\mathcal{E})$ over $B_3 \cong \mathbb{P}_{\Delta_3}$, defining the vector bundle as
\begin{equation}
    \mathcal{E} = \mathcal{O}_{B_3}(-K_{B_3}) \oplus \mathcal{O}_{B_3}(-K_{B_3}) \oplus \mathcal{O}_{B_3}(-K_{B_3})\coma
\end{equation}
where $K_{B_3}$ denotes the canonical class of the base $B_3$. Let $[x:y:z]$ denote the homogeneous coordinates of the $\mathbb{P}^2_{[1,1,2]}[4]$-fiber, 
in which the genus one curve is
realized as a quartic hypersurface. Under this bundle choice, the
divisor classes of $x$ and $y$ coincide, $[x]=[y]$, and the bisection
$S_0=\{x=0\}\cap\widetilde{\IY}_4$ has class
$[S_0]=[x]|_{\widetilde{\IY}_4}$~\cite{Klevers:2014bqa}.
Furthermore, the relevant Stanley--Reisner relations, together with
the hypersurface equation, imply that $x$ and $y$ cannot vanish
simultaneously on $\widetilde{\IY}_4$. Consequently,
    $[S_0]^2
    =
    \left.([x]\cdot[y])\right|_{\widetilde{\IY}_4}
    =0$
in the intersection ring of $\widetilde{\IY}_4$.

The intersection data  computation obtained by \texttt{Solver Agent} is shown in \cref{prmpt:ks_data,res:ks_data}. This example shows how to relate the data of a Calabi--Yau threefold $\IX_3$ in the trilayer construction to the data of the base $B_3 \cong \mathbb{P}_{\Delta_3}$ and the auxiliary genus one fibered Calabi--Yau fourfold $\widetilde{\IY}_4$. In particular, we show\footnote{In the GitHub repository \href{https://github.com/alexmininno/orientifold_databases}{\texttt{orientifold\_databases}}, the reader can find a script to verify \eqref{eq:IntB} for all 3d polytopes that give rise to a favorable trilayer Calabi--Yau as a toric hypersurface, together with other scripts to classify orientifold planes for Calabi--Yau in toric ambient space.} 
\begin{equation}
\scalebox{0.96}{$\displaystyle
\begin{aligned}
\int_{\IX_3} k_a \wedge k_b \wedge k_c &= \int_{\widetilde{\IY}_4} J_0 \wedge J_a \wedge J_b \wedge J_c  = 2 \int_{B_3} j_a \wedge j_b \wedge j_c = 2\kappa_{abc}\,, \\
\int_{\IX_3}c_2(T\IX_3)\wedge k_a &=  \int_{\widetilde{\IY}_4} c_2(T\widetilde{\IY}_4) \wedge J_0 \wedge J_a =  2 \int_{B_3} j_a  (c_1^2 (TB_3) + c_2(TB_3)) \,, \\
 \chi(\IX_3) & = \int_{\widetilde{\IY}_4} J_0 \wedge c_3(T\widetilde{\IY}_4) = 2\int_{B_3}  \left(c_3(TB_3)-c_1(TB_3) c_2(TB_3) -2c_1(TB_3)^3  \right)\,,
\end{aligned}$}
\label{eq:IntB}
\end{equation} 
where $k_a$ are the K\"ahler cone generators on $\IX_3$ induced by its embedding into $\widetilde{\IY}_4$, and $B_3= \mathbb{P}_{\Delta_3}$, with $j_a$, $a=1,\ldots,h^{1,1}(B_3)$, being its Kähler cone generators, and $J_0$ is the Kähler cone generator of $[S_0]$. This has been shown using the adjunction formula, assuming that $\IX_3= \widetilde{\IY}_4\cap \{x=0\}$ and $c_1(T\IX_3) = 0$, with $\IX_3$ being a Calabi--Yau manifold. To establish that $\IX_3$ is indeed the Calabi--Yau orientifold associated with the trilayer construction determined by $\Delta_3$, we invoke C.~T.~C. Wall's theorem and explicitly compare our formulas with the corresponding intersection data, which may be obtained, for instance, from the Kreuzer--Skarke database or \texttt{CYTools}. In this way, by embedding the Calabi--Yau orientifold $\IX_3$ into $\widetilde{\IY}_4$, we obtain an alternative F-theory uplift. In the conventional Weierstrass construction, the zero section $S_0 \cong B_3$ is identified with the orientifold quotient and typically develops terminal singularities in codimension-three at the orientifold projection images of the O3-planes. By contrast, when $\IX_3$ plays the role of bisection for $\widetilde{\IY}_4$, such singularities are not manifest in the geometry. Nevertheless, F-theory compactified on $\widetilde{\IY}_4$ describes a four-dimensional $\mathcal{N}=1$ supergravity theory with a discrete $\mathbb{Z}_2$ gauge symmetry~\cite{Mayrhofer:2014laa, Lin:2015qsa}. We leave a detailed investigation of the physics of this alternative uplift to future work.

\begin{saPromptBox}[label={prmpt:ks_data}]
Consider a three-dimensional lattice polytope $\Delta_3$ such that
$0\in\operatorname{int}(\Delta_3)$ and $2\Delta_3^*$ is a lattice
polytope, where $\Delta_3^*$ denotes the polar dual of $\Delta_3$, and
let
\begin{equation*}
B_3=\mathbb{P}_{\Delta_3}
\end{equation*}
be the associated toric variety, with a smooth toric resolution
understood
when necessary.

Now consider the family of smooth genus one fibered Calabi--Yau
fourfolds
\begin{equation*}
\varpi:\widetilde{\IY}_4\to B_3
\end{equation*}
with no section but with a bisection. Assume that $\varpi$ has no
codimension-one Kodaira fibers giving rise to non-Abelian gauge
enhancement.
The fourfolds are realized as hypersurfaces in a
$\mathbb{P}_{[1,1,2]}[4]$-fibered toric ambient space over $B_3$, with
fiber
coordinates $[x:y:z]$ of weights $(1,1,2)$. The fiber ambient space is
to
be understood as the toric fiber used in the $\mathbb{P}_{[1,1,2]}[4]$
construction, with its toric Stanley--Reisner ideal and divisor linear
equivalences.

The projective coordinates $[x:y:z]$ descend to those of the torus fiber
at
the Calabi--Yau locus, and the divisors associated with $x$ and $y$ are
linearly equivalent. The divisor class of $x$ may be identified with the
bisection divisor.

Let $j_a$, $a=1,\ldots,h^{1,1}(B_3)$, denote Kähler cone generators of
the
base $B_3$, and let their pullbacks to $\widetilde{\IY}_4$ be denoted by
\begin{equation*}
J_a=\varpi^{-1}(j_a).
\end{equation*}

Determine the divisor structure, a basis of Kähler cone generators, the
intersection ring, the relevant Chern-class intersection data, and the
Euler
characteristic of $\widetilde{\IY}_4$, expressing all results in terms
of the
topological and intersection data of the base $B_3$.
\end{saPromptBox}

\begin{saResultBox}[label=res:ks_data, prompt=prmpt:ks_data]Let $c_i \equiv c_i(TB_3)$ denote the Chern classes of the base, where we omit explicit pullbacks to $\widetilde{\IY}_4$ for notational clarity. We define the bisection divisor $S_0$ as the restriction of the ambient coordinates
\begin{equation}
S_0 \equiv [x]|_{\widetilde{\IY}_4} = [y]|_{\widetilde{\IY}_4}\coma
\end{equation}
and we retain the notation $J_a = \varpi^*j_a$ for the vertical divisors.

In the standard $\mathbb P_{[1,1,2]}[4]$ toric fibration, the fiber divisor relations are
\begin{equation}
[x]=[y]=S_0\coma \qquad [z]=2S_0+c_1\coma
\end{equation}
and the Calabi--Yau hypersurface class is determined to be
\begin{equation}
[\widetilde{\IY}_4]=4S_0+2c_1=2[z]\fstop
\end{equation}
The toric Stanley--Reisner relation $xyz=0$ yields the relation $S_0^2(2S_0+c_1)=0$ in divisor classes. Restricting this relation to $\widetilde{\IY}_4$ enforces the vanishing of all divisor intersections containing $S_0^2$. We identify the generic divisor and K\"ahler basis as $\{S_0,\varpi^*j_a\}$, which leads to
\begin{equation}
\label{eq:h11_Y4}
h^{1,1}(\widetilde{\IY}_4) = h^{1,1}(B_3) + 1\fstop
\end{equation}

By defining the triple intersection numbers on the base as $\kappa_{abc} \equiv \int_{B_3} j_a \wedge j_b \wedge j_c$, we evaluate the nonzero quadruple intersections on the fourfold exactly:
\begin{equation}
\label{eq:quad_ints_Y4}
\int_{\widetilde{\IY}_4} S_0 \wedge J_a \wedge J_b \wedge J_c = 2\kappa_{abc}\fstop
\end{equation}
All other combinations evaluate to zero, i.e.,
\begin{equation}
\int_{\widetilde{\IY}_4}J_a \wedge J_b \wedge J_c \wedge J_d = 0\coma
\qquad
\int_{\widetilde{\IY}_4}S_0^2 \wedge J_a \wedge J_b = \int_{\widetilde{\IY}_4}S_0^3 \wedge J_a = \int_{\widetilde{\IY}_4}S_0^4 = 0\fstop
\end{equation}
This indicates that the numerical divisor intersection ring is generated by $S_0$ and $J_a$ with the relations $S_0^2=0$ and all products of four vertical base divisors equal to zero.

We apply the adjunction formula from the ambient space $\mathcal{A}$ to compute the Chern classes. Given $c(T\mathcal{A})=c(TB_3)(1+S_0)^2(1+2S_0+c_1)$ and $[\widetilde{\IY}_4]=4S_0+2c_1$, we obtain
\begin{equation}
\scalebox{1}{$\displaystyle
\begin{aligned}
c_2(T\widetilde{\IY}_4) &=  6S_0 c_1 + c_1^2 + c_2\coma\\
c_3(T\widetilde{\IY}_4) &=  c_3  - c_1 c_2 - 2c_1^3 - 14S_0 c_1^2  \coma\\
c_4(T\widetilde{\IY}_4) &=  6S_0 (6c_1^3 +  c_1 c_2) \fstop
\end{aligned}$}
\end{equation}
The projection formula yields the useful Chern intersections:
\begin{align}
\label{eq:c2_ints_Y4}
\int_{\widetilde{\IY}_4}c_2(T\widetilde{\IY}_4) \wedge J_a \wedge J_b &= 12\int_{B_3} c_1 \wedge j_a \wedge j_b\coma\nonumber\\
\int_{\widetilde{\IY}_4}c_2(T\widetilde{\IY}_4) \wedge S \wedge J_a &= 2\int_{B_3}(c_1^2+c_2) \wedge j_a\coma\\
\int_{\widetilde{\IY}_4}c_2(T\widetilde{\IY}_4) \wedge S_0^2 &= 0\coma\nonumber
\end{align}
as well as the third Chern class intersections
\begin{equation}
\label{eq:c3_ints_Y4}
\begin{split}
    \int_{\widetilde{\IY}_4}c_3(T\widetilde{\IY}_4) \wedge J_a &= -28\int_{B_3}c_1^2 \wedge j_a\coma\\
\int_{\widetilde{\IY}_4}c_3(T\widetilde{\IY}_4) \wedge S_0 &= 2\int_{B_3}(c_3-c_1 c_2-2c_1^3)\fstop
\end{split}
\end{equation}
We also record the self-intersection of the second Chern class:
\begin{equation}
\int_{\widetilde{\IY}_4}c_2(T\widetilde{\IY}_4)^2 = 24\int_{B_3}(c_1^3+c_1 c_2)\fstop
\end{equation}

Finally, integrating the top Chern class provides the Euler characteristic of the fourfold:
\begin{equation}
\label{eq:chi_Y4}
\chi(\widetilde{\IY}_4) = \int_{\widetilde{\IY}_4} c_4(T\widetilde{\IY}_4) = \int_{B_3}\left(72c_1^3+12c_1 c_2\right)\fstop
\end{equation}
We evaluate all base integrals on the chosen smooth toric resolution of $B_3=\mathbb P_{\Delta_3}$.
\end{saResultBox}

\subsection{F-theory Uplift Tadpole Cancellation Conditions Examples}
\label{sec:ExampleO3}

\begin{table}[!htp]
\centering
\begin{tabular}{c |c| c| c |c| c| c}
$\IX_3$ & $B_3$ & $\kappa$ & $c_2(T\IX_3) \cdot J$ & $\chi(\IX_3)$   & $n_{\mathrm{sing}}(B_3)$ \\
\hhline{=|=|=|=|=|=}
$\mathbf{X}_8$ & $\mathbb{P}^3$ & 2 & 44 & $-296$   & 0\\
$\mathbf{X}_{10}$ & $\mathbb{P}^3_{[1,1,1,2]}$  & 1& 34 & $-288$  & 1\\
\end{tabular}
\caption{Here $J$ is the K\"ahler class in $\IX_3$, $\kappa = \int_X J^3$, and $n_{\mathrm{sing}}(B_3)$ is the number of terminal quotient singularities associated to the base $B_3$ of a given F-theory uplift for $\IX_3$.}
\label{tab:SimplestModels}
\end{table}

\noindent Following the discussion in Sections~\ref{sec:trilayer1} and~\ref{sec:orientifold-in-genus1}, we consider two of the simplest trilayer constructions, whose orientifold quotient bases are $\mathbb{P}^3$ and $\mathbb{P}^3_{[1,1,1,2]}$. The corresponding three-dimensional polytopes are
\begin{align}
\begin{split}
\Delta_3
&=\operatorname{Conv}\{
(1,0,0),(0,1,0),(0,0,1),(-1,-1,-1)\},\ \\
\Delta_3'
&=\operatorname{Conv}\{
(1,0,0),(0,1,0),(0,0,1),(-1,-1,-2)
\},
\end{split}
\end{align}
respectively. Applying the trilayer construction gives the four-dimensional reflexive polytopes
\begin{align}
\Delta_{\mathrm{HM}}
&=\operatorname{Conv}\bigl\{(v,1),(0,-1)
\mid v\in\operatorname{Vert}(\Delta_3)\bigr\},\\
\Delta_{\mathrm{HM}}'
&=\operatorname{Conv}\bigl\{(v,1),(0,-1)
\mid v\in\operatorname{Vert}(\Delta_3')\bigr\}.
\end{align}
The corresponding Calabi--Yau double covers are realized as anticanonical hypersurfaces through the Batyrev construction. We denote them by $\mathbf{X}_8$ and $\mathbf{X}_{10}$, respectively, since their intersection data match those of the geometries with these labels in Table~1 of~\cite{Alexandrov:2026rra}. We display these data in Table~\ref{tab:SimplestModels}; they also agree with our general formulas in~\eqref{eq:IntB}.
Moreover, the base $\mathbb{P}^3_{[1,1,1,2]}$ has an isolated terminal $\mathbb{Z}_2$ quotient singularity, whereas $\mathbb{P}^3$ is smooth. 
Correspondingly, the orientifold of $\mathbf{X}_{10}$ contains an O3-plane, while that of $\mathbf{X}_8$ contains none~\cite{Collinucci:2008zs}. We now describe the Type IIB orientifolds of these two threefolds in more detail.

\begin{table}[!htp]
\centering
\begin{tabular}{c |c| c| c |c| c| c}
$\IY_4$ & $B_3$ & $ N_{\mathrm{D3}}$ & $\chi(S_{\mathrm{O7}})$ & $\chi(S_{\mathrm{D7}_{\mathrm{WU}}})$  & $N_{\mathrm{O3}}$ & $\chi(Y_4)$ \\
\hhline{=|=|=|=|=|=|=}
$\mathbf{Y}_{24}$ & $\mathbb{P}^3$ & 1944 & 304 & 45440 & 0  & 23328\\
$\mathbf{Y}_{30}$ & $\mathbb{P}^3_{[1,1,1,2]}$  & 1898 & 295 & 44360 & 1 & 22776\\
\end{tabular}
\caption{Relevant tadpole cancellation condition data.}
\label{tab:SimplestModels-Y4}
\end{table}

\paragraph{Example: $\mathbb{P}_{[1,1,1,1,4]}[8]$}

Let us consider the Calabi--Yau threefold $\mathbf{X}_{8} \equiv \mathbb{P}_{[1,1,1,1,4]}[8]$. 
The toric rays of the ambient space are generated by the vertices
\begin{equation}
\begin{blockarray}{crrrrr}
	&&&&&l^1 \\
\begin{block}{c(rrrr|r)}
	v_1 & -1 & -1 & -1 & 1 & 1 \\
	v_2 & 0 & 0 & 0 & -1 & 4 \\
	v_3 & 0 & 0 & 1 & 1 & 1 \\
	v_4 & 0 & 1 & 0 & 1 & 1 \\
	v_5 & 1 & 0 & 0 & 1 & 1 \\
\end{block}
\end{blockarray}\,
\end{equation}
Here we follow the $P\vert Q$ convention of~\cite[Section~7.4]{Hori:2003ic}. The rows of $P$ are the primitive generators $v_i$ of the one-dimensional cones of the toric fan, while the columns $l^a$ of $Q$ generate the lattice $\Lambda$ of integral relations among them, satisfying $\sum_i l_i^a v_i=0$. Each row also labels a chiral field $\Phi_i$ in the GLSM, whose charge under the $a$th $U(1)$ gauge factor is $l_i^a$. In this example, the single column $l^1=(1,4,1,1,1)^{t}$ gives the charges of the five fields.

This Calabi--Yau threefold has Hodge numbers $h^{1,1} = 1$ and $h^{2,1} = 149$. For this space, $x_2=0$ defines a divisor $D_2$ with Hodge numbers $h^{2,0}(D_2) = 35$ and $h^{1,1}(D_2) = 232$. The other prime toric divisors have $h^{2,0} = 3$ and $h^{1,1} = 38$. 

We consider the orientifold involution leaving $x_2 = 0$ fixed. The fixed locus $D_2 = \{x_2 = 0\}$ wraps an O7-plane with Euler characteristic $\chi(S_{O7}) = 304$. Since $h^{2,0}(D_2) \neq 0$, the O7-plane tadpole is canceled by a Whitney umbrella D7-brane configuration with Euler characteristic $\chi(S_{D7_{\text{WU}}}) = 45440$. There are no O3-planes in this configuration, so $N_{O3} = 0$. 
The geometric contribution to the D3-brane tadpole evaluates to 
\begin{equation}
    N_{D3} = \frac{N_{O3}}{2} + \frac{\chi(S_{O7})}{6} + \frac{\chi(S_{D7_{\text{WU}}})}{24} = 0 + \frac{304}{6} + \frac{45440}{24} = 1944\,.
\end{equation}
This matches one-twelfth of the Euler characteristic of the corresponding F-theory uplift, given by the smooth elliptically fibered Calabi--Yau  fourfold $\pi: \mathbf{Y}_{24} \to  \mathbb{P}^3$. This geometry has been discussed in detail, for example, in~\cite{Klemm:1996ts, Klemm:2007in, Collinucci:2008pf,  Cota:2017aal}, where $\mathbf{Y}_{24}\equiv \mathbb{P}_{[1,1,1,1,8,12]}[24]$. Since there are no O3-planes, $\chi_{\mathrm{st}}(\IY_4)$ receives no singular point corrections, yielding $\chi_{\mathrm{st}}(\IY_4) = \int c_4(T\IY_4) = 23328$~\cite{Klemm:1996ts, Klemm:2007in, Collinucci:2008pf,  Cota:2017aal}, which satisfies the relation $\chi_{\mathrm{st}}(\IY_4) / 12 = 1944$.

\paragraph{Example: $\mathbb{P}_{[1,1,1,2,5]}[10]$}

As another relevant example, let us consider the Calabi--Yau threefold $\mathbf{X}_{10} \equiv \mathbb{P}_{[1,1,1,2,5]}[10]$. 
with the toric rays generated by 
\begin{equation}
\begin{blockarray}{crrrrr}
	&&&&&l^1 \\
\begin{block}{c(rrrr|r)}
	v_1 & -1 & -1 & -2 & 1 & 1 \\
	v_2 & 0 & 0 & 0 & -1 & 5 \\
	v_3 & 0 & 0 & 1 & 1 & 2 \\
	v_4 & 0 & 1 & 0 & 1 & 1 \\
	v_5 & 1 & 0 & 0 & 1 & 1 \\
\end{block}
\end{blockarray}\,
\end{equation}
The Hodge numbers are $h^{1,1} = 1$ and $h^{2,1} = 145$. The divisor $D_2 = \{x_2 = 0\}$ has Hodge numbers $h^{2,0}(D_2) = 34$ and $h^{1,1}(D_2) = 225$. 

Under the orientifold involution leaving $x_2 = 0$ fixed (defined by the trivial permutation $\rho = \ID$ and $b_2 = -1$, following the notation of Appendix \ref{sec:systematic_F_theory_uplifts}), 
the fixed locus gives a single O7-plane wrapping the divisor $D_2$ and one O3-plane ($N_{O3} = 1$). 
Because $D_2$ is not rigid, its tadpole is canceled by a Whitney umbrella with $\chi(S_{D7_{\text{WU}}}) = 44360$. 
The total D3-brane tadpole evaluates to
\begin{equation}
    N_{D3} = \frac{N_{O3}}{2} + \frac{\chi(S_{O7})}{6} + \frac{\chi(S_{D7_{\text{WU}}})}{24} = \frac{1}{2} + \frac{295}{6} + \frac{44360}{24} = 1898\,.
    \label{eq:1898}
\end{equation}
This tadpole must match $\chi_{\mathrm{st}}(\IY_4)/12$ for the F-theory fourfold $\IY_4$ built over $B_3 = \mathbb{P}^3_{[1,1,1,2]}$.

\paragraph{Uplift example:} We now proceed to construct the F-theory uplift of $\mathbf{X}_{10}$, namely the elliptic fibration
$\pi:\mathbf{Y}_{30}\to \mathbb{P}_{[1,1,1,2]}$, 
where $\mathbf{Y}_{30}$ is realized as a degree $30$ hypersurface in the weighted projective space $\mathbb{P}_{[1,1,1,2,10,15]}$. Following the methods of~\cite{Cota:2017aal}, this toric hypersurface can be obtained from the following polytope data:
\begin{align}
\begin{blockarray}{crrrrrrrl}
	&&&&&&l^1&l^2 \\
\begin{block}{c(rrrrr|rr)l}
         v_x & 0 & 0 &  0 & 1 & 0 & 2 & 0   \\ 
         v_y & 0&  0 &  0 & 0  &1  &3 & 0  \\ 
         v_z & 0&  0 &  0 & -2 & -3 &1 &-5  \\
            v_0 &-1 &-1&-2 & -2 &-3& 0 & 1 \\ 
	v_1 &  1&  0 &  0&  -2 &-3 & 0 &1 \\
	v_2 & 0 & 1& 0 &-2 & -3&0 &1 \\
	v_3 & 0 & 0& 1 & -2 &-3& 0 & 2 \\
\end{block}
\end{blockarray}\,
\end{align}
Here $x$, $y$, $z$, and $u_i$ are the homogeneous toric coordinates associated with their respective ray generator points $v_x$, $v_y$, $v_z$, and $v_i$; in particular, we identify $x$, $y$, and $z$ with the homogeneous coordinates of the fiber of the Weierstrass model, and $u_i$ with the base coordinates.  

Since $\mathbb{P}^3_{[1,1,1,2]}$ has an isolated terminal quotient singularity, as obtained in Result~\ref{res:terminal_p31112}, the theorem in \cref{prmpt:weierstrass_proof} implies that $\mathbf{Y}_{30}$ likewise has isolated terminal quotient singularities. We indeed encounter three isolated terminal quotient singularities upon intersecting the Calabi--Yau locus on the toric patch $U_\sigma = \mathbb{C}^5/\mathbb{Z}_2$ with $\sigma = \langle v_0, v_1, v_2, v_x, v_y\rangle$, where the $\mathbb{Z}_2$ generator acts on the affine coordinates as
\begin{equation}
    \sigma_2 : (u_0, u_1, u_2, x,y) \mapsto (-u_0, -u_1, -u_2, x,-y) \fstop 
\end{equation}
Its fixed locus is therefore the $x$-line 
$u_0=u_1=u_2=y=0$.
Restricting the Weierstrass equation 
to this locus gives
\begin{equation}
x^3+f_0x+g_0=0\coma
\end{equation}
where $f_0$ and $g_0$ denote the values of the Weierstrass coefficients $f$ and $g$ at
$u_0=u_1=u_2=0$. For a smooth fiber over the singular point of the base, $\Delta_W\neq0$, this cubic has three distinct roots.
It follows that the hypersurface intersects the fixed locus of the
ambient $\mathbb Z_2$ action in three isolated points. Near each such point, $x$ can be eliminated locally by the
Weierstrass equation, so that the residual $\mathbb Z_2$ action on the
hypersurface is
\begin{equation}
(u_0,u_1,u_2,y)
\mapsto
(-u_0,-u_1,-u_2,-y)\fstop
\label{eq:isoY30}
\end{equation}
Consequently, each singularity is locally analytically of type $\frac{1}{2}(1,1,1,1)$ and is, therefore, an isolated Gorenstein terminal quotient singularity.

The zero section contributes one further singular point over $p=[0:0:0:1]$. There the Weierstrass equation reduces to $y^2=x^3$, giving the point represented by $x=y=u_3=1$. The scalings encoded by $l^1,l^2$ preserve this representative when $\lambda^2=\lambda^3=1$ and $\mu^2=1$. Thus, the nontrivial residual action has $\lambda=1$ and $\mu=-1$. Since $u_0,u_1,u_2$ have $\mu$-weight $1$ and $z$ has $\mu$-weight $-5$, it acts on local coordinates of the hypersurface as
\begin{equation}
(u_0,u_1,u_2,z)\mapsto(-u_0,-u_1,-u_2,-z)\fstop
\end{equation}
Hence, this fourth point is an isolated Gorenstein terminal singularity of type $\frac12(1,1,1,1)$.

Using \texttt{CYTools} and the adjunction formula, we obtain that the uncorrected fourth Chern class intersection number is $22770$. Each of the four isolated quotient singularities contributes $2-\tfrac{1}{2}=\tfrac{3}{2}$ to the correction in~\eqref{eq:stringyTerminal}, giving a total of $6$. Equivalently, this is the correction associated with the single $\mathbb Z_2$ terminal quotient singularity of the base, as discussed in the correction formula~\eqref{eq:Euler4fold}. We therefore find
$\chi_{\mathrm{st}}(\mathbf{Y}_{30})=22770+6=22776$,
which perfectly matches the tadpole relation
$\frac{1}{12}\chi_{\mathrm{st}}(\mathbf{Y}_{30})=1898$,
in agreement with the tadpole relation~\eqref{eq:1898}.

This accounts for the O3-plane contribution that is absent from the naive fourfold Euler calculation discussed in~\cite{Collinucci:2008zs}.

\subsection{F-theory Uplift with S-folds Example}
\label{sec:Ftheory-SfoldUplifts}

Let us now discuss the construction of a global F-theory example containing both O3- and S-fold point singularities.
Concretely, we consider the base $B_3=\mathbb{P}^3_{[1,1,2,3]}$, discussed in \cref{prmpt:terminal_weighted,prmpt:local_groups_weighted}, which contains both a $\mathbb{Z}_2$ and a $\mathbb{Z}_3$ terminal quotient singularity.
The corresponding Weierstrass model over this toric threefold is realized as a toric hypersurface Calabi--Yau fourfold determined by the Batyrev construction associated with the polytope given by the convex hull of the following set of lattice points:
\begin{align}
\begin{blockarray}{crrrrrrrl}
	&&&&&&l^1&l^2 \\
\begin{block}{c(rrrrr|rr)l}
         v_x & 0 & 0 &  0 & 1 & 0 & 2 & 0   \\ 
         v_y & 0&  0 &  0 & 0  &1  &3 & 0  \\ 
         v_z & 0&  0 &  0 & -2 & -3 &1 &-7  \\
            v_0 &-1 & -2&-3 & -2 &-3& 0 & 1 \\ 
	v_1 &  1&  0 &  0&  -2 &-3 & 0 &1 \\
	v_2 & 0 & 1& 0 &-2 & -3&0 &2 \\
	v_3 & 0 & 0& 1 & -2 &-3& 0 & 3\\
\end{block}
\end{blockarray}\,
\label{eq:SfoldUplift}
\end{align}
As in the previous example, we denote by $x$, $y$, $z$, and $u_i$ the homogeneous toric coordinates associated with the respective ray generators $v_x$, $v_y$, $v_z$, and $v_i$. 
In particular, we identify $x$, $y$, and $z$ with the homogeneous coordinates of the Weierstrass fiber, 
while the $u_i$ are identified with the coordinates of the base.
The resulting geometry is therefore an elliptically fibered Calabi--Yau fourfold 
$ \pi:\mathbf{Y}_{42} \to \mathbb{P}^3_{[1,1,2,3]}$,
which can be realized as a degree 42 hypersurface in the ambient weighted projective space
$\mathbb{P}_{[1,1,2,3,14,21]}$.
By the theorem in \cref{prmpt:weierstrass_proof}, $\mathbf{Y}_{42}$ is $\mathbb{Q}$-factorial and contains isolated Gorenstein terminal singularities.
In what follows, we describe these singular points more explicitly.

Following \cref{prmpt:terminal_weighted,prmpt:local_groups_weighted}, the terminal singularities in $\mathbb{P}_{[1,1,2,3]}$ are given by the points 
\begin{equation}
p_2 =[0:0:1:0]\coma \quad p_3=[0:0:0:1]
\end{equation}
and are of the type $\frac{1}{2}(1,1,1)$ and $\frac{1}{3}(1,1,2)$, respectively. 
An analysis analogous to that of Section~\ref{sec:ExampleO3} for $\mathbf{Y}_{30}$ reveals three isolated $\mathbb{Z}_2$ terminal singularities located at the intersection of $\mathbf{Y}_{42}$ with the patch
$U_{\sigma}=\mathbb{C}^5/\mathbb{Z}_2$,
where $\sigma=\langle v_0,v_1,v_3,v_x,v_y\rangle$. In other words, these singularities correspond to the F-theory uplift of O3-plane points. Let us now turn to the singularities associated with the $\mathbb{Z}_3$ quotient. 

For the base point $p_3$, the relevant cone containing it is given by $\tau = \langle v_0, v_1, v_2 ,v_x, v_y\rangle$ 
with the associated toric patch $U_\tau = \mathbb{C}^5/\mathbb{Z}_3$, where the $\mathbb{Z}_3$ generator 
acts on the affine coordinates as 
\begin{equation}
\sigma_3 : (u_0,u_1,u_2, x,y) \mapsto (\zeta u_0, \zeta u_1 , \zeta^{2} u_2, \zeta^{2} x, y) \coma \quad \zeta = e^{2\pi i/3} \fstop 
\end{equation}
Its fixed locus is therefore the $y$-line $u_0 = u_1 = u_2 = x = 0$. Restricting the Weierstrass equation to this locus gives
\begin{equation}
y^2 = g_0 \coma
\end{equation}
where $g_0$ denotes the value of the Weierstrass coefficient $g$ at $u_0 = u_1 = u_2=0$. Thus, the hypersurface intersects the fixed locus of the
ambient $\mathbb Z_3$ action in two isolated points. Near each such point, $y$ can be eliminated locally by the
Weierstrass equation, so that the residual $\mathbb Z_3$ action on the hypersurface is
\begin{equation}
\sigma_3: (u_0,u_1,u_2,x)
\mapsto
(\zeta u_0,\zeta u_1,\zeta^{-1} u_2,\zeta^{-1} x)\fstop
\label{eq:isoY42}
\end{equation}
Thus, this gives rise to a pair of S-fold points that are locally analytically of type $\frac{1}{3}(1,1,2,2)$ and therefore correspond to isolated Gorenstein terminal quotient singularities. To our knowledge, this constitutes the first appearance in the literature of S-folds within a compact F-theory geometry.

We also encounter two further isolated terminal quotient singularities at the zero section, arising from residual finite group actions on the patches containing $z=0$. Consider first the $\mathbb Z_3$ singular point of the base $B_3=\mathbb P^3_{[1,1,2,3]}$ and the patch associated with
$\langle v_z,v_x,v_0,v_1,v_2\rangle$,
for which the complementary coordinates $y$ and $u_3$ are non-vanishing. The residual action can be read directly from the Mori vectors $l^1$ and $l^2$ in~(\ref{eq:SfoldUplift}), 
which define the $(\mathbb{C}^\ast)^2$ scaling action
\begin{equation}
(x,y,z,u_0,u_1,u_2,u_3)
\sim
(\lambda^2x,\lambda^3y,\lambda\mu^{-7}z,
\mu u_0,\mu u_1,\mu^2u_2,\mu^3u_3)\coma
\end{equation}
where $(\lambda, \mu) \in (\mathbb{C}^\ast)^2$. 
Fixing $y=u_3=1$, the singular point $p_3$ on the zero section is represented by
\begin{equation}
z=0\coma x=1\coma u_0=u_1=u_2=0\fstop
\end{equation}
Preserving these gauge choices requires
$\lambda^3=\lambda^2=1$, $\mu^3=1$, 
so that $\lambda=1$ and $\mu=\zeta=e^{2\pi i/3}$. The resulting action on the local coordinates is therefore
\begin{equation}
(u_0,u_1,u_2,z)
\mapsto
(\zeta u_0,\zeta u_1,\zeta^{-1}u_2,\zeta^{-1}z)\fstop
\end{equation} 
The residual $\mathbb Z_3$ action therefore acts on the four local coordinates as in~(\ref{eq:Spoint}), and the singularity is locally analytically of type
$\frac{1}{3}(1,1,2,2)$, hence it is an isolated Gorenstein terminal quotient singularity. 
Similarly, over the cone $\langle v_z, v_x, v_0, v_1, v_3\rangle$, 
the zero section point above the $p_2$ singularity of the base is locally of type $\frac{1}{2}(1,1,1,1)$. 

Although this example has no independent tadpole calculation, its polytope data provide a check of the local stringy correction formula. The Batyrev formulas implemented in \texttt{PALP} \cite{Kreuzer:2002uu} give $h^{1,1}_{\mathrm{st}}(Y_4)=2$, $h^{2,1}_{\mathrm{st}}(Y_4)=0$, and $h^{3,1}_{\mathrm{st}}(Y_4)=3462$. The Calabi--Yau fourfold identities then yield
\begin{equation}
h^{2,2}_{\mathrm{st}}(Y_4)=13900\coma
\chi_{\mathrm{st}}(Y_4)
=6(8+h^{1,1}_{\mathrm{st}}-h^{2,1}_{\mathrm{st}}+h^{3,1}_{\mathrm{st}})
=20832.
\end{equation}
Independently, toric adjunction gives $\int_{\mathcal Y_4}c_4(T\mathcal Y_4)=20818$ for the associated smooth orbifold stack. The single $\mathbb Z_2$ singularity on the base lifts to four isolated terminal points on $Y_4$, while the single $\mathbb Z_3$ singularity lifts to three isolated terminal points. Our  formula~\eqref{eq:stringyTerminal} therefore gives
\begin{equation}
\chi_{\mathrm{st}}(Y_4)
=20818
+4\left(2-\frac12\right)
+3\left(3-\frac13\right)
=20818+6+8
=20832\coma
\end{equation}
in agreement with the polytope computation.

\subsection{F-theory Uplift to Non-Higgsable Cluster  Example}
\label{sec:X18Uplift}

As a concrete application, let us revisit the Calabi--Yau threefold $\IX_3 = \mathbb{P}_{[1,1,1,6,9]}[18]$. While this geometry was previously studied in the literature (e.g., \cite{Candelas:1994hw}), its orientifold was recently analyzed from the Type IIB perspective in \cite{Jefferson:2022ssj}. Here, we report its toric data and GLSM following the conventions of our database, which differ from the presentation in \cite{Jefferson:2022ssj} by a permutation of the coordinates. 
The toric rays of the ambient space are generated by the vertices
\begin{equation}
\begin{blockarray}{crrrrrr}
	&&&&&l^1&l^2 \\
\begin{block}{c(rrrr|rr)}
	v_1 & -1 & -6 & -8 & 1 & 1 & 0 \\
	v_2 & 0 & 0 & 0 & -1 & 9 & 3 \\
	v_3 & 0 & 0 & 1 & 1 & 1 & 0 \\
	v_4 & 0 & 1 & 1 & 1 & 6 & 2 \\
	v_5 & 1 & 0 & 1 & 1 & 1 & 0 \\
	v_6 & 0 & -2 & -2 & 1 & 0 & 1 \\
\end{block}
\end{blockarray}\,
\label{eq:trilayerX18}
\end{equation}
This Calabi--Yau threefold has Hodge numbers $h^{1,1} = 2$ and $h^{2,1} = 272$. The Stanley--Reisner ideal for this specific triangulation is $\text{SRI} = \langle x_1 x_3 x_5, x_2 x_4 x_6 \rangle$. The prime toric divisors possess the following Hodge numbers:
\begin{equation}
\begin{array}{|c|c|c|c|c|}
\hline
& h^{0,0} & h^{1,0} & h^{2,0} & h^{1,1} \\\hline
D_1, D_3, D_5 & 1 & 0 & 2 & 30 \\\hline
D_2 & 1 & 0 & 65 & 417 \\\hline
D_4 & 1 & 0 & 28 & 218 \\\hline
D_6 & 1 & 0 & 0 & 1 \\\hline
\end{array}
\end{equation}

We consider the orientifold involution leaving $x_2 = 0$ and $x_6 = 0$ fixed. In the formalism introduced in Appendix \ref{sec:systematic_F_theory_uplifts}, this corresponds to a coordinate reflection defined by a trivial permutation $\rho = \ID$ and $b_2 = -1$ and $b_6 = -1$. 
In this configuration, the fixed locus $D_2 = \{x_2 = 0\}$ wraps an O7-plane that is canceled by a recombined Whitney umbrella D7-brane configuration. In contrast, the fixed locus $D_6 = \{x_6 = 0\}$ is completely rigid ($h^{2,0}(D_6) = 0$ and $h^{1,0}(D_6)=0$), so the associated O7-plane is canceled by an $\mathrm{SO}(8)$ stack of four D7-branes and their four orientifold images on top of it.
The total D7-brane Euler characteristic contributing to the tadpole is exactly the sum of the Whitney umbrella and the $\mathrm{SO}(8)$ stack contributions,
\begin{equation}
    \chi(S_{D7_{\text{tot}}}) = \chi(S_{D7_{\text{WU}}}) + \chi(S_{D7_{\mathrm{SO}(8)}}) = 86040 + 24 = 86064\,.
\end{equation}
The total O7-plane Euler characteristic evaluates to $\chi(S_{O7_{\text{tot}}}) = 552$. Because there are no O3-planes in this configuration ($N_{O3} = 0$), the geometric contribution to the D3-brane tadpole evaluates to 
\begin{equation}
    N_{D3} = \frac{N_{O3}}{2} + \frac{\chi(S_{O7_{\text{tot}}})}{6} + \frac{\chi(S_{D7_{\text{tot}}})}{24} = 3678\,.
    \label{eq:tadpoleX18}
\end{equation}
This means that the resolved F-theory fourfold $\IY_4$ constructed as an elliptic fibration over the orientifold base $B_3 = \IX_3 / \mathbb{Z}_2$, must have $\chi_{\mathrm{st}}(\IY_4) = 44136$. As there are no O3-planes, $\chi_{\mathrm{st}}(\IY_4)$ receives no singular point corrections, thus $\chi_{\mathrm{st}}(\IY_4) = \int c_4(T\IY_4) = 44136$. 

As already pointed out in~\cite{Kim:2022uni,Hassfeld:2026rzd}, the F-theory uplift of this particular orientifold gives rise to an elliptic fibration over a non-Higgsable cluster supporting an $\mathfrak{so}(8)$ gauge algebra. Such a base geometry was classified in~\cite{Morrison:2014lca} and corresponds to the non-Fano threefold
\begin{equation}
B_3 = \mathbb{P}_{\mathbb{P}^2}\left(\mathcal{O} \oplus \mathcal{O}(6)\right)\coma
\label{eq:NHC}
\end{equation}
which is a $\mathbb{P}^1$-fibration over $\mathbb{P}^2$ with a non-trivial twist, analogous to the non-Higgsable cluster in six dimensions  associated with the Hirzebruch surface $\mathbb{F}_4$~\cite{Morrison:2012np}.
Indeed, from the trilayer construction in~(\ref{eq:trilayerX18}), we observe that the toric fan associated with the polytope
\begin{equation}
\Delta_3'
=
\mathrm{Conv}\left(
(-1,-6,-8),
(0,0,1),
(0,1,1),
(1,0,1),
(0,-2,-2)
\right)
\end{equation}
is related, by an appropriate $\mathrm{GL}(3,\mathbb{Z})$ transformation and upon passing to primitive generators of its one-dimensional cones, to the toric fan associated with
\begin{equation}
\Delta_3
=
\mathrm{Conv}\left(
(1,0,0),
(0,1,0),
(-1,-1,-6),
(0,0,1),
(0,0,-1)
\right)\fstop
\label{eq:B3Polytope}
\end{equation}
The latter three-dimensional polytope realizes the toric projective bundle~(\ref{eq:NHC}). Thus, the corresponding F-theory uplift of the orientifold under consideration is given by the Weierstrass model
$\pi:Y_4\to
B_3=
\mathbb{P}_{\mathbb{P}^2}\left(\mathcal{O}\oplus\mathcal{O}(6)\right)$.
In the following, we proceed to explicitly construct this elliptically fibered Calabi--Yau fourfold and its resolution with the aid of \texttt{Solver Agent}. For completeness, we provide detailed documentation of the construction, including the intermediate calculations, in a separate file available \href{https://storage.googleapis.com/solver-agent-sessions/solver-agent-f-theory-uplifts/index.html}{here}.

We now follow the approach of~\cite{DelZotto:2017mee} for the toric hypersurface construction of an elliptically fibered Calabi--Yau fourfold whose base realizes a non-Higgsable cluster in F-theory. To this end, let us introduce the lattice $N\cong\mathbb{Z}^{5}$ and write a lattice point as
\begin{equation}
  v=(a,b\,;c,d,e)\coma
  \label{eq:5dPoint}
\end{equation}
such that $N_F=\{c=d=e=0\}\subset N$ plays the role of the fiber lattice and
\begin{equation}
  \varphi:N \to N_B\cong\mathbb{Z}^{3},
  \qquad (a,b\,;c,d,e)\mapsto (c,d,e)
  \label{eq:ToricProj}
\end{equation}
is the projection to the base lattice. We then consider the reflexive polytope $\Delta\subset N_{\mathbb{R}}$ with the following boundary lattice points:
\begin{align*}
\begin{array}{rclrclrcl}
v_x&= & (1,0\,;0,0,0)\coma &
  v_y & =& (0,1\,;0,0,0)\coma &
    v_z & = & (-2,-3\,;0,0,0)\,, \\  
  u_1& = & (-2,-3\,;1,0,0)\coma  &
  u_2 & = & (-2,-3\,;0,1,0)\coma &
  u_3 & = & (-2,-3\,;-1,-1,-6)\coma\\
  \multicolumn{9}{c}{\begin{array}{rclrcl} u_4& = &(-2,-3\,;0,0, 1)\coma &
      u_5 & = & (-2,-3\,;0,0, -1)\fstop \end{array}}
\end{array}
\end{align*}
With this toric data in place, we proceed to carry out the explicit construction of the elliptic fibration $\pi:Y_4\to B_3$ and the corresponding computations through \texttt{Solver Agent} as follows:

\begin{saPromptBox}[label={prmpt:NHC1}]
Show that $\varphi$ maps $\{u_1,u_2,u_3,u_4, u_5\}$ onto the rays of the fan of
a smooth non-Fano threefold $B_3$ that is given by the projective bundle $\mathbb{P}_{\mathbb{P}^2}(\mathcal{O} \oplus \mathcal{O}(6))$. 
\end{saPromptBox}

\begin{saResultBox}[label=res:NHC1, prompt=prmpt:NHC1]
Under the projection $\varphi:N\rightarrow N_B$, the five rays $u_i$ are mapped to
the ray generators associated to $\Delta_3$ in~(\ref{eq:B3Polytope}). 
These rays generate a complete fan whose maximal cones are
\begin{equation}
\langle r_i,r_j,s_{\pm}\rangle\coma 1\leq i<j\leq 3\coma
\end{equation}
where $r_1,r_2,r_3$ denote the first three rays and $s_{\pm}$ the two vertical rays. Each maximal cone is unimodular, and hence the associated toric threefold $B_3$ is smooth. Modulo the vertical direction, the rays $r_1,r_2,r_3$ reduce to the standard fan of $\mathbb{P}^2$. Moreover, the ray relations are
\begin{equation}
s_+ +s_-=0,
\coma
r_1+r_2+r_3=6s_-\coma
\label{eq:LinearB3}
\end{equation}
which identifies the corresponding Cox quotient with the projective bundle
$
\mathbb{P}_{\mathbb{P}^2}
(\mathcal{O}\oplus\mathcal{O}(6)).
$
Finally, the second primitive relation in~(\ref{eq:LinearB3}) 
has anticanonical degree $3-6=-3$. Equivalently, the invariant curves associated with the walls $\langle r_i,s_-\rangle$ satisfy
$(-K_{B_3})\cdot C=-3$. 
Thus $-K_{B_3}$ is not nef, and in particular not ample. We therefore conclude that the projection $\varphi$ realizes the base as the smooth, complete, non-Fano toric threefold
\begin{equation}
\boxed{
B_3\cong
\mathbb{P}_{\mathbb{P}^2}
\bigl(\mathcal{O}\oplus\mathcal{O}(6)\bigr)
}\fstop
\end{equation}
\end{saResultBox}

\begin{saPromptBox}[label={prmpt:NHC2}]
Using the eight rays $\{v_x,v_y,v_z,u_1, u_2, u_3, u_4, u_5\}$, write down the fan $\Sigma_{W}$ whose
maximal cones are of the form $\sigma_{F}+\sigma_{B_3}$, with $\sigma_{F}$ a two-dimensional cone of
the $\mathbb{P}_{[2,3,1]}^2$ fan and $\sigma_{B_3}$ the lift of a three-dimensional cone of the fan $\Sigma_{B_3}$ associated to $B_3$. Let $\\mathbb{P}_{\\Sigma_W}$ be the toric variety associated to the toric fan $\\Sigma_W$. Show that
\begin{equation}
  \mathbb{P}_{\Sigma_{W}} \cong 
  \mathbb{P}\bigl(K_{B_3}^{-2}\oplus K_{B_3}^{-3} \oplus \mathcal{O}_{B_3}\bigr),
\end{equation}
and that the generic anticanonical hypersurface is the Weierstrass model $\pi:Y_4\to B_3$,
\begin{equation}
  y^2=x^3+f x z^4+gz^6.
\end{equation}
Determine $\operatorname{ord}_{S}(f,g,\Delta_{W})$, where $\Delta_W = 4 f^3 +27g^2$ is the discriminant of the Weierstrass model, and identify the Kodaira fiber over a divisor $S \subset B_3$. Also determine $S$ in terms of toric divisors of $B_3$.
\end{saPromptBox}

\begin{saResultBox}[label=res:NHC2, prompt=prmpt:NHC2]
The fiber rays satisfy
\begin{equation}
2v_x+3v_y+v_z=0\coma
\end{equation}
and hence generate the fan of $\mathbb{P}_{[2,3,1]}^2$. Combining its three two-dimensional cones with the six lifted maximal cones of $\Sigma_{B_3}$ gives the fan $\Sigma_W$ with maximal cones
\begin{equation}
\Sigma_W^{(5)}
=
\left\{
\langle v_a,v_b,u_i,u_j,u_k\rangle
:
ab\in\{xy,yz,zx\}\,,
ijk\in\{124,125,234,235,314,315\}
\right\} \fstop
\end{equation}
The corresponding Cox grading shows that, if $H$ denotes the relative weighted hyperplane class,
\begin{equation}
[x]=2H+2c_1(TB_3),\qquad
[y]=3H+3c_1(TB_3), \qquad
[z]=H,
\end{equation}
so that
$$
\mathbb{P}_{\Sigma_W}
\cong
\mathbb{P}_{[2,3,1]}
\bigl(
 K_{B_3}^{-2}\oplus K_{B_3}^{-3} \oplus  \mathcal{O}
\bigr).
$$
Its generic anticanonical hypersurface is therefore the usual Tate model with coefficients $a_i\in \Gamma(B_3,K_{B_3}^{-i})$, which can be brought to Weierstrass form
$$
y^2=x^3+fxz^4+gz^6,
$$
where $f\in \Gamma (B_3,K_{B_3}^{-4})$ and $g\in \Gamma(B_3,K_{B_3}^{-6})$.

To identify the non-Higgsable divisor, let $D_{r_i}$ and $D_{s_\pm}$ denote the toric divisors of $B_3$ associated with the projected rays
$r_i=\varphi(u_i)$,
$s_+=\varphi(u_{4})$, and $s_- = \varphi(u_5)$. 
We denote by 
$S:=D_{s_-}\subset B_3$ the toric divisor associated with $s_-$. By a slight abuse of notation, from now on, we use the same symbols for divisors and their divisor classes whenever no confusion can arise. The divisor relations are then
\begin{equation}
L:=D_{r_1}=D_{r_2}=D_{r_3},
\qquad
D_{s_+}=S+6L,
\qquad
-K_{B_3}=9L+2S.
\end{equation}
Since the rays $s_+$ and $s_-$ do not span a cone of the fan of $B_3$, one has
$D_{s_+}D_{s_-}=0$. 
It follows that $S\cong\mathbb P^2$ and
$N_{S/B_3}\cong\mathcal O_{\mathbb P^2}(-6)$. 
The toric monomial structure of the sections of $-4K_{B_3}$ and $-6K_{B_3}$ forces
\begin{equation}
\operatorname{ord}_S(f,g,\Delta_W)=(2,3,6)\fstop
\end{equation}
Hence the generic fiber over $S$ is of Kodaira type $I_0^*$. Since the leading coefficients of $f$ and $g$ restrict to constants on $S$, the associated monodromy cubic has three distinct constant roots for generic complex structure. The $I_0^*$ fiber is therefore split, corresponding to the non-Higgsable gauge algebra
\begin{equation}
\boxed{\mathfrak{g} = \mathfrak{so}(8)}\fstop
\end{equation}

\end{saResultBox}

\begin{saPromptBox}[label={prmpt:NHC3}]
Show that, apart from the seven points listed above, $\Delta$ contains further lattice
points that do not lie in the interior of a facet, and identify them. Express each as a positive
integer combination of ray points at $\partial \Delta$, and interpret the two corresponding star
subdivisions as successive weighted blow-ups over $S$. Read off the multiplicity of the associated
fiber components from $\varphi$.
\end{saPromptBox}

\begin{saResultBox}[label=res:NHC3, prompt=prmpt:NHC3]
Besides the originally displayed boundary points, the polytope $\Delta$ contains precisely two further lattice points that do not lie in the relative interior of a facet,
\begin{equation}
p_A=(-2,-3;0,0,-2),
\qquad
p_B=(-1,-1;0,0,-1).
\end{equation}
The remaining boundary lattice points lie in the relative interiors of facets and are therefore not relevant for the present identification of the additional non-facet-interior points. Both $p_A$ and $p_B$ lie in the cone $\langle v_x,v_y,u_5\rangle$ and satisfy
\begin{equation}
p_A=2v_x+3v_y+2u_5,
\qquad
2p_B=p_A+v_y.
\end{equation}
The first star subdivision inserts $p_A$ in the interior of $\langle v_x,v_y,u_5\rangle$ and corresponds to a weighted blow-up of the toric locus $V(x,y,u_5)$ with weights $(2,3,2)$. After this subdivision, $p_B$ is the primitive midpoint of the edge $\langle p_A,v_y\rangle$, so that a second star subdivision inserts $p_B$ along this edge and corresponds to a further index-two weighted blow-up. In the following, we denote by $D_v$ the toric divisor associated with a ray generator $v$ of the corresponding toric fan. Both refinements are supported over the toric divisor $D_{u_5}$, which projects onto the divisor $S\subset B_3$.
The multiplicities of the corresponding fiber components follow directly from the projection,
\begin{equation}
\varphi(u_5)=s_-,
\qquad
\varphi(p_A)=2s_-,
\qquad
\varphi(p_B)=s_-,
\end{equation}
and hence
\begin{equation}
\varphi^*S
=
D_{u_5}+2D_{p_A}+D_{p_B}+\cdots .
\end{equation}
Thus, the exceptional divisors associated with $p_A$ and $p_B$ appear in the fiber over $S$ with multiplicities $2$ and $1$, respectively.

\end{saResultBox}

\begin{saPromptBox}[label={prmpt:NHC4}]
Determine a fine, regular, star triangulation $\hat{\mathcal{T}}$ of $\Delta$ that is compatible with the
fibration, i.e.\ such that every cone of $\Sigma_{\hat{\mathcal{T}}}$ is mapped by $\varphi$ into a cone of
$\Sigma_{B_3}$. Verify that the resulting hypersurface $\hat{Y}_4$ is smooth and that
$\rho:\hat{Y}_4\to Y_4$ is a crepant resolution. Show that the
fiber over a generic point of $S$ has five components.  
Draw the resulting
dual graph, identify the affine Dynkin type, and read off the gauge algebra $\mathfrak{g}$.
\end{saPromptBox}

\begin{saResultBox}[label=res:NHC4, prompt=prmpt:NHC4]

Including the additional boundary lattice points $p_A,p_B$, together with the four boundary lattice points lying in the relative interiors of facets
\begin{equation}
\begin{array}{rclrcl}
q_1 & = & (-1,-2;0,0,-1)\coma & 
q_2 & = & (-1,-2;0,0,0)\coma  \\
q_3 & = & (-1,-1;0,0,0)\coma &
q_4 & = & (0,-1;0,0,0)\coma 
\end{array}
\end{equation}
 one finds a fine, regular, star triangulation $\hat{\mathcal T}$ of $\Delta$ with $54$ maximal cones. The triangulation is compatible with the elliptic fibration in the sense that every cone of $\Sigma_{\hat{\mathcal T}}$ is mapped by $\varphi$ into a cone of $\Sigma_{B_3}$. Moreover, all maximal cones are unimodular, so that the associated toric ambient space is smooth, and regularity of $\hat{\mathcal T}$ ensures that the induced birational toric morphism is projective. The proper transform $\hat{Y}_4$ of the generic anticanonical hypersurface is smooth. For each exceptional ray, the vanishing order of the anticanonical hypersurface precisely cancels the corresponding ambient discrepancy, and hence
$
K_{\hat{Y}_4}=\rho^*K_{Y_4}$.
Thus,
$\rho:\hat{Y}_4 \to Y_4$
defines a projective crepant resolution.

Over a generic point $\eta_S$ of the divisor $S\subset B_3$, the intersections of $D_{u_5}$ and $D_{p_A}$ with $\hat{Y}_4$ are irreducible, while $D_{p_B}\cap\hat{Y}_4$ splits into three irreducible fiber components $C_i$, with $1\leq i \leq3$. Denoting these components by
\begin{equation}
U:=D_{u_5}\cap\hat{Y}_4\coma
A:=D_{p_A}\cap\hat{Y}_4\coma
D_{p_B}\cap\hat{Y}_4=C_1+C_2+C_3\coma
\end{equation}
the generic fiber takes the form
\begin{equation}
\hat{\pi}^{-1}(\eta_S)
=
U+2A+C_1+C_2+C_3\fstop
\end{equation}
The remaining divisor $D_{q_1}$ appearing in the toric pullback of $S$ does not intersect $\hat{Y}_4$. The only non-trivial intersections among the five fiber components are
\begin{equation}
U\cdot A=1\coma
A\cdot C_i=1\coma  1\leq i\leq 3\coma
\end{equation}
so that their dual graph is
\begin{equation}
\begin{tikzpicture}[baseline=0pt,scale = 1]
    \node (A) {$A$};
    \node (U) [above right=4mm of A] {$U$};
    \node (C1) [above left=4mm of A] {$C_1$};
    \node (C2) [below left=4mm of A] {$C_2$};
    \node (C3) [below right=4mm of A] {$C_3$};
    \draw (A) -- (U);
    \draw (A) -- (C1);
    \draw (A) -- (C2);
    \draw (A) -- (C3);
\end{tikzpicture}
\end{equation}
with multiplicities $(1,1,2,1,1)$ corresponding to the nodes $(C_1, C_2, A, C_3, U)$. This is the affine Dynkin diagram $\widetilde D_4$. Since the corresponding $I_0^*$ fiber is split, the non-Higgsable gauge algebra supported over $S$ is indeed $\mathfrak{so}(8)$.
\end{saResultBox}

\begin{saPromptBox}[label={prmpt:NHC5}]
Identify a basis of divisor classes of $\hat{Y}_4$ compatible with the Shioda--Tate decomposition, 
\begin{equation}
H^{1,1}(\hat{Y}_4) = \langle S_0,\bar{D}_1,\bar{D}_2,E_1,E_2,E_3,E_4\rangle \,, 
\end{equation}
 where $S_0$ is the zero section, $\bar{D}_1$ and $\bar{D}_2$ are pullback divisors of  $B_3$, respectively, and $E_1,\ldots,E_4$ are the Cartan divisors associated with the $\mathfrak{so}(8)$ resolution. Using the toric intersection ring of the resolved ambient space, compute the intersection numbers 
 \begin{equation}
  \kappa_{ABCD} = \int_{\hat{Y}_4} D_A\wedge D_B\wedge D_C\wedge D_D  \,.
  \end{equation}
\end{saPromptBox}

\begin{saResultBox}[label=res:NHC5, prompt=prmpt:NHC5]

Let $L$ and $S$ denote the divisor classes on the base introduced above, with
\begin{equation}
L^3=0\coma L^2\cdot S=1\coma L\cdot S^2=-6\coma S^3=36\coma
\end{equation}
and $c_1(TB_3)=9L+2S$. A basis of $H^{1,1}(\hat{Y}_4)$ compatible with the Shioda--Tate decomposition is given by
\begin{equation}
H^{1,1}(\hat{Y}_4)
= 
\left\langle
Z,\bar D_1,\bar D_2,E_1,E_2,E_3,E_4
\right\rangle \,,
\label{eq:ShiodaTate}
\end{equation}
where
\begin{equation}
Z=D_{v_z}\big|_{\hat{Y}_4}\coma
\bar D_1=\hat{\pi}^*L\coma
\bar D_2=\hat{\pi}^*S\fstop
\end{equation}
Here $E_2=D_{p_A}|_{\hat{Y}_4}$ is the Cartan divisor associated with the central node of the $D_4$ fiber, 
while the divisor $D_{p_B}|_{\hat{Y}_4}$ splits into the three outer Cartan divisors as follows
\begin{equation}
D_{p_B}\big|_{\hat{Y}_4}=E_1+E_3+E_4\fstop
\end{equation}
In particular, the pullback of the divisor $S$ decomposes as
\begin{equation}
\bar D_2
=
E_0+E_1+2E_2+E_3+E_4\coma
\end{equation}
where $E_0$ denotes the fibral divisor associated with the affine node of the resolved $I_0^*$ fiber.

The complete quadruple intersection tensor can be encoded compactly by introducing the Kähler parameters $(z,\ell,s,e_1,e_2,e_3,e_4)$ associated with the ordered basis~(\ref{eq:ShiodaTate}) and
$A_k:=e_1^k+e_3^k+e_4^k$. 
Then
\begin{align}
\begin{split}
I_4
 =&\,
\int_{\hat{Y}_4}
\left(
zZ+\ell\bar D_1+s\bar D_2+\sum_{i=1}^4 e_iE_i
\right)^4
\\
=&\,
-126z^4+48z^3\ell+36z^3s
-12z^2\ell^2+36z^2\ell s\\
&-108z^2s^2
+12z\ell^2s-72z\ell s^2+144zs^3
\\
&
+\ell^2\left(-12A_2+12e_2A_1-12e_2^2\right)
+\ell s\left(144A_2-144e_2A_1+144e_2^2\right)
\\
&
+s^2\left(-432A_2+432e_2A_1-432e_2^2\right)
+\ell\left(48A_3+48e_2^3-36e_2^2A_1\right)
\\
&
+s\left(-288A_3-288e_2^3+216e_2^2A_1\right)
-72A_4-54e_2^4+36e_2^3A_1 \fstop
\end{split}
\end{align}
For a monomial
\begin{equation}
z^{n_0}\ell^{n_1}s^{n_2}e_1^{n_3}e_2^{n_4}e_3^{n_5}e_4^{n_6}\coma
\sum_{i=0}^{6} n_i=4\coma
\end{equation}
the integers $n_0,\ldots,n_6$ denote the corresponding powers of the variables in the monomial. Its coefficient in $I_4$ is related to the associated intersection number by
\begin{equation}
\operatorname{coeff}(I_4)
=
\frac{4!}{n_0!\cdots n_6!}
\kappa_{ABCD}\coma
\end{equation}
so that this polynomial determines all symmetric quadruple intersections.

As checks, the section and base intersections include~\cite{Weigand:2018rez}
\begin{equation}
Z\cdot\bar D_1^2\cdot\bar D_2=1\coma
Z\cdot\bar D_1\cdot\bar D_2^2=-6\coma
Z\cdot\bar D_2^3=36\coma
\end{equation}
and obey the usual section relation 
$Z^2=-\hat{\pi}^*c_1(TB_3)\,Z$. 
Moreover, in the ordering $(E_1,E_2,E_3,E_4)$, the Cartan intersections satisfy~\cite{Weigand:2018rez}
\begin{equation}
\kappa_{E_iE_j\bar D_a\bar D_b}
=
-C^{D_4}_{ij}
\int_{B_3}S\wedge D_a\wedge D_b\coma
\end{equation}
confirming that the exceptional divisors reproduce the $D_4$ Cartan matrix associated with the $\mathfrak{so}(8)$ resolution.
\end{saResultBox}

\noindent As a final sanity check, we asked \texttt{Solver Agent} to compute the Euler characteristic of $\hat{Y}_4$ to check that it agrees with the one needed by \eqref{eq:tadpoleX18}.

\begin{saPromptBox}[label=prmpt:computechiofY4]

Using adjunction and the quadruple intersection numbers obtained above,
compute $c_4(T\hat{Y}_4)$ and hence the Euler characteristic
\begin{equation}
 \chi(\hat Y_4)
 =
 \int_{\hat Y_4}c_4(T\hat Y_4)\fstop
\end{equation}
\end{saPromptBox}

\begin{saResultBox}
[label=res:computechiofY4,prompt=prmpt:computechiofY4]

Let $X=X_{\mathcal T}$ be the smooth toric fivefold associated with
the 54-cone FRST, and write
\begin{equation}
S_X=[\hat Y_4]=-K_X=\sum_r D_r,
\qquad
c(TX)=\prod_r(1+D_r)=1+e_1+e_2+\cdots.
\end{equation}
Since $e_1=S_X$, adjunction gives
\begin{equation}
c_4(T\hat Y_4)
=\left(e_4-S_Xe_3+S_X^2e_2\right)\big|_{\hat Y_4}.
\end{equation}
Reducing the resulting degree-five intersections using the linear and
Stanley--Reisner relations of the resolved fan, we obtain
\begin{equation}
\chi(\hat Y_4)
=\int_X S_X\left(e_4-S_Xe_3+S_X^2e_2\right)
=234-5346+49248
=44136.
\end{equation}
As an independent check, Batyrev's formulas give
$(h^{1,1},h^{2,1},h^{3,1},h^{2,2})=(7,0,7341,29436)$, and hence
\begin{equation}
\chi(\hat Y_4)
=4+2h^{1,1}-4h^{2,1}+2h^{3,1}+h^{2,2}
=44136.
\end{equation}

\end{saResultBox}

Notice that Result~\ref{res:computechiofY4} agrees with both the tadpole cancellation relation
$\frac{1}{12}\chi(\hat{Y}_4)=3678$ in~\eqref{eq:tadpoleX18}
and the Euler characteristic previously reported
in~\cite[Table 6.3]{Klemm:1996ts}. 

Following the flux constraints reviewed in~\cite[Section 9]{Weigand:2018rez} and using the results of our previous computations, we now consider the following ansatz for a non-trivial non-abelian four-form flux $G_4$:

\begin{saPromptBox}[label=prmpt:FluxConstraints]

Consider the most general vertical flux generated by quadratic divisor products,
\begin{equation}
 G_4=\sum_{A\leq B}g_{AB}D_AD_B\in H^{2,2}_{\mathrm{vert}}(\hat Y_4)\coma
 D_A\in\{Z,\bar D_1,\bar D_2,E_1,E_2,E_3,E_4\}\fstop
\end{equation}
Use the intersection relations obtained above to reduce this ansatz to an independent basis.  Impose the F-theory transversality conditions
\begin{equation}
 \int_{\hat Y_4}G_4Z\bar D_\alpha=0\coma
 \int_{\hat Y_4}G_4\bar D_\alpha\bar D_\beta=0\coma
\end{equation}
and the cohomological conditions for preserving the full
non-abelian gauge algebra $\mathfrak{so}(8)$,
\begin{equation}
 \int_{\hat Y_4}G_4E_i\bar D_\alpha=0\coma
\end{equation}
where $\alpha,\beta=1,2$ and $i=1,\ldots,4$.

Require in addition
\begin{equation}
 G_4+\frac{1}{2}c_2(T\hat Y_4)
 \in H^4(\hat Y_4,\mathbb Z)\coma
 J\wedge G_4=0\coma
 n_{\mathrm{D3}}
 +\frac{1}{2}
 \int_{\hat Y_4}G_4\wedge G_4
 =
 \frac{\chi(\hat Y_4)}{24}\coma
\end{equation}
where $J$ is the Kähler form of $\hat Y_4$ and
$n_{\mathrm{D3}}\in\mathbb Z_{\geq0}$. Determine whether there exists a
non-zero supersymmetric, properly quantized vertical flux satisfying
all these conditions.

\end{saPromptBox}

\begin{saResultBox}
[label=res:FluxConstraints,prompt=prmpt:FluxConstraints]

Write $L=\bar D_1$ and $S=\bar D_2$. The intersection relations reduce
the general vertical flux to
\begin{equation}
G_4=AZ^2+BZL+CL^2+DLS+L\sum_{i=1}^4 f_iE_i\fstop
\end{equation}
The F-theory transversality conditions give
\begin{equation}
-2A=0\coma 3A+B=0\coma
12A-2B+D=0\coma 9A+3B+C-6D=0\fstop
\end{equation}
Thus $A=B=C=D=0$, leaving $G_4=L\sum_i f_iE_i$. The
$\mathfrak{so}(8)$-preserving conditions already force this remaining
flux to vanish: using $\bar D_1=L$, they read
\begin{equation}
0=\int_{\hat Y_4}G_4E_jL
=-\sum_{i=1}^4(C_{D_4})_{ji}f_i\coma
C_{D_4}=
\begin{pmatrix}
 2&-1& 0& 0\\
-1& 2&-1&-1\\
 0&-1& 2& 0\\
 0&-1& 0& 2
\end{pmatrix}.
\end{equation}
Since $C_{D_4}$ is invertible, all $f_i$ vanish.

In fact, dropping gauge preservation does not produce a nonzero
supersymmetric flux. For $f=(f_1,\ldots,f_4)^{\mathsf T}$, the remaining
classes satisfy
\begin{equation}
\int_{\hat Y_4}G_4\wedge G_4=-f^{\mathsf T}C_{D_4}f<0
\qquad\text{if }f\neq0\fstop
\end{equation}
This contradicts the Hodge--Riemann bilinear relations, which require positive self-intersection for a nonzero real $(2,2)$-class satisfying $J\wedge G_4=0$ on the smooth projective fourfold $\hat Y_4$. Hence there is no nonzero supersymmetric vertical flux within this ansatz, irrespective of shifted quantization and the D3-brane tadpole. The zero flux requires, then, that the tadpole constraint is satisfied by
\begin{equation}
n_{\mathrm{D3}}=\frac{\chi(\hat Y_4)}{24}
=\frac{44136}{24}=1839\fstop
\end{equation}

\end{saResultBox}

The absence of non-zero gauge-preserving vertical flux in our $\mathfrak{so}(8)$ non-Higgsable model raises questions about its  geometric and physical significance. The absence of codimension-two fiber enhancements along the gauge divisor suggests that there is no localized charged matter~\cite{Morrison:2014lca}, as in its six-dimensional counterpart~\cite{Morrison:2021wuv}. Although this is consistent with the absence of an associated chiral index, it does not, by itself, explain why the allowed vertical flux vanishes. Relaxing gauge preservation to allow Cartan fluxes leaves only the trivial class once primitivity is imposed in the Kähler chamber considered. We leave a deeper understanding of this restriction, and whether it extends to other non-Higgsable models, for future work.

\subsection{Calabi--Yau Example with S-folds and Gauge Enhancement}
\label{sec:CYExamples-SFoldsGauge}

In this final section, we focus on a Calabi--Yau fourfold geometry that simultaneously incorporates O3-plane and S-fold point singularities, together with additional Kodaira singularities giving rise to a non-abelian gauge enhancement. This provides an explicit example of the scenarios described in points 3. and 4. of Theorem~\ref{thm:weierstrass_proof}.

Let us start by analyzing the interesting base geometry $\mathbb{P}^3_{[1,1,3,4]}$, which possesses a $\mathbb{Z}_4$ terminal quotient singularity and, by Theorem~\ref{thm:weierstrass_proof}, we expect terminal quotient singularities in its Weierstrass model. However, this threefold also contains a canonical singularity at $p_3=[0:0:1:0]$. Indeed, on the patch $U_\sigma=\mathbb{C}^3/\mathbb{Z}_3$, with $\sigma=\langle u_0,u_1,u_3\rangle$, the generator $g$ of $\mathbb{Z}_3$ acts as
\begin{equation}
g:(u_0,u_1,u_3)\mapsto
(\zeta u_0,\zeta u_1,\zeta^4u_3)
=
(\zeta u_0,\zeta u_1,\zeta u_3)\coma
\zeta=e^{2\pi i/3}\fstop
\end{equation}
Thus, $\mathrm{age}(g)=1$, while $\mathrm{age}(g^2)=2$, and hence $p_3$ is a canonical but non-terminal quotient singularity. 
Since this singularity admits a crepant toric resolution whose exceptional divisor is a $\mathbb{P}^2$, we consider the geometry
\begin{equation}
B_3=\mathrm{Bl}_{p_3}(\mathbb{P}^3_{[1,1,3,4]})\fstop
\label{eq:BossBase}
\end{equation}
The resulting threefold retains only the $\mathbb{Z}_4$ terminal quotient singularity at $p_4=[0:0:0:1]\in\mathbb{P}^3_{[1,1,3,4]}$, 
and therefore provides a suitable working example for Theorem~\ref{thm:weierstrass_proof}.

Before proceeding with \texttt{Solver Agent}, let us summarize our main
results and make a few remarks.

We initially followed the approach to constructing an elliptic
fibration over $B_3$ described in Section~\ref{sec:X18Uplift}, but \texttt{Solver Agent}
revealed that some of our assumptions required correction. In
particular, the generic Weierstrass model over $B_3$ does not admit a toric realization. We therefore chose a tuned
Weierstrass model that gives a Calabi--Yau fourfold with a toric
realization. For its associated polytope, we could not find an FRST triangulation that is also compatible with the
elliptic fibration, as in some toric constructions discussed
in~\cite{Huang:2019pne}. We instead decided to preserve the fibration structure and
realize the Calabi--Yau hypersurface using a non-FRST triangulation.

In this tuned Weierstrass model $\pi:Y_4\to B_3$, we find an $I_6$
gauge enhancement along the blowup divisor $S\cong\mathbb{P}^2$.
The divisor $S$ does not intersect $p_4$, and the discriminant
does not meet the singular locus of the base
$\{\Delta_W=0\}\cap\operatorname{Sing}(B_3)=\varnothing$.
Moreover, $Y_4$ has two isolated $\mathbb{Z}_4$ terminal quotient
singularities and one isolated $\mathbb{Z}_2$ terminal quotient
singularity. The partial crepant resolution
$\rho:\hat Y_4\to Y_4$ resolves the $I_6$ singularity while leaving
these isolated Gorenstein terminal singularities unchanged, as expected from Theorem~\ref{thm:weierstrass_proof}.

Following a similar  formulation introduced in Section~\ref{sec:X18Uplift}, 
we use~\texttt{Solver Agent} to construct the corresponding F-theory uplift for~(\ref{eq:BossBase}):

\begin{saPromptBox}[label={prmpt:Boss1}]
Let $N \simeq \mathbb{Z}^5$ and write a lattice point as
\begin{equation}
v=(a,b\,;c,d,e),
\end{equation}
so that
\begin{equation}
N_F=\{c=d=e=0\}\subset N
\end{equation}
plays the role of the fiber lattice and
\begin{equation}
\varphi:N\longrightarrow N_B\simeq\mathbb{Z}^3,
\qquad
(a,b\,;c,d,e)\longmapsto(c,d,e)
\end{equation}
is the projection to the base lattice. Let $\Delta\subset N_{\mathbb{R}}$ be the reflexive
polytope with the following distinguished ray points on $\partial\Delta$:
\begin{align*}
v_x&=(1,0\,;0,0,0),&
v_y&=(0,1\,;0,0,0),&
v_z&=(-2,-3\,;0,0,0),\\
u_0&=(-2,-3\,;-1,-3,-4),&
u_1&=(-2,-3\,;1,0,0),&
u_2&=(-2,-3\,;0,1,0),\\
u_3&=(-2,-3\,;0,0,1),&
e_b&=(1,0\,;0,-1,-1).
\end{align*}

Show that $\varphi$ maps
$\{u_0,u_1,u_2,u_3,e_b\}$
onto the rays of the fan of a projective threefold $B_3$ obtained as the toric crepant blow-up
\begin{equation}
B_3=
\operatorname{Bl}_{p_3}\!\left(
\mathbb{P}^3_{[1,1,3,4]}
\right),
\qquad
p_3=[0:0:1:0]\in\mathbb{P}^3_{[1,1,3,4]}.
\end{equation}
Show that the $\mathbb{Z}_3$ canonical quotient singularity at $p_3$ is resolved and identify the remaining isolated singular point of $B_3$. Identify the local group structure around the latter.
\end{saPromptBox}

\begin{saResultBox}[label=res:Boss1,prompt=prmpt:Boss1]
Let
$r_i=\varphi(u_i)$ and 
$q=\varphi(e_b)$, 
so that
\begin{equation}
\begin{aligned}
r_0&=(-1,-3,-4),&
r_1&=(1,0,0),&
r_2&=(0,1,0),\\
r_3&=(0,0,1),&
q&=(0,-1,-1).
\end{aligned}
\end{equation}
The four rays $r_0,r_1,r_2,r_3$ satisfy the primitive positive relation
\begin{equation}
r_0+r_1+3r_2+4r_3=0,
\end{equation}
and therefore generate the fan of the weighted projective space
$\mathbb{P}^3_{[1,1,3,4]}$.

The maximal cone
$\sigma=\langle r_0,r_1,r_3\rangle$ corresponds to the coordinate point
$p_3=[0:0:1:0]$. 
Its lattice index is three, and the associated local quotient singularity is
$\frac{1}{3}(1,1,1)$. 
In particular, $p_3$ is a canonical Gorenstein quotient singularity.

The additional ray $q$ satisfies
\begin{equation}
q=\frac{1}{3}\left(r_0+r_1+r_3\right),
\end{equation}
and hence lies in the relative interior of $\sigma$. The corresponding
star subdivision replaces $\sigma$ by the three cones
$\langle q,r_1,r_3\rangle$, 
$\langle r_0,q,r_3\rangle$, 
and  
$\langle r_0,r_1,q\rangle$, 
while leaving the remaining maximal cones unchanged. Each of the new
cones is unimodular, so the singularity at $p_3$ is completely resolved.

The subdivision is moreover crepant. Indeed, the local canonical support
covector
$m=(1,-2,1)$ 
satisfies
\begin{equation}
\langle m,r_0\rangle
=
\langle m,r_1\rangle
=
\langle m,r_3\rangle
=
\langle m,q\rangle
=1.
\end{equation}
The exceptional discrepancy is therefore
$\langle m,q\rangle-1=0$. 
Since a toric star subdivision is projective over the original toric
variety, the resulting base is the projective crepant toric blow-up
\begin{equation}
B_3=
\operatorname{Bl}_{p_3}\!\left(
\mathbb{P}^3_{[1,1,3,4]}
\right),
\end{equation}
where the blow-up is understood in the normalized toric sense.

After this subdivision, the only non-smooth maximal cone is
$\tau=\langle r_0,r_1,r_2\rangle$,
which corresponds to
$p_4=[0:0:0:1]$. 
This cone has lattice index four, with local analytic quotient type
$\frac{1}{4}(1,1,3)$. 
Furthermore, all proper faces of $\tau$ are smooth, so the
singularity at $p_4$ is isolated. Consequently,
\begin{equation}
\operatorname{Sing}(B_3)=\{p_4\},
\qquad
(B_3,p_4)\cong
\mathbb{C}^3/\mathbb{Z}_4 \text{ of type }\tfrac{1}{4}(1,1,3).
\end{equation}

\end{saResultBox}

\begin{saPromptBox}[label={prmpt:Boss2}]
Show that the lattice point
\begin{equation}
\widetilde e_b=(-2,-3\,;0,-1,-1)
\end{equation}
lies on $\partial\Delta$ and compute $\varphi(\widetilde e_b)$.
Using the eight rays
$\{v_x,v_y,v_z,u_0,u_1,u_2,u_3,\widetilde e_b\}$, 
write down a fan $\Sigma_W$ compatible with $\varphi$, whose maximal cones are of the form
$\sigma_F+\widetilde{\sigma}_B$,
where $\sigma_F$ is a two-dimensional cone of the fan of
$\mathbb{P}^2_{[2,3,1]}$
and $\widetilde{\sigma}_B$ is the lift of a three-dimensional cone
$\sigma_B\in\Sigma_{B_3}$.

Let $\mathbb{P}_{\Sigma_W}$ be the toric variety associated to the fan $\Sigma_W$.
Show that
\begin{equation}
\mathbb{P}_{\Sigma_W}
\cong
\mathbb{P}^2_{[2,3,1]}
\Bigl(
\mathcal{O}_{B_3}(-2K_{B_3})
\oplus
\mathcal{O}_{B_3}(-3K_{B_3})
\oplus
\mathcal{O}_{B_3}
\Bigr),
\end{equation}
where $\mathcal{O}_{B_3}(-mK_{B_3})$ is understood as the corresponding
rank-one reflexive sheaf whenever $-mK_{B_3}$ is not Cartier. 
Show furthermore that a generic anticanonical hypersurface in
$\mathbb{P}_{\Sigma_W}$ gives the untuned Weierstrass model
\begin{equation}
y^2=x^3+fxz^4+gz^6,
\end{equation}
and verify if Kodaira degeneration is generically forced along the exceptional divisor of $B_3$.

Now consider the additional ray
\begin{equation}
e_b=(1,0\,;0,-1,-1).
\end{equation}
Express $e_b$ in terms of $v_x,v_y,v_z,u_0,u_1,u_2,u_3,\widetilde e_b$,
so that adding $e_b$ defines a toric star subdivision of the toric ambient space of the untuned Weierstrass model. 
Denote the corresponding toric modification by
$\mathbb{P}_{\Sigma_{W}'}
\to 
\mathbb{P}_{\Sigma_W}$.
Show that a generic anticanonical hypersurface in
$\mathbb{P}_{\Sigma_{W}'}$ is the proper transform of a tuned
Weierstrass model
$\pi:Y_4\to B_3$.
Determine the vanishing orders
\begin{equation}
\operatorname{ord}_S(a_1,a_2,a_3,a_4,a_6)
\end{equation}
of the corresponding Tate coefficients along the divisor
$S\subset B_3$, and hence determine
\begin{equation}
\operatorname{ord}_S(f,g,\Delta_W),
\end{equation}
where $(f,g,\Delta_W)$ are the Weierstrass coefficients and discriminant of the tuned Weierstrass model. 
Identify the Kodaira fiber over $S$ and determine $S$ in terms of the toric divisors of $B_3$.
\end{saPromptBox}

\begin{saResultBox}[label=res:Boss2,prompt=prmpt:Boss2]
For the polytope
$\Delta=\operatorname{conv}\{v_x,v_y,v_z,u_0,u_1,u_2,u_3,e_b\}$,
the lattice point
\begin{equation}
\widetilde e_b
=
\frac{1}{3}\left(u_0+u_1+u_3\right)
=
(-2,-3\,;0,-1,-1)
\end{equation}
lies on $\partial\Delta$. Indeed, the supporting functional
$\ell(a,b\,;c,d,e)=c-2d+e$ satisfies $\ell\leq1$ on $\Delta$,
with equality at $u_0,u_1,u_3$ and $\widetilde e_b$. Moreover,
\begin{equation}
\varphi(\widetilde e_b)=\varphi(e_b)=q=(0,-1,-1).
\end{equation}

To construct the untuned Weierstrass ambient space, lift the six maximal
cones of $\Sigma_{B_3}$ by replacing $r_i$ with $u_i$ and $q$ with
$\widetilde e_b$. Since $2v_x+3v_y+v_z=0$, the fiber fan is that of
$\mathbb{P}^2_{[2,3,1]}$, with maximal cones
$\langle v_x,v_y\rangle$, 
$\langle v_y,v_z\rangle$,
and $\langle v_z,v_x\rangle$.
Taking the sums of these fiber cones with the lifted base cones gives
a complete fan $\Sigma_W$ with eighteen maximal cones. Each such cone
satisfies
$\varphi(\sigma_F+\widetilde{\sigma}_B)=\sigma_B$,
so the resulting toric variety admits a morphism to $B_3$.

Writing $H=[D_z]$ and suppressing pullbacks of base divisor classes,
the principal divisor relations give
\begin{equation}
[D_x]=2H-2K_{B_3},
\qquad
[D_y]=3H-3K_{B_3}.
\end{equation}
Consequently,
\begin{equation}
\mathbb{P}_{\Sigma_W}
\cong
\mathbb{P}^2_{[2,3,1]}
\Bigl(
\mathcal{O}_{B_3}(-2K_{B_3})
\oplus
\mathcal{O}_{B_3}(-3K_{B_3})
\oplus
\mathcal{O}_{B_3}
\Bigr),
\end{equation}
where the sheaves are understood as rank-one reflexive sheaves whenever
the corresponding divisors are not Cartier. The ambient anticanonical
class is $6H-6K_{B_3}$, and a generic anticanonical hypersurface takes
the Tate form
\begin{equation}
y^2+a_1xyz+a_3yz^3
=
x^3+a_2x^2z^2+a_4xz^4+a_6z^6,
\qquad
a_i\in \Gamma\!\left(B_3,\mathcal{O}_{B_3}(-iK_{B_3})\right).
\end{equation}
Over characteristic zero, this becomes
\begin{equation}
y^2=x^3+fxz^4+gz^6,
\qquad
f\in \Gamma\!\left(B_3,\mathcal{O}_{B_3}(-4K_{B_3})\right),
\quad
g\in \Gamma\!\left(B_3,\mathcal{O}_{B_3}(-6K_{B_3})\right).
\end{equation}

No degeneration is forced along the exceptional divisor $D_q$ in this
untuned model. Indeed, for $n=4,6$, the character $(-n,2n,-n)$ gives a
section of $\mathcal{O}_{B_3}(-nK_{B_3})$ with vanishing order zero
along $D_q$. Thus, for generic coefficients,
\begin{equation}
\operatorname{ord}_{D_q}(f,g,\Delta_W)=(0,0,0),
\end{equation}
and the generic fiber over $D_q$ is smooth.

We now introduce the additional ray $e_b$, which satisfies
\begin{equation}
e_b=\widetilde e_b+3v_x+3v_y.
\end{equation}
It lies in the relative interior of
$\langle v_x,v_y,\widetilde e_b\rangle$ and defines a star subdivision
$\mu:\mathbb{P}_{\Sigma'_W}\to\mathbb{P}_{\Sigma_W}$.
The new ambient divisor has discrepancy $3+3+1-1=6$.
Locally, with $w_{\mathrm{old}}=0$ defining $D_q$, the coordinate
substitution is
\begin{equation}
x_{\mathrm{old}}=xe^3,\qquad
y_{\mathrm{old}}=ye^3,\qquad
w_{\mathrm{old}}=we,
\end{equation}
where $e$ is the exceptional coordinate. After division by $e^6$,
an original monomial
$x_{\mathrm{old}}^{n_x}y_{\mathrm{old}}^{n_y}
w_{\mathrm{old}}^{n_w}$
has exceptional exponent
\begin{equation}
3n_x+3n_y+n_w-6.
\end{equation}
Requiring this exponent to be nonnegative imposes the Tate vanishing
orders
\begin{equation}
\operatorname{ord}_S(a_1,a_2,a_3,a_4,a_6)
=
(0,0,3,3,6),
\qquad
S=D_q.
\end{equation}
These bounds are attained by global toric monomials. Hence a generic
anticanonical hypersurface in $\mathbb{P}_{\Sigma'_W}$ is the strict
transform of a tuned Weierstrass model $\pi:Y_4\to B_3$, whose local
Tate coefficients have the form
\begin{equation}
a_1=\alpha_1\coma
a_2=\alpha_2\coma
a_3=w^3\alpha_3\coma
a_4=w^3\alpha_4\coma
a_6=w^6\alpha_6\coma
\end{equation}
where the $\alpha_i$ are regular coefficient functions and $S=\{w=0\}$.

Substitution into the Weierstrass formulas yields
\begin{equation}
\operatorname{ord}_S(f,g,\Delta_W)=(0,0,6),
\end{equation}
where $\Delta_W$ now denotes the discriminant of the tuned model.
Thus the generic fiber over $S$ has Kodaira type $I_6$. The supporting
divisor is the exceptional divisor of the base blow-up, with class
\begin{equation}
[S]=[D_q]
=
[D_{r_2}]-3[D_{r_0}]
=
[D_{r_3}]-4[D_{r_0}].
\end{equation}
The degeneration is therefore induced by the additional toric ray
$e_b$, rather than being forced in the generic Weierstrass model
over the same base.
\end{saResultBox}

\begin{saPromptBox}[label={prmpt:Boss3}]
Compute $\{\Delta_W=0\}\cap\operatorname{Sing}(B_3)$.
Determine also the singular locus $\operatorname{Sing}(Y_4)$ of the contracted Weierstrass model.
\end{saPromptBox}

\begin{saResultBox}[label=res:Boss3,prompt=prmpt:Boss3]
The unique singular point of $B_3$ is $p_4=[0:0:0:1]$, with
local quotient type $\frac14(1,1,3)$. Since its maximal cone
$\langle r_0,r_1,r_2\rangle$ does not contain $q$, this point
does not belong to the divisor $S=D_q$ supporting the $I_6$
degeneration.

On the index-one cover of a neighborhood of $p_4$, the group
$\mu_4$ acts on the base coordinates as
\begin{equation}
\mu_4: (s_0,s_1,s_2)
\mapsto
(\zeta s_0,\zeta s_1,\zeta^3s_2)\coma
\zeta=e^{2\pi i/4}\fstop
\end{equation}
A local representative of a section of
$\mathcal{O}_{B_3}(-mK_{B_3})$ has character $m$ modulo four.
Consequently, at the origin of the cover,
\begin{equation}
a_1=a_2=a_3=a_6=0\coma
A:=a_4(p_4)\neq0
\end{equation}
for generic coefficients. The Weierstrass formulas therefore give
\begin{equation}
f(p_4)=A\coma
g(p_4)=0\coma
\Delta_W(p_4)=4A^3\neq0\coma
\end{equation}
where $\Delta_W=4f^3+27g^2$. It follows that
\begin{equation}
\{\Delta_W=0\}\cap\operatorname{Sing}(B_3)=\varnothing\fstop
\end{equation}

We next determine the singularities of the total space.
In the affine Tate chart $z=1$, with $S=\{w=0\}$, the tuned
equation reads
\begin{equation}
F
=
y^2+\alpha_1xy+w^3\alpha_3y
-x^3-\alpha_2x^2-w^3\alpha_4x-w^6\alpha_6
=0\coma
\end{equation}
where the $\alpha_i$ are regular coefficient functions.
The equation and all its first derivatives vanish along
\begin{equation}
\Gamma=\{w=x=y=0\}\cong S\fstop
\end{equation}
Conversely, restricting $F=F_x=F_y=0$ to $w=0$ forces
$x=y=0$. Hence, set-theoretically,
\begin{equation}
\operatorname{Sing}(Y_4)\cap\pi^{-1}(S)=\Gamma\fstop
\end{equation}
At a generic point of $\Gamma$, the transverse analytic equation
takes the form
$uv=\lambda w^6$,
where $\lambda$ is a local holomorphic unit. Thus $\Gamma$
is a singular surface whose generic transverse singularity is of Du Val type $A_5$.

Although the discriminant is nonzero at $p_4$, the total space
has additional quotient singularities above this point.
On the cover, the fiber is the smooth elliptic curve
\begin{equation}
E_4:\qquad y^2=x^3+Axz^4\coma
\end{equation}
with induced action
\begin{equation}
[x:y:z]\mapsto[\zeta^2x:\zeta^3y:z]\fstop
\end{equation}
The points
\begin{equation}
O=[1:1:0]\coma
T_0=[0:0:1]
\end{equation}
are fixed by $\mu_4$, whereas
\begin{equation}
T_\pm=[\pm\alpha:0:1]\coma
\alpha^2=-A\coma
\end{equation}
form a single $\mu_4$-orbit with stabilizer $\mu_2$.

At both $O$ and $T_0$, the fiber tangent has weight three.
Together with the base tangent weights $(1,1,3)$, this gives
two quotient singularities of type $\frac14(1,1,3,3)$.
At $T_\pm$, the stabilizer $\mu_2$ acts by $-1$ on all four
tangent directions, giving one quotient singularity of type
$\frac12(1,1,1,1)$. Denoting their images in $Y_4$ by
$Q_O,Q_T,Q_2$, respectively, the local analytic types are
\begin{equation}
\begin{aligned}
(Y_4,Q_O)&\cong(Y_4,Q_T)
\cong\mathbb{C}^4/\mathbb{Z}_4 \text{ of type }\tfrac14(1,1,3,3)\coma\\
(Y_4,Q_2)&\cong
\mathbb{C}^4/\mathbb{Z}_2 \text{ of type }\tfrac12(1,1,1,1)\fstop
\end{aligned}
\end{equation}
These singularities are isolated, since no nonidentity
stabilizer element has an invariant tangent direction.

For generic coefficients, Bertini and Jacobian transversality
exclude further singularities away from $S\cup\{p_4\}$.
Since $p_4\notin S$, the three quotient points are disjoint
from $\Gamma$. Consequently, the singular locus of the tuned
Weierstrass model is
\begin{equation}
\operatorname{Sing}(Y_4)
=
\Gamma\sqcup\{Q_O,Q_T,Q_2\}\coma
\qquad
\Gamma\cong S\fstop
\end{equation}

\end{saResultBox}

\begin{saPromptBox}[label={prmpt:Boss4}]
Show that, apart from the eight distinguished ray points listed above,
$\Delta$ contains further lattice points that do not lie in the interior of a facet,
including the point $\widetilde e_b$ introduced above, and identify all of them.
For each such point, find a primitive positive integral relation with the ray
generators of $\Delta$, and interpret the corresponding star subdivisions as
successive weighted blow-ups over $S$. Read off from $\varphi$ the multiplicities
of the associated fiber components.
\end{saPromptBox}

\begin{saResultBox}[label=res:Boss4,prompt=prmpt:Boss4]
For the polytope 
$\Delta
=
\operatorname{conv}
\{v_x,v_y,v_z,u_0,u_1,u_2,u_3,e_b\}$, 
exact lattice enumeration gives fifteen lattice points, with the origin
as the unique interior point. Besides the eight distinguished points,
the remaining boundary points are
\begin{equation}
\begin{aligned}
w_1&=\widetilde{e}_b=(-2,-3\,;0,-1,-1),&
s_1&=(-1,-2\,;0,0,0),\\
w_2&=(-1,-2\,;0,-1,-1),&
s_2&=(-1,-1\,;0,0,0),\\
w_3&=(0,-1\,;0,-1,-1),&
s_3&=(0,-1\,;0,0,0).
\end{aligned}
\end{equation}
The points $s_1,s_3$ lie in the relative interior of the facet
$-a+b+1=0$, while $s_2$ lies in the relative interior of the facet
$2a-b+1=0$. By contrast, the smallest faces containing $w_1$ and
$w_2,w_3$ have dimensions two and three, respectively. Thus
$w_1,w_2,w_3$ are precisely the additional boundary lattice points
outside facet interiors.

These points satisfy the primitive positive integral relations
\begin{equation}
\begin{aligned}
3w_1&=u_0+u_1+u_3\coma\\
9w_2&=2u_0+2u_1+2u_3+3e_b\coma\\
9w_3&=u_0+u_1+u_3+6e_b\fstop
\end{aligned}
\end{equation}
After introducing $w_1$, the latter two relations can be written
in the form
\begin{equation}
3w_2=2w_1+e_b\coma
2w_3=w_2+e_b\fstop
\end{equation}
The coefficients in each relation have greatest common divisor one,
and their positivity places the new ray in the relative interior
of the corresponding cone.

The relation for $w_1$ describes a star subdivision of
$\langle u_0,u_1,u_3\rangle$, with center 
$D_{u_0}\cap D_{u_1}\cap D_{u_3}$.
The transverse lattice index is three, and the local transverse
quotient has type $\frac13(1,1,1)$. The subdivision is therefore
induced by an ordinary weight-$(1,1,1)$ blow-up on the covering space.
Its center maps to the original base point $p_3$, whereas the
new divisor $D_{w_1}$ dominates the exceptional base divisor
$S=D_q$.

The next subdivision introduces $w_2$ into
$\langle w_1,e_b\rangle$. Its center is
$D_{w_1}\cap D_{e_b}$, with transverse quotient type
$\frac13(2,1)$, and it is induced by a weight-$(2,1)$ blow-up
on the covering space. Finally, introducing $w_3$ into
$\langle w_2,e_b\rangle$ gives a subdivision with center
$D_{w_2}\cap D_{e_b}$ and transverse quotient type
$\frac12(1,1)$, induced by an ordinary weight-$(1,1)$ blow-up.
Both of these centers lie over the generic orbit of $S$.

The multiplicities over $S$ follow directly from the projection:
\begin{equation}
\varphi(e_b)
=
\varphi(w_1)
=
\varphi(w_2)
=
\varphi(w_3)
=
q=(0,-1,-1)\fstop
\end{equation}
For a primitive vertical ray $v$ satisfying $\varphi(v)=mq$,
the associated generic toric fiber multiplicity is 
$m_v
=
[\mathbb{Z}q:\varphi(\mathbb{Z}v)]
=
m$.
Since all four rays map to the primitive base ray $q$, one obtains
$m_{e_b}=m_{w_1}=m_{w_2}=m_{w_3}=1$. 
Thus the corresponding toric divisors occur with multiplicity one
over $S$. The integers $3,3,2$ appearing in the subdivision relations
are transverse quotient indices, not fiber multiplicities.
\end{saResultBox}

\begin{saPromptBox}[label={prmpt:Boss5}]

Determine whether there exists a fine, regular, star triangulation
$\hat{\mathcal T}$ of $\Delta$ such that every cone of
$\Sigma_{\hat{\mathcal T}}$ is mapped by $\varphi$ into a cone of
$\Sigma_{B_3}$. If no such triangulation exists, identify a
combinatorial obstruction.

Independently, starting from the fibration-compatible fan
$\Sigma'_W$ constructed in Prompt \ref{prmpt:Boss2}, consider projective simplicial
refinements
$\hat\Sigma_W\to\Sigma'_W$ 
obtained by toric star subdivisions using the boundary lattice points
identified in Result \ref{res:Boss4}. These refinements are not required to arise from
an FRST of $\Delta$.

Let
$\hat Y_4\subset\mathbb{P}_{\hat\Sigma_W}$
be the strict transform of the tuned Weierstrass model in the resulting
toric ambient space. Verify that its defining equation is a section of
the anticanonical reflexive sheaf
\begin{equation}
\mathcal{O}_{\mathbb{P}_{\hat\Sigma_W}}
\!\left(-K_{\mathbb{P}_{\hat\Sigma_W}}\right)
\end{equation}
and that its Newton polytope is $\Delta^\ast$, where $\Delta^\ast$ is the polar dual polytope of the reflexive polytope $\Delta$. Thus,
$\hat Y_4$ is an anticanonical toric hypersurface described by the
Batyrev polynomial data associated with the reflexive polytope $\Delta$,
although $\hat\Sigma_W$ is not required to arise from an FRST of
$\Delta$.

Determine whether the induced morphism
$\rho:\hat Y_4\to Y_4$ 
is a crepant partial resolution and whether $\hat Y_4$
is $\mathbb{Q}$-factorial with at most isolated terminal quotient
singularities. If further modifications are needed, identify the
remaining singular loci.

Determine the irreducible components and multiplicities of the
fiber over a generic point of $S$, taking into account possible
splitting of toric divisors upon restriction to the hypersurface.
Determine the monodromy cover, identify the geometric fiber
graph and its affine Dynkin type, and read off the gauge algebra
$\mathfrak g$.

\end{saPromptBox}

\begin{saResultBox}[label=res:Boss5,prompt=prmpt:Boss5]

\paragraph{Obstruction to a compatible FRST.}
The segment $[e_b,u_2]$ is a one-dimensional face of $\Delta$,
and its direction
\begin{equation}
e_b-u_2=(3,3\,;0,-2,-1)
\end{equation}
is primitive. Hence this edge contains no further lattice point
and occurs in every triangulation of $\partial\Delta$. The
corresponding star fan therefore necessarily contains the cone
$\langle e_b,u_2\rangle$. However,
\begin{equation}
\varphi(e_b)=q\coma
\varphi(u_2)=r_2\coma
\end{equation}
and no cone of $\Sigma_{B_3}$ contains both $q$ and $r_2$.
Consequently, no star triangulation of $\Delta$ is compatible
with the prescribed base fan. In particular, a
fibration-compatible FRST does not exist.

\paragraph{A separate toric refinement.}
Instead, starting from the fibration-compatible fan $\Sigma'_W$,
we perform the successive star subdivisions determined by
\begin{equation}
3w_2=2w_1+e_b\coma
2w_3=w_2+e_b\coma
w_1=\widetilde e_b\fstop
\end{equation}
The resulting fan $\hat\Sigma_W$ has thirty-six maximal
cones and is complete, simplicial, and compatible with
$\varphi$. It is also projective: the original fan $\Sigma_W$
admits a strictly convex integral support function, and each
subsequent star subdivision is projective over the preceding
toric variety.

The subdivisions introducing $w_2$ and $w_3$ are crepant
on the ambient space, since their discrepancies are
\begin{equation}
a(w_2)=\frac23+\frac13-1=0\coma
a(w_3)=\frac12+\frac12-1=0\fstop
\end{equation}
The earlier subdivision introducing $e_b$ has ambient
discrepancy six, but the pullback of the tuned hypersurface
has multiplicity six along its exceptional divisor.
These contributions cancel by adjunction. Thus, denoting
the strict transform by
\begin{equation}
\hat Y_4\subset
X:=\mathbb{P}_{\hat\Sigma_W},
\end{equation}
the induced morphism
$\rho:\hat Y_4\to Y_4$ 
is projective, birational, and crepant.

\paragraph{Anticanonical equation and Newton polytope.}
For the primitive ray generators $v_\rho$ of
$\hat\Sigma_W$, the anticanonical monomial polytope is
\begin{equation}
P_{-K_X}
=
\left\{
m\in M_{\mathbb R}
\ \middle|\
\langle m,v_\rho\rangle\geq-1
\text{ for every }\rho
\right\}
=
\Delta^\ast.
\end{equation}
Indeed, the inequalities associated with $w_1,w_2,w_3$
are redundant, since these points are convex combinations
of the original distinguished rays. Consequently, the
generic defining polynomial $\hat P$ satisfies
\begin{equation}
\hat P\in
H^0\!\left(X,\mathcal{O}_X(-K_X)\right),
\qquad
\operatorname{Newt}(\hat P)=\Delta^\ast.
\end{equation}
The dual polytope contains $3920$ lattice points, which label
the anticanonical monomials. Thus $\hat Y_4$ retains the
Batyrev polynomial data associated with $\Delta$, although
its ambient fan does not arise from an FRST of $\Delta$.

\paragraph{Residual singularities.}
The refinement resolves the positive-dimensional singular
locus $\Gamma\cong S$ of the tuned Weierstrass model.
For generic coefficients, $\hat Y_4$ is smooth except
at three points above $p_4$, with quotient types
\begin{equation}
2\times\frac14(1,1,3,3),
\qquad
1\times\frac12(1,1,1,1).
\end{equation}
Their local divisor class groups are finite, respectively
$\mathbb{Z}_4$ and $\mathbb{Z}_2$, so $\hat Y_4$ is
$\mathbb{Q}$-factorial. Moreover, every nontrivial stabilizer
element has age two. The Reid--Tai criterion therefore
shows that all three quotient singularities are terminal.
Hence $\rho$ is a crepant partial resolution with only
isolated terminal quotient singularities remaining.

\paragraph{The resolved fiber and its monodromy.}
Over the function field $K=\mathbb C(S)$, the divisors
$D_{w_1}$ and $D_{e_b}$ each contribute one geometrically
irreducible fiber component, denoted $C_0$ and $C_3$.
On both $D_{w_2}$ and $D_{w_3}$, the hypersurface equation
reduces to
\begin{equation}
Q=y^2+\bar a_1xyz-\bar a_2x^2z^2=0\coma
\bar a_i=a_i|_S\fstop
\end{equation}
The irreducibility of these intersections is therefore
controlled by the monodromy cover
\begin{equation}
\psi^2+\bar a_1\psi-\bar a_2=0\fstop
\end{equation}

The actual monomial set of $\Delta^\ast$ imposes
\begin{equation}
\bar a_1=c_1h\coma
\bar a_2=c_2h^2\coma
h=t_2^3\coma
\end{equation}
where $c_1,c_2\in\mathbb C$ are generic constants and $t_2$
denotes the base Cox coordinate associated with $r_2$.
In particular, $h$ is nonzero at the generic point of $S$.
Choosing $\gamma\in\mathbb C$ with
$\gamma^2=c_1^2+4c_2\neq0$, we obtain
\begin{equation}
b_2|_S
=
\bar a_1^2+4\bar a_2
=
(\gamma h)^2\fstop
\end{equation}
The change of variable $\xi=2\psi+\bar a_1$ therefore
puts the monodromy cover in the factorized form
\begin{equation}
(\xi-\gamma h)(\xi+\gamma h)=0\fstop
\end{equation}
Thus the cover splits and the monodromy is trivial.

Correspondingly,
\begin{equation}
Q=(y-r_+xz)(y-r_-xz)\coma
r_\pm=\frac{h}{2}(-c_1\pm\gamma)\fstop
\end{equation}
The intersections with $D_{w_2}$ and $D_{w_3}$ each split
into two geometrically irreducible components, denoted
$C_{1,\pm}$ and $C_{2,\pm}$, respectively. All components
are already defined over $K$ and occur with multiplicity
one. The generic fiber is therefore reduced, with
\begin{equation}
F_K
=
C_0+C_{1,+}+C_{2,+}+C_3+C_{2,-}+C_{1,-}\fstop
\end{equation}
Its components form the cycle
\begin{equation}
C_0
\;-\;
C_{1,+}
\;-\;
C_{2,+}
\;-\;
C_3
\;-\;
C_{2,-}
\;-\;
C_{1,-}
\;-\;
C_0\fstop
\end{equation}
Adjacent components meet transversely at one point,
while all other distinct pairs are disjoint. The dual
graph is consequently the affine Dynkin diagram
$A_5^{(1)}$. Since there is no monodromy folding, the
fiber is split Kodaira type $I_6^s$, with gauge algebra
\begin{equation}
\mathfrak g=\mathfrak{sl}_6,
\end{equation}
whose compact real form is $\mathfrak{su}(6)$.

\end{saResultBox}

\begin{saPromptBox}[label={prmpt:Boss6}]
Determine $\operatorname{Sing}(\hat{Y}_4)$.
\end{saPromptBox}

\begin{saResultBox}[label=res:Boss6,prompt=prmpt:Boss6]
The subdivisions by $w_2$ and $w_3$ resolve the singular surface
$\Gamma\cong S$ of the tuned Weierstrass model. Since their centers
lie over $S$ and $p_4\notin S$, the local quotient geometry above
$p_4$ remains unchanged. Consequently, for generic coefficients,
\begin{equation}
\operatorname{Sing}(\hat Y_4)
=
\{Q_O,Q_T,Q_2\}\coma
\end{equation}
where all three points lie above $p_4$, with local analytic types
\begin{equation}
\begin{aligned}
(\hat Y_4,Q_O)
&\cong
(\hat Y_4,Q_T)
\cong
\mathbb{C}^4/\mathbb{Z}_4 \text{ of type }\tfrac14(1,1,3,3),\\
(\hat Y_4,Q_2)
&\cong
\mathbb{C}^4/\mathbb{Z}_2 \text{ of type }\tfrac12(1,1,1,1)\fstop
\end{aligned}
\end{equation}
These are isolated terminal cyclic quotient singularities.
No positive-dimensional singular locus remains.
\end{saResultBox}

\section{Conclusions}
\label{sec:conclusions}

In this work, we have introduced \texttt{Solver Agent} as an agentic framework for research in mathematics and theoretical physics. The purpose of this system is to substantially accelerate and facilitate proofs and computations, and we have applied it to the study of F-theory uplifts of Type IIB orientifold compactifications, particularly in improving our understanding of terminal singularities in Calabi--Yau fourfolds. \texttt{Solver Agent} was not only used to perform and verify computations, but it was also directly involved in the proof of Theorem \ref{thm:weierstrass_proof}. Noting that isolated fixed points of O3 orientifold involutions descend to terminal $\mathbb{Z}_2$ quotient singularities, this theorem states that terminal quotient singularities in Type IIB Calabi--Yau orientifold compactifications uplift to an elliptically fibered Calabi--Yau fourfold carrying $\mathbb{Q}$-factorial terminal quotient singularities over fixed points of groups $\ZZ_k$ with $k \in \{2,3,4,6\}$. Moreover, \texttt{Solver Agent} computed the fixed-point contributions to the Hodge data and Euler characteristics, demonstrating that the resulting Euler characteristic correction explicitly fixes the geometric contribution of these defects. We further employed the system to systematically analyze toric trilayer constructions, showing how a single three-dimensional reflexive polytope dictates the geometry of both the Type IIB Calabi--Yau threefold and the F-theory base. In this framework, the orientifold double cover is realized as a bisection of an alternative genus-one-fibered uplift carrying a discrete $\mathbb{Z}_2$ gauge symmetry that holds independently of a Sen limit. Ultimately, the framework provided a practical approach to compute explicit intersection data and systematically analyze four-form fluxes for four-dimensional $\mathcal{N}=1$ compactifications with non-abelian gauge sectors.

\texttt{Solver Agent} has proven to be a powerful framework for making concrete progress in research and demonstrates how AI can be integrated into traditional research beyond the out-of-the-box LLMs present in the market. In fact, developing these agentic frameworks for specialized research may be the way to overcome the limitations of current AI models, whose trustworthiness is sometimes in doubt. Nevertheless, human oversight remains essential for guiding the development of the project, validating results, and carrying out independent checks. In our experience, physical intuition `guided` the course of the work, while critical thinking was indispensable for assessing the consistency and reliability of the results. In fact, this is another positive outcome from using \texttt{Solver Agent} to assist us in the research process, where we are able to delegate to it most of the `cumbersome and tedious tasks`, while human researchers can focus on the physical interpretation of the outcomes and the broader trajectory of the projects. 

\texttt{Solver Agent} is publicly available at the following \href{https://github.com/starrfree/solver-agent}{GitHub repository},\footnote{Public Git repository under \href{https://github.com/starrfree/solver-agent/blob/main/LICENSE}{AGPL-3.0} license: \href{https://github.com/starrfree/solver-agent}{github.com/starrfree/solver-agent}} along with an \href{https://github.com/starrfree/solver-agent\#installation}{installation guide} and a \href{https://github.com/starrfree/solver-agent\#using-solver-agent}{user guide}. We advise anyone using this tool to export all sessions and make them publicly available along with the publication of their work. All sessions used to conduct the work presented in this paper are available \href{https://storage.googleapis.com/solver-agent-sessions/solver-agent-f-theory-uplifts/index.html}{here}.\footnote{\href{https://storage.googleapis.com/solver-agent-sessions/solver-agent-f-theory-uplifts/index.html}{storage.googleapis.com/solver-agent-sessions/solver-agent-f-theory-uplifts/index.html}} We also encourage users who develop new tools, integrations, or extensions that may be useful to the broader research community to contribute them back to the public repository.

With the tools currently available to the agent, several directions can be pursued immediately, including the following:
\begin{itemize}
\item A deeper understanding of the enumerative geometry of Donaldson--Thomas invariants for Calabi--Yau fourfolds, their relation to four-form fluxes, and their physical interpretation would be highly desirable~\cite{Bae:2022pif}. Topological string theory on elliptically fibered Calabi--Yau fourfolds with multiple Kähler moduli remains largely unexplored, despite its close connection to the study of gauge fluxes in F-theory~\cite{Cota:2017aal, Lee:2019tst, Lee:2020gvu, Lee:2020blx}.

\item A natural candidate for a BPS index associated with S-folds is provided by the generalized Donaldson--Thomas partition function of the quotient stack $[\mathbb{C}^4/\mathbb{Z}_k]$, extended to terminal quotient singularities with $k\in\{2,3,4,6\}$~\cite{Cao:2023gvn}. It would be interesting to make this connection more precise and understand its physical interpretation.

\item \begingroup\emergencystretch=3em Our computations capture stringy orbifold corrections arising from terminal quotient singularities that modify the results of classical intersection computations \cite{batyrev1996strong, Batyrev:1997hj}. More broadly, it would be interesting to investigate how such singularities affect Gromov--Witten theory \cite{Klemm:2007in} and to deepen our understanding of topological string theory on Calabi--Yau geometries with terminal singularities \cite{Schimannek:2021pau, Katz:2022lyl}. \par\endgroup

\item The methods developed here could also facilitate computationally demanding tasks, including determining topological string partition functions on compact Calabi--Yau threefolds~\cite{Huang:2006hq, Alexandrov:2023zjb, Alexandrov:2023ltz}, proving related conjectures~\cite{Huang:2015sta,Cota:2019cjx, Duque:2025kaa, Pioline:2025uov, FierroCota:2025pkp, Huang:2025xkc}, and extending such computations to multi-parameter compact Calabi--Yau geometries~\cite{Alim:2012ss, Kuusela:2023vgi, Doran:2024kcb} and to non-perturbative regimes~\cite{Gu:2023mgf, Douaud:2024khu, Douaud:2026qfo}.

\item A further application resides in the systematic study and classification of Calabi--Yau geometries and their fibration structures. While substantial progress has been made in mapping the allowed bases for F-theory compactifications, identifying generic fibers, and locating non-Higgsable clusters (see, e.g., \cite{Morrison:2012js,Morrison:2012np,Taylor:2015isa,Abbasi:2025lvn}), fully characterizing the geometry requires an exact determination of the singularity structure. Building on the algebraic computations and geometric proofs utilized in this work, \texttt{Solver Agent} can be employed to systematically construct explicit global Weierstrass models across Calabi--Yau datasets. 
\item A further application for \texttt{Solver Agent} is the Swampland program~\cite{Vafa:2005ui}. The verification of Swampland conjectures typically requires the determination of the geometric properties of string compactifications. For instance, evaluating the Swampland Distance Conjecture~\cite{Ooguri:2006in} and identifying geometrical obstructions to infinite distance limits in Calabi--Yau moduli spaces entail analysis of singularity structures and asymptotic Hodge theory (see, e.g., \cite{Blumenhagen:2018nts,Corvilain:2018lgw,Grimm:2019bey,Joshi:2019nzi}). Furthermore, completing the proof of the Minimal Weak Gravity Conjecture~\cite{Arkani-Hamed:2006emk,FierroCota:2023bsp} or extending the Weak Gravity Conjecture to AdS space~\cite{Lin:2025wfe,Lin:2025gco} requires explicit computations of massive string spectra and black hole extremality bounds in specific geometries. Following the methodology used to prove \cref{thm:weierstrass_proof}, \texttt{Solver Agent} can be employed to automate the underlying algebraic derivations, allowing for a systematic verification of Swampland criteria across broad classes of Calabi--Yau fourfolds.
\end{itemize}
However, the flexibility of the \texttt{Solver Agent} resides precisely in the choice of tools that the user can decide to provide to it. For our purposes, \texttt{Solver Agent} has made use of \texttt{CYTools} \cite{Demirtas:2022hqf} as a tool for the computations, but we see interesting and possible applications in other fields, should different tools be provided. For instance, 
\begin{itemize}
    \item In the context of Seiberg dualities \cite{Seiberg:1994pq} applied to quiver gauge theories, it would be interesting to check how \texttt{Solver Agent} performs when compared to recent applications of ML to determine if two SCFTs are dual \cite{Heckman:2026xsi}. 
    \item In the context of 6d $\mathcal{N}=(1,0)$ SCFTs, there is much evidence that their compactifications lead to large classes of known 4d $\mathcal{N}=2$ SCFTs \cite{Heckman:2022suy,Giacomelli:2025zqn,Giacomelli:2024ycb,Giacomelli:2024dbd}. In particular, the so-called A-type orbi-instanton theories \cite{Aspinwall:1997ye,DelZotto:2014hpa,Heckman:2015bfa,Mekareeya:2017jgc}, i.e., 6d $\mathcal{N}=(1,0)$ theories realized on M5-branes probing an M9-brane \cite{Horava:1996ma} on a $\CC^2/\ZZ_k$ singularity, were recently shown to provide the origin of class $\mathcal{S}$ \cite{Gaiotto:2009we,Gaiotto:2009gz} theories of type A with both regular and irregular untwisted punctures \cite{Giacomelli:2024ycb,Giacomelli:2025zqn}. It would be nice to use \texttt{Solver Agent} either to extend the statement to other kinds of 6d theories or to formally prove these relations.
    \item Another application in SCFTs is in the context of 3d mirror symmetry \cite{Intriligator:1996ex} of Argyres--Douglas (AD) theories \cite{Argyres:1995jj}. AD theories are $\mathcal{N}=2$ 4d SCFTs for  mutually nonlocal dyons that are massless, and one cannot 
go to a duality frame in which these dyons carry only electric charge, leading to intrinsically non-Lagrangian theories.\footnote{See, e.g., \cite{Akhond:2021xio,Argyres:2022mnu}, for recent reviews of these theories.} However, a lot of effort has been put in over the past years to propose Lagrangian 3d magnetic quivers corresponding to such AD theories \cite{Giacomelli:2020ryy,Carta:2021whq,Carta:2021dyx,Carta:2022spy,Carta:2022fxc}. It would be interesting if \texttt{Solver Agent} could be used to find the missing 3d magnetic quivers or, related to the previous point, provide a general proof of principles that relate the various theories. 
\end{itemize}

The use of LLMs has become an undeniable part of everyday life. This work demonstrates how agentic frameworks can support research by increasing efficiency and enabling access to computations that would otherwise be too cumbersome to perform. We believe that embracing this technological innovation should become part of future research practices. Importantly, this new way of working is not incompatible with --- and should not replace --- the traditional aspects that remain essential to research, namely exploration, curiosity, patience, understanding, and experimentation. In particular, LLMs should be incorporated into research by researchers themselves, in accordance with the scientific principles of transparency, reproducibility, and openness, notably through the use and development of open-source software; while remaining closely aligned with the established practices, norms, and workflows of the respective research communities.

\paragraph{AI Disclosure:} The code produced for this work was written with the help of \texttt{GPT 5.6 Sol} (OpenAI), \texttt{Gemini 3.1 Pro} (Google), \texttt{Claude 5 Opus} (Anthropic), and \texttt{Claude Fable 5} (Anthropic). The authors have used LLMs to improve the language and grammar of the paper.

\subsubsection*{Acknowledgments}

We thank P. Balavoine, F. Carta, M. Danese, M. R. Douglas, B. Hassfeld, C. Lawrie, J. Moritz, B. Pioline, F. Ruehle, G. Shiu and T. Weigand for helpful comments and discussions. The research of C.~F.~C. is supported by the Initiative Physique des Infinis at Sorbonne Universit\'e. A.~M. is supported in part by DOE (HEP) Awards DE-SC0017647 and DE-SC0023719. A.~M. thanks the ESI program ``The Unreasonable Effectiveness of Toric Geometry: Bridging Mathematics, Computation, and String Theory" and the Simons Center for Geometry and Physics during the ``23rd Simons Physics Summer Workshop: Theory, Experiment and the Emerging New Physics" for their hospitality.

\appendix

\section{Calabi--Yau Hypersurfaces in Toric Ambient Spaces}
\label{sec:KS}

A toric ambient space $\mathcal{A}$ of dimension $d$ is a complex algebraic variety that contains an algebraic torus $T \simeq (\CC^*)^d$ as a subset \cite{cox1995homogeneous, cox2011toric, Fulton}. Geometrically, one constructs this space by considering $k$ homogeneous coordinates $x_i$ as
\begin{equation}
    \mathcal{A} \simeq \left(\CC^k \setminus Z\right)/(\CC^*)^r\coma
\end{equation}
where $r=k-d$ is the rank of the toric variety, giving the dimension of the Picard group.  The torus $T$ defines a lattice $N = \mathrm{Hom}(\CC^*, T) \simeq \ZZ^d$ and a dual character lattice $M = \mathrm{Hom}(T, \CC^*) \simeq \ZZ^d$, alongside their real vector spaces $N_\RR = N\otimes_\ZZ \RR$ and $M_\RR = M\otimes_\ZZ \RR$. The topological data of the toric variety is captured by a fan $\Sigma_F \subset N_\RR$, defined as a collection of strongly convex rational polyhedral cones $\sigma$ intersecting along shared faces $\tau$. The homogeneous coordinates $x_i$ are in one-to-one correspondence with the rays $\nu_i\in \Sigma_F$. Consequently, the exclusion set $Z$ is the zero locus of the Stanley--Reisner (SR) ideal
\begin{equation}
    \text{SR} = \left\langle \prod_{\nu_i\notin \sigma} x_i \mid \sigma \in \Sigma_F \right\rangle\coma
\end{equation}
eliminating the coordinate combinations that do not share a common cone in the fan.

The construction of a Calabi--Yau threefold hypersurface $\IX_3$ in toric ambient space uses precisely this correspondence between the algebraic torus $T$ defining the toric variety and the lattices $N$ and $M$ to associate each Calabi--Yau with a polytope $\Delta\subset M_\RR$ and its dual $\Delta^\circ\subset N_\RR$  \cite{Batyrev:1993oya,Kreuzer:2000xy}. The geometry is captured by the normal fan $\Sigma_\Delta$, constructed by considering the cones $\sigma$ over the proper faces $\tau$ of $\Delta^\circ$ with their apexes at the origin \cite{de2010loera}. This normal fan $\Sigma_\Delta$ defines a toric fourfold $\mathcal{A}_{\Sigma_\Delta}$ containing a potentially singular Calabi--Yau hypersurface \cite{cox2011toric, Fulton, reid1983decomposition}.

The Calabi--Yau hypersurface can be made smooth via crepant resolutions, which translates to refining the normal fan $\Sigma_\Delta$ into a simplicial fan $\Sigma_F(\mathcal{T})$ through a FRST $\mathcal{T}$ of $\Delta^\circ$ \cite{de2010loera,Altman:2014bfa}. Such triangulation leads to a smooth ambient toric fourfold $\mathcal{A} \equiv \mathcal{A}_{\Delta^\circ,\mathcal{T}}$ where a generic hypersurface is the zero locus of a polynomial formed by associating a monomial to every point $m\in \Delta$ \cite{Batyrev:1993oya}:
\begin{equation}
    x^{[m]} = \prod_{i=1}^{n_\text{div}} x_i^{\langle m,\nu_i\rangle+1}\,,
\end{equation}
where $\nu_i$ are the edge vectors of the refined fan $\Sigma_F(\mathcal{T})$, and $x_i$ are the corresponding homogeneous coordinates \cite{cox1995homogeneous}. By choosing this defining polynomial to be a section of the anticanonical line bundle $\mathcal{O}_\mathcal{A}(-K)$, where $K = -\sum_{i=1}^{n_\text{div}} D_i$ is the canonical bundle of $\mathcal{A}$ and $D_i \equiv \{x_i = 0\}$ are the prime toric divisors, the resulting hypersurface is Calabi--Yau with normal bundle $N_{X_3} = \mathcal{O}_\mathcal{A}(-K)$.

The non-trivial Hodge numbers of $\IX_n$ are extracted from the data of the polytope \cite{Batyrev:1993oya}:
\begin{equation}
\label{eq:h11h21eqs}
\begin{split}
    h^{1,1}(\IX_3)&=\ell(\Delta^\circ)-(n+2)-\sum_{\Gamma^\circ}\ell^*(\Gamma^\circ)+\sum_{\Theta^\circ}\ell^*(\Theta^\circ)\ell^*(\hat{\Theta}^\circ)\,,\\
    h^{n-1,1}(\IX_3)&=\ell(\Delta)-(n+2)-\sum_\Gamma\ell^*(\Gamma)+\sum_\Theta\ell^*(\Theta)\ell^*(\hat{\Theta})\,,
\end{split}
\end{equation}
where $\ell(\alpha)$ and $\ell^*(\alpha)$ are the number of lattice points and interior lattice points of a face $\alpha$, respectively. The sets $\Gamma$ ($\Gamma^\circ$) and $\Theta$ ($\Theta^\circ$) represent codimension-$1$ and codimension-$2$ faces of $\Delta$ ($\Delta^\circ$), while $\hat{\Theta}$ ($\hat{\Theta}^\circ$) is the dual face. For favorable Calabi--Yau manifolds, $h^{1,1}(\IX_3)$ coincides with $h^{1,1}(\mathcal{A})$, and one can directly expand the K\"ahler form $J$ of the Calabi--Yau in terms of the toric divisors of the ambient space. The linear equivalence relations among the prime toric divisors, i.e. $\sum_i \langle m, \nu_i \rangle D_i = 0$, correspond to the linear relations among the fan rays, $\sum_i Q_i^a \nu_i = 0$. This defines the gauged linear sigma model (GLSM) \cite{Witten:1993yc,hori2003mirror} charge matrix $Q_i^a$, which identifies a basis of divisors for $H^{1,1}(\mathcal{A})$ that descends to $\IX_3$. The intersection numbers $\kappa_{abc}$ of the Calabi--Yau threefold are then evaluated within the ambient fourfold $\mathcal{A}$ by uplifting the integral via the first Chern class of the normal bundle, $c_1(N_{X_3}) = \sum_{i=1}^{n_\text{div}} D_i$:
\begin{equation}
    \kappa_{abc} = \int_{\IX_3} D_a \wedge D_b \wedge D_c = \int_{\mathcal{A}} D_a \wedge D_b \wedge D_c \wedge c_1(N_{X_3})\,.
\end{equation}

\subsection{Orientifolding Kreuzer--Skarke}
\label{sec:systematic_F_theory_uplifts}

The values of $h^{1,1}_+$ and $h^{1,1}_-$ can be extracted independently by projecting the orientifold involution onto the Picard lattice of the Calabi--Yau \cite{Crino:2022zjk,Jefferson:2022ssj,Hassfeld:2026rzd}, as we will explain below. 

In practice, constructing an orientifold requires finding the locations that are left invariant under the involution, meaning we must identify the fixed loci in the Calabi--Yau manifold. To do this, for a Calabi--Yau hypersurface in toric ambient space, one can use the formalism introduced in Section \ref{sec:KS} and translate the involutions to actions on the homogeneous coordinates of the ambient space. A general geometric involution $\iota$ takes the form $x_i \mapsto b_i x_{\rho(i)}$ \cite{Moritz:2023jdb,Hassfeld:2026rzd}, combining a coordinate permutation $\rho$ with a diagonal sign shift characterized by $b_i$. The components of these actions can be of three main types: coordinate reflections $x_i \mapsto -x_i$ (see, e.g., \cite{Crino:2022zjk,Jefferson:2022ssj}), coordinate permutations $x_i \mapsto x_{\rho(i)}$ (see, e.g., \cite{Gao:2013pra,Altman:2021pyc,Gao:2021xbs,Collinucci:2009uh,Gao:2022fdi}), and generalized diagonal shifts \cite{Carta:2020ohw,Moritz:2023jdb,Hassfeld:2026rzd}.

The permutation $\rho\in S_{n_\text{div}}$, with $S_{n_\text{div}}$ being the symmetric group of the $n_\text{div}$ homogeneous coordinates, is described as the action of a matrix $P_\rho\in \mathrm{GL}(4,\ZZ)$ on the lattice $N$ such that the normal fan $\Sigma_F(\mathcal{T})$ is preserved, i.e., $P_\rho \nu_i = \nu_{\rho(i)}$. On the other hand, $b_i\in\{-1,1\}$ defines the $\ZZ_2$ lattice shifts, which are characterized by half-lattice torsion points $\xi \in \frac{1}{2}M / M$ on the dual lattice. Any vector $\xi$ can be represented as a four-dimensional vector with entries being $\{0,\frac{1}{2}\}$. For each of these 16 distinct vectors, we can compute the scalar product with the rays $\nu_i$ of the normal fan $\Sigma_F(\mathcal{T})$, and the sign vector on the corresponding homogeneous coordinate is given by \cite{Moritz:2023jdb,Hassfeld:2026rzd}
\begin{equation}
    b_i = (-1)^{2\langle \xi, \nu_i \rangle}\fstop
\end{equation}

With the involution action on the coordinates defined, the fixed loci determine the location of the orientifold planes. If the coordinates are mapped to minus themselves, the fixed locus requires setting them to zero. On the other hand, for coordinates that are exchanged by a permutation, one instead defines the invariant and anti-invariant combinations \cite{Gao:2021xbs, Altman:2021pyc}.  If the vanishing of these coordinates is forbidden by the SR ideal, that locus cannot admit O-planes \cite{Crino:2022zjk}. All allowed fixed loci are then classified by their complex codimension with respect to the Calabi--Yau threefold \cite{Gimon:1996rq}: codimension-one loci correspond to O7-planes, and isolated points correspond to O3-planes \cite{Garcia-Etxebarria:2015wns}. Similarly, codimension-two loci lead to O5-planes, and space-filling fixed loci (codimension-zero) correspond to O9-planes.\footnote{Depending on the choice of discrete torsion, the orientifold planes can appear as either $O^-$- or $O^+$-planes, which carry opposite RR charges and tensions \cite{Polchinski:1995mt}. In this work, we will restrict our attention entirely to the standard case of $O^-$-planes, which carry negative D-brane charges and thus require the introduction of D-branes and fluxes for tadpole cancellation.}

Finally, we can use the involution action to determine the splitting of the K\"ahler moduli. The splitting depends on the permutation $\rho$ and is computed by projecting its matrix representation $\mathbf{P}=\{P_{ij}\}$ with $P_{ij} = \delta_{i,\rho(j)}$ onto the Picard lattice. By considering the GLSM charge matrix $\mathbf{Q} = \{Q_i^a\}$, one defines $\Lambda = \mathbf{Q} \mathbf{P}\mathbf{Q}^+$,\footnote{Because $\mathbf{Q}$ is a rectangular $h^{1,1} \times n_\text{div}$ matrix, it does not possess a standard inverse. The pseudo-inverse $\mathbf{Q}^+ = \mathbf{Q}^T(\mathbf{Q}\mathbf{Q}^T)^{-1}$ allows us to map the coordinate permutations back down to the K\"ahler moduli.} and 
\begin{equation}
    h^{1,1}_\pm = \frac{1}{2}(h^{1,1} \pm \mathrm{Tr}(\Lambda))\fstop
\end{equation}

As established in Section~\ref{sec:tadpole_fluxes}, rigid divisors intersecting the O7-plane prevent the D7-branes from recombining into a smooth Whitney umbrella~\cite{Collinucci:2008pf,Braun:2008ua,Collinucci:2008sq}. Because of this geometric rigidity, the sections $\eta$ and $\psi$ cannot deform away from zero, forcing every section of $f$ and $g$ to vanish to degrees 2 and 3 along the divisor. This forms a non-Higgsable cluster~\cite{Morrison:2012np,Morrison:2012js,Morrison:2014lca,Halverson:2015jua,Halverson:2016vwx,Halverson:2017vde}. The polynomial $\Delta$ vanishes to degree six, and the elliptic fiber is forced into an $I_0^*$ degeneration~\cite{Sen:1996vd,Sen:1997gv}, rendering $\IY_4^s$ singular along the rigid divisor.

A crepant K\"ahler resolution of these $I_0^*$ singularities smooths the Calabi--Yau fourfold. Physically, this geometric resolution corresponds to moving onto the Coulomb branch of the effective field theory, breaking the non-abelian gauge symmetry to its Cartan subgroup~\cite{Morrison:1996xf,Intriligator:1997pq}. The resolution is achievable via toric blowups in the ambient space. For every prime toric divisor $D_i$ hosting an NHC, we refine the fan $\Sigma_{V_6^s}$ of the ambient space by introducing two exceptional rays:
\begin{equation}
    \nu_{e_1} = \nu_i + \nu_x + 2\nu_y\,, \qquad \nu_{e_2} = 2\nu_i + 2\nu_x + 3\nu_y\coma
\end{equation}
where $\nu_x$ and $\nu_y$ correspond to the homogeneous coordinates $x$ and $y$ of the fiber.\footnote{This specific refinement resolves the $I_0^*$ singularity by performing a local star subdivision on the cones containing $(\nu_i, \nu_x, \nu_y)$.} These rays correspond to the exceptional divisors that blow up the singular strata. The smooth toric sixfold $V_6$ allows the resolved Calabi--Yau fourfold $\IY_4 \subset V_6$ to be expressed in terms of the blown-up homogeneous coordinates, exposing the exceptional divisors of the Cartan algebra. The resulting $\IY_4$ is smooth away from point-like terminal $\mathbb{Z}_2$ orbifold singularities inherited from the base $\widetilde{\mathcal{A}}$.

\paragraph{Example: A Two-Parameter Toric Hypersurface}

The toric rays of the ambient space are generated by the vertices
\begin{equation}
\begin{blockarray}{crrrrrr}
	&&&&&l^1&l^2 \\
\begin{block}{c(rrrr|rr)}
	v_1 & -1 & -1 & 0 & 0 & 1 & 0 \\
	v_2 & -1 & -1 & 0 & 1 & 0 & 1 \\
	v_3 & -1 & -1 & 1 & 0 & 0 & 1 \\
	v_4 & -1 & 7 & -1 & -1 & 0 & 1 \\
	v_5 & 1 & -1 & 0 & 0 & 1 & 3 \\
	v_6 & 0 & -1 & 0 & 0 & -2 & 2 \\
\end{block}
\end{blockarray}\,
\end{equation}
This Calabi--Yau threefold has Hodge numbers $h^{1,1} = 2$ and $h^{2,1} = 106$. The Stanley--Reisner ideal for this triangulation is $\text{SRI} = \langle x_1 x_5, x_2 x_3 x_4 x_6 \rangle$. The prime toric divisors possess the following Hodge numbers:
\begin{equation}
\begin{array}{|c|c|c|c|c|}
\hline
& h^{0,0} & h^{1,0} & h^{2,0} & h^{1,1} \\\hline
D_1 & 1 & 0 & 0 & 1 \\\hline
D_2, D_3, D_4 & 1 & 0 & 2 & 29 \\\hline
D_5 & 1 & 0 & 13 & 104 \\\hline
D_6 & 1 & 21 & 0 & 2 \\\hline
\end{array}
\end{equation}

The orientifold involution is defined by a trivial permutation $\rho = \ID$ and $b_3 = -1$ (i.e., $x_3 \mapsto -x_3$). This implies that the K\"ahler moduli do not split ($h^{1,1}_- = 0$). 
This involution generates one O7-plane and one O3-plane. The fixed locus $D_5 = \{x_5 = 0\}$ wraps an O7-plane with $\chi(S_{O7}) = 132$. Because $h^{2,0}(D_5) = 13 \neq 0$, the generic D7-brane configuration canceling its tadpole is a recombined Whitney umbrella with Euler characteristic $\chi(S_{D7_{\text{WU}}}) = 13152$. The O7-plane at the fixed locus $D_1 = \{x_1 = 0\}$ wraps a completely rigid divisor ($h^{2,0}(D_1) = 0$ and $h^{1,0}(D_1) = 0$) with $\chi(S_{O7}) = 3$, which an $\mathrm{SO}(8)$ stack of four D7-branes and their four orientifold images cancels, contributing $\chi(S_{D7_{\mathrm{SO}(8)}}) = 24$. The invariant locus $\{x_2=x_3=x_4 = 0\}$ defines $N_{O3} = 1$ O3-plane.

The total D7-brane Euler characteristic contributing to the tadpole is the sum of the Whitney umbrella and the $\mathrm{SO}(8)$ stack contributions:
\begin{equation}
    \chi(S_{D7_{\text{tot}}}) = \chi(S_{D7_{\text{WU}}}) + \chi(S_{D7_{\mathrm{SO}(8)}}) = 13152 + 24 = 13176\,.
\end{equation}
The total O7-plane Euler characteristic evaluates to $\chi(S_{O7_{\text{tot}}}) = 132 + 3 = 135$. The geometric contribution to the D3-brane tadpole evaluates to 
\begin{equation}
    N_{D3} = \frac{N_{O3}}{2} + \frac{\chi(S_{O7_{\text{tot}}})}{6} + \frac{\chi(S_{D7_{\text{tot}}})}{24} = \frac{1}{2} + \frac{135}{6} + \frac{13176}{24} = 572\,.
\end{equation}
This tadpole, then, predicts $\chi_{\mathrm{st}}(\IY_4) =6864$.

\paragraph{Example: $\mathbb{P}_{[1,1,1,6,9]}[18]$ -- 2}

We now consider another involution of the examples considered in Section \ref{sec:X18Uplift}. All the toric data are the same, but the orientifold involution leaves $x_5 = 0$ fixed ($x_5 \mapsto -x_5$). This involution is defined by a trivial permutation $\rho = \ID$ and $b_5 = -1$, meaning that the K\"ahler moduli do not split and $h^{1,1}_- = 0$.  However, unlike the previous example, this involution gives rise to both O7-planes and O3-planes. The fixed locus of codimension one is $D_5 = \{x_5 = 0\}$, which wraps an O7-plane with Hodge numbers $h^{2,0}(D_5) = 2$ and $h^{1,1}(D_5)=30$, and an Euler characteristic $\chi(D_5) = 36$. The fixed loci of codimension three give rise to O3-planes. Specifically, the invariant locus $\{x_1=x_2=x_3=0\}$ represents the intersection of three divisors in the four-dimensional ambient toric space, defining a curve. When intersected with the Calabi--Yau hypersurface, this curve punctures the geometry at exactly $\int_{\IX_3} [D_1] \wedge [D_2] \wedge [D_3] = 3$ distinct, physically separated points. Consequently, this single algebraic locus corresponds to three distinct O3-planes. Similarly, the locus $\{x_1=x_3=x_6=0\}$ evaluates to an intersection number of $1$, corresponding to a single isolated O3-plane. Together, they yield a total of $N_{O3} = 4$ O3-planes.

Because $D_5$ is not rigid, the generic D7-brane configuration canceling its tadpole is a recombined Whitney umbrella rather than a coincident $\mathrm{SO}(8)$ stack. However, because the self-intersection of $D_5$ vanishes ($D_5^3 = 0$), the Whitney umbrella has no self-intersection points. Consequently, both configurations would degenerate to the same topology and yield the same D7-brane Euler characteristic:
\begin{equation}
    \chi(S_{D7_{\text{tot}}}) = 288\,.
\end{equation}
The total O7-plane Euler characteristic is $\chi(S_{O7_{\text{tot}}}) = 36$. The geometric contribution to the D3-brane tadpole evaluates to 
\begin{equation}
    N_{D3} = \frac{N_{O3}}{2} + \frac{\chi(S_{O7_{\text{tot}}})}{6} + \frac{\chi(S_{D7_{\text{tot}}})}{24} = \frac{4}{2} + \frac{36}{6} + \frac{288}{24} = 2 + 6 + 12 = 20\,.
\end{equation}
The stringy Euler characteristic for the Calabi--Yau fourfold that corresponds to the F-theory uplift of this orientifold configuration is then $\chi_{\mathrm{st}}(\IY_4) =240$, where $\int c_4(T\IY_4) = 216$, with the 4 O3-planes giving the correction of $+24$ needed to match the tadpole.

\bibliographystyle{JHEP}
\bibliography{ref}

@article{Cao:2023gvn,
    author = "Cao, Yalong and Kool, Martijn and Monavari, Sergej",
    title = "{A Donaldson-Thomas crepant resolution conjecture on Calabi-Yau 4-folds}",
    eprint = "2301.11629",
    archivePrefix = "arXiv",
    primaryClass = "math.AG",
    reportNumber = "RIKEN-iTHEMS-Report-23",
    doi = "10.1090/tran/9027",
    journal = "Trans. Am. Math. Soc.",
    volume = "376",
    number = "11",
    pages = "8225--8268",
    year = "2023"
}

@article{Grassi:2026ept,
    author = "Grassi, Antonella and Miranda, Rick and Paranjape, Kapil and Srinivas, Vasudevan and Weigand, Timo",
    title = "{Bounds on the Mordell-Weil rank of elliptic fibrations}",
    eprint = "2603.24666",
    archivePrefix = "arXiv",
    primaryClass = "math.AG",
    month = "3",
    year = "2026"
}

@article{Apruzzi:2020pmv,
    author = {Apruzzi, Fabio and Giacomelli, Simone and Sch{\"a}fer-Nameki, Sakura},
    title = "{4d $\mathcal{N}=2$ S-folds}",
    eprint = "2001.00533",
    archivePrefix = "arXiv",
    primaryClass = "hep-th",
    doi = "10.1103/PhysRevD.101.106008",
    journal = "Phys. Rev. D",
    volume = "101",
    number = "10",
    pages = "106008",
    year = "2020"
}

@article{Kreuzer:1998vb,
    author = "Kreuzer, Maximilian and Skarke, Harald",
    title = "{Classification of reflexive polyhedra in three-dimensions}",
    eprint = "hep-th/9805190",
    archivePrefix = "arXiv",
    reportNumber = "UTTG-07-98, TUW-98-13",
    doi = "10.4310/ATMP.1998.v2.n4.a5",
    journal = "Adv. Theor. Math. Phys.",
    volume = "2",
    pages = "853--871",
    year = "1998"
}

@incollection{NAKAYAMA1988405,
title = {On Weierstrass Models},
editor = {Hiroaki Hijikata and Heisuke Hironaka and Masaki Maruyama and Hideyuki Matsumura and Masayoshi Miyanishi and Tadao Oda and Kenji Ueno},
booktitle = {Algebraic Geometry and Commutative Algebra},
publisher = {Academic Press},
pages = {405-431},
year = {1988},
isbn = {978-0-12-348032-3},
doi = {https://doi.org/10.1016/B978-0-12-348032-3.50004-9},
url = {https://www.sciencedirect.com/science/article/pii/B9780123480323500049},
author = {Noboru Nakayama}
}

@article{Nakayama2002Global,
  author  = {Nakayama, Noboru},
  title   = {Global structure of an elliptic fibration},
  journal = {Publications of the Research Institute for Mathematical Sciences},
  volume  = {38},
  number  = {3},
  pages   = {451--649},
  year    = {2002},
  doi     = {10.2977/prims/1145476270},
  url     = {https://ems.press/journals/prims/articles/2598}
}

@misc{StacksProject,
  author       = {The Stacks Project Authors},
  title        = {The Stacks Project},
  howpublished = {\url{https://stacks.math.columbia.edu}},
  year         = {2026}
}

@book{hartshorne1977algebraic,
  title     = {Algebraic Geometry},
  author    = {Hartshorne, Robin},
  isbn      = {978-0-387-90244-9},
  series    = {Graduate Texts in Mathematics},
  volume    = {52},
  year      = {1977},
  publisher = {Springer},
  address   = {New York}
}

@book{Bartocci:2009tcf,
    author = "Bartocci, Claudio and Bruzzo, Ugo and Hern{\'a}ndez Ruip{\'e}rez, Daniel",
    title = "{Fourier-Mukai and Nahm Transforms in Geometry and Mathematical Physics}",
    doi = "10.1007/b11801",
    isbn = "978-0-8176-3246-5, 978-0-8176-4663-9",
    publisher = {Birkh{\"a}user Boston},
    address = "Boston",
    year = "2009"
}

@article{Schuett:2009pej,
    author = "Schuett, Matthias and Shioda, Tetsuji",
    title = "{Elliptic Surfaces}",
    eprint = "0907.0298",
    archivePrefix = "arXiv",
    primaryClass = "math.AG",
    month = "7",
    year = "2009"
}

@article{Weigand:2018rez,
    author = "Weigand, Timo",
    title = "{F-theory}",
    eprint = "1806.01854",
    archivePrefix = "arXiv",
    primaryClass = "hep-th",
    reportNumber = "CERN-TH-2018-126",
    journal = "PoS",
    volume = "TASI2017",
    pages = "016",
    year = "2018"
}

@article{Hassfeld:2026rzd,
    author = "Hassfeld, Bjoern and Moritz, Jakob",
    title = "{Calabi-Yau Orientifold Hypersurfaces and their F-theory Uplifts}",
    eprint = "2606.19423",
    archivePrefix = "arXiv",
    primaryClass = "hep-th",
    month = "6",
    year = "2026"
}

@article{Alexandrov:2026rra,
    author = "Alexandrov, Sergei and Klemm, Albrecht and Pioline, Boris",
    title = "{Large Order Enumerative Geometry, Black Holes and Black Rings}",
    eprint = "2605.19552",
    archivePrefix = "arXiv",
    primaryClass = "hep-th",
    month = "5",
    year = "2026"
}

@inproceedings{Morrison2008TerminalQS,
  title={Terminal Quotient Singularities in Dimensions Three and Four Author ( s ) :},
  author={David R. Morrison and Glenn Stevens},
  year={2008},
  url={https://api.semanticscholar.org/CorpusID:268524629},
  booktitle = {""},
}

@article{Kreuzer:2002uu,
    author = "Kreuzer, Maximilian and Skarke, Harald",
    title = "{PALP: A Package for analyzing lattice polytopes with applications to toric geometry}",
    eprint = "math/0204356",
    archivePrefix = "arXiv",
    reportNumber = "TUW-02-10",
    doi = "10.1016/S0010-4655(03)00491-0",
    journal = "Comput. Phys. Commun.",
    volume = "157",
    pages = "87--106",
    year = "2004"
}

@article{Altman:2014bfa,
    author = "Altman, Ross and Gray, James and He, Yang-Hui and Jejjala, Vishnu and Nelson, Brent D.",
    title = "{A Calabi-Yau Database: Threefolds Constructed from the Kreuzer-Skarke List}",
    eprint = "1411.1418",
    archivePrefix = "arXiv",
    primaryClass = "hep-th",
    doi = "10.1007/JHEP02(2015)158",
    journal = "JHEP",
    volume = "02",
    pages = "158",
    year = "2015"
}

@article{Sethi:1996es,
    author = "Sethi, S. and Vafa, C. and Witten, Edward",
    title = "{Constraints on low dimensional string compactifications}",
    eprint = "hep-th/9606122",
    archivePrefix = "arXiv",
    reportNumber = "HUTP-96-A025, IASSNS-HEP-96-60",
    doi = "10.1016/S0550-3213(96)00483-X",
    journal = "Nucl. Phys. B",
    volume = "480",
    pages = "213--224",
    year = "1996"
}

@article{Blumenhagen:2009up,
    author = "Blumenhagen, Ralph and Grimm, Thomas W. and Jurke, Benjamin and Weigand, Timo",
    title = "{F-theory uplifts and GUTs}",
    eprint = "0906.0013",
    archivePrefix = "arXiv",
    primaryClass = "hep-th",
    reportNumber = "MPP-2009-66, SLAC-PUB-13654",
    doi = "10.1088/1126-6708/2009/09/053",
    journal = "JHEP",
    volume = "09",
    pages = "053",
    year = "2009"
}

@article{Collinucci:2008zs,
    author = "Collinucci, Andres",
    title = "{New F-theory lifts}",
    eprint = "0812.0175",
    archivePrefix = "arXiv",
    primaryClass = "hep-th",
    doi = "10.1088/1126-6708/2009/08/076",
    journal = "JHEP",
    volume = "08",
    pages = "076",
    year = "2009"
}

@article{Collinucci:2009uh,
    author = "Collinucci, Andres",
    title = "{New F-theory lifts. II. Permutation orientifolds and enhanced singularities}",
    eprint = "0906.0003",
    archivePrefix = "arXiv",
    primaryClass = "hep-th",
    doi = "10.1007/JHEP04(2010)076",
    journal = "JHEP",
    volume = "04",
    pages = "076",
    year = "2010"
}

@article{Carta:2020ohw,
    author = "Carta, Federico and Moritz, Jakob and Westphal, Alexander",
    title = "{A landscape of orientifold vacua}",
    eprint = "2003.04902",
    archivePrefix = "arXiv",
    primaryClass = "hep-th",
    reportNumber = "DESY-20-046",
    doi = "10.1007/JHEP05(2020)107",
    journal = "JHEP",
    volume = "05",
    pages = "107",
    year = "2020"
}

@article{Klemm:2007in,
	archiveprefix = {arXiv},
	author = {Klemm, A. and Pandharipande, R.},
	doi = {10.1007/s00220-008-0490-9},
	eprint = {math/0702189},
	journal = {Commun. Math. Phys.},
	pages = {621--653},
	title = {{Enumerative geometry of Calabi-Yau 4-folds}},
	volume = {281},
	year = {2008}}

@article{DelZotto:2017mee,
    author = "Del Zotto, Michele and Gu, Jie and Huang, Min-Xin and Kashani-Poor, Amir-Kian and Klemm, Albrecht and Lockhart, Guglielmo",
    title = "{Topological Strings on Singular Elliptic Calabi-Yau 3-folds and Minimal 6d SCFTs}",
    eprint = "1712.07017",
    archivePrefix = "arXiv",
    primaryClass = "hep-th",
    doi = "10.1007/JHEP03(2018)156",
    journal = "JHEP",
    volume = "03",
    pages = "156",
    year = "2018"
}

@article{Bae:2022pif,
    author = "Bae, Younghan and Kool, Martijn and Park, Hyeonjun",
    title = "{Counting surfaces on Calabi-Yau 4-folds I: Foundations}",
    eprint = "2208.09474",
    archivePrefix = "arXiv",
    primaryClass = "math.AG",
    doi = "10.2140/gt.2026.30.2431",
    journal = "Geom. Topol.",
    volume = "30",
    pages = "2431--2564",
    year = "2026"
}

@article{Klevers:2014bqa,
    author = "Klevers, Denis and Mayorga Pena, Damian Kaloni and Oehlmann, Paul-Konstantin and Piragua, Hernan and Reuter, Jonas",
    title = "{F-Theory on all Toric Hypersurface Fibrations and its Higgs Branches}",
    eprint = "1408.4808",
    archivePrefix = "arXiv",
    primaryClass = "hep-th",
    reportNumber = "UPR-1264-T, CERN-PH-TH-2014-171, Bonn-TH-2014-13",
    doi = "10.1007/JHEP01(2015)142",
    journal = "JHEP",
    volume = "01",
    pages = "142",
    year = "2015"
}

@article{Arras:2016evy,
    author = "Arras, Philipp and Grassi, Antonella and Weigand, Timo",
    title = "{Terminal Singularities, Milnor Numbers, and Matter in F-theory}",
    eprint = "1612.05646",
    archivePrefix = "arXiv",
    primaryClass = "hep-th",
    doi = "10.1016/j.geomphys.2017.09.001",
    journal = "J. Geom. Phys.",
    volume = "123",
    pages = "71--97",
    year = "2018"
}

@article{Taylor:2025gnp,
    author = "Taylor, Washington and Wang, Yi-Nan and Yu, Yihang",
    title = "{Statistics of base polytopes in F-theory}",
    eprint = "2509.13252",
    archivePrefix = "arXiv",
    primaryClass = "hep-th",
    reportNumber = "MIT-CTP/5924",
    doi = "10.1007/JHEP03(2026)032",
    journal = "JHEP",
    volume = "03",
    pages = "032",
    year = "2026"
}

@article{Heckman:2020svr,
    author = "Heckman, Jonathan J. and Lawrie, Craig and Rochais, Thomas B. and Zhang, Hao Y. and Zoccarato, Gianluca",
    title = "{$S$-folds, string junctions, and $\mathcal{N} = 2$ SCFTs}",
    eprint = "2009.10090",
    archivePrefix = "arXiv",
    primaryClass = "hep-th",
    doi = "10.1103/PhysRevD.103.086013",
    journal = "Phys. Rev. D",
    volume = "103",
    number = "8",
    pages = "086013",
    year = "2021"
}

@article{Kollr2008ExercisesIT,
  author = {{Koll{\'a}r}, J{\'a}nos},
        title = "{Exercises in the birational geometry of algebraic varieties}",
      journal = {arXiv e-prints},
         year = 2008,
        month = sep,
          eid = {arXiv:0809.2579},
        pages = {arXiv:0809.2579},
          doi = {10.48550/arXiv.0809.2579},
archivePrefix = {arXiv},
       eprint = {0809.2579},
 primaryClass = {math.AG},
       adsurl = {https://ui.adsabs.harvard.edu/abs/2008arXiv0809.2579K}
}

@article{Grassi:2018rva,
    author = "Grassi, Antonella and Weigand, Timo",
    title = "{On topological invariants of algebraic threefolds with ($\mathbb Q$-factorial) singularities}",
    eprint = "1804.02424",
    archivePrefix = "arXiv",
    primaryClass = "math.AG",
    reportNumber = "CERN-TH-2018-013",
    month = "4",
    year = "2018"
}

@article{Mayrhofer:2014laa,
    author = "Mayrhofer, Christoph and Palti, Eran and Till, Oskar and Weigand, Timo",
    title = "{On Discrete Symmetries and Torsion Homology in F-Theory}",
    eprint = "1410.7814",
    archivePrefix = "arXiv",
    primaryClass = "hep-th",
    doi = "10.1007/JHEP06(2015)029",
    journal = "JHEP",
    volume = "06",
    pages = "029",
    year = "2015"
}

@article{Lin:2015qsa,
    author = "Lin, Ling and Mayrhofer, Christoph and Till, Oskar and Weigand, Timo",
    title = "{Fluxes in F-theory Compactifications on Genus-One Fibrations}",
    eprint = "1508.00162",
    archivePrefix = "arXiv",
    primaryClass = "hep-th",
    doi = "10.1007/JHEP01(2016)098",
    journal = "JHEP",
    volume = "01",
    pages = "098",
    year = "2016"
}

@article{Klemm:1996ts,
    author = "Klemm, A. and Lian, B. and Roan, S. S. and Yau, Shing-Tung",
    title = "{Calabi-Yau fourfolds for M theory and F theory compactifications}",
    eprint = "hep-th/9701023",
    archivePrefix = "arXiv",
    reportNumber = "EFI-97-01",
    doi = "10.1016/S0550-3213(97)00798-0",
    journal = "Nucl. Phys. B",
    volume = "518",
    pages = "515--574",
    year = "1998"
}

@article{Lee:2020gvu,
	archiveprefix = {arXiv},
	author = {Lee, Seung-Joo and Lerche, Wolfgang and Lockhart, Guglielmo and Weigand, Timo},
	doi = {10.1007/JHEP01(2021)162},
	eprint = {2005.10837},
	journal = {JHEP},
	pages = {162},
	primaryclass = {hep-th},
	title = {{Quasi-Jacobi forms, elliptic genera and strings in four dimensions}},
	volume = {01},
	year = {2021}}

@article{Lee:2019tst,
	archiveprefix = {arXiv},
	author = {Lee, Seung-Joo and Lerche, Wolfgang and Weigand, Timo},
	doi = {10.1007/JHEP08(2019)104},
	eprint = {1901.08065},
	journal = {JHEP},
	pages = {104},
	primaryclass = {hep-th},
	title = {{Modular Fluxes, Elliptic Genera, and Weak Gravity Conjectures in Four Dimensions}},
	volume = {08},
	year = {2019}}

@misc{douglas2026formalizationqft,
      title={Formalization of QFT}, 
      author={Michael R. Douglas and Sarah Hoback and Anna Mei and Ron Nissim},
      year={2026},
      eprint={2603.15770},
      archivePrefix={arXiv},
      primaryClass={hep-th},
      url={https://arxiv.org/abs/2603.15770}, 
}

@article{Bergman:2009zh,
    author = "Bergman, Oren and Hirano, Shinji",
    title = "{Anomalous radius shift in AdS(4)/CFT(3)}",
    eprint = "0902.1743",
    archivePrefix = "arXiv",
    primaryClass = "hep-th",
    doi = "10.1088/1126-6708/2009/07/016",
    journal = "JHEP",
    volume = "07",
    pages = "016",
    year = "2009"
}

@article{Altman:2021pyc,
    author = "Altman, Ross and Carifio, Jonathan and Gao, Xin and Nelson, Brent D.",
    title = "{Orientifold Calabi-Yau threefolds with divisor involutions and string landscape}",
    eprint = "2111.03078",
    archivePrefix = "arXiv",
    primaryClass = "hep-th",
    doi = "10.1007/JHEP03(2022)087",
    journal = "JHEP",
    volume = "03",
    pages = "087",
    year = "2022"
}

@article{Hanany:2000fq,
    author = "Hanany, Amihay and Kol, Barak",
    title = "{On orientifolds, discrete torsion, branes and M theory}",
    eprint = "hep-th/0003025",
    archivePrefix = "arXiv",
    reportNumber = "MIT-CTP-2946, TAUP-2609-2000, NSF-ITP-00-07",
    doi = "10.1088/1126-6708/2000/06/013",
    journal = "JHEP",
    volume = "06",
    pages = "013",
    year = "2000"
}

@article{Esole:2017kyr,
	archiveprefix = {arXiv},
	author = {Esole, Mboyo and Jefferson, Patrick and Kang, Monica Jinwoo},
	doi = {10.1007/s00220-019-03517-1},
	eprint = {1703.00905},
	journal = {Commun. Math. Phys.},
	number = {1},
	pages = {99--144},
	primaryclass = {math.AG},
	title = {{Euler Characteristics of Crepant Resolutions of Weierstrass Models}},
	volume = {371},
	year = {2019}}

@article{Witten:1996bn,
	archiveprefix = {arXiv},
	author = {Witten, Edward},
	doi = {10.1016/0550-3213(96)00283-0},
	eprint = {hep-th/9604030},
	journal = {Nucl. Phys. B},
	pages = {343--360},
	reportnumber = {IASSNS-HEP-96-29},
	title = {{Nonperturbative superpotentials in string theory}},
	volume = {474},
	year = {1996}}

@article{Demirtas:2019sip,
	archiveprefix = {arXiv},
	author = {Demirtas, Mehmet and Kim, Manki and Mcallister, Liam and Moritz, Jakob},
	doi = {10.1103/PhysRevLett.124.211603},
	eprint = {1912.10047},
	journal = {Phys. Rev. Lett.},
	number = {21},
	pages = {211603},
	primaryclass = {hep-th},
	title = {{Vacua with Small Flux Superpotential}},
	volume = {124},
	year = {2020}}

@article{cox1995homogeneous,
 author = {{Cox}, David A.},
        title = "{Erratum to ``The Homogeneous Coordinate Ring of a Toric Variety'', along with the original paper}",
      journal = {arXiv e-prints},
         year = 1992,
        month = oct,
          eid = {alg-geom/9210008},
        pages = {alg-geom/9210008},
          doi = {10.48550/arXiv.alg-geom/9210008},
archivePrefix = {arXiv},
       eprint = {alg-geom/9210008},
 primaryClass = {math.AG},
       adsurl = {https://ui.adsabs.harvard.edu/abs/1992alg.geom.10008C}
    }

@article{Witten:1993yc,
	archiveprefix = {arXiv},
	author = {Witten, Edward},
	doi = {10.1016/0550-3213(93)90033-L},
	editor = {Greene, B. and Yau, Shing-Tung},
	eprint = {hep-th/9301042},
	journal = {Nucl. Phys. B},
	pages = {159--222},
	reportnumber = {IASSNS-HEP-93-3},
	title = {{Phases of N=2 theories in two-dimensions}},
	volume = {403},
	year = {1993}}

@book{hori2003mirror,
	author = {Hori, K. and Katz, S. and Vafa, C. and Thomas, R. and American Mathematical Society and Clay Mathematics Institute and Pandharipande, R. and Klemm, A.},
	isbn = {9780821829554},
	lccn = {03052414},
	publisher = {American Mathematical Society},
	series = {Clay mathematics monographs},
	title = {Mirror Symmetry},
	url = {https://books.google.com/books?id=WFsAqFDRqHgC},
	year = {2003}}

@article{Huang:2019pne,
    author = "Huang, Yu-Chien and Taylor, Washington",
    title = "{Fibration structure in toric hypersurface Calabi-Yau threefolds}",
    eprint = "1907.09482",
    archivePrefix = "arXiv",
    primaryClass = "hep-th",
    reportNumber = "MIT-CTP-5132",
    doi = "10.1007/JHEP03(2020)172",
    journal = "JHEP",
    volume = "03",
    pages = "172",
    year = "2020"
}

@book{Fulton,
SBN = {9780691000497},
 URL = {http://www.jstor.org/stable/j.ctt1b7x7vc},
 author = {William Fulton},
 publisher = {Princeton University Press},
 title = {Introduction to Toric Varieties. (AM-131)},
 urldate = {2026-09-21},
 year = {1993}
    }

@article{Garcia-Etxebarria:2015wns,
	archiveprefix = {arXiv},
	author = {Garc\'\i{}a-Etxebarria, I\~naki and Regalado, Diego},
	doi = {10.1007/JHEP03(2016)083},
	eprint = {1512.06434},
	journal = {JHEP},
	pages = {083},
	primaryclass = {hep-th},
	reportnumber = {MPP-2015-307},
	title = {{$ \mathcal{N}=3 $ four dimensional field theories}},
	volume = {03},
	year = {2016}}

@article{Candelas:1994hw,
	archiveprefix = {arXiv},
	author = {Candelas, Philip and Font, Anamaria and Katz, Sheldon H. and Morrison, David R.},
	doi = {10.1016/0550-3213(94)90155-4},
	eprint = {hep-th/9403187},
	journal = {Nucl. Phys. B},
	pages = {626--674},
	reportnumber = {UTTG-25-93, IASSNS-HEP-94-12, OSU-M-94-1},
	title = {{Mirror symmetry for two parameter models. 2.}},
	volume = {429},
	year = {1994}}

@article{Polchinski:1995mt,
	archiveprefix = {arXiv},
	author = {Polchinski, Joseph},
	doi = {10.1103/PhysRevLett.75.4724},
	eprint = {hep-th/9510017},
	journal = {Phys. Rev. Lett.},
	pages = {4724--4727},
	reportnumber = {NSF-ITP-95-122},
	title = {{Dirichlet Branes and Ramond-Ramond charges}},
	volume = {75},
	year = {1995}}

@article{Gimon:1996rq,
	archiveprefix = {arXiv},
	author = {Gimon, Eric G. and Polchinski, Joseph},
	doi = {10.1103/PhysRevD.54.1667},
	eprint = {hep-th/9601038},
	journal = {Phys. Rev. D},
	pages = {1667--1676},
	reportnumber = {NSF-ITP-96-01},
	title = {{Consistency conditions for orientifolds and D-manifolds}},
	volume = {54},
	year = {1996}}

@book{Polchinski:1998rq,
	author = {Polchinski, J.},
	doi = {10.1017/CBO9780511816079},
	isbn = {978-0-511-25227-3, 978-0-521-67227-6, 978-0-521-63303-1},
	month = {12},
	publisher = {Cambridge University Press},
	series = {Cambridge Monographs on Mathematical Physics},
	title = {{String theory. Vol. 1: An introduction to the bosonic string}},
	year = {2007}}

@book{Polchinski:1998rr,
	author = {Polchinski, J.},
	doi = {10.1017/CBO9780511618123},
	isbn = {978-0-511-25228-0, 978-0-521-63304-8, 978-0-521-67228-3},
	month = {12},
	publisher = {Cambridge University Press},
	series = {Cambridge Monographs on Mathematical Physics},
	title = {{String theory. Vol. 2: Superstring theory and beyond}},
	year = {2007}}

@article{morrison1984terminal,
ISSN = {00029939, 10886826},
 URL = {http://www.jstor.org/stable/2044659},
 author = {David R. Morrison and Glenn Stevens},
 journal = {Proceedings of the American Mathematical Society},
 number = {1},
 pages = {15--20},
 publisher = {American Mathematical Society},
 title = {Terminal Quotient Singularities in Dimensions Three and Four},
 urldate = {2026-09-21},
 volume = {90},
 year = {1984}
    }

@article{anno2003four,
	author = {Anno, Rina E},
	journal = {Mathematical Notes},
	number = {5},
	pages = {769--776},
	publisher = {Springer},
	title = {Four-dimensional terminal Gorenstein quotient singularities},
	volume = {73},
	year = {2003}}

@article{Esole:2017qeh,
	archiveprefix = {arXiv},
	author = {Esole, Mboyo and Jagadeesan, Ravi and Kang, Monica Jinwoo},
	eprint = {1709.04913},
	month = {9},
	primaryclass = {hep-th},
	title = {{The Geometry of G$_2$, Spin(7), and Spin(8)-models}},
	year = {2017}}

@article{Esole:2020alo,
	archiveprefix = {arXiv},
	author = {Esole, Mboyo and Pasterski, Sabrina},
	eprint = {2004.06104},
	month = {4},
	primaryclass = {hep-th},
	title = {{Flops and Fibral Geometry of E$_7$-models}},
	year = {2020}}

@article{Esole:2019ocl,
	archiveprefix = {arXiv},
	author = {Esole, Mboyo and Jefferson, Patrick},
	eprint = {1910.09536},
	month = {10},
	primaryclass = {hep-th},
	title = {{USp(4)-models}},
	year = {2019}}

@article{Esole:2019hgr,
	archiveprefix = {arXiv},
	author = {Esole, Mboyo and Jefferson, Patrick},
	eprint = {1905.12620},
	month = {5},
	primaryclass = {hep-th},
	title = {{The Geometry of SO(3), SO(5), and SO(6) models}},
	year = {2019}}

@article{Esole:2019asj,
	archiveprefix = {arXiv},
	author = {Esole, Mboyo and Jagadeesan, Ravi and Kang, Monica Jinwoo},
	eprint = {1905.05174},
	month = {5},
	primaryclass = {hep-th},
	title = {{48 Crepant Paths to $\text{SU}(2)\!\times\!\text{SU}(3)$}},
	year = {2019}}

@article{Esole:2018bmf,
	archiveprefix = {arXiv},
	author = {Esole, Mboyo and Kang, Monica Jinwoo},
	eprint = {1808.07054},
	month = {8},
	primaryclass = {hep-th},
	title = {{Characteristic numbers of elliptic fibrations with non-trivial Mordell-Weil groups}},
	year = {2018}}

@article{Esole:2018tuz,
	archiveprefix = {arXiv},
	author = {Esole, Mboyo and Kang, Monica Jinwoo},
	eprint = {1807.08755},
	month = {7},
	primaryclass = {hep-th},
	title = {{Characteristic numbers of crepant resolutions of Weierstrass models}},
	year = {2018}}

@article{Sen:1996vd,
	archiveprefix = {arXiv},
	author = {Sen, Ashoke},
	doi = {10.1016/0550-3213(96)00347-1},
	eprint = {hep-th/9605150},
	journal = {Nucl. Phys. B},
	pages = {562--578},
	reportnumber = {MRI-PHY-96-14},
	title = {{F theory and orientifolds}},
	volume = {475},
	year = {1996}}

@article{Morrison:2014lca,
	archiveprefix = {arXiv},
	author = {Morrison, David R. and Taylor, Washington},
	doi = {10.1007/JHEP05(2015)080},
	eprint = {1412.6112},
	journal = {JHEP},
	pages = {080},
	primaryclass = {hep-th},
	reportnumber = {UCSB-MATH-2014-38, MIT-CTP-4582},
	title = {{Non-Higgsable clusters for 4D F-theory models}},
	volume = {05},
	year = {2015}}

@article{batyrev1996strong,
	author = {Batyrev, Victor V and Dais, Dimitrios I},
	journal = {Topology},
	number = {4},
	pages = {901--929},
	publisher = {Elsevier},
	title = {Strong McKay correspondence, string-theoretic Hodge numbers and mirror symmetry},
	volume = {35},
	year = {1996}}

@article{Kachru:2003aw,
	archiveprefix = {arXiv},
	author = {Kachru, Shamit and Kallosh, Renata and Linde, Andrei D. and Trivedi, Sandip P.},
	doi = {10.1103/PhysRevD.68.046005},
	eprint = {hep-th/0301240},
	journal = {Phys. Rev. D},
	pages = {046005},
	reportnumber = {SLAC-PUB-9630, SU-ITP-03-01, TIFR-TH-03-03},
	title = {{De Sitter vacua in string theory}},
	volume = {68},
	year = {2003}}

@article{Katz:1996th,
	archiveprefix = {arXiv},
	author = {Katz, Sheldon H. and Vafa, Cumrun},
	doi = {10.1016/S0550-3213(97)00283-6},
	eprint = {hep-th/9611090},
	journal = {Nucl. Phys. B},
	pages = {196--204},
	reportnumber = {HUTP-96-A051, OSU-M-96-25, OSU-MATH-1996-25},
	title = {{Geometric engineering of N=1 quantum field theories}},
	volume = {497},
	year = {1997}}

@article{Diaconescu:1998ua,
	archiveprefix = {arXiv},
	author = {Diaconescu, Duiliu-Emanuel and Gukov, Sergei},
	doi = {10.1016/S0550-3213(98)00597-5},
	eprint = {hep-th/9804059},
	journal = {Nucl. Phys. B},
	pages = {171--196},
	reportnumber = {RU-98-11, PUPT-1780, ITEP-TH-52-97, LANDAU-97-TMP-4},
	title = {{Three-dimensional N=2 gauge theories and degenerations of Calabi-Yau four folds}},
	volume = {535},
	year = {1998}}

@article{Kim:2022uni,
	archiveprefix = {arXiv},
	author = {Kim, Manki},
	eprint = {2207.01440},
	month = {7},
	primaryclass = {hep-th},
	reportnumber = {MIT-CTP/5447},
	title = {{On D3-brane Superpotential}},
	year = {2022}}

@article{Grimm:2011dj,
	archiveprefix = {arXiv},
	author = {Grimm, Thomas W. and Kerstan, Max and Palti, Eran and Weigand, Timo},
	doi = {10.1103/PhysRevD.84.066001},
	eprint = {1105.3193},
	journal = {Phys. Rev. D},
	pages = {066001},
	primaryclass = {hep-th},
	title = {{On Fluxed Instantons and Moduli Stabilisation in IIB Orientifolds and F-theory}},
	volume = {84},
	year = {2011}}

@article{Bianchi:2011qh,
	archiveprefix = {arXiv},
	author = {Bianchi, Massimo and Collinucci, Andres and Martucci, Luca},
	doi = {10.1007/JHEP12(2011)045},
	eprint = {1107.3732},
	journal = {JHEP},
	pages = {045},
	primaryclass = {hep-th},
	title = {{Magnetized E3-brane instantons in F-theory}},
	volume = {12},
	year = {2011}}

@article{Kaufmann:2026fli,
    author = "Kaufmann, Lukas and Monnee, Jeroen and Weigand, Timo and Wiesner, Max",
    title = "{Quantum obstructions for $N=1$ infinite distance limits -- Part I: $g_s$ obstructions}",
    eprint = "2603.12315",
    archivePrefix = "arXiv",
    primaryClass = "hep-th",
    doi = "10.1103/blb9-hrwd",
    journal = "Phys. Rev. D",
    volume = "113",
    number = "12",
    pages = "126027",
    year = "2026"
}

@article{Kaufmann:2026tsy,
    author = "Kaufmann, Lukas and Weigand, Timo and Wiesner, Max",
    title = "{On Quantum Obstructions in Type IIA Orientifolds}",
    eprint = "2604.25988",
    archivePrefix = "arXiv",
    primaryClass = "hep-th",
    month = "4",
    year = "2026"
}

@article{Kaufmann:2026mha,
    author = "Kaufmann, Lukas and Monnee, Jeroen and Weigand, Timo and Wiesner, Max",
    title = {{Quantum obstructions for $N=1$ infinite distance limits -- Part II: K{\"a}hler obstructions}},
    eprint = "2603.13470",
    archivePrefix = "arXiv",
    primaryClass = "hep-th",
    doi = "10.1103/ypyx-mg6r",
    journal = "Phys. Rev. D",
    volume = "113",
    number = "12",
    pages = "126028",
    year = "2026"
}

@article{Blumenhagen:2006xt,
	archiveprefix = {arXiv},
	author = {Blumenhagen, Ralph and Cvetic, Mirjam and Weigand, Timo},
	doi = {10.1016/j.nuclphysb.2007.02.016},
	eprint = {hep-th/0609191},
	journal = {Nucl. Phys. B},
	pages = {113--142},
	reportnumber = {MPP-2006-119, NSF-KITP-06-73, UPR-1162-T},
	title = {{Spacetime instanton corrections in 4D string vacua: The Seesaw mechanism for D-Brane models}},
	volume = {771},
	year = {2007}}

@article{Ibanez:2006da,
	archiveprefix = {arXiv},
	author = {Ibanez, L. E. and Uranga, A. M.},
	doi = {10.1088/1126-6708/2007/03/052},
	eprint = {hep-th/0609213},
	journal = {JHEP},
	pages = {052},
	reportnumber = {IFT-UAM-CSIC-06-45, CERN-PH-TH-2006-199},
	title = {{Neutrino Majorana Masses from String Theory Instanton Effects}},
	volume = {03},
	year = {2007}}

@article{Beasley:2008dc,
	archiveprefix = {arXiv},
	author = {Beasley, Chris and Heckman, Jonathan J. and Vafa, Cumrun},
	doi = {10.1088/1126-6708/2009/01/058},
	eprint = {0802.3391},
	journal = {JHEP},
	pages = {058},
	primaryclass = {hep-th},
	title = {{GUTs and Exceptional Branes in F-theory - I}},
	volume = {01},
	year = {2009}}

@article{Blumenhagen:2012kz,
	archiveprefix = {arXiv},
	author = {Blumenhagen, Ralph and Gao, Xin and Rahn, Thorsten and Shukla, Pramod},
	doi = {10.1007/JHEP06(2012)162},
	eprint = {1205.2485},
	journal = {JHEP},
	pages = {162},
	primaryclass = {hep-th},
	reportnumber = {MPP-2012-87},
	title = {{A Note on Poly-Instanton Effects in Type IIB Orientifolds on Calabi-Yau Threefolds}},
	volume = {06},
	year = {2012}}

@article{Gao:2013pra,
	archiveprefix = {arXiv},
	author = {Gao, Xin and Shukla, Pramod},
	doi = {10.1007/JHEP11(2013)170},
	eprint = {1307.1139},
	journal = {JHEP},
	pages = {170},
	primaryclass = {hep-th},
	reportnumber = {MPP-2013-183},
	title = {{On Classifying the Divisor Involutions in Calabi-Yau Threefolds}},
	volume = {11},
	year = {2013}}

@article{Gao:2021xbs,
	archiveprefix = {arXiv},
	author = {Gao, Xin and Zou, Hao},
	doi = {10.1103/PhysRevD.105.046017},
	eprint = {2112.04950},
	journal = {Phys. Rev. D},
	number = {4},
	pages = {046017},
	primaryclass = {hep-th},
	title = {{Applying machine learning to the Calabi-Yau orientifolds with string vacua}},
	volume = {105},
	year = {2022}}

@article{Palti:2020qlc,
    author = "Palti, Eran and Vafa, Cumrun and Weigand, Timo",
    title = "{Supersymmetric Protection and the Swampland}",
    eprint = "2003.10452",
    archivePrefix = "arXiv",
    primaryClass = "hep-th",
    doi = "10.1007/JHEP06(2020)168",
    journal = "JHEP",
    volume = "06",
    pages = "168",
    year = "2020"
}

@book{cox2011toric,
	author = {Cox, David A and Little, John B and Schenck, Henry K},
	publisher = {American Mathematical Soc.},
	title = {Toric varieties},
	volume = {124},
	year = {2011}}

@incollection{reid1983decomposition,
author    = {Reid, Miles},
    title     = {Decomposition of toric morphisms},
    booktitle = {Arithmetic and Geometry},
    volume    = {II: Geometry},
    pages     = {395--418},
    publisher = {Birkh{\"a}user Boston},
    address   = {Boston, MA},
    year      = {1983},
    editor    = {Artin, M. and Tate, J.},
    series    = {Progress in Mathematics},
    isbn      = {978-1-4757-9288-1}
    }

@article{Morrison:2012js,
	archiveprefix = {arXiv},
	author = {Morrison, David R. and Taylor, Washington},
	doi = {10.1002/prop.201200086},
	eprint = {1204.0283},
	journal = {Fortsch. Phys.},
	pages = {1187--1216},
	primaryclass = {hep-th},
	reportnumber = {UCSB-MATH-2012-14, MIT-CTP-4356},
	title = {{Toric bases for 6D F-theory models}},
	volume = {60},
	year = {2012}}

@article{Morrison:2021wuv,
    author = "Morrison, David R. and Taylor, Washington",
    title = "{Charge completeness and the massless charge lattice in F-theory models of supergravity}",
    eprint = "2108.02309",
    archivePrefix = "arXiv",
    primaryClass = "hep-th",
    reportNumber = "MIT-CTP-5172, UCSB-MATH-2021-03",
    doi = "10.1007/JHEP12(2021)040",
    journal = "JHEP",
    volume = "12",
    pages = "040",
    year = "2021"
}

@article{Morrison:2012np,
	archiveprefix = {arXiv},
	author = {Morrison, David R. and Taylor, Washington},
	doi = {10.2478/s11534-012-0065-4},
	eprint = {1201.1943},
	journal = {Central Eur. J. Phys.},
	pages = {1072--1088},
	primaryclass = {hep-th},
	reportnumber = {UCSB-MATH-2012-01, MIT-CTP-4339},
	title = {{Classifying bases for 6D F-theory models}},
	volume = {10},
	year = {2012}}

@article{Halverson:2015jua,
	archiveprefix = {arXiv},
	author = {Halverson, James and Taylor, Washington},
	doi = {10.1007/JHEP09(2015)086},
	eprint = {1506.03204},
	journal = {JHEP},
	pages = {086},
	primaryclass = {hep-th},
	reportnumber = {NSF-KITP-15-068, MIT-CTP-4677},
	title = {{$ {\mathrm{\mathbb{P}}}^1 $-bundle bases and the prevalence of non-Higgsable structure in 4D F-theory models}},
	volume = {09},
	year = {2015}}

@misc{CYTools,
  author = {Demirtas, Mehmet and Rios-Tascon, Andres and McAllister, Liam},
  title = {{CYTools}: A Software Package for Analyzing Calabi-Yau Manifolds},
  howpublished = {\url{https://cy.tools}},
  eprint = {2211.03823},
  archivePrefix = {arXiv},
  note = {Repository: \url{https://github.com/LiamMcAllisterGroup/cytools} (License: GPL-3.0)}
}

@article{Seiberg:1994pq,
    author = "Seiberg, N.",
    title = "{Electric - Magnetic Duality in Supersymmetric Non-Abelian Gauge Theories}",
    eprint = "hep-th/9411149",
    archivePrefix = "arXiv",
    reportNumber = "RU-94-82, IASSNS-HEP-94-98",
    doi = "10.1016/0550-3213(94)00023-8",
    journal = "Nucl. Phys. B",
    volume = "435",
    pages = "129--146",
    year = "1995"
}

@article{Collinucci:2008pf,
	archiveprefix = {arXiv},
	author = {Collinucci, Andres and Denef, Frederik and Esole, Mboyo},
	doi = {10.1088/1126-6708/2009/02/005},
	eprint = {0805.1573},
	journal = {JHEP},
	pages = {005},
	primaryclass = {hep-th},
	title = {{D-brane Deconstructions in IIB Orientifolds}},
	volume = {02},
	year = {2009}}

@article{Jefferson:2022ssj,
    author = "Jefferson, Patrick and Kim, Manki",
    title = "{On the intermediate Jacobian of M5-branes}",
    eprint = "2211.00210",
    archivePrefix = "arXiv",
    primaryClass = "hep-th",
    reportNumber = "MIT-CTP-5470",
    doi = "10.1007/JHEP05(2024)180",
    journal = "JHEP",
    volume = "05",
    pages = "180",
    year = "2024"
}

@article{Crino:2022zjk,
    author = "Crin{\`o}, Chiara and Quevedo, Fernando and Schachner, Andreas and Valandro, Roberto",
    title = "{A database of Calabi-Yau orientifolds and the size of D3-tadpoles}",
    eprint = "2204.13115",
    archivePrefix = "arXiv",
    primaryClass = "hep-th",
    doi = "10.1007/JHEP08(2022)050",
    journal = "JHEP",
    volume = "08",
    pages = "050",
    year = "2022"
}

@article{Kreuzer:2000xy,
    author = "Kreuzer, Maximilian and Skarke, Harald",
    title = "{Complete classification of reflexive polyhedra in four-dimensions}",
    eprint = "hep-th/0002240",
    archivePrefix = "arXiv",
    reportNumber = "HUB-EP-00-13, TUW-00-07",
    doi = "10.4310/ATMP.2000.v4.n6.a2",
    journal = "Adv. Theor. Math. Phys.",
    volume = "4",
    pages = "1209--1230",
    year = "2000"
}

@article{Batyrev:1993oya,
    author = "Batyrev, Victor V.",
    title = "{Dual polyhedra and mirror symmetry for Calabi-Yau hypersurfaces in toric varieties}",
    eprint = "alg-geom/9310003",
    archivePrefix = "arXiv",
    journal = "J. Alg. Geom.",
    volume = "3",
    pages = "493--545",
    year = "1994"
}

@article{Cicoli:2021dhg,
    author = "Cicoli, Michele and Etxebarria, I{\~n}aki Garc{\'\i}a and Quevedo, Fernando and Schachner, Andreas and Shukla, Pramod and Valandro, Roberto",
    title = "{The Standard Model quiver in de Sitter string compactifications}",
    eprint = "2106.11964",
    archivePrefix = "arXiv",
    primaryClass = "hep-th",
    doi = "10.1007/JHEP08(2021)109",
    journal = "JHEP",
    volume = "08",
    pages = "109",
    year = "2021"
}

@article{Gendler:2022qof,
	archiveprefix = {arXiv},
	author = {Gendler, Naomi and Kim, Manki and McAllister, Liam and Moritz, Jakob and Stillman, Mike},
	eprint = {2204.06566},
	month = {4},
	primaryclass = {hep-th},
	reportnumber = {MIT-CTP/5388},
	title = {{Superpotentials from Singular Divisors}},
	year = {2022}}

@article{Freed:1999vc,
    author = "Freed, Daniel S. and Witten, Edward",
    title = "{Anomalies in string theory with D-branes}",
    eprint = "hep-th/9907189",
    archivePrefix = "arXiv",
    journal = "Asian J. Math.",
    volume = "3",
    pages = "819",
    year = "1999"
}

@article{Moritz:2023jdb,
    author = "Moritz, Jakob",
    title = "{Orientifolding Kreuzer-Skarke}",
    eprint = "2305.06363",
    archivePrefix = "arXiv",
    primaryClass = "hep-th",
    month = "5",
    year = "2023"
}

@article{Blumenhagen:2008zz,
    author = "Blumenhagen, Ralph and Braun, Volker and Grimm, Thomas W. and Weigand, Timo",
    title = "{GUTs in Type IIB Orientifold Compactifications}",
    eprint = "0811.2936",
    archivePrefix = "arXiv",
    primaryClass = "hep-th",
    reportNumber = "MPP-2008-144, DIAS-STP-08-15, SLAC-PUB-13466",
    doi = "10.1016/j.nuclphysb.2009.02.011",
    journal = "Nucl. Phys. B",
    volume = "815",
    pages = "1--94",
    year = "2009"
}

@book{de2010loera,
title={Triangulations: Structures for Algorithms and Applications},
  author={De Loera, Jesús A. and Rambau, Jörg and Santos, Francisco},
  volume={25},
  year={2010},
  publisher={Springer Science \& Business Media},
  series={Algorithms and Computation in Mathematics},
  isbn={978-3-642-12970-3},
  doi={10.1007/978-3-642-12971-1}
}

@article{Braun:2011zm,
    author = "Braun, Andreas P. and Collinucci, Andres and Valandro, Roberto",
    title = "{G-flux in F-theory and algebraic cycles}",
    eprint = "1107.5337",
    archivePrefix = "arXiv",
    primaryClass = "hep-th",
    reportNumber = "TUW-11-19, LMU-ASC-33-11, ZMP-HH-11-13",
    doi = "10.1016/j.nuclphysb.2011.10.034",
    journal = "Nucl. Phys. B",
    volume = "856",
    pages = "129--179",
    year = "2012"
}

@article{Demirtas:2022hqf,
    author = "Demirtas, Mehmet and Rios-Tascon, Andres and McAllister, Liam",
    title = "{CYTools: A Software Package for Analyzing Calabi-Yau Manifolds}",
    eprint = "2211.03823",
    archivePrefix = "arXiv",
    primaryClass = "hep-th",
    month = "11",
    year = "2022"
}

@article{OpenAI:2023ktj,
    author = "OpenAI and others",
    title = "{GPT-4 Technical Report}",
    eprint = "2303.08774",
    archivePrefix = "arXiv",
    primaryClass = "cs.CL",
    month = "3",
    year = "2023"
}

@article{Yao:2022react,
     author = {{Yao}, Shunyu and {Zhao}, Jeffrey and {Yu}, Dian and {Du}, Nan and {Shafran}, Izhak and {Narasimhan}, Karthik and {Cao}, Yuan},
        title = "{ReAct: Synergizing Reasoning and Acting in Language Models}",
      journal = {arXiv e-prints},
         year = 2022,
        month = oct,
          eid = {arXiv:2210.03629},
        pages = {arXiv:2210.03629},
          doi = {10.48550/arXiv.2210.03629},
archivePrefix = {arXiv},
       eprint = {2210.03629},
 primaryClass = {cs.CL},
       adsurl = {https://ui.adsabs.harvard.edu/abs/2022arXiv221003629Y}
}

@article{Wu:2023autogen,
author = {{Wu}, Qingyun and {Bansal}, Gagan and {Zhang}, Jieyu and {Wu}, Yiran and {Li}, Beibin and {Zhu}, Erkang and {Jiang}, Li and {Zhang}, Xiaoyun and {Zhang}, Shaokun and {Liu}, Jiale and {Awadallah}, Ahmed Hassan and {White}, Ryen W and {Burger}, Doug and {Wang}, Chi},
        title = "{AutoGen: Enabling Next-Gen LLM Applications via Multi-Agent Conversation}",
      journal = {arXiv e-prints},
         year = 2023,
        month = aug,
          eid = {arXiv:2308.08155},
        pages = {arXiv:2308.08155},
          doi = {10.48550/arXiv.2308.08155},
archivePrefix = {arXiv},
       eprint = {2308.08155},
 primaryClass = {cs.AI},
       adsurl = {https://ui.adsabs.harvard.edu/abs/2023arXiv230808155W}
}

@inproceedings{deMoura:2021lean4,
  author = {de Moura, Leonardo and Ullrich, Sebastian},
  title = {The Lean 4 Theorem Prover and Programming Language},
  booktitle = {Automated Deduction -- CADE 28},
  year = {2021},
  publisher = {Springer International Publishing},
  pages = {625--635},
  doi = {10.1007/978-3-030-79876-5_37}
}

@inproceedings{mathlib:2020,
  author = {{The mathlib Community}},
  title = {The Lean Mathematical Library},
  booktitle = {Proceedings of the 9th ACM SIGPLAN International Conference on Certified Programs and Proofs},
  year = {2020},
  publisher = {Association for Computing Machinery},
  pages = {367--381},
  doi = {10.1145/3372885.3373824}
}

@article{Bena:2020xrh,
    author = {Bena, Iosif and Bl\r{a}b\"ack, Johan and Gra\~na, Mariana and L\"ust, Severin},
    title = "{The tadpole problem}",
    eprint = "2010.10519",
    archivePrefix = "arXiv",
    primaryClass = "hep-th",
    doi = "10.1007/JHEP11(2021)223",
    journal = "JHEP",
    volume = "11",
    pages = "223",
    year = "2021"
}

@article{Cota:2017aal,
    author = "Cota, Cesar Fierro and Klemm, Albrecht and Schimannek, Thorsten",
    title = "{Modular Amplitudes and Flux-Superpotentials on elliptic Calabi-Yau fourfolds}",
    eprint = "1709.02820",
    archivePrefix = "arXiv",
    primaryClass = "hep-th",
    doi = "10.1007/JHEP01(2018)086",
    journal = "JHEP",
    volume = "01",
    pages = "086",
    year = "2018"
}

@article{Balasubramanian:2005zx,
	archiveprefix = {arXiv},
	author = {Balasubramanian, Vijay and Berglund, Per and Conlon, Joseph P. and Quevedo, Fernando},
	doi = {10.1088/1126-6708/2005/03/007},
	eprint = {hep-th/0502058},
	journal = {JHEP},
	pages = {007},
	reportnumber = {DAMTP-2005-10, UNH-05-01, UPR-1109-T},
	title = {{Systematics of moduli stabilisation in Calabi-Yau flux compactifications}},
	volume = {03},
	year = {2005}}

@article{Katz:1996fh,
	archiveprefix = {arXiv},
	author = {Katz, Sheldon H. and Klemm, Albrecht and Vafa, Cumrun},
	doi = {10.1016/S0550-3213(97)00282-4},
	eprint = {hep-th/9609239},
	journal = {Nucl. Phys. B},
	pages = {173--195},
	reportnumber = {EFI-96-37, HUTP-96-A046, OSU-M-96-24},
	title = {{Geometric engineering of quantum field theories}},
	volume = {497},
	year = {1997}}

@article{Denef:2008wq,
	archiveprefix = {arXiv},
	author = {Denef, Frederik},
	editor = {Bachas, Costas and Baulieu, Laurent and Douglas, Michael and Kiritsis, Elias and Rabinovici, Eliezer and Vanhove, Pierre and Windey, Paul and Cugliandolo, Leticia F.},
	eprint = {0803.1194},
	journal = {Les Houches},
	pages = {483--610},
	primaryclass = {hep-th},
	title = {{Les Houches Lectures on Constructing String Vacua}},
	volume = {87},
	year = {2008}}

@article{Anderson:2015yzz,
	archiveprefix = {arXiv},
	author = {Anderson, Lara B. and Apruzzi, Fabio and Gao, Xin and Gray, James and Lee, Seung-Joo},
	doi = {10.1103/PhysRevD.93.086001},
	eprint = {1511.05188},
	journal = {Phys. Rev. D},
	number = {8},
	pages = {086001},
	primaryclass = {hep-th},
	title = {{Instanton superpotentials, Calabi-Yau geometry, and fibrations}},
	volume = {93},
	year = {2016}}

@article{Blumenhagen:2006ci,
    author = "Blumenhagen, Ralph and Kors, Boris and Lust, Dieter and Stieberger, Stephan",
    title = "{Four-dimensional String Compactifications with D-Branes, Orientifolds and Fluxes}",
    eprint = "hep-th/0610327",
    archivePrefix = "arXiv",
    reportNumber = "CERN-PH-TH-2006-218",
    doi = "10.1016/j.physrep.2007.04.003",
    journal = "Phys. Rept.",
    volume = "445",
    pages = "1--193",
    year = "2007"
}

@article{Maharana:2012tu,
    author = "Maharana, Anshuman and Palti, Eran",
    title = "{Models of Particle Physics from Type IIB String Theory and F-theory: A Review}",
    eprint = "1212.0555",
    archivePrefix = "arXiv",
    primaryClass = "hep-th",
    reportNumber = "HRI-ST-1212, CPHT-RR088.1112",
    doi = "10.1142/S0217751X13300056",
    journal = "Int. J. Mod. Phys. A",
    volume = "28",
    pages = "1330005",
    year = "2013"
}

@article{Weigand:2010wm,
    author = "Weigand, Timo",
    editor = "Walcher, J.",
    title = "{Lectures on F-theory compactifications and model building}",
    eprint = "1009.3497",
    archivePrefix = "arXiv",
    primaryClass = "hep-th",
    doi = "10.1088/0264-9381/27/21/214004",
    journal = "Class. Quant. Grav.",
    volume = "27",
    pages = "214004",
    year = "2010"
}

@article{Braun:2008ua,
    author = "Braun, A. P. and Hebecker, Arthur and Triendl, H.",
    title = "{D7-Brane Motion from M-Theory Cycles and Obstructions in the Weak Coupling Limit}",
    eprint = "0801.2163",
    archivePrefix = "arXiv",
    primaryClass = "hep-th",
    reportNumber = "HD-THEP-08-1",
    doi = "10.1016/j.nuclphysb.2008.03.021",
    journal = "Nucl. Phys. B",
    volume = "800",
    pages = "298--329",
    year = "2008"
}

@article{Collinucci:2008sq,
    author = "Collinucci, Andres and Kreuzer, Maximilian and Mayrhofer, Christoph and Walliser, Nils-Ole",
    title = "{Four-modulus 'Swiss Cheese' chiral models}",
    eprint = "0811.4599",
    archivePrefix = "arXiv",
    primaryClass = "hep-th",
    doi = "10.1088/1126-6708/2009/07/074",
    journal = "JHEP",
    volume = "07",
    pages = "074",
    year = "2009"
}

@article{Braun:2020jrx,
    author = "Braun, Andreas P. and Valandro, Roberto",
    title = "{$G_{4}$ flux, algebraic cycles and complex structure moduli stabilization}",
    eprint = "2009.11873",
    archivePrefix = "arXiv",
    primaryClass = "hep-th",
    doi = "10.1007/JHEP01(2021)207",
    journal = "JHEP",
    volume = "01",
    pages = "207",
    year = "2021"
}

@article{Gao:2022fdi,
    author = "Gao, Xin and Hebecker, Arthur and Schreyer, Simon and Venken, Victoria",
    title = "{The LVS parametric tadpole constraint}",
    eprint = "2202.04087",
    archivePrefix = "arXiv",
    primaryClass = "hep-th",
    doi = "10.1007/JHEP07(2022)056",
    journal = "JHEP",
    volume = "07",
    pages = "056",
    year = "2022"
}

@article{Aluffi:2009tm,
    author = "Aluffi, Paolo and Esole, Mboyo",
    title = "{New Orientifold Weak Coupling Limits in F-theory}",
    eprint = "0908.1572",
    archivePrefix = "arXiv",
    primaryClass = "hep-th",
    reportNumber = "R",
    doi = "10.1007/JHEP02(2010)020",
    journal = "JHEP",
    volume = "02",
    pages = "020",
    year = "2010"
}

@article{Esole:2012tf,
    author = "Esole, Mboyo and Savelli, Raffaele",
    title = "{Tate Form and Weak Coupling Limits in F-theory}",
    eprint = "1209.1633",
    archivePrefix = "arXiv",
    primaryClass = "hep-th",
    reportNumber = "MPP-2012-128",
    doi = "10.1007/JHEP06(2013)027",
    journal = "JHEP",
    volume = "06",
    pages = "027",
    year = "2013"
}

@article{Dasgupta:1996yh,
    author = "Dasgupta, Keshav and Mukhi, Sunil",
    title = "{A Note on low dimensional string compactifications}",
    eprint = "hep-th/9612188",
    archivePrefix = "arXiv",
    reportNumber = "TIFR-TH-96-61",
    doi = "10.1016/S0370-2693(97)00216-5",
    journal = "Phys. Lett. B",
    volume = "398",
    pages = "285--290",
    year = "1997"
}

@article{Morrison:1996xf,
    author = "Morrison, David R. and Seiberg, Nathan",
    title = "{Extremal transitions and five-dimensional supersymmetric field theories}",
    eprint = "hep-th/9609070",
    archivePrefix = "arXiv",
    reportNumber = "DUKE-TH-96-130, RU-96-80",
    doi = "10.1016/S0550-3213(96)00592-5",
    journal = "Nucl. Phys. B",
    volume = "483",
    pages = "229--247",
    year = "1997"
}

@article{Intriligator:1997pq,
    author = "Intriligator, Kenneth A. and Morrison, David R. and Seiberg, Nathan",
    title = "{Five-dimensional supersymmetric gauge theories and degenerations of Calabi-Yau spaces}",
    eprint = "hep-th/9702198",
    archivePrefix = "arXiv",
    reportNumber = "RU-96-99, IASSNS-HEP-96-112",
    doi = "10.1016/S0550-3213(97)00279-4",
    journal = "Nucl. Phys. B",
    volume = "497",
    pages = "56--100",
    year = "1997"
}

@article{Sen:1997gv,
	archiveprefix = {arXiv},
	author = {Sen, Ashoke},
	doi = {10.1103/PhysRevD.55.R7345},
	eprint = {hep-th/9702165},
	journal = {Phys. Rev. D},
	pages = {R7345--R7349},
	reportnumber = {MRI-PHY-P970202, NI-97013},
	title = {{Orientifold limit of F theory vacua}},
	volume = {55},
	year = {1997}}

@article{Morrison:1996na,
    author = "Morrison, David R. and Vafa, Cumrun",
    title = "{Compactifications of F theory on Calabi-Yau threefolds. 1}",
    eprint = "hep-th/9602114",
    archivePrefix = "arXiv",
    reportNumber = "DUKE-TH-96-106, HUTP-96-A007",
    doi = "10.1016/0550-3213(96)00242-8",
    journal = "Nucl. Phys. B",
    volume = "473",
    pages = "74--92",
    year = "1996"
}

@article{Morrison:1996pp,
    author = "Morrison, David R. and Vafa, Cumrun",
    title = "{Compactifications of F theory on Calabi-Yau threefolds. 2.}",
    eprint = "hep-th/9603161",
    archivePrefix = "arXiv",
    reportNumber = "DUKE-TH-96-107, HUTP-96-A012",
    doi = "10.1016/0550-3213(96)00369-0",
    journal = "Nucl. Phys. B",
    volume = "476",
    pages = "437--469",
    year = "1996"
}

@article{Taylor:2015isa,
    author = "Taylor, Washington and Wang, Yi-Nan",
    title = "{Non-toric bases for elliptic Calabi{\textendash}Yau threefolds and 6D F-theory vacua}",
    eprint = "1504.07689",
    archivePrefix = "arXiv",
    primaryClass = "hep-th",
    reportNumber = "MIT-CTP-4629",
    doi = "10.4310/ATMP.2017.v21.n4.a6",
    journal = "Adv. Theor. Math. Phys.",
    volume = "21",
    pages = "1063--1114",
    year = "2017"
}

@article{Abbasi:2025lvn,
    author = "Abbasi, Fatima and Nally, Richard and Taylor, Washington",
    title = "{Classifying Fibers and Bases in Toric Hypersurface Calabi-Yau Threefolds}",
    eprint = "2511.10601",
    archivePrefix = "arXiv",
    primaryClass = "hep-th",
    reportNumber = "MIT-CTP/5958",
    month = "11",
    year = "2025"
}

@article{Morrison:1991cd,
    author = "Morrison, David R.",
    editor = "Yau, Shing-Tung",
    title = "{Picard-Fuchs equations and mirror maps for hypersurfaces}",
    eprint = "hep-th/9111025",
    archivePrefix = "arXiv",
    reportNumber = "DUK-M-91-14",
    journal = "AMS/IP Stud. Adv. Math.",
    volume = "9",
    pages = "185--199",
    year = "1998"
}

@article{DelZotto:2014hpa,
    author = "Del Zotto, Michele and Heckman, Jonathan J. and Tomasiello, Alessandro and Vafa, Cumrun",
    title = "{6d Conformal Matter}",
    eprint = "1407.6359",
    archivePrefix = "arXiv",
    primaryClass = "hep-th",
    doi = "10.1007/JHEP02(2015)054",
    journal = "JHEP",
    volume = "02",
    pages = "054",
    year = "2015"
}

@article{DenefLoeserMcKay,
  author        = {Denef, Jan and Loeser, Fran{\c{c}}ois},
  title         = {Motivic Integration, Quotient Singularities and the
                   {McKay} Correspondence},
  journal       = {Compositio Mathematica},
  volume        = {131},
  number        = {3},
  pages         = {267--290},
  year          = {2002},
  doi           = {10.1023/A:1015565912485},
  eprint        = {math/9903187},
  archivePrefix = {arXiv},
  primaryClass  = {math.AG}
}

@article{Hosono:1993qy,
    author = "Hosono, S. and Klemm, A. and Theisen, S. and Yau, Shing-Tung",
    title = "{Mirror symmetry, mirror map and applications to Calabi-Yau hypersurfaces}",
    eprint = "hep-th/9308122",
    archivePrefix = "arXiv",
    reportNumber = "HUTMP-93-0801, LMU-TPW-93-22",
    doi = "10.1007/BF02100589",
    journal = "Commun. Math. Phys.",
    volume = "167",
    pages = "301--350",
    year = "1995"
}

@article{Vafa:1996xn,
    author = "Vafa, Cumrun",
    title = "{Evidence for F theory}",
    eprint = "hep-th/9602022",
    archivePrefix = "arXiv",
    reportNumber = "HUTP-96-A004",
    doi = "10.1016/0550-3213(96)00172-1",
    journal = "Nucl. Phys. B",
    volume = "469",
    pages = "403--418",
    year = "1996"
}

@article{Halverson:2016vwx,
    author = "Halverson, James",
    title = "{Strong Coupling in F-theory and Geometrically Non-Higgsable Seven-branes}",
    eprint = "1603.01639",
    archivePrefix = "arXiv",
    primaryClass = "hep-th",
    reportNumber = "NSF-KITP-16-023",
    doi = "10.1016/j.nuclphysb.2017.02.014",
    journal = "Nucl. Phys. B",
    volume = "919",
    pages = "267--296",
    year = "2017"
}

@article{Halverson:2017vde,
    author = "Halverson, James and Long, Cody and Sung, Benjamin",
    title = "{On the Scarcity of Weak Coupling in the String Landscape}",
    eprint = "1710.09374",
    archivePrefix = "arXiv",
    primaryClass = "hep-th",
    doi = "10.1007/JHEP02(2018)113",
    journal = "JHEP",
    volume = "02",
    pages = "113",
    year = "2018"
}

@article{Demirtas:2023als,
    author = "Demirtas, Mehmet and Kim, Manki and McAllister, Liam and Moritz, Jakob and Rios-Tascon, Andres",
    title = "{Computational Mirror Symmetry}",
    eprint = "2303.00757",
    archivePrefix = "arXiv",
    primaryClass = "hep-th",
    reportNumber = "MIT-CTP/5528",
    doi = "10.1007/JHEP01(2024)184",
    journal = "JHEP",
    volume = "01",
    pages = "184",
    year = "2024"
}

@article{Schimannek:2021pau,
    author = "Schimannek, Thorsten",
    title = "{Modular curves, the Tate-Shafarevich group and Gopakumar-Vafa invariants with discrete charges}",
    eprint = "2108.09311",
    archivePrefix = "arXiv",
    primaryClass = "hep-th",
    reportNumber = "UWThPh-2021-13",
    doi = "10.1007/JHEP02(2022)007",
    journal = "JHEP",
    volume = "02",
    pages = "007",
    year = "2022"
}

@article{Katz:2022lyl,
    author = "Katz, Sheldon and Klemm, Albrecht and Schimannek, Thorsten and Sharpe, Eric",
    title = "{Topological Strings on Non-commutative Resolutions}",
    eprint = "2212.08655",
    archivePrefix = "arXiv",
    primaryClass = "hep-th",
    doi = "10.1007/s00220-023-04896-2",
    journal = "Commun. Math. Phys.",
    volume = "405",
    number = "3",
    pages = "62",
    year = "2024"
}

@article{Huang:2006hq,
    author = "Huang, Min-xin and Klemm, Albrecht and Quackenbush, Seth",
    title = "{Topological string theory on compact Calabi-Yau: Modularity and boundary conditions}",
    eprint = "hep-th/0612125",
    archivePrefix = "arXiv",
    reportNumber = "MAD-TH-06-12",
    doi = "10.1007/978-3-540-68030-7_3",
    journal = "Lect. Notes Phys.",
    volume = "757",
    pages = "45--102",
    year = "2009"
}

@article{Huang:2015sta,
    author = "Huang, Min-xin and Katz, Sheldon and Klemm, Albrecht",
    title = "{Topological String on elliptic CY 3-folds and the ring of Jacobi forms}",
    eprint = "1501.04891",
    archivePrefix = "arXiv",
    primaryClass = "hep-th",
    reportNumber = "USTC-ICTS-15-02, BONN-TH-2015-01",
    doi = "10.1007/JHEP10(2015)125",
    journal = "JHEP",
    volume = "10",
    pages = "125",
    year = "2015"
}

@article{Alexandrov:2023zjb,
    author = "Alexandrov, Sergei and Feyzbakhsh, Soheyla and Klemm, Albrecht and Pioline, Boris and Schimannek, Thorsten",
    title = "{Quantum geometry, stability and modularity}",
    eprint = "2301.08066",
    archivePrefix = "arXiv",
    primaryClass = "hep-th",
    doi = "10.4310/CNTP.2024.v18.n1.a2",
    journal = "Commun. Num. Theor. Phys.",
    volume = "18",
    number = "1",
    pages = "49--151",
    year = "2024"
}

@article{Alexandrov:2023ltz,
    author = "Alexandrov, Sergei and Feyzbakhsh, Soheyla and Klemm, Albrecht and Pioline, Boris",
    title = "{Quantum geometry and mock modularity}",
    eprint = "2312.12629",
    archivePrefix = "arXiv",
    primaryClass = "hep-th",
    doi = "10.4310/cntp.260401215658",
    journal = "Commun. Num. Theor. Phys.",
    volume = "20",
    number = "1",
    pages = "97--148",
    year = "2026"
}

@article{Cota:2019cjx,
    author = "Cota, Cesar Fierro and Klemm, Albrecht and Schimannek, Thorsten",
    title = "{Topological strings on genus one fibered Calabi-Yau 3-folds and string dualities}",
    eprint = "1910.01988",
    archivePrefix = "arXiv",
    primaryClass = "hep-th",
    reportNumber = "BONN-TH-2019-05, UWThPh-2019-29",
    doi = "10.1007/JHEP11(2019)170",
    journal = "JHEP",
    volume = "11",
    pages = "170",
    year = "2019"
}

@article{Duque:2025kaa,
    author = "Duque, David Jaramillo and Kashani-Poor, Amir-Kian and Schimannek, Thorsten",
    title = "{The twisted geometry of 6d F-theory vacua with discrete gauge symmetries}",
    eprint = "2508.16500",
    archivePrefix = "arXiv",
    primaryClass = "hep-th",
    month = "8",
    year = "2025"
}

@article{Pioline:2025uov,
    author = "Pioline, Boris and Schimannek, Thorsten",
    title = "{Revisiting the Quantum Geometry of Torus-fibered Calabi-Yau Threefolds}",
    eprint = "2510.23722",
    archivePrefix = "arXiv",
    primaryClass = "hep-th",
    doi = "10.21468/SciPostPhys.21.2.037",
    journal = "SciPost Phys.",
    volume = "21",
    pages = "037",
    year = "2026"
}

@article{FierroCota:2025pkp,
    author = "Fierro Cota, Cesar",
    title = "{Supergravity anomaly equations from modularity of Calabi--Yau threefolds}",
    eprint = "2512.18151",
    archivePrefix = "arXiv",
    primaryClass = "hep-th",
    month = "12",
    year = "2025"
}

@article{Huang:2025xkc,
    author = "Huang, Min-xin and Katz, Sheldon and Klemm, Albrecht and Wang, Xin",
    title = "{Refined BPS numbers on compact Calabi-Yau threefolds from Wilson loops}",
    eprint = "2503.16270",
    archivePrefix = "arXiv",
    primaryClass = "hep-th",
    reportNumber = "USTC-ICTS/PCFT-25-03, USTC-ICTS/PCFT-25-03,MPIM-Bonn-2024",
    doi = "10.1007/JHEP08(2025)178",
    journal = "JHEP",
    volume = "08",
    pages = "178",
    year = "2025"
}

@article{Alim:2012ss,
    author = "Alim, Murad and Scheidegger, Emanuel",
    title = "{Topological Strings on Elliptic Fibrations}",
    eprint = "1205.1784",
    archivePrefix = "arXiv",
    primaryClass = "hep-th",
    doi = "10.4310/CNTP.2014.v8.n4.a4",
    journal = "Commun. Num. Theor. Phys.",
    volume = "08",
    pages = "729--800",
    year = "2014"
}

@article{Gu:2023mgf,
    author = "Gu, Jie and Kashani-Poor, Amir-Kian and Klemm, Albrecht and Marino, Marcos",
    title = "{Non-perturbative topological string theory on compact Calabi-Yau 3-folds}",
    eprint = "2305.19916",
    archivePrefix = "arXiv",
    primaryClass = "hep-th",
    doi = "10.21468/SciPostPhys.16.3.079",
    journal = "SciPost Phys.",
    volume = "16",
    number = "3",
    pages = "079",
    year = "2024"
}

@article{Douaud:2024khu,
    author = "Douaud, Simon and Kashani-Poor, Amir-Kian",
    title = "{Borel singularities and Stokes constants of the topological string free energy on one-parameter Calabi-Yau threefolds}",
    eprint = "2412.16140",
    archivePrefix = "arXiv",
    primaryClass = "hep-th",
    doi = "10.1007/JHEP06(2025)253",
    journal = "JHEP",
    volume = "06",
    pages = "253",
    year = "2025"
}

@article{Douaud:2026qfo,
    author = "Douaud, Simon and Kashani-Poor, Amir-Kian",
    title = "{The non-perturbative topological string: from resurgence to wall-crossing of DT invariants}",
    eprint = "2604.19731",
    archivePrefix = "arXiv",
    primaryClass = "hep-th",
    doi = "10.1007/JHEP08(2026)041",
    journal = "JHEP",
    volume = "08",
    pages = "041",
    year = "2026"
}

@article{Kuusela:2023vgi,
    author = "Kuusela, Pyry and McGovern, Joseph",
    title = "{Reflections in the mirror: Studying infinite Coxeter symmetries of GV-invariants}",
    eprint = "2312.06753",
    archivePrefix = "arXiv",
    primaryClass = "hep-th",
    doi = "10.1142/S0217751X24460163",
    journal = "Int. J. Mod. Phys. A",
    volume = "39",
    number = "33",
    pages = "2446016",
    year = "2024"
}

@article{Doran:2024kcb,
    author = "Doran, Charles and Pioline, Boris and Schimannek, Thorsten",
    title = "{Enumerative geometry and modularity in two-modulus K3-fibered Calabi-Yau threefolds}",
    eprint = "2408.02994",
    archivePrefix = "arXiv",
    primaryClass = "hep-th",
    doi = "10.4310/ATMP.260522000535",
    month = "8",
    year = "2024"
}

@article{Lee:2020blx,
    author = "Lee, Seung-Joo and Lerche, Wolfgang and Lockhart, Guglielmo and Weigand, Timo",
    title = "{Holomorphic anomalies, fourfolds and fluxes}",
    eprint = "2012.00766",
    archivePrefix = "arXiv",
    primaryClass = "hep-th",
    doi = "10.1007/JHEP03(2022)072",
    journal = "JHEP",
    volume = "03",
    pages = "072",
    year = "2022"
}

@article{Sharpe:2001bs,
    author = "Sharpe, Eric R.",
    title = "{String orbifolds and quotient stacks}",
    eprint = "hep-th/0102211",
    archivePrefix = "arXiv",
    reportNumber = "DUKE-CGTP-2001-03",
    doi = "10.1016/S0550-3213(02)00039-1",
    journal = "Nucl. Phys. B",
    volume = "627",
    pages = "445--505",
    year = "2002"
}

@article{Behrend:2005wzg,
    author = "Behrend, Kai",
    title = "{Donaldson-Thomas invariants via microlocal geometry}",
    eprint = "math/0507523",
    archivePrefix = "arXiv",
    month = "12",
    year = "2005"
}

@article{AbramovichGraberVistoli,
  author        = {Abramovich, Dan and Graber, Tom and Vistoli, Angelo},
  title         = {Gromov--{W}itten Theory of {D}eligne--{M}umford Stacks},
  journal       = {American Journal of Mathematics},
  volume        = {130},
  number        = {5},
  pages         = {1337--1398},
  year          = {2008},
  doi           = {10.1353/ajm.0.0017},
  eprint        = {math/0603151},
  archivePrefix = {arXiv},
  primaryClass  = {math.AG}
}

@article{SatakeGaussBonnet,
  author  = {Satake, Ichir{\^o}},
  title   = {The {Gauss--Bonnet} Theorem for {$V$}-Manifolds},
  journal = {Journal of the Mathematical Society of Japan},
  volume  = {9},
  number  = {4},
  pages   = {464--492},
  year    = {1957},
  doi     = {10.2969/jmsj/00940464}
}

@article{Giddings:2001yu,
    author = "Giddings, Steven B. and Kachru, Shamit and Polchinski, Joseph",
    title = "{Hierarchies from fluxes in string compactifications}",
    eprint = "hep-th/0105097",
    archivePrefix = "arXiv",
    reportNumber = "SLAC-PUB-8807, NSF-ITP-01-37, SU-ITP-01-16",
    doi = "10.1103/PhysRevD.66.106006",
    journal = "Phys. Rev. D",
    volume = "66",
    pages = "106006",
    year = "2002"
}

@article{Grimm:2004uq,
    author = "Grimm, Thomas W. and Louis, Jan",
    title = "{The Effective action of N = 1 Calabi-Yau orientifolds}",
    eprint = "hep-th/0403067",
    archivePrefix = "arXiv",
    reportNumber = "LPTENS-04-14",
    doi = "10.1016/j.nuclphysb.2004.08.005",
    journal = "Nucl. Phys. B",
    volume = "699",
    pages = "387--426",
    year = "2004"
}

@article{Marchesano:2024gul,
    author = "Marchesano, Fernando and Shiu, Gary and Weigand, Timo",
    title = "{The Standard Model from String Theory: What Have We Learned?}",
    eprint = "2401.01939",
    archivePrefix = "arXiv",
    primaryClass = "hep-th",
    reportNumber = "IFT-UAM/CSIC-24-01, ZMP-HH/24-01",
    doi = "10.1146/annurev-nucl-102622-012235",
    journal = "Ann. Rev. Nucl. Part. Sci.",
    volume = "74",
    number = "1",
    pages = "113--140",
    year = "2024"
}

@inbook{Marchesano:2022qbx,
    author="Marchesano, Fernando
and Schellekens, Bert
and Weigand, Timo",
editor="Bambi, Cosimo
and Modesto, Leonardo
and Shapiro, Ilya",
title="D-brane and F-theory Model Building",
bookTitle="Handbook of Quantum Gravity",
year="2023",
publisher="Springer Nature Singapore",
address="Singapore",
pages="1--68",
isbn="978-981-19-3079-9",
doi="10.1007/978-981-19-3079-9_57-1",
url="https://doi.org/10.1007/978-981-19-3079-9_57-1"
}

@article{Cvetic:2022fnv,
    author = "Cvetic, Mirjam and Halverson, James and Shiu, Gary and Taylor, Washington",
    title = "{Snowmass White Paper: String Theory and Particle Physics}",
    eprint = "2204.01742",
    archivePrefix = "arXiv",
    primaryClass = "hep-th",
    reportNumber = "CERN-TH-2022-054, UPR-1318-T, MIT-CTP-5419",
    month = "4",
    year = "2022"
}

@article{Ibanez:2001nd,
    author = "Ibanez, Luis E. and Marchesano, F. and Rabadan, R.",
    title = "{Getting just the standard model at intersecting branes}",
    eprint = "hep-th/0105155",
    archivePrefix = "arXiv",
    reportNumber = "FTUAM-01-09, IFT-UAM-CSIC-01-15",
    doi = "10.1088/1126-6708/2001/11/002",
    journal = "JHEP",
    volume = "11",
    pages = "002",
    year = "2001"
}

@article{Cvetic:2019gnh,
    author = "Cveti{\v{c}}, Mirjam and Halverson, James and Lin, Ling and Liu, Muyang and Tian, Jiahua",
    title = "{Quadrillion $F$-Theory Compactifications with the Exact Chiral Spectrum of the Standard Model}",
    eprint = "1903.00009",
    archivePrefix = "arXiv",
    primaryClass = "hep-th",
    reportNumber = "UPR-1297-T",
    doi = "10.1103/PhysRevLett.123.101601",
    journal = "Phys. Rev. Lett.",
    volume = "123",
    number = "10",
    pages = "101601",
    year = "2019"
}

@article{McAllister:2023vgy,
    author = "McAllister, Liam and Quevedo, Fernando",
    title = "{Moduli Stabilization in String Theory}",
    eprint = "2310.20559",
    archivePrefix = "arXiv",
    primaryClass = "hep-th",
    month = "10",
    year = "2023"
}

@article{Gukov:1999ya,
    author = "Gukov, Sergei and Vafa, Cumrun and Witten, Edward",
    title = "{CFT's from Calabi-Yau four folds}",
    eprint = "hep-th/9906070",
    archivePrefix = "arXiv",
    reportNumber = "HUTP-99-A034, IASSNS-HEP-99-52, PUPT-1864",
    doi = "10.1016/S0550-3213(00)00373-4",
    journal = "Nucl. Phys. B",
    volume = "584",
    pages = "69--108",
    year = "2000",
    note = "[Erratum: Nucl.Phys.B 608, 477--478 (2001)]"
}

@article{Dasgupta:1999ss,
    author = "Dasgupta, Keshav and Rajesh, Govindan and Sethi, Savdeep",
    title = "{M theory, orientifolds and G - flux}",
    eprint = "hep-th/9908088",
    archivePrefix = "arXiv",
    reportNumber = "IASSNS-HEP-99-75, NSF-ITP-99-095",
    doi = "10.1088/1126-6708/1999/08/023",
    journal = "JHEP",
    volume = "08",
    pages = "023",
    year = "1999"
}

@misc{anthropic2024claude,
      title={The Claude 3 Model Family: Opus, Sonnet, Haiku}, 
      author={{Anthropic}},
      year={2024},
      howpublished={\url{https://www-cdn.anthropic.com/de8ba9b01c9ab7cbabf5c33b80b7bbc618857627/Model_Card_Claude_3.pdf}}
}

@article{Benander_2025,
    title={Nexic Reasoning: Defining a Generalized Calculus Over Anthropic Parameters},
    author={Ajax Benander},
    journal={OSF Preprints},
    doi={10.31219/osf.io/jvkcp},
    year={2025}
}

@misc{OpenAI_Navier_2026,
    title={On the Navier–Stokes Millennium Prize Problem},
    author={{OpenAI}},
    year={2026},
    howpublished={\url{https://openai.com/index/navier-stokes-solution/}}
}

@article{Lefkowitz_2026,
    title={Comparative Micro-Study: Behavioral Reasoning Differences Between Gemini-3-Pro and Grok-4.1-Thinking},
    author={Dana Lefkowitz},
    journal={SSRN Electronic Journal},
    doi={10.2139/ssrn.6123586},
    year={2026}
}

@inproceedings{de2015lean,
  title={The Lean theorem prover (system description)},
  author={de Moura, Leonardo and Kong, Soonho and Avigad, Jeremy and van Doorn, Floris and von Raumer, Jakob},
  booktitle={International Conference on Automated Deduction},
  pages={378--388},
  year={2015},
  organization={Springer}
}

@article{Yip:2026jhw,
    author = "Yip, Jacky H. T. and Mininno, Alessandro and Shiu, Gary",
    title = "{Exploring Line Bundle Standard Models with Transformers}",
    eprint = "2607.00078",
    archivePrefix = "arXiv",
    primaryClass = "hep-th",
    month = "6",
    year = "2026"
}

@inproceedings{Arnal:2026zyo,
    author = "Arnal, Charles and Yip, Jacky H. T. and Charton, Fran{\c{c}}ois and Shiu, Gary",
    title = "{Generating Special Triangulations with Transformers}",
    eprint = "2606.26660",
    archivePrefix = "arXiv",
    primaryClass = "hep-th",
    month = "6",
    year = "2026",
    booktitle = "{}"
}

@article{Heckman:2026xsi,
    author = "Heckman, Jonathan J. and Meynet, Shani and Mininno, Alessandro and Shiu, Gary",
    title = "{Learning to Trace Seiberg Dualities}",
    eprint = "2607.28628",
    archivePrefix = "arXiv",
    primaryClass = "hep-th",
    month = "7",
    year = "2026"
}

@article{Loges:2021hvn,
    author = "Loges, Gregory J. and Shiu, Gary",
    title = "{Breeding Realistic D-Brane Models}",
    eprint = "2112.08391",
    archivePrefix = "arXiv",
    primaryClass = "hep-th",
    doi = "10.1002/prop.202200038",
    journal = "Fortsch. Phys.",
    volume = "70",
    number = "5",
    pages = "2200038",
    year = "2022"
}

@article{Loges:2022mao,
    author = "Loges, Gregory J. and Shiu, Gary",
    title = "{134 billion intersecting brane models}",
    eprint = "2206.03506",
    archivePrefix = "arXiv",
    primaryClass = "hep-th",
    doi = "10.1007/JHEP12(2022)097",
    journal = "JHEP",
    volume = "12",
    pages = "097",
    year = "2022"
}

@article{Halverson:2019tkf,
    author = "Halverson, James and Nelson, Brent and Ruehle, Fabian",
    title = "{Branes with Brains: Exploring String Vacua with Deep Reinforcement Learning}",
    eprint = "1903.11616",
    archivePrefix = "arXiv",
    primaryClass = "hep-th",
    doi = "10.1007/JHEP06(2019)003",
    journal = "JHEP",
    volume = "06",
    pages = "003",
    year = "2019"
}

@article{Yip:2025hon,
    author = "Yip, Jacky H. T. and Arnal, Charles and Charton, Fran{\c{c}}ois and Shiu, Gary",
    title = "{Transforming Calabi-Yau Constructions: Generating New Calabi-Yau Manifolds with Transformers}",
    eprint = "2507.03732",
    archivePrefix = "arXiv",
    primaryClass = "hep-th",
    month = "7",
    year = "2025"
}

@article{Walden:2025cpf,
    author = "Walden, Moritz and Larfors, Magdalena",
    title = "{Sampling string vacua using generative models}",
    eprint = "2509.16029",
    archivePrefix = "arXiv",
    primaryClass = "hep-th",
    doi = "10.1088/2632-2153/ae32dc",
    journal = "Mach. Learn. Sci. Tech.",
    volume = "7",
    number = "1",
    pages = "015039",
    year = "2026"
}

@article{Heckman:2022suy,
    author = "Heckman, Jonathan J. and Lawrie, Craig and Lin, Ling and Zhang, Hao Y. and Zoccarato, Gianluca",
    title = "{6D SCFTs, center-flavor symmetries, and Stiefel-Whitney compactifications}",
    eprint = "2205.03411",
    archivePrefix = "arXiv",
    primaryClass = "hep-th",
    reportNumber = "CERN-TH 2022-074, DESY-22-069",
    doi = "10.1103/PhysRevD.106.066003",
    journal = "Phys. Rev. D",
    volume = "106",
    number = "6",
    pages = "066003",
    year = "2022"
}

@article{Giacomelli:2025zqn,
    author = "Giacomelli, Simone and Harding, William and Mekareeya, Noppadol and Mininno, Alessandro",
    title = "{From regular to irregular: a unified origin for Argyres-Douglas theories}",
    eprint = "2507.13434",
    archivePrefix = "arXiv",
    primaryClass = "hep-th",
    doi = "10.1007/JHEP10(2025)155",
    journal = "JHEP",
    volume = "10",
    pages = "155",
    year = "2025"
}

@article{Giacomelli:2024ycb,
    author = "Giacomelli, Simone and Harding, William and Mekareeya, Noppadol and Mininno, Alessandro",
    title = "{All class $ \mathcal{S} $ theories of type-A originate from orbi-instantons}",
    eprint = "2411.03425",
    archivePrefix = "arXiv",
    primaryClass = "hep-th",
    reportNumber = "ZMP-HH/24-25",
    doi = "10.1007/JHEP02(2025)185",
    journal = "JHEP",
    volume = "02",
    pages = "185",
    year = "2025"
}

@article{Giacomelli:2024dbd,
    author = "Giacomelli, Simone and Savelli, Raffaele and Zoccarato, Gianluca",
    title = "{$ \mathcal{N} $ = 2 Orbi-S-Folds}",
    eprint = "2405.00101",
    archivePrefix = "arXiv",
    primaryClass = "hep-th",
    doi = "10.1007/JHEP01(2025)059",
    journal = "JHEP",
    volume = "01",
    pages = "059",
    year = "2025"
}

@article{Aspinwall:1997ye,
    author = "Aspinwall, Paul S. and Morrison, David R.",
    title = "{Point - like instantons on K3 orbifolds}",
    eprint = "hep-th/9705104",
    archivePrefix = "arXiv",
    reportNumber = "RU-97-29, IASSNS-HEP-97-46",
    doi = "10.1016/S0550-3213(97)00516-6",
    journal = "Nucl. Phys. B",
    volume = "503",
    pages = "533--564",
    year = "1997"
}

@article{Heckman:2015bfa,
    author = "Heckman, Jonathan J. and Morrison, David R. and Rudelius, Tom and Vafa, Cumrun",
    title = "{Atomic Classification of 6D SCFTs}",
    eprint = "1502.05405",
    archivePrefix = "arXiv",
    primaryClass = "hep-th",
    doi = "10.1002/prop.201500024",
    journal = "Fortsch. Phys.",
    volume = "63",
    pages = "468--530",
    year = "2015"
}

@article{Mekareeya:2017jgc,
    author = "Mekareeya, Noppadol and Ohmori, Kantaro and Tachikawa, Yuji and Zafrir, Gabi",
    title = "{E$_{8}$ instantons on type-A ALE spaces and supersymmetric field theories}",
    eprint = "1707.04370",
    archivePrefix = "arXiv",
    primaryClass = "hep-th",
    reportNumber = "IPMU-17-0097",
    doi = "10.1007/JHEP09(2017)144",
    journal = "JHEP",
    volume = "09",
    pages = "144",
    year = "2017"
}

@article{Horava:1996ma,
    author = "Horava, Petr and Witten, Edward",
    title = "{Eleven-dimensional supergravity on a manifold with boundary}",
    eprint = "hep-th/9603142",
    archivePrefix = "arXiv",
    reportNumber = "IASSNS-HEP-96-17, PUPT-1597",
    doi = "10.1016/0550-3213(96)00308-2",
    journal = "Nucl. Phys. B",
    volume = "475",
    pages = "94--114",
    year = "1996"
}

@article{Gaiotto:2009we,
    author = "Gaiotto, Davide",
    title = "{N=2 dualities}",
    eprint = "0904.2715",
    archivePrefix = "arXiv",
    primaryClass = "hep-th",
    doi = "10.1007/JHEP08(2012)034",
    journal = "JHEP",
    volume = "08",
    pages = "034",
    year = "2012"
}

@article{Gaiotto:2009gz,
    author = "Gaiotto, Davide and Maldacena, Juan",
    title = "{The Gravity duals of N=2 superconformal field theories}",
    eprint = "0904.4466",
    archivePrefix = "arXiv",
    primaryClass = "hep-th",
    doi = "10.1007/JHEP10(2012)189",
    journal = "JHEP",
    volume = "10",
    pages = "189",
    year = "2012"
}

@article{Intriligator:1996ex,
    author = "Intriligator, Kenneth A. and Seiberg, N.",
    title = "{Mirror symmetry in three-dimensional gauge theories}",
    eprint = "hep-th/9607207",
    archivePrefix = "arXiv",
    reportNumber = "RU-96-63, IASSNS-HEP-96-80",
    doi = "10.1016/0370-2693(96)01088-X",
    journal = "Phys. Lett. B",
    volume = "387",
    pages = "513--519",
    year = "1996"
}

@article{Argyres:1995jj,
    author = "Argyres, Philip C. and Douglas, Michael R.",
    title = "{New phenomena in SU(3) supersymmetric gauge theory}",
    eprint = "hep-th/9505062",
    archivePrefix = "arXiv",
    reportNumber = "IASSNS-HEP-95-31, RU-95-28",
    doi = "10.1016/0550-3213(95)00281-V",
    journal = "Nucl. Phys. B",
    volume = "448",
    pages = "93--126",
    year = "1995"
}

@article{Akhond:2021xio,

  archiveprefix = {arXiv},
  author = {Akhond, Mohammad and Arias-Tamargo, Guillermo and Mininno, Alessandro and Sun, Hao-Yu and Sun, Zhengdi and Wang, Yifan and Xu, Fengjun},
  doi = {10.21468/SciPostPhysLectNotes.64},
  eprint = {2112.14764},
  journal = {SciPost Phys. Lect. Notes},
  pages = {64},
  primaryclass = {hep-th},
  publisher = {SciPost},
  reportnumber = {IFT-UAM/CSIC-21-151, ZMP-HH/21-28},
  title = {{The Hitchhiker's Guide to 4d $\mathcal{N}=2$ Superconformal Field Theories}},
  year = {2022},
}

@article{Argyres:2022mnu,
  archiveprefix = {arXiv},
  author = {Argyres, Philip C. and Heckman, Jonathan J. and Intriligator, Kenneth and Martone, Mario},
  eprint = {2202.07683},
  month = {2},
  primaryclass = {hep-th},
  title = {{Snowmass White Paper on SCFTs}},
  year = {2022},
}

@article{Giacomelli:2020ryy,
  archiveprefix = {arXiv},
  author = {Giacomelli, Simone and Mekareeya, Noppadol and Sacchi, Matteo},
  doi = {10.1007/JHEP03(2021)242},
  eprint = {2012.12852},
  journal = {JHEP},
  pages = {242},
  primaryclass = {hep-th},
  title = {{New aspects of Argyres--Douglas theories and their dimensional reduction}},
  volume = {03},
  year = {2021},
}

@article{Carta:2021whq,
    author = "Carta, Federico and Giacomelli, Simone and Mekareeya, Noppadol and Mininno, Alessandro",
    title = "{Conformal manifolds and 3d mirrors of Argyres-Douglas theories}",
    eprint = "2105.08064",
    archivePrefix = "arXiv",
    primaryClass = "hep-th",
    reportNumber = "IFT-UAM/CSIC-21-55",
    doi = "10.1007/JHEP08(2021)015",
    journal = "JHEP",
    volume = "08",
    pages = "015",
    year = "2021"
}

@article{Carta:2021dyx,

  archiveprefix = {arXiv},
  author = {Carta, Federico and Giacomelli, Simone and Mekareeya, Noppadol and Mininno, Alessandro},
  doi = {10.1007/JHEP02(2022)014},
  eprint = {2110.06940},
  journal = {JHEP},
  pages = {14},
  primaryclass = {hep-th},
  reportnumber = {IFT-UAM/CSIC-21-109, ZMP-HH/21-20},
  title = {{Conformal manifolds and 3d mirrors of (D$_{n}$, D$_{m}$) theories}},
  volume = {02},
  year = {2022},
}

@article{Carta:2022spy,

  archiveprefix = {arXiv},
  author = {Carta, Federico and Giacomelli, Simone and Mekareeya, Noppadol and Mininno, Alessandro},
  doi = {10.1007/JHEP06(2022)059},
  eprint = {2203.16550},
  journal = {JHEP},
  pages = {59},
  primaryclass = {hep-th},
  reportnumber = {ZMP-HH/22-7},
  title = {{Dynamical consequences of 1-form symmetries and the exceptional Argyres-Douglas theories}},
  volume = {06},
  year = {2022},
}

@article{Carta:2022fxc,

  archiveprefix = {arXiv},
  author = {Carta, Federico and Giacomelli, Simone and Mekareeya, Noppadol and Mininno, Alessandro},
  doi = {10.1007/JHEP06(2023)102},
  eprint = {2208.11130},
  journal = {JHEP},
  pages = {102},
  primaryclass = {hep-th},
  reportnumber = {ZMP-HH/22-16},
  title = {{A tale of 2-groups: D$_{p}$(USp(2N)) theories}},
  volume = {06},
  year = {2023},
}

@article{Vafa:2005ui,
  archiveprefix = {arXiv},
  author = {Vafa, Cumrun},
  eprint = {hep-th/0509212},
  month = {9},
  primaryclass = {hep-th},
  reportnumber = {HUTP-05-A043},
  title = {{The String landscape and the swampland}},
  year = {2005},
}

@inproceedings{Batyrev:1997hj,
    author = "Batyrev, V. V.",
    title = "{Stringy hodge numbers of varieties with Gorenstein canonical singularities}",
    booktitle = "{Taniguchi Symposium on Integrable Systems and Algebraic Geometry}",
    pages = "1--32",
    year = "1997"
}

@article{Dixon:1985jw,
    author = "Dixon, L. J. and Harvey, J. A. and Vafa, C. and Witten, E.",
    title = "{Strings on Orbifolds}",
    doi = "10.1016/0550-3213(85)90593-0",
    journal = "Nucl. Phys. B",
    volume = "261",
    pages = "678--686",
    year = "1985"
}

@article{Aluffi:2010cbu,
	author = {Aluffi, Paolo},
	title = {Chern classes of blow-ups},
	journal = {Math. Proc. Cambridge Philos. Soc.},
	volume = {148},
	number = {2},
	pages = {227--242},
	year = {2010}
}

@article{guevara2026singleminus,
    author = "Guevara, Alfredo and Lupsasca, Alexandru and Skinner, David and Strominger, Andrew and Weil, Kevin",
    title = "{Single-minus gluon tree amplitudes are nonzero}",
    eprint = "2602.12176",
    archivePrefix = "arXiv",
    primaryClass = "hep-th",
    month = "2",
    year = "2026"
}

@article{Shih:2026jfe,
    author = "Shih, David",
    title = "{Learning to Unscramble Feynman Loop Integrals with SAILIR}",
    eprint = "2604.05034",
    archivePrefix = "arXiv",
    primaryClass = "hep-ph",
    month = "4",
    year = "2026"
}

@article{Pan_2025,
   title={Quantum many-body physics calculations with large language models},
   volume={8},
   ISSN={2399-3650},
   url={http://dx.doi.org/10.1038/s42005-025-01956-y},
   DOI={10.1038/s42005-025-01956-y},
   number={1},
   journal={Communications Physics},
   publisher={Springer Science and Business Media LLC},
   author={Pan, Haining and Mudur, Nayantara and Taranto, William and Tikhanovskaya, Maria and Venugopalan, Subhashini and Bahri, Yasaman and Brenner, Michael P. and Kim, Eun-Ah},
   year={2025},
   month=jan }

@article{lu2025languageagentsphysics,
   author = {Lu, Sirui and Jin, Zhijing and Zhang, Terry Jingchen and Kos, Pavel and Cirac, J. Ignacio and Sch{\"o}lkopf, Bernhard},
    title = "{Can Theoretical Physics Research Benefit from Language Agents?}",
    eprint = "2506.06214",
    archivePrefix = "arXiv",
    primaryClass = "cs.CL",
    month = "6",
    year = "2025"
}

@article{Agrawal:2026lvg,
    author = "Agrawal, Prateek and Craig, Nathaniel and Madden, Amalia and Lombera, I{\~n}igo Valenzuela",
    title = "{The FERMIACC: Agents for Particle Theory}",
    eprint = "2603.22538",
    archivePrefix = "arXiv",
    primaryClass = "hep-ph",
    month = "3",
    year = "2026"
}

@article{Schwartz:2026ekw,
    author = "Schwartz, Matthew D.",
    title = "{Resummation of the C-Parameter Sudakov Shoulder Using Effective Field Theory}",
    eprint = "2601.02484",
    archivePrefix = "arXiv",
    primaryClass = "hep-ph",
    month = "1",
    year = "2026"
}

@article{Cai__2025,
   title={Learning-at-Criticality in Large Language Models for Quantum Field Theory and Beyond},
   volume={42},
   ISSN={1741-3540},
   url={http://dx.doi.org/10.1088/0256-307X/42/12/120002},
   DOI={10.1088/0256-307x/42/12/120002},
   number={12},
   journal={Chinese Physics Letters},
   publisher={IOP Publishing},
   author={Cai, Xiansheng and Hu, Sihan  and Wang, Tao and Huang, Yuan and Zhang, Pan and Deng, Youjin and Chen, Kun},
   year={2025},
   month=nov, pages={120002} }

@article{jaiswal2024mora,
author = {{Jaiswal}, Raj and {Jain}, Dhruv and {Parimal Popat}, Harsh and {Anand}, Avinash and {Dharmadhikari}, Abhishek and {Marathe}, Atharva and {Ratn Shah}, Rajiv},
        title = "{Improving Physics Reasoning in Large Language Models Using Mixture of Refinement Agents}",
      journal = {arXiv e-prints},
         year = 2024,
        month = dec,
          eid = {arXiv:2412.00821},
        pages = {arXiv:2412.00821},
          doi = {10.48550/arXiv.2412.00821},
archivePrefix = {arXiv},
       eprint = {2412.00821},
 primaryClass = {cs.AI},
       adsurl = {https://ui.adsabs.harvard.edu/abs/2024arXiv241200821J}
}

@article{Shih:2026lmy,
    author = "Shih, David",
    title = "{Learning to Unscramble: Simplifying Symbolic Expressions via Self-Supervised Oracle Trajectories}",
    eprint = "2603.11164",
    archivePrefix = "arXiv",
    primaryClass = "hep-th",
    month = "3",
    year = "2026"
}

@article{xu2025multiagentphysicist,
 author = {{Xu}, Yinggan and {Kimlee}, Hana and {Xiao}, Yijia and {Luo}, Di},
        title = "{Advancing AI-Scientist Understanding: Multi-Agent LLMs with Interpretable Physics Reasoning}",
      journal = {arXiv e-prints},
         year = 2025,
        month = apr,
          eid = {arXiv:2504.01911},
        pages = {arXiv:2504.01911},
          doi = {10.48550/arXiv.2504.01911},
archivePrefix = {arXiv},
       eprint = {2504.01911},
 primaryClass = {cs.AI},
       adsurl = {https://ui.adsabs.harvard.edu/abs/2025arXiv250401911X}
}

@article{Plehn:2026gxv,
    author = "Plehn, Tilman and Schiller, Daniel and Schmal, Nikita",
    title = "{MadAgents}",
    eprint = "2601.21015",
    archivePrefix = "arXiv",
    primaryClass = "hep-ph",
    month = "1",
    year = "2026"
}

@article{Woodward:2026abc,
    author = {Woodward, Nathaniel S. and Gao, Zhiqi and Kvasiuk, Yurii and Smith, Kendrick M. and Sala, Frederic and M{\"u}nchmeyer, Moritz},
    title = "{Fine-Tuning Small Reasoning Models for Quantum Field Theory}",
    eprint = "2604.18936",
    archivePrefix = "arXiv",
    primaryClass = "cs.LG",
    month = "4",
    year = "2026"
}

@article{Niarchos:2026abc,
    author = "Niarchos, Vasilis and Papageorgakis, Constantinos and Stapleton, Alexander G. and Trifinopoulos, Sokratis",
    title = "{When Does Critique Improve AI-Assisted Theoretical Physics? SCALAR: Structured Critic--Actor Loop for Agentic Reasoning}",
    eprint = "2605.06772",
    archivePrefix = "arXiv",
    primaryClass = "cs.AI",
    reportNumber = "CCTP-2026-7; ITCP-2026-7; CERN-TH-2026-097; QMUL-PH-26-15, CCTP-2026-7, ITCP-2026-7, CERN-TH-2026-097, QMUL-PH-26-15",
    month = "5",
    year = "2026"
}

@article{Cvetic:2001nr,
    author = "Cvetic, Mirjam and Shiu, Gary and Uranga, Angel M.",
    title = "{Chiral four-dimensional N=1 supersymmetric type 2A orientifolds from intersecting D6 branes}",
    eprint = "hep-th/0107166",
    archivePrefix = "arXiv",
    reportNumber = "UPR-943-T, CERN-TH-2001-182, CAMTP-01-8",
    doi = "10.1016/S0550-3213(01)00427-8",
    journal = "Nucl. Phys. B",
    volume = "615",
    pages = "3--32",
    year = "2001"
}

@article{Cvetic:2001tj,
    author = "Cvetic, Mirjam and Shiu, Gary and Uranga, Angel M.",
    title = "{Three family supersymmetric standard - like models from intersecting brane worlds}",
    eprint = "hep-th/0107143",
    archivePrefix = "arXiv",
    reportNumber = "UPR-948-T, CERN-TH-2001-190",
    doi = "10.1103/PhysRevLett.87.201801",
    journal = "Phys. Rev. Lett.",
    volume = "87",
    pages = "201801",
    year = "2001"
}

@article{Blumenhagen:2005mu,
    author = "Blumenhagen, Ralph and Cvetic, Mirjam and Langacker, Paul and Shiu, Gary",
    title = "{Toward realistic intersecting D-brane models}",
    eprint = "hep-th/0502005",
    archivePrefix = "arXiv",
    reportNumber = "MPP-2005-7, UPR-1093-T, MAD-TH-05-02",
    doi = "10.1146/annurev.nucl.55.090704.151541",
    journal = "Ann. Rev. Nucl. Part. Sci.",
    volume = "55",
    pages = "71--139",
    year = "2005"
}

@article{Marchesano:2004xz,
    author = "Marchesano, Fernando and Shiu, Gary",
    title = "{Building MSSM flux vacua}",
    eprint = "hep-th/0409132",
    archivePrefix = "arXiv",
    reportNumber = "MAD-TH-04-9",
    doi = "10.1088/1126-6708/2004/11/041",
    journal = "JHEP",
    volume = "11",
    pages = "041",
    year = "2004"
}

@article{Blumenhagen:2009yv,
    author = "Blumenhagen, Ralph and Grimm, Thomas W. and Jurke, Benjamin and Weigand, Timo",
    title = "{Global F-theory GUTs}",
    eprint = "0908.1784",
    archivePrefix = "arXiv",
    primaryClass = "hep-th",
    reportNumber = "MPP-2009-148, SLAC-PUB-13751",
    doi = "10.1016/j.nuclphysb.2009.12.013",
    journal = "Nucl. Phys. B",
    volume = "829",
    pages = "325--369",
    year = "2010"
}

@article{Donagi:2008ca,
    author = "Donagi, Ron and Wijnholt, Martijn",
    title = "{Model Building with F-Theory}",
    eprint = "0802.2969",
    archivePrefix = "arXiv",
    primaryClass = "hep-th",
    reportNumber = "AEI-2007-174",
    doi = "10.4310/ATMP.2011.v15.n5.a2",
    journal = "Adv. Theor. Math. Phys.",
    volume = "15",
    number = "5",
    pages = "1237--1317",
    year = "2011"
}

@article{Beasley:2008kw,
    author = "Beasley, Chris and Heckman, Jonathan J. and Vafa, Cumrun",
    title = "{GUTs and Exceptional Branes in F-theory - II: Experimental Predictions}",
    eprint = "0806.0102",
    archivePrefix = "arXiv",
    primaryClass = "hep-th",
    doi = "10.1088/1126-6708/2009/01/059",
    journal = "JHEP",
    volume = "01",
    pages = "059",
    year = "2009"
}

@article{Grimm:2009yu,
    author = "Grimm, Thomas W. and Krause, Sven and Weigand, Timo",
    title = "{F-Theory GUT Vacua on Compact Calabi-Yau Fourfolds}",
    eprint = "0912.3524",
    archivePrefix = "arXiv",
    primaryClass = "hep-th",
    reportNumber = "HD-THEP-09-30",
    doi = "10.1007/JHEP07(2010)037",
    journal = "JHEP",
    volume = "07",
    pages = "037",
    year = "2010"
}

@article{Marsano:2009gv,
    author = "Marsano, Joseph and Saulina, Natalia and Schafer-Nameki, Sakura",
    title = "{Monodromies, Fluxes, and Compact Three-Generation F-theory GUTs}",
    eprint = "0906.4672",
    archivePrefix = "arXiv",
    primaryClass = "hep-th",
    reportNumber = "CALT-68-2733",
    doi = "10.1088/1126-6708/2009/08/046",
    journal = "JHEP",
    volume = "08",
    pages = "046",
    year = "2009"
}

@article{Grana:2005jc,
    author = "Grana, Mariana",
    title = "{Flux compactifications in string theory: A Comprehensive review}",
    eprint = "hep-th/0509003",
    archivePrefix = "arXiv",
    reportNumber = "LPTENS-05-26, CPHT-RR-049-0805",
    doi = "10.1016/j.physrep.2005.10.008",
    journal = "Phys. Rept.",
    volume = "423",
    pages = "91--158",
    year = "2006"
}

@article{Douglas:2006es,
    author = "Douglas, Michael R. and Kachru, Shamit",
    title = "{Flux compactification}",
    eprint = "hep-th/0610102",
    archivePrefix = "arXiv",
    reportNumber = "SLAC-PUB-12131",
    doi = "10.1103/RevModPhys.79.733",
    journal = "Rev. Mod. Phys.",
    volume = "79",
    pages = "733--796",
    year = "2007"
}

@book{Ibanez:2012zz,
    author = "Ibanez, Luis E. and Uranga, Angel M.",
    title = "{String theory and particle physics: An introduction to string phenomenology}",
    isbn = "978-0-521-51752-2, 978-1-139-22742-1",
    publisher = "Cambridge University Press",
    month = "2",
    year = "2012"
}

@article{Aluffi:2007sx,
    author = "Aluffi, Paolo and Esole, Mboyo",
    title = "{Chern class identities from tadpole matching in type IIB and F-theory}",
    eprint = "0710.2544",
    archivePrefix = "arXiv",
    primaryClass = "hep-th",
    reportNumber = "MPIM2007-126, FSU07-23, KUL-TF-07-22",
    doi = "10.1088/1126-6708/2009/03/032",
    journal = "JHEP",
    volume = "03",
    pages = "032",
    year = "2009"
}

@article{Caraffi:2026wzk,
    author = "Caraffi, Luca and Carta, Federico and Cicoli, Michele",
    title = "{New Type IIB Poly-instanton Effects and Axion Quintessence}",
    eprint = "2607.20613",
    archivePrefix = "arXiv",
    primaryClass = "hep-th",
    month = "7",
    year = "2026"
}

@article{Becker:1996gj,
    author = "Becker, Katrin and Becker, Melanie",
    title = "{M theory on eight manifolds}",
    eprint = "hep-th/9605053",
    archivePrefix = "arXiv",
    reportNumber = "NSF-ITP-96-19",
    doi = "10.1016/0550-3213(96)00367-7",
    journal = "Nucl. Phys. B",
    volume = "477",
    pages = "155--167",
    year = "1996"
}

@article{Berasaluce-Gonzalez:2012awn,
    author = "Berasaluce-Gonzalez, M. and Camara, P. G. and Marchesano, F. and Uranga, A. M.",
    title = "{Zp charged branes in flux compactifications}",
    eprint = "1211.5317",
    archivePrefix = "arXiv",
    primaryClass = "hep-th",
    reportNumber = "IFT-UAM-CSIC-12-108",
    doi = "10.1007/JHEP04(2013)138",
    journal = "JHEP",
    volume = "04",
    pages = "138",
    year = "2013"
}

@article{Bershadsky:1996nh,
    author = "Bershadsky, M. and Intriligator, Kenneth A. and Kachru, S. and Morrison, David R. and Sadov, V. and Vafa, Cumrun",
    title = "{Geometric singularities and enhanced gauge symmetries}",
    eprint = "hep-th/9605200",
    archivePrefix = "arXiv",
    reportNumber = "HUTP-96-A017, IASSNS-HEP-96-49, RU-96-40",
    doi = "10.1016/S0550-3213(96)90131-5",
    journal = "Nucl. Phys. B",
    volume = "481",
    pages = "215--252",
    year = "1996"
}

@article{Blumenhagen:2007bn,
    author = "Blumenhagen, Ralph and Cvetic, Mirjam and Richter, Robert and Weigand, Timo",
    title = "{Lifting D-Instanton Zero Modes by Recombination and Background Fluxes}",
    eprint = "0708.0403",
    archivePrefix = "arXiv",
    primaryClass = "hep-th",
    doi = "10.1088/1126-6708/2007/10/098",
    journal = "JHEP",
    volume = "10",
    pages = "098",
    year = "2007"
}

@article{Blumenhagen:2009qh,
    author = "Blumenhagen, Ralph and Cvetic, Mirjam and Kachru, Shamit and Weigand, Timo",
    title = "{D-Brane Instantons in Type II Orientifolds}",
    eprint = "0902.3251",
    archivePrefix = "arXiv",
    primaryClass = "hep-th",
    reportNumber = "MPP-2009-15, UPR-1205-T, SLAC-PUB-13531",
    doi = "10.1146/annurev.nucl.010909.083113",
    journal = "Ann. Rev. Nucl. Part. Sci.",
    volume = "59",
    pages = "269--296",
    year = "2009"
}

@book{Blumenhagen:2013fgp,
    author = {Blumenhagen, Ralph and L{\"u}st, Dieter and Theisen, Stefan},
    title = "{Basic concepts of string theory}",
    doi = "10.1007/978-3-642-29497-6",
    isbn = "978-3-642-29496-9",
    publisher = "Springer",
    address = "Heidelberg, Germany",
    series = "Theoretical and Mathematical Physics",
    year = "2013"
}

@article{Camara:2003ku,
    author = "Camara, Pablo G. and Ibanez, L. E. and Uranga, A. M.",
    title = "{Flux induced SUSY breaking soft terms}",
    eprint = "hep-th/0311241",
    archivePrefix = "arXiv",
    reportNumber = "IFT-UAM-CSIC-03-42",
    doi = "10.1016/j.nuclphysb.2004.04.013",
    journal = "Nucl. Phys. B",
    volume = "689",
    pages = "195--242",
    year = "2004"
}

@article{Camara:2004jj,
    author = "Camara, Pablo G. and Ibanez, L. E. and Uranga, A. M.",
    title = "{Flux-induced SUSY-breaking soft terms on D7-D3 brane systems}",
    eprint = "hep-th/0408036",
    archivePrefix = "arXiv",
    reportNumber = "IFT-UAM-CSIC-04-36",
    doi = "10.1016/j.nuclphysb.2004.11.035",
    journal = "Nucl. Phys. B",
    volume = "708",
    pages = "268--316",
    year = "2005"
}

@article{Cicoli:2011qg,
    author = "Cicoli, Michele and Mayrhofer, Christoph and Valandro, Roberto",
    title = "{Moduli Stabilisation for Chiral Global Models}",
    eprint = "1110.3333",
    archivePrefix = "arXiv",
    primaryClass = "hep-th",
    reportNumber = "DESY-11-179, ZMP-HH-11-15",
    doi = "10.1007/JHEP02(2012)062",
    journal = "JHEP",
    volume = "02",
    pages = "062",
    year = "2012"
}

@article{Cicoli:2023opf,
    author = "Cicoli, Michele and Conlon, Joseph P. and Maharana, Anshuman and Parameswaran, Susha and Quevedo, Fernando and Zavala, Ivonne",
    title = "{String cosmology: From the early universe to today}",
    eprint = "2303.04819",
    archivePrefix = "arXiv",
    primaryClass = "hep-th",
    doi = "10.1016/j.physrep.2024.01.002",
    journal = "Phys. Rept.",
    volume = "1059",
    pages = "1--155",
    year = "2024"
}

@article{Collinucci:2014taa,
    author = "Collinucci, Andres and Savelli, Raffaele",
    title = "{F-theory on singular spaces}",
    eprint = "1410.4867",
    archivePrefix = "arXiv",
    primaryClass = "hep-th",
    doi = "10.1007/JHEP09(2015)100",
    journal = "JHEP",
    volume = "09",
    pages = "100",
    year = "2015"
}

@article{Douglas:1995bn,
    author = "Douglas, Michael R.",
    editor = "Baulieu, L. and Kazakov, V. and Picco, M. and Windey, Paul and Di Francesco, P. and Douglas, Michael R.",
    title = "{Branes within branes}",
    eprint = "hep-th/9512077",
    archivePrefix = "arXiv",
    reportNumber = "RU-95-92",
    journal = "NATO Sci. Ser. C",
    volume = "520",
    pages = "267--275",
    year = "1999"
}

@article{Gomis:2005wc,
    author = "Gomis, Jaume and Marchesano, Fernando and Mateos, David",
    title = "{An Open string landscape}",
    eprint = "hep-th/0506179",
    archivePrefix = "arXiv",
    reportNumber = "MAD-TH-05-4",
    doi = "10.1088/1126-6708/2005/11/021",
    journal = "JHEP",
    volume = "11",
    pages = "021",
    year = "2005"
}

@article{Grana:2003ek,
    author = "Grana, Mariana and Grimm, Thomas W. and Jockers, Hans and Louis, Jan",
    title = "{Soft supersymmetry breaking in Calabi-Yau orientifolds with D-branes and fluxes}",
    eprint = "hep-th/0312232",
    archivePrefix = "arXiv",
    reportNumber = "CPHT-RR-116-1203",
    doi = "10.1016/j.nuclphysb.2004.04.021",
    journal = "Nucl. Phys. B",
    volume = "690",
    pages = "21--61",
    year = "2004"
}

@article{Grana:2022dfw,
    author = "Gra{\~n}a, Mariana and Grimm, Thomas W. and van de Heisteeg, Damian and Herraez, Alvaro and Plauschinn, Erik",
    title = "{The tadpole conjecture in asymptotic limits}",
    eprint = "2204.05331",
    archivePrefix = "arXiv",
    primaryClass = "hep-th",
    doi = "10.1007/JHEP08(2022)237",
    journal = "JHEP",
    volume = "08",
    pages = "237",
    year = "2022"
}

@article{Green:1996dd,
    author = "Green, Michael B. and Harvey, Jeffrey A. and Moore, Gregory W.",
    title = "{I-brane inflow and anomalous couplings on d-branes}",
    eprint = "hep-th/9605033",
    archivePrefix = "arXiv",
    reportNumber = "DAMTP-96-40, EFI-96-13, YCTP-P8-96, RU-96-29",
    doi = "10.1088/0264-9381/14/1/008",
    journal = "Class. Quant. Grav.",
    volume = "14",
    pages = "47--52",
    year = "1997"
}

@book{Hori:2003ic,
    author = "Hori, K. and Katz, S. and Klemm, A. and Pandharipande, R. and Thomas, R. and Vafa, C. and Vakil, R. and Zaslow, E.",
    title = "{Mirror symmetry}",
    publisher = "AMS",
    address = "Providence, USA",
    series = "Clay mathematics monographs",
    volume = "1",
    year = "2003"
}

@article{Jockers:2005zy,
    author = "Jockers, Hans and Louis, Jan",
    title = "{D-terms and F-terms from D7-brane fluxes}",
    eprint = "hep-th/0502059",
    archivePrefix = "arXiv",
    doi = "10.1016/j.nuclphysb.2005.04.011",
    journal = "Nucl. Phys. B",
    volume = "718",
    pages = "203--246",
    year = "2005"
}

@article{Lust:2022mhk,
    author = {L{\"u}st, Severin and Wiesner, Max},
    title = "{The tadpole conjecture in the interior of moduli space}",
    eprint = "2211.05128",
    archivePrefix = "arXiv",
    primaryClass = "hep-th",
    doi = "10.1007/JHEP12(2023)029",
    journal = "JHEP",
    volume = "12",
    pages = "029",
    year = "2023"
}

@article{Marchesano:2004yq,
    author = "Marchesano, Fernando and Shiu, Gary",
    title = "{MSSM vacua from flux compactifications}",
    eprint = "hep-th/0408059",
    archivePrefix = "arXiv",
    reportNumber = "MAD-TH-04-8",
    doi = "10.1103/PhysRevD.71.011701",
    journal = "Phys. Rev. D",
    volume = "71",
    pages = "011701",
    year = "2005"
}

@article{Minasian:1997mm,
    author = "Minasian, Ruben and Moore, Gregory W.",
    title = "{K theory and Ramond-Ramond charge}",
    eprint = "hep-th/9710230",
    archivePrefix = "arXiv",
    reportNumber = "YCTP-P21-97",
    doi = "10.1088/1126-6708/1997/11/002",
    journal = "JHEP",
    volume = "11",
    pages = "002",
    year = "1997"
}

@article{Plauschinn:2021hkp,
    author = "Plauschinn, Erik",
    title = "{The tadpole conjecture at large complex-structure}",
    eprint = "2109.00029",
    archivePrefix = "arXiv",
    primaryClass = "hep-th",
    doi = "10.1007/JHEP02(2022)206",
    journal = "JHEP",
    volume = "02",
    pages = "206",
    year = "2022"
}

@article{Uranga:2000xp,
    author = "Uranga, Angel M.",
    title = "{D-brane probes, RR tadpole cancellation and K-theory charge}",
    eprint = "hep-th/0011048",
    archivePrefix = "arXiv",
    reportNumber = "CERN-TH-2000-331",
    doi = "10.1016/S0550-3213(00)00787-2",
    journal = "Nucl. Phys. B",
    volume = "598",
    pages = "225--246",
    year = "2001"
}

@article{He:2017aed,
    author = "He, Yang-Hui",
    title = "{Deep-Learning the Landscape}",
    eprint = "1706.02714",
    archivePrefix = "arXiv",
    primaryClass = "hep-th",
    month = "6",
    year = "2017"
}

@article{He:2020lbz,
    author = "He, Yang-Hui and Lukas, Andre",
    title = "{Machine Learning Calabi-Yau Four-folds}",
    eprint = "2009.02544",
    archivePrefix = "arXiv",
    primaryClass = "hep-th",
    doi = "10.1016/j.physletb.2021.136139",
    journal = "Phys. Lett. B",
    volume = "815",
    pages = "136139",
    year = "2021"
}

@article{Yu:2026ogu,
    author = "Yu, Xingyang",
    title = "{DualityCert: Verifier-Gated Language-Model Repair of Broken Duality Claims in Quantum Field Theory}",
    eprint = "2607.23614",
    archivePrefix = "arXiv",
    primaryClass = "cs.CR",
    month = "7",
    year = "2026"
}

@article{Sympy,
     title = {SymPy: symbolic computing in Python},
     author = {Meurer, Aaron and Smith, Christopher P. and Paprocki, Mateusz and \v{C}ert\'{i}k, Ond\v{r}ej and Kirpichev, Sergey B. and Rocklin, Matthew and Kumar, AMiT and Ivanov, Sergiu and Moore, Jason K. and Singh, Sartaj and Rathnayake, Thilina and Vig, Sean and Granger, Brian E. and Muller, Richard P. and Bonazzi, Francesco and Gupta, Harsh and Vats, Shivam and Johansson, Fredrik and Pedregosa, Fabian and Curry, Matthew J. and Terrel, Andy R. and Rou\v{c}ka, \v{S}t\v{e}p\'{a}n and Saboo, Ashutosh and Fernando, Isuru and Kulal, Sumith and Cimrman, Robert and Scopatz, Anthony},
     year = 2017,
     month = jan,
     volume = 3,
     pages = {e103},
     journal = {PeerJ Computer Science},
     issn = {2376-5992},
     url = {https://doi.org/10.7717/peerj-cs.103},
     doi = {10.7717/peerj-cs.103}
    }

@article{Ruehle:2020jrk,
    author = "Ruehle, Fabian",
    title = "{Data science applications to string theory}",
    doi = "10.1016/j.physrep.2019.09.005",
    journal = "Phys. Rept.",
    volume = "839",
    pages = "1--117",
    year = "2020"
}

@inproceedings{Cole:2021nnt,
    author = "Cole, Alex and Krippendorf, Sven and Schachner, Andreas and Shiu, Gary",
    title = "{Probing the Structure of String Theory Vacua with Genetic Algorithms and Reinforcement Learning}",
    booktitle = "{35th Conference on Neural Information Processing Systems}",
    eprint = "2111.11466",
    archivePrefix = "arXiv",
    primaryClass = "hep-th",
    month = "11",
    year = "2021"
}

@inproceedings{Anderson:2023viv,
    author = "Anderson, Lara B. and Gray, James and Larfors, Magdalena",
    title = "{Lectures on Numerical and Machine Learning Methods for Approximating Ricci-flat Calabi-Yau Metrics}",
    eprint = "2312.17125",
    archivePrefix = "arXiv",
    primaryClass = "hep-th",
    month = "12",
    year = "2023",
    booktitle = "{}",
}

@article{Douglas:2021ces,
    author = "Douglas, Michael R.",
    title = "{From Algebraic Geometry to Machine Learning}",
    eprint = "2107.14387",
    archivePrefix = "arXiv",
    primaryClass = "math.HO",
    month = "7",
    year = "2021"
}

@article{Corvilain:2018lgw,
    author = "Corvilain, Pierre and Grimm, Thomas W. and Valenzuela, Irene",
    title = {{The Swampland Distance Conjecture for K{\"a}hler moduli}},
    eprint = "1812.07548",
    archivePrefix = "arXiv",
    primaryClass = "hep-th",
    doi = "10.1007/JHEP08(2019)075",
    journal = "JHEP",
    volume = "08",
    pages = "075",
    year = "2019"
}

@article{Grimm:2019bey,
    author = "Grimm, Thomas W. and Ruehle, Fabian and van de Heisteeg, Damian",
    title = "{Classifying Calabi{\textendash}Yau Threefolds Using Infinite Distance Limits}",
    eprint = "1910.02963",
    archivePrefix = "arXiv",
    primaryClass = "hep-th",
    doi = "10.1007/s00220-021-03972-9",
    journal = "Commun. Math. Phys.",
    volume = "382",
    number = "1",
    pages = "239--275",
    year = "2021"
}

@article{Ooguri:2006in,
    author = "Ooguri, Hirosi and Vafa, Cumrun",
    title = "{On the Geometry of the String Landscape and the Swampland}",
    eprint = "hep-th/0605264",
    archivePrefix = "arXiv",
    reportNumber = "CALT-68-2600, HUTP-06-A017",
    doi = "10.1016/j.nuclphysb.2006.10.033",
    journal = "Nucl. Phys. B",
    volume = "766",
    pages = "21--33",
    year = "2007"
}

@article{Arkani-Hamed:2006emk,
    author = "Arkani-Hamed, Nima and Motl, Lubos and Nicolis, Alberto and Vafa, Cumrun",
    title = "{The String landscape, black holes and gravity as the weakest force}",
    eprint = "hep-th/0601001",
    archivePrefix = "arXiv",
    reportNumber = "HUTP-05-A0057",
    doi = "10.1088/1126-6708/2007/06/060",
    journal = "JHEP",
    volume = "06",
    pages = "060",
    year = "2007"
}

@article{Blumenhagen:2018nts,
    author = {Blumenhagen, Ralph and Kl{\"a}wer, Daniel and Schlechter, Lorenz and Wolf, Florian},
    title = "{The Refined Swampland Distance Conjecture in Calabi-Yau Moduli Spaces}",
    eprint = "1803.04989",
    archivePrefix = "arXiv",
    primaryClass = "hep-th",
    reportNumber = "MPP-2018-34",
    doi = "10.1007/JHEP06(2018)052",
    journal = "JHEP",
    volume = "06",
    pages = "052",
    year = "2018"
}

@article{FierroCota:2023bsp,
    author = "Fierro Cota, Cesar and Mininno, Alessandro and Weigand, Timo and Wiesner, Max",
    title = "{The minimal weak gravity conjecture}",
    eprint = "2312.04619",
    archivePrefix = "arXiv",
    primaryClass = "hep-th",
    reportNumber = "ZMP-HH/23-21",
    doi = "10.1007/JHEP05(2024)285",
    journal = "JHEP",
    volume = "05",
    pages = "285",
    year = "2024"
}

@article{Lin:2025wfe,
    author = "Lin, Puxin and Mininno, Alessandro and Shiu, Gary",
    title = "{Formulating the Weak Gravity Conjecture in AdS space}",
    eprint = "2503.05862",
    archivePrefix = "arXiv",
    primaryClass = "hep-th",
    doi = "10.1007/JHEP06(2025)100",
    journal = "JHEP",
    volume = "06",
    pages = "100",
    year = "2025"
}

@article{Ebelt:2023clh,
    author = "Ebelt, Julian and Krippendorf, Sven and Schachner, Andreas",
    title = "{W0{\_}sample = np.random.normal(0,1)?}",
    eprint = "2307.15749",
    archivePrefix = "arXiv",
    primaryClass = "hep-th",
    doi = "10.1016/j.physletb.2024.138786",
    journal = "Phys. Lett. B",
    volume = "855",
    pages = "138786",
    year = "2024"
}

@article{Anderson:2020hux,
    author = "Anderson, Lara B. and Gerdes, Mathis and Gray, James and Krippendorf, Sven and Raghuram, Nikhil and Ruehle, Fabian",
    title = "{Moduli-dependent Calabi-Yau and SU(3)-structure metrics from Machine Learning}",
    eprint = "2012.04656",
    archivePrefix = "arXiv",
    primaryClass = "hep-th",
    reportNumber = "CERN-TH-2020-205",
    doi = "10.1007/JHEP05(2021)013",
    journal = "JHEP",
    volume = "05",
    pages = "013",
    year = "2021"
}

@article{Larfors:2021pbb,
    author = "Larfors, Magdalena and Lukas, Andre and Ruehle, Fabian and Schneider, Robin",
    title = "{Learning Size and Shape of Calabi-Yau Spaces}",
    eprint = "2111.01436",
    archivePrefix = "arXiv",
    primaryClass = "hep-th",
    reportNumber = "UUITP-53/21",
    month = "11",
    year = "2021"
}

@article{Larfors:2022nep,
    author = "Larfors, Magdalena and Lukas, Andre and Ruehle, Fabian and Schneider, Robin",
    title = "{Numerical metrics for complete intersection and Kreuzer{\textendash}Skarke Calabi{\textendash}Yau manifolds}",
    eprint = "2205.13408",
    archivePrefix = "arXiv",
    primaryClass = "hep-th",
    reportNumber = "UUITP-25/22",
    doi = "10.1088/2632-2153/ac8e4e",
    journal = "Mach. Learn. Sci. Tech.",
    volume = "3",
    number = "3",
    pages = "035014",
    year = "2022"
}

@article{Halverson:2023ndu,
    author = "Halverson, James and Ruehle, Fabian",
    title = "{Metric flows with neural networks}",
    eprint = "2310.19870",
    archivePrefix = "arXiv",
    primaryClass = "hep-th",
    doi = "10.1088/2632-2153/ad8533",
    journal = "Mach. Learn. Sci. Tech.",
    volume = "5",
    number = "4",
    pages = "045020",
    year = "2024"
}

@article{Lust:2026mys,
    author = {L{\"u}st, Severin and Ruehle, Fabian and Schreyer, Simon},
    title = "{Warped Numerical Calabi-Yau Metrics}",
    eprint = "2607.18402",
    archivePrefix = "arXiv",
    primaryClass = "hep-th",
    month = "7",
    year = "2026"
}

@article{Krippendorf:2026pou,
    author = "Krippendorf, Sven and Tooby-Smith, Joseph",
    title = "{Physics as Code: From Scans to Theorems with ITP APIs in $SU(5)$ Model Building}",
    eprint = "2603.28406",
    archivePrefix = "arXiv",
    primaryClass = "hep-th",
    month = "3",
    year = "2026"
}

@article{Gu:2026zbf,
    author = "Gu, Yi and Krippendorf, Sven",
    title = "{Scattering Amplitudes as Programs: Self-Evolving Search for Theory and Event Generation}",
    eprint = "2607.21629",
    archivePrefix = "arXiv",
    primaryClass = "hep-ph",
    month = "7",
    year = "2026"
}

@article{Krippendorf:2025mhp,
    author = "Krippendorf, Sven and Liu, Zhimei",
    title = "{Solving inverse problems of Type IIB flux vacua with conditional generative models}",
    eprint = "2506.22551",
    archivePrefix = "arXiv",
    primaryClass = "hep-th",
    doi = "10.1007/JHEP07(2026)103",
    journal = "JHEP",
    volume = "07",
    pages = "103",
    year = "2026"
}

@article{Brodie:2019dfx,
    author = "Brodie, Callum R. and Constantin, Andrei and Deen, Rehan and Lukas, Andre",
    title = "{Machine Learning Line Bundle Cohomology}",
    eprint = "1906.08730",
    archivePrefix = "arXiv",
    primaryClass = "hep-th",
    doi = "10.1002/prop.201900087",
    journal = "Fortsch. Phys.",
    volume = "68",
    number = "1",
    pages = "1900087",
    year = "2020"
}

@article{Constantin:2021for,
    author = "Constantin, Andrei and Harvey, Thomas R. and Lukas, Andre",
    title = "{Heterotic String Model Building with Monad Bundles and Reinforcement Learning}",
    eprint = "2108.07316",
    archivePrefix = "arXiv",
    primaryClass = "hep-th",
    doi = "10.1002/prop.202100186",
    journal = "Fortsch. Phys.",
    volume = "70",
    number = "2-3",
    pages = "2100186",
    year = "2022"
}

@article{Lin:2025gco,
    author = "Lin, Puxin and Mininno, Alessandro and Shiu, Gary",
    title = "{Expanding the weak gravity conjecture in AdS space}",
    eprint = "2511.21809",
    archivePrefix = "arXiv",
    primaryClass = "hep-th",
    doi = "10.1007/JHEP04(2026)208",
    journal = "JHEP",
    volume = "04",
    pages = "208",
    year = "2026"
}

@article{Joshi:2019nzi,
    author = "Joshi, Abhinav and Klemm, Albrecht",
    title = "{Swampland Distance Conjecture for One-Parameter Calabi-Yau Threefolds}",
    eprint = "1903.00596",
    archivePrefix = "arXiv",
    primaryClass = "hep-th",
    doi = "10.1007/JHEP08(2019)086",
    journal = "JHEP",
    volume = "08",
    pages = "086",
    year = "2019"
}

\end{document}